\documentclass[longauth]{aa}  

\usepackage{graphicx}
\usepackage{txfonts}
\usepackage{multirow}
\usepackage{amsmath}
\usepackage{amssymb}

\newcommand{\twosixtycgpsepPpone}{\ensuremath{13.475853^{+0.000013}_{-0.000011}}}

\newcommand{\twosixtycgpseptzeropone}{\ensuremath{2458392.2942^{+0.0015}_{-0.0017}}}
\newcommand{\twosixtycgpsepppone}{\ensuremath{0.0258^{+0.0011}_{-0.0001}}}
\newcommand{\twosixtycgpsepbpone}{\ensuremath{0.679^{+0.095}_{-0.081}}}
\newcommand{\twosixtycgpseprho}{\ensuremath{6957.2^{+2114.6}_{-2427.2}}}
\newcommand{\twosixtycgpsepqoneTESS}{\ensuremath{0.39^{+0.23}_{-0.19}}}
\newcommand{\twosixtycgpsepqtwoTESS}{\ensuremath{0.34^{+0.28}_{-0.23}}}
\newcommand{\twosixtycgpsepmuESPRESSOeighteen}{\ensuremath{-10943.2^{+2.9}_{-2.6}}}
\newcommand{\twosixtycgpsepsigmawESPRESSOeighteen}{\ensuremath{0.03^{+0.34}_{-0.03}}}
\newcommand{\twosixtycgpsepmuESPRESSOnineteen}{\ensuremath{-10945.42 \pm 0.92}}
\newcommand{\twosixtycgpsepsigmawESPRESSOnineteen}{\ensuremath{1.51^{+0.32}_{-0.28}}}
\newcommand{\twosixtycgpsepmuHIRES}{\ensuremath{-0.4^{+1.0}_{-1.1}}}
\newcommand{\twosixtycgpsepsigmawHIRES}{\ensuremath{0.2^{+1.5}_{-0.2}}}
\newcommand{\twosixtycgpsepKpone}{\ensuremath{1.57^{+0.59}_{-0.57}}}
\newcommand{\twosixtycgpsepGPsigmaESPRESSOeighteenESPRESSOnineteen}{\ensuremath{3.95^{+0.59}_{-0.52}}}
\newcommand{\twosixtycgpsepGPalphaESPRESSOeighteenESPRESSOnineteen}{\ensuremath{0.026^{+0.011}_{-0.010}}}
\newcommand{\twosixtycgpsepGPGammaESPRESSOeighteenESPRESSOnineteen}{\ensuremath{1.25^{+0.82}_{-0.64}}}
\newcommand{\twosixtycgpsepGPProtESPRESSOeighteenESPRESSOnineteenHIRES}{\ensuremath{23.7^{+5.1}_{-5.0}}}
\newcommand{\twosixtycgpsepGPsigmaHIRES}{\ensuremath{5.382^{+0.018}_{-0.015}}}
\newcommand{\twosixtycgpsepGPalphaHIRES}{\ensuremath{0.106^{+0.061}_{-0.051}}}
\newcommand{\twosixtycgpsepGPGammaHIRES}{\ensuremath{5.2 \pm 2.3}}
\newcommand{\twosixtycgpsepmfluxTESSthree}{\ensuremath{(-1.5^{+3.1}_{-3.4})\times10^{-5}}}
\newcommand{\twosixtycgpsepsigmawTESSthree}{\ensuremath{5.2^{+21.9}_{-4.6}}}
\newcommand{\twosixtycgpsepGPsigmaTESSthree}{\ensuremath{(25.5 \pm 1.4)\times10^{-5}}}
\newcommand{\twosixtycgpsepGPrhoTESSthree}{\ensuremath{0.1968 \pm 0.0048}}
\newcommand{\twosixtycgpsepmfluxTESSfortytwo}{\ensuremath{-0.6^{+1.6}_{-1.7})\times10^{-5}}}
\newcommand{\twosixtycgpsepsigmawTESSfortytwo}{\ensuremath{197.7^{+6.9}_{-7.5}}}
\newcommand{\twosixtycgpsepGPsigmaTESSfortytwo}{\ensuremath{(8.63^{+0.65}_{-0.69})\times10^{-5}}}
\newcommand{\twosixtycgpsepGPrhoTESSfortytwo}{\ensuremath{0.250^{+0.033}_{-0.027}}}
\newcommand{\twosixtycgpsepmfluxTESSseventy}{\ensuremath{(0.4 \pm 6.2)\times10^{-5}}}
\newcommand{\twosixtycgpsepsigmawTESSseventy}{\ensuremath{179.2^{+7.4}_{-7.6}}}
\newcommand{\twosixtycgpsepGPsigmaTESSseventy}{\ensuremath{(24.7^{+2.8}_{-2.5})\times10^{-5}}}
\newcommand{\twosixtycgpsepGPrhoTESSseventy}{\ensuremath{0.91^{+0.12}_{-0.11}}}
\newcommand{\twosixtycgpsepeccpone}{\ensuremath{0.0}}
\newcommand{\twosixtycgpsepomegapone}{\ensuremath{90.0}}
\newcommand{\twosixtycgpsepmdilutionTESSthree}{\ensuremath{1.0}}
\newcommand{\twosixtycgpsepmdilutionTESSfortytwo}{\ensuremath{1.0}}
\newcommand{\twosixtycgpsepmdilutionTESSseventy}{\ensuremath{1.0}}

\newcommand{\twosixtyesevenPpone}{\ensuremath{13.475853^{+0.000011}_{-0.000013}}}

\newcommand{\twosixtyeseventzeropone}{\ensuremath{2458392.2945^{+0.0017}_{-0.0014}}}
\newcommand{\twosixtyesevenppone}{\ensuremath{0.02458^{+0.00073}_{-0.00081}}}
\newcommand{\twosixtyesevenbpone}{\ensuremath{0.33 \pm 0.16}}
\newcommand{\twosixtyesevensesinomegapone}{\ensuremath{0.39^{+0.17}_{-0.13}}}
\newcommand{\twosixtyesevensecosomegapone}{\ensuremath{0.72^{+0.06}_{-0.16}}}
\newcommand{\twosixtyeseveneccpone}{\ensuremath{0.678^{+0.017}_{-0.048}}}
\newcommand{\twosixtyesevenomegapone}{\ensuremath{28.3^{+17.4}_{-10.4}}}
\newcommand{\twosixtyesevenrho}{\ensuremath{2609.8^{+1068.1}_{-694.2}}}
\newcommand{\twosixtyesevenqoneTESS}{\ensuremath{0.34^{+0.24}_{-0.17}}}
\newcommand{\twosixtyesevenqtwoTESS}{\ensuremath{0.37^{+0.25}_{-0.22}}}
\newcommand{\twosixtyesevenmuESPRESSOeighteen}{\ensuremath{-10943.2^{+2.3}_{-2.4}}}
\newcommand{\twosixtyesevensigmawESPRESSOeighteen}{\ensuremath{0.04^{+0.48}_{-0.03}}}
\newcommand{\twosixtyesevenmuESPRESSOnineteen}{\ensuremath{-10945.24^{+0.81}_{-0.83}}}
\newcommand{\twosixtyesevensigmawESPRESSOnineteen}{\ensuremath{1.50^{+0.32}_{-0.29}}}
\newcommand{\twosixtyesevenmuHIRES}{\ensuremath{-0.36^{+0.95}_{-0.86}}}
\newcommand{\twosixtyesevensigmawHIRES}{\ensuremath{0.05^{+1.07}_{-0.04}}}
\newcommand{\twosixtyesevenKpone}{\ensuremath{3.49^{+0.86}_{-1.09}}}
\newcommand{\twosixtyesevenGPsigmaESPRESSOeighteenESPRESSOnineteen}{\ensuremath{3.57^{+0.44}_{-0.39}}}
\newcommand{\twosixtyesevenGPalphaESPRESSOeighteenESPRESSOnineteen}{\ensuremath{0.027^{+0.012}_{-0.009}}}
\newcommand{\twosixtyesevenGPGammaESPRESSOeighteenESPRESSOnineteen}{\ensuremath{1.19^{+0.67}_{-0.49}}}
\newcommand{\twosixtyesevenGPProtESPRESSOeighteenESPRESSOnineteenHIRES}{\ensuremath{23.8^{+4.9}_{-5.1}}}
\newcommand{\twosixtyesevenGPsigmaHIRES}{\ensuremath{4.82^{+0.54}_{-0.56}}}
\newcommand{\twosixtyesevenGPalphaHIRES}{\ensuremath{0.091^{+0.054}_{-0.041}}}
\newcommand{\twosixtyesevenGPGammaHIRES}{\ensuremath{4.4^{+2.3}_{-1.8}}}
\newcommand{\twosixtyesevenmfluxTESSthree}{\ensuremath{(-1.7^{+3.3}_{-3.2})\times10^{-5}}}
\newcommand{\twosixtyesevensigmawTESSthree}{\ensuremath{3.1^{+13.6}_{-2.6}}}
\newcommand{\twosixtyesevenGPsigmaTESSthree}{\ensuremath{(25.3^{+1.3}_{-1.2})\times10^{-5}}}
\newcommand{\twosixtyesevenGPrhoTESSthree}{\ensuremath{0.1996^{+0.0042}_{-0.0056}}}
\newcommand{\twosixtyesevenmfluxTESSfortytwo}{\ensuremath{-0.4 \pm 1.4)\times10^{-5}}}
\newcommand{\twosixtyesevensigmawTESSfortytwo}{\ensuremath{195.3^{+7.1}_{-7.6}}}
\newcommand{\twosixtyesevenGPsigmaTESSfortytwo}{\ensuremath{(8.92^{+0.64}_{-0.63})\times10^{-5}}}
\newcommand{\twosixtyesevenGPrhoTESSfortytwo}{\ensuremath{0.244^{+0.032}_{-0.026}}}
\newcommand{\twosixtyesevenmfluxTESSseventy}{\ensuremath{(-0.6^{+6.8}_{-6.3})\times10^{-5}}}
\newcommand{\twosixtyesevensigmawTESSseventy}{\ensuremath{180.5^{+7.0}_{-6.1}}}
\newcommand{\twosixtyesevenGPsigmaTESSseventy}{\ensuremath{(23.9\pm 2.2)\times10^{-5}}}
\newcommand{\twosixtyesevenGPrhoTESSseventy}{\ensuremath{0.88^{+0.11}_{-0.09}}}
\newcommand{\twosixtyesevenmdilutionTESSthree}{\ensuremath{1.0}}
\newcommand{\twosixtyesevenmdilutionTESSfortytwo}{\ensuremath{1.0}}
\newcommand{\twosixtyesevenmdilutionTESSseventy}{\ensuremath{1.0}}

\newcommand{\twoeightysixTESSmfluxTESSone}{\ensuremath{(-2.1^{+1.4}_{-1.5})\times10^{-5}}}
\newcommand{\twoeightysixTESSsigmawTESSone}{\ensuremath{194.9^{+11.8}_{-12.5}}}
\newcommand{\twoeightysixTESSGPsigmaTESSone}{\ensuremath{(8.2^{+1.1}_{-0.9})\times10^{-5}}}
\newcommand{\twoeightysixTESSGPrhoTESSone}{\ensuremath{0.33^{+0.10}_{-0.08}}}
\newcommand{\twoeightysixTESSmdilutionTESSone}{\ensuremath{1.0}}
\newcommand{\twoeightysixTESSmfluxTESStwo}{\ensuremath{(-1.0^{+1.2}_{-1.1})\times10^{-5}}}
\newcommand{\twoeightysixTESSsigmawTESStwo}{\ensuremath{45.5^{+50.4}_{-44.7}}}
\newcommand{\twoeightysixTESSGPsigmaTESStwo}{\ensuremath{(6.74 \pm 0.91)\times10^{-5}}}
\newcommand{\twoeightysixTESSGPrhoTESStwo}{\ensuremath{0.21^{+0.36}_{-0.07}}}
\newcommand{\twoeightysixTESSmdilutionTESStwo}{\ensuremath{1.0}}
\newcommand{\twoeightysixTESSmfluxTESSthree}{\ensuremath{(-3.9^{+3.2}_{-6.6})\times10^{-5}}}
\newcommand{\twoeightysixTESSsigmawTESSthree}{\ensuremath{5.3^{+48.8}_{-4.9}}}
\newcommand{\twoeightysixTESSGPsigmaTESSthree}{\ensuremath{(8.9^{+8.6}_{-2.3})\times10^{-5}}}
\newcommand{\twoeightysixTESSGPrhoTESSthree}{\ensuremath{1.3^{+2.3}_{-0.6}}}
\newcommand{\twoeightysixTESSmdilutionTESSthree}{\ensuremath{1.0}}
\newcommand{\twoeightysixTESSmfluxTESSfour}{\ensuremath{(-2.6^{+3.1}_{-3.3})\times10^{-5}}}
\newcommand{\twoeightysixTESSsigmawTESSfour}{\ensuremath{186.0^{+12.6}_{-13.1}}}
\newcommand{\twoeightysixTESSGPsigmaTESSfour}{\ensuremath{(21.7^{+2.2}_{-1.9})\times10^{-5}}}
\newcommand{\twoeightysixTESSGPrhoTESSfour}{\ensuremath{0.206^{+0.044}_{-0.031}}}
\newcommand{\twoeightysixTESSmdilutionTESSfour}{\ensuremath{1.0}}
\newcommand{\twoeightysixTESSmfluxTESSfive}{\ensuremath{(-3.2^{+5.0}_{-5.4})\times10^{-5}}}
\newcommand{\twoeightysixTESSsigmawTESSfive}{\ensuremath{226.1^{+10.7}_{-11.5}}}
\newcommand{\twoeightysixTESSGPsigmaTESSfive}{\ensuremath{(15.1^{+5.2}_{-2.9})\times10^{-5}}}
\newcommand{\twoeightysixTESSGPrhoTESSfive}{\ensuremath{1.28^{+0.60}_{-0.40}}}
\newcommand{\twoeightysixTESSmdilutionTESSfive}{\ensuremath{1.0}}
\newcommand{\twoeightysixTESSmfluxTESSseven}{\ensuremath{(-1.7\pm 1.6)\times10^{-5}}}
\newcommand{\twoeightysixTESSsigmawTESSseven}{\ensuremath{183.3^{+12.4}_{-13.2}}}
\newcommand{\twoeightysixTESSGPsigmaTESSseven}{\ensuremath{(6.8^{+1.2}_{-1.0})\times10^{-5}}}
\newcommand{\twoeightysixTESSGPrhoTESSseven}{\ensuremath{0.56^{+0.23}_{-0.16}}}
\newcommand{\twoeightysixTESSmdilutionTESSseven}{\ensuremath{1.0}}
\newcommand{\twoeightysixTESSmfluxTESSeight}{\ensuremath{-2.28323^{+2.3}_{-2.4})\times10^{-5}}}
\newcommand{\twoeightysixTESSsigmawTESSeight}{\ensuremath{271.3^{+10.3}_{-11.0}}}
\newcommand{\twoeightysixTESSGPsigmaTESSeight}{\ensuremath{(9.4^{+1.8}_{-1.3})\times10^{-5}}}
\newcommand{\twoeightysixTESSGPrhoTESSeight}{\ensuremath{0.50^{+0.16}_{-0.13}}}
\newcommand{\twoeightysixTESSmdilutionTESSeight}{\ensuremath{1.0}}
\newcommand{\twoeightysixTESSmfluxTESSnine}{\ensuremath{(-3.1^{+1.9}_{-2.0})\times10^{-5}}}
\newcommand{\twoeightysixTESSsigmawTESSnine}{\ensuremath{233.5^{+10.5}_{-11.8}}}
\newcommand{\twoeightysixTESSGPsigmaTESSnine}{\ensuremath{(8.4^{+1.6}_{-1.2})\times10^{-5}}}
\newcommand{\twoeightysixTESSGPrhoTESSnine}{\ensuremath{0.48^{+0.16}_{-0.12}}}
\newcommand{\twoeightysixTESSmdilutionTESSnine}{\ensuremath{1.0}}
\newcommand{\twoeightysixTESSmfluxTESSten}{\ensuremath{(-1.9 \pm 2.1)\times10^{-5}}}
\newcommand{\twoeightysixTESSsigmawTESSten}{\ensuremath{311.2^{+9.9}_{-9.6}}}
\newcommand{\twoeightysixTESSGPsigmaTESSten}{\ensuremath{(10.0^{+1.5}_{-1.2})\times10^{-5}}}
\newcommand{\twoeightysixTESSGPrhoTESSten}{\ensuremath{0.35^{+0.11}_{-0.08}}}
\newcommand{\twoeightysixTESSmdilutionTESSten}{\ensuremath{1.0}}
\newcommand{\twoeightysixTESSmfluxTESSeleven}{\ensuremath{(-3.3^{+3.8}_{-4.1})\times10^{-5}}}
\newcommand{\twoeightysixTESSsigmawTESSeleven}{\ensuremath{214.2^{+13.1}_{-13.8}}}
\newcommand{\twoeightysixTESSGPsigmaTESSeleven}{\ensuremath{(13.7^{+3.1}_{-2.2})\times10^{-5}}}
\newcommand{\twoeightysixTESSGPrhoTESSeleven}{\ensuremath{0.68^{+0.23}_{-0.15}}}
\newcommand{\twoeightysixTESSmdilutionTESSeleven}{\ensuremath{1.0}}
\newcommand{\twoeightysixTESSmfluxTESStwelve}{\ensuremath{(-1.6 \pm 4.9)\times10^{-5}}}
\newcommand{\twoeightysixTESSsigmawTESStwelve}{\ensuremath{222.8^{+14.4}_{-13.6}}}
\newcommand{\twoeightysixTESSGPsigmaTESStwelve}{\ensuremath{(15.7^{+4.0}_{-2.4})\times10^{-5}}}
\newcommand{\twoeightysixTESSGPrhoTESStwelve}{\ensuremath{0.81^{+0.32}_{-0.22}}}
\newcommand{\twoeightysixTESSmdilutionTESStwelve}{\ensuremath{1.0}}
\newcommand{\twoeightysixTESSmfluxTESSthirteen}{\ensuremath{(-2.1 \pm 1.3)\times10^{-5}}}
\newcommand{\twoeightysixTESSsigmawTESSthirteen}{\ensuremath{253.1^{+10.5}_{-10.5}}}
\newcommand{\twoeightysixTESSGPsigmaTESSthirteen}{\ensuremath{(9.3^{+1.0}_{-0.9})\times10^{-5}}}
\newcommand{\twoeightysixTESSGPrhoTESSthirteen}{\ensuremath{0.187^{+0.057}_{-0.041}}}
\newcommand{\twoeightysixTESSmdilutionTESSthirteen}{\ensuremath{1.0}}
\newcommand{\twoeightysixTESSmfluxTESStwentyseven}{\ensuremath{(-2.0 \pm 3.0)\times10^{-5}}}
\newcommand{\twoeightysixTESSsigmawTESStwentyseven}{\ensuremath{205.5^{+13.8}_{-13.6}}}
\newcommand{\twoeightysixTESSGPsigmaTESStwentyseven}{\ensuremath{(10.0^{+2.2}_{-1.6})\times10^{-5}}}
\newcommand{\twoeightysixTESSGPrhoTESStwentyseven}{\ensuremath{0.85^{+0.31}_{-0.22}}}
\newcommand{\twoeightysixTESSmdilutionTESStwentyseven}{\ensuremath{1.0}}
\newcommand{\twoeightysixTESSmfluxTESStwentyeight}{\ensuremath{(-1.6^{+1.8}_{-1.7})\times10^{-5}}}
\newcommand{\twoeightysixTESSsigmawTESStwentyeight}{\ensuremath{229.5^{+11.7}_{-12.9}}}
\newcommand{\twoeightysixTESSGPsigmaTESStwentyeight}{\ensuremath{(9.4^{+1.3}_{-1.1})\times10^{-5}}}
\newcommand{\twoeightysixTESSGPrhoTESStwentyeight}{\ensuremath{0.248^{+0.069}_{-0.053}}}
\newcommand{\twoeightysixTESSmdilutionTESStwentyeight}{\ensuremath{1.0}}
\newcommand{\twoeightysixTESSmfluxTESStwentynine}{\ensuremath{(-1.5^{+3.0}_{-2.8})\times10^{-5}}}
\newcommand{\twoeightysixTESSsigmawTESStwentynine}{\ensuremath{266.5^{+10.3}_{-10.2}}}
\newcommand{\twoeightysixTESSGPsigmaTESStwentynine}{\ensuremath{(12.2^{+2.2}_{-1.6})\times10^{-5}}}
\newcommand{\twoeightysixTESSGPrhoTESStwentynine}{\ensuremath{0.48^{+0.15}_{-0.11}}}
\newcommand{\twoeightysixTESSmdilutionTESStwentynine}{\ensuremath{1.0}}
\newcommand{\twoeightysixTESSmfluxTESSthirty}{\ensuremath{(-2.4^{+6.0}_{-6.5})\times10^{-5}}}
\newcommand{\twoeightysixTESSsigmawTESSthirty}{\ensuremath{298.7^{+9.5}_{-9.2}}}
\newcommand{\twoeightysixTESSGPsigmaTESSthirty}{\ensuremath{(26.8^{+5.3}_{-3.6})\times10^{-5}}}
\newcommand{\twoeightysixTESSGPrhoTESSthirty}{\ensuremath{0.59^{+0.17}_{-0.11}}}
\newcommand{\twoeightysixTESSmdilutionTESSthirty}{\ensuremath{1.0}}
\newcommand{\twoeightysixTESSmfluxTESSthirtyone}{\ensuremath{(-1.1^{+3.0}_{-2.7})\times10^{-5}}}
\newcommand{\twoeightysixTESSsigmawTESSthirtyone}{\ensuremath{238.2^{+10.3}_{-10.2}}}
\newcommand{\twoeightysixTESSGPsigmaTESSthirtyone}{\ensuremath{(13.0^{+2.0}_{-1.6})\times10^{-5}}}
\newcommand{\twoeightysixTESSGPrhoTESSthirtyone}{\ensuremath{0.50^{+0.14}_{-0.10}}}
\newcommand{\twoeightysixTESSmdilutionTESSthirtyone}{\ensuremath{1.0}}
\newcommand{\twoeightysixTESSmfluxTESSthirtytwo}{\ensuremath{(3.7^{+9.5}_{-9.1})\times10^{-5}}}
\newcommand{\twoeightysixTESSsigmawTESSthirtytwo}{\ensuremath{257.0^{+10.0}_{-9.6}}}
\newcommand{\twoeightysixTESSGPsigmaTESSthirtytwo}{\ensuremath{(29.1^{+8.5}_{-5.3})\times10^{-5}}}
\newcommand{\twoeightysixTESSGPrhoTESSthirtytwo}{\ensuremath{1.08^{+0.43}_{-0.28}}}
\newcommand{\twoeightysixTESSmdilutionTESSthirtytwo}{\ensuremath{1.0}}
\newcommand{\twoeightysixTESSmfluxTESSthirtythree}{\ensuremath{(-2.3^{+2.1}_{-2.5})\times10^{-5}}}
\newcommand{\twoeightysixTESSsigmawTESSthirtythree}{\ensuremath{226.9^{+10.6}_{-11.0}}}
\newcommand{\twoeightysixTESSGPsigmaTESSthirtythree}{\ensuremath{(12.5^{+1.7}_{-1.3})\times10^{-5}}}
\newcommand{\twoeightysixTESSGPrhoTESSthirtythree}{\ensuremath{0.35^{+0.11}_{-0.08}}}
\newcommand{\twoeightysixTESSmdilutionTESSthirtythree}{\ensuremath{1.0}}
\newcommand{\twoeightysixTESSmfluxTESSthirtyfour}{\ensuremath{(-1.7^{+7.1}_{-7.2})\times10^{-5}}}
\newcommand{\twoeightysixTESSsigmawTESSthirtyfour}{\ensuremath{210.1^{+11.2}_{-11.0}}}
\newcommand{\twoeightysixTESSGPsigmaTESSthirtyfour}{\ensuremath{(26.4^{+5.0}_{-3.6})\times10^{-5}}}
\newcommand{\twoeightysixTESSGPrhoTESSthirtyfour}{\ensuremath{0.84^{+0.25}_{-0.19}}}
\newcommand{\twoeightysixTESSmdilutionTESSthirtyfour}{\ensuremath{1.0}}
\newcommand{\twoeightysixTESSmfluxTESSthirtyfive}{\ensuremath{(-3.9^{+11.1}_{-11.6})\times10^{-5}}}
\newcommand{\twoeightysixTESSsigmawTESSthirtyfive}{\ensuremath{262.5^{+10.2}_{-10.5}}}
\newcommand{\twoeightysixTESSGPsigmaTESSthirtyfive}{\ensuremath{(29.5^{+10.1}_{-5.9})\times10^{-5}}}
\newcommand{\twoeightysixTESSGPrhoTESSthirtyfive}{\ensuremath{1.43^{+0.56}_{-0.36}}}
\newcommand{\twoeightysixTESSmdilutionTESSthirtyfive}{\ensuremath{1.0}}
\newcommand{\twoeightysixTESSmfluxTESSthirtysix}{\ensuremath{(-2.0^{+2.7}_{-2.6})\times10^{-5}}}
\newcommand{\twoeightysixTESSsigmawTESSthirtysix}{\ensuremath{230.7^{+11.3}_{-11.0}}}
\newcommand{\twoeightysixTESSGPsigmaTESSthirtysix}{\ensuremath{(11.2^{+1.7}_{-1.4})\times10^{-5}}}
\newcommand{\twoeightysixTESSGPrhoTESSthirtysix}{\ensuremath{0.50^{+0.19}_{-0.13}}}
\newcommand{\twoeightysixTESSmdilutionTESSthirtysix}{\ensuremath{1.0}}
\newcommand{\twoeightysixTESSmfluxTESSthirtyseven}{\ensuremath{(-0.2^{+7.6}_{-7.1})\times10^{-5}}}
\newcommand{\twoeightysixTESSsigmawTESSthirtyseven}{\ensuremath{279.0^{+10.5}_{-9.9}}}
\newcommand{\twoeightysixTESSGPsigmaTESSthirtyseven}{\ensuremath{(23.2^{+5.9}_{-4.1})\times10^{-5}}}
\newcommand{\twoeightysixTESSGPrhoTESSthirtyseven}{\ensuremath{1.04^{+0.28}_{-0.20}}}
\newcommand{\twoeightysixTESSmdilutionTESSthirtyseven}{\ensuremath{1.0}}
\newcommand{\twoeightysixTESSmfluxTESSthirtyeight}{\ensuremath{(3.3^{+8.5}_{-7.3})\times10^{-5}}}
\newcommand{\twoeightysixTESSsigmawTESSthirtyeight}{\ensuremath{316.8^{+8.8}_{-8.7}}}
\newcommand{\twoeightysixTESSGPsigmaTESSthirtyeight}{\ensuremath{(27.6^{+7.3}_{-4.3})\times10^{-5}}}
\newcommand{\twoeightysixTESSGPrhoTESSthirtyeight}{\ensuremath{0.96^{+0.38}_{-0.23}}}
\newcommand{\twoeightysixTESSmdilutionTESSthirtyeight}{\ensuremath{1.0}}
\newcommand{\twoeightysixTESSmfluxTESSthirtynine}{\ensuremath{(-3.6^{+4.8}_{-5.0})\times10^{-5}}}
\newcommand{\twoeightysixTESSsigmawTESSthirtynine}{\ensuremath{216.5^{+12.4}_{-12.0}}}
\newcommand{\twoeightysixTESSGPsigmaTESSthirtynine}{\ensuremath{(23.2^{+3.1}_{-2.6})\times10^{-5}}}
\newcommand{\twoeightysixTESSGPrhoTESSthirtynine}{\ensuremath{0.53^{+0.10}_{-0.08}}}
\newcommand{\twoeightysixTESSmdilutionTESSthirtynine}{\ensuremath{1.0}}
\newcommand{\twoeightysixTESSmfluxTESSsixtyone}{\ensuremath{(-0.6^{+2.8}_{-3.0})\times10^{-5}}}
\newcommand{\twoeightysixTESSsigmawTESSsixtyone}{\ensuremath{254.4 \pm 10.1}}
\newcommand{\twoeightysixTESSGPsigmaTESSsixtyone}{\ensuremath{(11.6^{+2.1}_{-1.6})\times10^{-5}}}
\newcommand{\twoeightysixTESSGPrhoTESSsixtyone}{\ensuremath{0.63^{+0.21}_{-0.15}}}
\newcommand{\twoeightysixTESSmdilutionTESSsixtyone}{\ensuremath{1.0}}
\newcommand{\twoeightysixTESSmfluxTESSsixtytwo}{\ensuremath{(-1.3^{+2.2}_{-2.0})\times10^{-5}}}
\newcommand{\twoeightysixTESSsigmawTESSsixtytwo}{\ensuremath{2.9^{+22.8}_{-2.6}}}
\newcommand{\twoeightysixTESSGPsigmaTESSsixtytwo}{\ensuremath{(13.6^{+1.4}_{-1.3})\times10^{-5}}}
\newcommand{\twoeightysixTESSGPrhoTESSsixtytwo}{\ensuremath{0.20^{+0.28}_{-0.08}}}
\newcommand{\twoeightysixTESSmdilutionTESSsixtytwo}{\ensuremath{1.0}}
\newcommand{\twoeightysixTESSmfluxTESSsixtythree}{\ensuremath{(-2.8^{+3.5}_{-3.9})\times10^{-5}}}
\newcommand{\twoeightysixTESSsigmawTESSsixtythree}{\ensuremath{171.7^{+13.0}_{-12.5}}}
\newcommand{\twoeightysixTESSGPsigmaTESSsixtythree}{\ensuremath{(16.7^{+2.2}_{-1.9})\times10^{-5}}}
\newcommand{\twoeightysixTESSGPrhoTESSsixtythree}{\ensuremath{0.53^{+0.12}_{-0.09}}}
\newcommand{\twoeightysixTESSmdilutionTESSsixtythree}{\ensuremath{1.0}}
\newcommand{\twoeightysixTESSmfluxTESSsixtyfour}{\ensuremath{(-0.7^{+5.0}_{-4.7})\times10^{-5}}}
\newcommand{\twoeightysixTESSsigmawTESSsixtyfour}{\ensuremath{215.3^{+11.2}_{-11.9}}}
\newcommand{\twoeightysixTESSGPsigmaTESSsixtyfour}{\ensuremath{(17.6^{+3.3}_{-2.4})\times10^{-5}}}
\newcommand{\twoeightysixTESSGPrhoTESSsixtyfour}{\ensuremath{0.90^{+0.28}_{-0.19}}}
\newcommand{\twoeightysixTESSmdilutionTESSsixtyfour}{\ensuremath{1.0}}
\newcommand{\twoeightysixTESSmfluxTESSsixtyfive}{\ensuremath{(-1.8^{+6.2}_{-6.1})\times10^{-5}}}
\newcommand{\twoeightysixTESSsigmawTESSsixtyfive}{\ensuremath{237.3^{+11.3}_{-12.1}}}
\newcommand{\twoeightysixTESSGPsigmaTESSsixtyfive}{\ensuremath{(22.7^{+4.7}_{-3.4})\times10^{-5}}}
\newcommand{\twoeightysixTESSGPrhoTESSsixtyfive}{\ensuremath{0.73^{+0.19}_{-0.14}}}
\newcommand{\twoeightysixTESSmdilutionTESSsixtyfive}{\ensuremath{1.0}}
\newcommand{\twoeightysixTESSmfluxTESSsixtysix}{\ensuremath{(-2.1^{+4.0}_{-4.2})\times10^{-5}}}
\newcommand{\twoeightysixTESSsigmawTESSsixtysix}{\ensuremath{170.2 \pm 16.8}}
\newcommand{\twoeightysixTESSGPsigmaTESSsixtysix}{\ensuremath{(14.0^{+3.4}_{-2.2})\times10^{-5}}}
\newcommand{\twoeightysixTESSGPrhoTESSsixtysix}{\ensuremath{0.74^{+0.24}_{-0.18}}}
\newcommand{\twoeightysixTESSmdilutionTESSsixtysix}{\ensuremath{1.0}}
\newcommand{\twoeightysixTESSmfluxTESSsixtyseven}{\ensuremath{(-0.9^{+2.1}_{-2.3})\times10^{-5}}}
\newcommand{\twoeightysixTESSsigmawTESSsixtyseven}{\ensuremath{93.8^{+31.0}_{-87.4}}}
\newcommand{\twoeightysixTESSGPsigmaTESSsixtyseven}{\ensuremath{(7.4^{+1.9}_{-1.4})\times10^{-5}}}
\newcommand{\twoeightysixTESSGPrhoTESSsixtyseven}{\ensuremath{0.68^{+0.64}_{-0.32}}}
\newcommand{\twoeightysixTESSmdilutionTESSsixtyseven}{\ensuremath{1.0}}
\newcommand{\twoeightysixTESSmfluxTESSsixtyeight}{\ensuremath{(-2.2 \pm 2.5}}
\newcommand{\twoeightysixTESSsigmawTESSsixtyeight}{\ensuremath{214.1^{+11.3}_{-13.7}}}
\newcommand{\twoeightysixTESSGPsigmaTESSsixtyeight}{\ensuremath{(12.8^{+2.0}_{-1.5})\times10^{-5}}}
\newcommand{\twoeightysixTESSGPrhoTESSsixtyeight}{\ensuremath{0.31^{+0.10}_{-0.07}}}
\newcommand{\twoeightysixTESSmdilutionTESSsixtyeight}{\ensuremath{1.0}}
\newcommand{\twoeightysixTESSmfluxTESSsixtynine}{\ensuremath{(-2.3 \pm 1.6)\times10^{-5}}}
\newcommand{\twoeightysixTESSsigmawTESSsixtynine}{\ensuremath{121.0^{+19.5}_{-24.5}}}
\newcommand{\twoeightysixTESSGPsigmaTESSsixtynine}{\ensuremath{(7.1^{+1.2}_{-0.9})\times10^{-5}}}
\newcommand{\twoeightysixTESSGPrhoTESSsixtynine}{\ensuremath{0.42^{+0.15}_{-0.11}}}
\newcommand{\twoeightysixTESSmdilutionTESSsixtynine}{\ensuremath{1.0}}

\newcommand{\twoeightysixoneplPpone}{\ensuremath{4.5117240^{+0.0000029}_{-0.0000025}}}

\newcommand{\twoeightysixonepltzeropone}{\ensuremath{2460186.64900^{+0.00070}_{-0.00084}}}
\newcommand{\twoeightysixoneplppone}{\ensuremath{0.0170^{+0.0012}_{-0.0008}}}
\newcommand{\twoeightysixoneplbpone}{\ensuremath{0.62^{+0.17}_{-0.33}}}
\newcommand{\twoeightysixoneplrho}{\ensuremath{2866.4^{+2328.4}_{-1422.7}}}
\newcommand{\twoeightysixoneplqoneTESS}{\ensuremath{0.38^{+0.23}_{-0.13}}}
\newcommand{\twoeightysixoneplqtwoTESS}{\ensuremath{0.42^{+0.36}_{-0.27}}}
\newcommand{\twoeightysixoneplmuESPRESSOeighteen}{\ensuremath{17780.8^{+1.5}_{-1.8}}}
\newcommand{\twoeightysixoneplsigmawESPRESSOeighteen}{\ensuremath{1.45^{+0.55}_{-0.37}}}
\newcommand{\twoeightysixoneplmuESPRESSOnineteen}{\ensuremath{17779.5840865174 \pm 1.4}}
\newcommand{\twoeightysixoneplsigmawESPRESSOnineteen}{\ensuremath{1.34^{+0.37}_{-0.42}}}
\newcommand{\twoeightysixoneplmuHARPS}{\ensuremath{1.5^{+1.1}_{-1.4}}}
\newcommand{\twoeightysixoneplsigmawHARPS}{\ensuremath{0.05^{+1.19}_{-0.05}}}
\newcommand{\twoeightysixoneplKpone}{\ensuremath{1.89^{+0.45}_{-0.41}}}
\newcommand{\twoeightysixoneplGPsigmarv}{\ensuremath{3.29^{+0.61}_{-0.78}}}
\newcommand{\twoeightysixoneplGPalpharv}{\ensuremath{(7.4^{+3.9}_{-4.1}) \times 10^{-4}}}
\newcommand{\twoeightysixoneplGPGammarv}{\ensuremath{0.90^{+0.55}_{-0.51}}}
\newcommand{\twoeightysixoneplGPProtrv}{\ensuremath{36.7 \pm 1.7}}
\newcommand{\twoeightysixonepleccpone}{\ensuremath{0.0}}
\newcommand{\twoeightysixoneplomegapone}{\ensuremath{90.0}}
\newcommand{\twoeightysixtwopcPpone}{\ensuremath{4.5117244^{+0.0000031}_{-0.0000027}}}

\newcommand{\twoeightysixtwopctzeropone}{\ensuremath{2460186.64911 \pm 0.00076}}
\newcommand{\twoeightysixtwopcppone}{\ensuremath{0.01673^{+0.00086}_{-0.00036}}}
\newcommand{\twoeightysixtwopcbpone}{\ensuremath{0.60^{+0.14}_{-0.06}}}
\newcommand{\twoeightysixtwopcrho}{\ensuremath{3040.5^{+462.4}_{-1223.2}}}
\newcommand{\twoeightysixtwopcPptwo}{\ensuremath{39.361826^{+0.000070}_{-0.000081}}}

\newcommand{\twoeightysixtwopctzeroptwo}{\ensuremath{2460155.0251^{+0.0018}_{-0.0020}}}
\newcommand{\twoeightysixtwopcpptwo}{\ensuremath{0.02206^{+0.00097}_{-0.00046}}}
\newcommand{\twoeightysixtwopcbptwo}{\ensuremath{0.32^{+0.29}_{-0.17}}}
\newcommand{\twoeightysixtwopcqoneTESS}{\ensuremath{0.3847784499^{+0.18}_{-0.13}}}
\newcommand{\twoeightysixtwopcqtwoTESS}{\ensuremath{0.36^{+0.26}_{-0.22}}}
\newcommand{\twoeightysixtwopcmuESPRESSOeighteen}{\ensuremath{17780.0^{+1.2}_{-1.1}}}
\newcommand{\twoeightysixtwopcsigmawESPRESSOeighteen}{\ensuremath{1.70^{+0.60}_{-0.52}}}
\newcommand{\twoeightysixtwopcmuESPRESSOnineteen}{\ensuremath{17779.8 \pm 1.0}}
\newcommand{\twoeightysixtwopcsigmawESPRESSOnineteen}{\ensuremath{1.42^{+0.37}_{-0.36}}}
\newcommand{\twoeightysixtwopcmuHARPS}{\ensuremath{0.70^{+1.02}_{-0.83}}}
\newcommand{\twoeightysixtwopcsigmawHARPS}{\ensuremath{0.75^{+0.91}_{-0.74}}}
\newcommand{\twoeightysixtwopcKpone}{\ensuremath{1.98 \pm 0.33}}
\newcommand{\twoeightysixtwopcKptwo}{\ensuremath{0.79^{+0.64}_{-0.47}}}
\newcommand{\twoeightysixtwopcGPsigmarv}{\ensuremath{2.01^{+0.94}_{-0.81}}}
\newcommand{\twoeightysixtwopcGPalpharv}{\ensuremath{(10.7^{+4.8}_{-4.3})\times 10^{-4}}}
\newcommand{\twoeightysixtwopcGPGammarv}{\ensuremath{0.93^{+0.44}_{-0.39}}}
\newcommand{\twoeightysixtwopcGPProtrv}{\ensuremath{36.7 \pm 2.5}}
\newcommand{\twoeightysixtwopceccpone}{\ensuremath{0.0}}
\newcommand{\twoeightysixtwopcomegapone}{\ensuremath{90.0}}
\newcommand{\twoeightysixtwopceccptwo}{\ensuremath{0.0}}
\newcommand{\twoeightysixtwopcomegaptwo}{\ensuremath{90.0}}
\newcommand{\twoeightysixtwopePpone}{\ensuremath{4.5117244^{+0.0000028}_{-0.0000027}}}

\newcommand{\twoeightysixtwopetzeropone}{\ensuremath{2460186.64917^{+0.00068}_{-0.000689}}}
\newcommand{\twoeightysixtwopeppone}{\ensuremath{0.01656^{+0.00052}_{-0.00043}}}
\newcommand{\twoeightysixtwopebpone}{\ensuremath{0.56^{+0.11}_{-0.15}}}
\newcommand{\twoeightysixtwopesesinomegapone}{\ensuremath{0.24^{+0.16}_{-0.15}}}
\newcommand{\twoeightysixtwopesecosomegapone}{\ensuremath{0.24\pm0.13}}
\newcommand{\twoeightysixtwopeeccpone}{\ensuremath{0.144^{+0.093}_{-0.073}}}
\newcommand{\twoeightysixtwopeomegapone}{\ensuremath{45.1^{+24.7}_{-28.7}}}
\newcommand{\twoeightysixtwoperho}{\ensuremath{2465.8^{+495.8}_{-612.6}}}
\newcommand{\twoeightysixtwopePptwo}{\ensuremath{39.361812^{+0.000058}_{-0.000063}}}

\newcommand{\twoeightysixtwopetzeroptwo}{\ensuremath{2460155.0247^{+0.0017}_{-0.0018}}}
\newcommand{\twoeightysixtwopepptwo}{\ensuremath{0.02196^{+0.00043}_{-0.00040}}}
\newcommand{\twoeightysixtwopebptwo}{\ensuremath{0.29^{+0.18}_{-0.17}}}
\newcommand{\twoeightysixtwopesesinomegaptwo}{\ensuremath{0.16^{+0.13}_{-0.10}}}
\newcommand{\twoeightysixtwopesecosomegaptwo}{\ensuremath{0.23^{+0.20}_{-0.15}}}
\newcommand{\twoeightysixtwopeeccptwo}{\ensuremath{0.11^{+0.11}_{-0.07}}}
\newcommand{\twoeightysixtwopeomegaptwo}{\ensuremath{33.9^{+36.2}_{-23.6}}}
\newcommand{\twoeightysixtwopeqoneTESS}{\ensuremath{0.43^{+0.23}_{-0.18}}}
\newcommand{\twoeightysixtwopeqtwoTESS}{\ensuremath{0.32^{+0.31}_{-0.20}}}
\newcommand{\twoeightysixtwopemuESPRESSOeighteen}{\ensuremath{17780.2\pm1.2}}
\newcommand{\twoeightysixtwopesigmawESPRESSOeighteen}{\ensuremath{1.34^{+0.67}_{-0.45}}}
\newcommand{\twoeightysixtwopemuESPRESSOnineteen}{\ensuremath{17780.0^{+1.1}_{-1.2}}}
\newcommand{\twoeightysixtwopesigmawESPRESSOnineteen}{\ensuremath{1.40^{+0.30}_{-0.28}}}
\newcommand{\twoeightysixtwopemuHARPS}{\ensuremath{1.0\pm1.0}}
\newcommand{\twoeightysixtwopesigmawHARPS}{\ensuremath{0.13^{+0.97}_{-0.12}}}
\newcommand{\twoeightysixtwopeKpone}{\ensuremath{1.96^{+0.33}_{-0.29}}}
\newcommand{\twoeightysixtwopeKptwo}{\ensuremath{1.01^{+0.72}_{-0.64}}}
\newcommand{\twoeightysixtwopeGPsigmarv}{\ensuremath{2.94^{+0.52}_{-0.58}}}
\newcommand{\twoeightysixtwopeGPalpharv}{\ensuremath{0.00096^{+0.00047}_{-0.00040}}}
\newcommand{\twoeightysixtwopeGPGammarv}{\ensuremath{1.05^{+0.39}_{-0.34}}}
\newcommand{\twoeightysixtwopeGPProtrv}{\ensuremath{36.8\pm1.8}}
\newcommand{\onethirtyfourcPpone}{\ensuremath{1.40152604^{+0.00000074}_{-0.00000082}}}

\newcommand{\onethirtyfourctzeropone}{\ensuremath{2459082.85679^{+0.00041}_{-0.00037}}}
\newcommand{\onethirtyfourcppone}{\ensuremath{0.0247^{+0.0015}_{-0.0010}}}
\newcommand{\onethirtyfourcbpone}{\ensuremath{0.74^{+0.09}_{-0.12}}}
\newcommand{\onethirtyfourcrho}{\ensuremath{2236.4^{+1277.2}_{-928.3}}}
\newcommand{\onethirtyfourcqoneTESS}{\ensuremath{0.36^{+0.26}_{-0.19}}}
\newcommand{\onethirtyfourcqtwoTESS}{\ensuremath{0.36^{+0.33}_{-0.25}}}
\newcommand{\onethirtyfourcmuESPRESSOeighteen}{\ensuremath{29710.5 \pm 1.8}}
\newcommand{\onethirtyfourcsigmawESPRESSOeighteen}{\ensuremath{2.61^{+1.02}_{-0.77}}}
\newcommand{\onethirtyfourcmuESPRESSOnineteen}{\ensuremath{29715.2^{+1.7}_{-1.5}}}
\newcommand{\onethirtyfourcsigmawESPRESSOnineteen}{\ensuremath{1.68^{+0.49}_{-0.42}}}
\newcommand{\onethirtyfourcmuHARPS}{\ensuremath{29772.5^{+1.7}_{-1.6}}}
\newcommand{\onethirtyfourcsigmawHARPS}{\ensuremath{0.47^{+0.94}_{-0.46}}}
\newcommand{\onethirtyfourcmuPFS}{\ensuremath{4.1^{+1.7}_{-1.6}}}
\newcommand{\onethirtyfourcsigmawPFS}{\ensuremath{2.81^{+0.31}_{-0.30}}}
\newcommand{\onethirtyfourcKpone}{\ensuremath{3.20^{+0.29}_{-0.30}}}
\newcommand{\onethirtyfourcGPsigmarv}{\ensuremath{4.80^{+0.50}_{-0.46}}}
\newcommand{\onethirtyfourcGPalpharv}{\ensuremath{(9.8 \pm 1.4) \times 10^{-4}}}
\newcommand{\onethirtyfourcGPGammarv}{\ensuremath{3.16^{+0.80}_{-0.69}}}
\newcommand{\onethirtyfourcGPProtrv}{\ensuremath{31.08^{+0.73}_{-0.61}}}
\newcommand{\onethirtyfourcmfluxTESSone}{\ensuremath{(-3.8^{+4.3}_{-4.2})\times 10^{-5}}}
\newcommand{\onethirtyfourcsigmawTESSone}{\ensuremath{9.6^{+56.4}_{-9.0}}}
\newcommand{\onethirtyfourcGPsigmaTESSone}{\ensuremath{(24.2 \pm 1.7)\times 10^{-5}}}
\newcommand{\onethirtyfourcGPrhoTESSone}{\ensuremath{0.486^{+0.065}_{-0.057}}}
\newcommand{\onethirtyfourcmfluxTESStwentyeight}{\ensuremath{(-4.8^{+11.1}_{-9.9})\times 10^{-5}}}
\newcommand{\onethirtyfourcsigmawTESStwentyeight}{\ensuremath{218.0^{+15.4}_{-17.4}}}
\newcommand{\onethirtyfourcGPsigmaTESStwentyeight}{\ensuremath{(53.2^{+4.0}_{-4.1})\times 10^{-5}}}
\newcommand{\onethirtyfourcGPrhoTESStwentyeight}{\ensuremath{0.410^{+0.044}_{-0.041}}}
\newcommand{\onethirtyfourcmfluxTESSsixtyeight}{\ensuremath{(15.5^{+33.2}_{-28.6})\times 10^{-5}}}
\newcommand{\onethirtyfourcsigmawTESSsixtyeight}{\ensuremath{179.5^{+15.6}_{-18.3}}}
\newcommand{\onethirtyfourcGPsigmaTESSsixtyeight}{\ensuremath{(67.3^{+26.4}_{-18.5})\times 10^{-5}}}
\newcommand{\onethirtyfourcGPrhoTESSsixtyeight}{\ensuremath{2.89^{+0.94}_{-0.77}}}
\newcommand{\onethirtyfourceccpone}{\ensuremath{0.0}}
\newcommand{\onethirtyfourcomegapone}{\ensuremath{90.0}}
\newcommand{\onethirtyfourcmdilutionTESSone}{\ensuremath{1.0}}
\newcommand{\onethirtyfourcmdilutionTESStwentyeight}{\ensuremath{1.0}}
\newcommand{\onethirtyfourcmdilutionTESSsixtyeight}{\ensuremath{1.0}}
\newcommand{\onethirtyfourePpone}{\ensuremath{1.40152634^{+0.00000073}_{-0.00000094}}}

\newcommand{\onethirtyfouretzeropone}{\ensuremath{2459082.85677^{+0.00033}_{-0.00034}}}
\newcommand{\onethirtyfoureppone}{\ensuremath{0.02370^{+0.00083}_{-0.00087}}}
\newcommand{\onethirtyfourebpone}{\ensuremath{0.54^{+0.19}_{-0.24}}}
\newcommand{\onethirtyfouresesinomegapone}{\ensuremath{0.099^{+0.096}_{-0.067}}}
\newcommand{\onethirtyfouresecosomegapone}{\ensuremath{0.075^{+0.065}_{-0.048}}}
\newcommand{\onethirtyfoureeccpone}{\ensuremath{0.022^{+0.029}_{-0.015}}}
\newcommand{\onethirtyfoureomegapone}{\ensuremath{52.2^{+26.4}_{-32.8}}}
\newcommand{\onethirtyfourerho}{\ensuremath{4125.8^{+1858.6}_{-1816.1}}}
\newcommand{\onethirtyfoureqoneTESS}{\ensuremath{0.32^{+0.28}_{-0.20}}}
\newcommand{\onethirtyfoureqtwoTESS}{\ensuremath{0.37^{+0.33}_{-0.25}}}
\newcommand{\onethirtyfouremuESPRESSOeighteen}{\ensuremath{29709.8^{+1.9}_{-1.7}}}
\newcommand{\onethirtyfouresigmawESPRESSOeighteen}{\ensuremath{2.65^{+0.91}_{-0.74}}}
\newcommand{\onethirtyfouremuESPRESSOnineteen}{\ensuremath{29715.5^{+1.5}_{-1.6}}}
\newcommand{\onethirtyfouresigmawESPRESSOnineteen}{\ensuremath{1.70^{+0.55}_{-0.51}}}
\newcommand{\onethirtyfouremuHARPS}{\ensuremath{29771.7^{+1.8}_{-1.6}}}
\newcommand{\onethirtyfouresigmawHARPS}{\ensuremath{1.02^{+0.64}_{-0.93}}}
\newcommand{\onethirtyfouremuPFS}{\ensuremath{3.3^{+1.8}_{-1.7}}}
\newcommand{\onethirtyfouresigmawPFS}{\ensuremath{2.78^{+0.31}_{-0.30}}}
\newcommand{\onethirtyfoureKpone}{\ensuremath{3.17^{+0.29}_{-0.27}}}
\newcommand{\onethirtyfoureGPsigmarv}{\ensuremath{4.74^{+0.50}_{-0.44}}}
\newcommand{\onethirtyfoureGPalpharv}{\ensuremath{0.00103^{+0.00014}_{-0.00015}}}
\newcommand{\onethirtyfoureGPGammarv}{\ensuremath{3.24^{+0.78}_{-0.75}}}
\newcommand{\onethirtyfoureGPProtrv}{\ensuremath{31.14^{+0.68}_{-0.70}}}
\newcommand{\onethirtyfouremfluxTESSone}{\ensuremath{(-2.4^{+4.0}_{-4.5})\times 10^{-5}}}
\newcommand{\onethirtyfouresigmawTESSone}{\ensuremath{8.3^{+57.2}_{-7.8}}}
\newcommand{\onethirtyfoureGPsigmaTESSone}{\ensuremath{(24.0^{+1.6}_{-1.5})\times 10^{-5}}}
\newcommand{\onethirtyfoureGPrhoTESSone}{\ensuremath{0.497^{+0.060}_{-0.054}}}
\newcommand{\onethirtyfouremfluxTESStwentyeight}{\ensuremath{(-5.2^{+10.4}_{-10.2})\times 10^{-5}}}
\newcommand{\onethirtyfouresigmawTESStwentyeight}{\ensuremath{218.8^{+15.1}_{-17.4}}}
\newcommand{\onethirtyfoureGPsigmaTESStwentyeight}{\ensuremath{(52.9^{+3.9}_{-3.8})\times 10^{-5}}}
\newcommand{\onethirtyfoureGPrhoTESStwentyeight}{\ensuremath{0.407^{+0.043}_{-0.038}}}
\newcommand{\onethirtyfouremfluxTESSsixtyeight}{\ensuremath{(14.2^{+37.1}_{-30.7})\times 10^{-5}}}
\newcommand{\onethirtyfouresigmawTESSsixtyeight}{\ensuremath{179.2^{+16.1}_{-15.8}}}
\newcommand{\onethirtyfoureGPsigmaTESSsixtyeight}{\ensuremath{(68.4^{+22.2}_{-17.8})\times 10^{-5}}}
\newcommand{\onethirtyfoureGPrhoTESSsixtyeight}{\ensuremath{3.05^{+0.86}_{-0.75}}}
\newcommand{\onethirtyfouremdilutionTESSone}{\ensuremath{1}}
\newcommand{\onethirtyfouremdilutionTESStwentyeight}{\ensuremath{1}}
\newcommand{\onethirtyfouremdilutionTESSsixtyeight}{\ensuremath{1}}

\begin{document}

   \title{Three super-Earths and a possible water world from TESS and ESPRESSO}

   \subtitle{}

   \author{M.~J. Hobson
          \inst{\ref{unige}}
          \and
          F. Bouchy
          \inst{\ref{unige}}
          \and
          B. Lavie
          \inst{\ref{unige}}
          \and
          C. Lovis
          \inst{\ref{unige}}
          \and 
          V. Adibekyan
          \inst{\ref{IAporto}, \ref{uniporto}}
          \and
          C. Allende Prieto
          \inst{\ref{iac}}
          \and 
          Y. Alibert
          \inst{\ref{unibe1}, \ref{unibe2}}
          \and 
          S.~C.~C. Barros
          \inst{\ref{IAporto}, \ref{uniporto}}
          \and
          A. Castro-González
          \inst{\ref{CSIC}}
          \and
          S. Cristiani
          \inst{\ref{trieste}}
          \and
          V. D'Odorico
          \inst{\ref{trieste}}
          \and
          M. Damasso
          \inst{\ref{torino}}
          \and
          P. Di Marcantonio
          \inst{\ref{trieste}}
          \and
          X. Dumusque
          \inst{\ref{unige}}
          \and
          D. Ehrenreich
          \inst{\ref{unige}}
          \and
          P. Figueira
          \inst{\ref{IAporto},\ref{uniporto},\ref{ESO}}
          \and
          R. Génova Santos
          \inst{\ref{iac}}
          \and
          J.~I. Gonz\'alez Hern\'andez
          \inst{\ref{iac}, \ref{laguna}}
          \and
          J. Lillo-Box
          \inst{\ref{CSIC}}
          \and 
          G. Lo Curto
          \inst{\ref{ESO}}
          \and
          C.~J.~A.~P. Martins	
          \inst{\ref{CAporto},\ref{IAporto}}
          \and
          A. Mehner
          \inst{\ref{ESO}}
          \and
          G. Micela
          \inst{\ref{palermo}}
          \and
          P. Molaro
          \inst{\ref{trieste}}
          \and
          N.~J. Nunes
          \inst{\ref{lisboa}}
          \and
          E. Palle
          \inst{\ref{iac}, \ref{laguna}}
          \and 
          F. Pepe
          \inst{\ref{unige}}
          \and
          R. Rebolo
          \inst{\ref{iac}}
          \and
          J.\,Rodrigues
            \inst{\ref{IAporto}, \ref{uniporto},\ref{OFXB}}
          \and
          N. Santos
          \inst{\ref{IAporto}, \ref{uniporto}}
          \and
          S.~G. Sousa
          \inst{\ref{IAporto}, \ref{uniporto}}
          \and
          A. Sozzetti
          \inst{\ref{torino}}
          \and
          A. Suárez Mascareño
          \inst{\ref{iac},\ref{laguna}}
          \and
          H. M. Tabernero
          \inst{\ref{UCM}}
          \and
          S. Udry
          \inst{\ref{unige}}
          \and
          M.-R. Zapatero Osorio
          \inst{\ref{CSIC}}
          \and
          D.~J. Armstrong
          \inst{\ref{coventry1}, \ref{coventry2}}
          \and 
          D.~R. Ciardi
          \inst{\ref{caltech}}
          \and
          K.~A. Collins
          \inst{\ref{CFA}}
          \and
          K.~I. Collins
          \inst{\ref{GeorgeMason}}
          \and
          M. Everett
          \inst{\ref{NOIRLab}}
          \and
          D. Gandolfi
          \inst{\ref{unitorino}}
          \and
          S.~B. Howell
          \inst{\ref{NASA-Ames}}
          \and
          J.~M. Jenkins
          \inst{\ref{NASA-Ames}}
          \and
          J. Kielkopf
          \inst{\ref{louisville}}
          \and
          J.~H. Livingston
          \inst{\ref{abc-japan}, \ref{NAOJ}}
          \and
          M.~B. Lund
          \inst{\ref{caltech}}
          \and
          I. Mireles
          \inst{\ref{UNM}}
          \and
          G.~R. Ricker
          \inst{\ref{Kavli}}
          \and
          R.~P. Schwarz
          \inst{\ref{CFA}}
          \and
          S. Seager
          \inst{\ref{Kavli}, \ref{MIT1}, \ref{MIT2}}
          \and
          T.-G. Tan
          \inst{\ref{PEST}}
          \and
          E.~B. Ting
          \inst{\ref{NASA-Ames}}
          \and
          J.~N. Winn
          \inst{\ref{princeton}}
          }

   \institute{Observatoire de Genève, Département d'Astronomie, Université de Genève, Chemin Pegasi 51b, 1290 Versoix, Switzerland \label{unige}\\
              \email{melissa.hobson@unige.ch}
    \and
    Instituto de Astrofísica e Ciências do Espaço, CAUP, Universidade do Porto, Rua das Estrelas, 4150-762, Porto, Portugal\label{IAporto}
    \and
    Departamento de Física e Astronomia, Faculdade de Ciências, Universidade do Porto, Rua do Campo Alegre, 4169-007, Porto, Portugal\label{uniporto}
    \and
    Instituto de Astrof\'{\i}sica de Canarias, c/ V\'ia L\'actea s/n, 38205 La Laguna, Tenerife, Spain\label{iac}
    \and
    Physics Institute, University of Bern, Gesellsschaftstrasse 6, CH-3012 Bern, Switzerland\label{unibe1}
    \and
    Center for Space and Habitability, University of Bern, Gesellsschaftstrasse 6, CH-3012 Bern, Switzerland\label{unibe2}
    \and
    Centro de Astrobiolog\'ia (CAB), CSIC-INTA, Camino Bajo del Castillo s/n, 28692, Villanueva de la Ca\~nada (Madrid), Spain\label{CSIC}
    \and
    INAF – Osservatorio Astronomico di Trieste, via G. B. Tiepolo 11, I-34143, Trieste, Italy\label{trieste}
    \and
    INAF - Osservatorio Astrofisico di Torino, Strada Osservatorio 20, 10025, Pino Torinese (TO), Italy\label{torino}
    \and
    European Southern Observatory, Av. Alonso de Cordova, 3107, Vitacura, Santiago de Chile, Chile\label{ESO}
    \and
    Universidad de La Laguna, Dept. Astrof{\'\i}sica, E-38206 La Laguna, Tenerife, Spain\label{laguna}
    \and
    Centro de Astrof\'{\i}sica da Universidade do Porto, Rua das Estrelas, 4150-762 Porto, Portugal\label{CAporto}
    \and
    INAF - Osservatorio Astronomico di Palermo, Piazza del Parlamento 1, 90134, Palermo, Italy\label{palermo}
    \and
    Instituto de Astrof\'isica e Ci\^encias do Espa\c{c}o, Faculdade de Ci\^encias da Universidade de Lisboa, 1749-016 Lisboa, Portugal\label{lisboa}
     \and
     Observatoire François-Xavier Bagnoud -- OFXB, 3961 Saint-Luc, Switzerland\label{OFXB}
    \and
    Departamento de Física de la Tierra y Astrofísica \& IPARCOS-UCM (Instituto de Física de Partículas y del Cosmos de la UCM), Facultad de Ciencias Físicas, Universidad Complutense de Madrid, 28040, Madrid, Spain\label{UCM}
    \and
    Centre for Exoplanets and Habitability, University of Warwick, Coventry, CV4 7AL, UK\label{coventry1}
    \and
    Department of Physics, University of Warwick, Coventry, CV4 7AL, UK\label{coventry2}
    \and
    Caltech/IPAC-NASA Exoplanet Science Institute, 770 S. Wilson Avenue, Pasadena, CA 91106, USA\label{caltech}
    \and
    Center for Astrophysics \textbar \ Harvard \& Smithsonian, 60 Garden Street, Cambridge, MA 02138, USA\label{CFA}
    \and
    George Mason University, 4400 University Drive, Fairfax, VA, 22030 USA\label{GeorgeMason}
    \and
    NSF's National Optical-Infrared Astronomy Research Laboratory, 950 N. Cherry Avenue, Tucson, AZ 85719, USA\label{NOIRLab}
    \and
    Dipartimento di Fisica, Università degli Studi di Torino, Via Pietro Giuria, 1, 10125, Torino, Italy\label{unitorino}
    \and
    NASA Ames Research Center, Moffett Field, CA 94035 USA\label{NASA-Ames}
    \and
    Department of Physics and Astronomy, University of Louisville, Louisville, KY 40292, USA\label{louisville}
    \and
    Astrobiology Center, 2-21-1 Osawa, Mitaka, Tokyo 181-8588, Japan\label{abc-japan}
    \and
    National Astronomical Observatory of Japan, 2-21-1 Osawa, Mitaka, Tokyo 181-8588, Japan\label{NAOJ}
    \and
    Department of Physics and Astronomy, University of New Mexico, 210 Yale Blvd NE, Albuquerque, NM 87106, USA\label{UNM}
    \and
    Department of Physics and Kavli Institute for Astrophysics and Space Research, Massachusetts Institute of Technology, Cambridge, MA 02139, USA\label{Kavli}
    \and
    Department of Earth, Atmospheric and Planetary Sciences, Massachusetts Institute of Technology, Cambridge, MA 02139, USA\label{MIT1}
    \and
    Department of Aeronautics and Astronautics, MIT, 77 Massachusetts Avenue, Cambridge, MA 02139, USA\label{MIT2}
    \and
    Perth Exoplanet Survey Telescope, Perth, Western Australia, Australia\label{PEST}
    \and
    Department of Astrophysical Sciences, Princeton University, Princeton, NJ 08544, USA\label{princeton}
    }

   \date{Received 25 April 2024; revised 7 June 2024; accepted 10 June 2024}

 
  \abstract
   {Since 2018, the ESPRESSO spectrograph at the VLT has been hunting for planets in the Southern skies via the radial velocity (RV) method. One of its goals is to follow up candidate planets from transit surveys such as the TESS mission, with particular focus on small planets for which ESPRESSO's RV precision is vital.}
   {We aim to confirm and characterize in detail three super-Earth candidate transiting planets from TESS using precise RVs from ESPRESSO.}
   {We analyzed photometry from TESS and ground-based facilities, high-resolution imaging, and RVs from ESPRESSO, HARPS, and HIRES, to confirm and characterize three new planets: TOI-260~b, transiting a late K-dwarf, and TOI-286~b and c, orbiting an early K-dwarf. We also update parameters for the known super-Earth TOI-134~b (L 168-9 b), which is hosted by an M-dwarf.}
   {TOI-260~b has a \twosixtycgpsepPpone{} d period, $4.23 \pm1.60 \,\mathrm{M_\oplus}$ mass and $1.71\pm0.08\,\mathrm{R_\oplus}$ radius. For TOI-286~b we find a \twoeightysixtwopcPpone{} d period, $4.53\pm0.78\,\mathrm{M_\oplus}$ mass and $1.42\pm0.10\,\mathrm{R_\oplus}$ radius; for TOI-286~c, a \twoeightysixtwopcPptwo{} d period, $3.72\pm2.22\,\mathrm{M_\oplus}$ mass and $1.88\pm 0.12\,\mathrm{R_\oplus}$ radius. For TOI-134~b we obtain a \onethirtyfourcPpone{} d period, $4.07\pm0.45\,\mathrm{M_\oplus}$ mass, and $1.63\pm0.14\,\mathrm{R_\oplus}$ radius. Circular models are preferred for all the planets, although for TOI-260~b the eccentricity is not well-constrained. We compute bulk densities and place the planets in the context of composition models.}
   {TOI-260~b lies within the radius valley, and is most likely a rocky planet. However, the uncertainty on the eccentricity and thus on the mass renders its composition hard to determine. TOI-286~b and c span the radius valley, with TOI-286~b lying below it and having a likely rocky composition, while TOI-286~c is within the valley, close to the upper border, and probably has a significant water fraction. With our updated parameters for TOI-134~b, we obtain a lower density than previous findings, giving a rocky or Earth-like composition.}

   \maketitle
%

\section{Introduction}

Transiting planets orbiting around bright nearby stars are excellent targets for detailed characterization with follow-up high-resolution spectroscopy. By combining the transit photometry with the radial velocities (RVs) obtained from the spectra, both radius and true mass can be determined. This in turn allows for the determination of the planet's bulk density, which provides valuable constraints on its likely composition. 

The \textit{Transiting Exoplanet Survey Satellite} NASA mission \citep[\textit{TESS},][]{Ricker2015} has been conducting an all-sky survey to search for transiting planet candidates around nearby stars since 2018, through its prime mission (25 July 2018 - 4 July 2020), first extended mission (4 July 2020 - 1 September 2022), and second extended mission (1 September 2022 - present). One of the primary science requirements of the \textit{TESS} prime mission was to determine the masses for at least 50 planets with radii $\mathrm{<4\, R_\oplus}$. This has been amply achieved, with 128 such planets with masses determined to date \footnote{According to the NASA Exoplanet Archive, \cite{Akeson2013}, \url{https://exoplanetarchive.ipac.caltech.edu/}, accessed 16 February 2024}, though only 97 have masses measured to an accuracy of 25\% or better\footnote{According to the PlanetS catalog, \cite{Otegi2020}, Parc et al. in prep, available on the DACE platform, accessed 31 March 2024.}. In order to achieve this requirement, \textit{TESS} relies on extreme precision radial velocity (EPRV) facilities, which can confirm the \textit{TESS} candidates and provide mass measurements.

The Echelle SPectrograph for Rocky Exoplanets and Stable Spectroscopic Observations, (ESPRESSO, \citealt{Pepe2013, Pepe2021}) is an EPRV spectrograph at ESO's Very Large Telescope (VLT), Paranal, Chile. Achieving an RV precision of better than $\mathrm{25\, cm\, s^{-1}}$ during a single night, ESPRESSO is designed for and ideally suited to the precise mass measurement of low-mass planets. One of the main objectives of the ESPRESSO Guaranteed Time Observations \citep[GTO,][]{Pepe2021} is indeed the follow-up of such candidate planets from \textit{TESS} and \textit{Kepler}. The rocky planet population, especially, with masses typically below $\mathrm{6-8 M_\oplus}$, can only be characterized with instruments of ESPRESSO's RV precision. 

The population of planets with $\mathrm{R_p <4\, R_\oplus}$ encompasses the super-Earth and sub-Neptune groups, types of planets that are not found in the Solar System but which are common in exoplanetary systems, with occurrence rates of $\sim 30\%$ \citep{Fressin2013}. The two groups are separated by a paucity of planets in the $\mathrm{1.5-2R_\oplus}$ region, known as the radius valley or Fulton gap, which was predicted by \cite{Owen2013} and confirmed by \cite{Fulton2017}. The radius valley is thought to divide the smaller, rocky super-Earths from the larger, volatile-rich sub-Neptunes. The accurate characterization of planets in this regime allows us to probe their likely composition, and also gain insights into their formation and evolution \citep[e.g.][]{Luque2022,Burn2024}.

In this work, we present the confirmation and characterization of three new planets in the super-Earth and sub-Neptune domains: TOI-260~b, TOI-286~b, and TOI-286~c. TOI-260~b, which has a $1.71 \pm 0.11 \, \mathrm{R_\oplus}$ radius, orbits a late K-dwarf with a $\mathrm{\sim 13.5 \,d}$ period. TOI-286~b, and TOI-286~c are hosted by an early K dwarf; the inner planet, TOI-286~b, has a $1.41 \pm 0.07 \, \mathrm{R_\oplus}$ radius and a $\mathrm{\sim 4.5 \,d}$ period, while the outer planet, TOI-286~c, has a  $1.86 \pm 0.1  \, \mathrm{R_\oplus}$ radius and a $\mathrm{\sim 39.4 \,d}$ period. We also update the planetary parameters and characterization of the previously confirmed planet TOI-134~b, also known as L 168-9~b, which orbits an M dwarf at a $\mathrm{\sim 1.4 \,d}$ period. This planet was first described in \cite{Astudillo-Defru2020} and some of its parameters were updated by \cite{Patel2022}.

We describe the data in Section \ref{sect:obs}, and present our analysis of it in Section \ref{sect:analysis}. Our results are discussed in Section \ref{sect:discussion}, and we summarise and conclude in Section \ref{sect:conclusions}.

\section{Observations}\label{sect:obs}

In this section we describe the data used to identify and confirm the new planets we present in this paper, TOI-260~b, TOI-286~b, and TOI-286~c. For TOI-134~b, which is a known planet, we present only the \textit{TESS} data (some of which postdates the work of \citealt{Astudillo-Defru2020} and \citealt{Patel2022}) and the new ESPRESSO data which we use to refine the planetary mass.  

\subsection{\textit{TESS} photometry}
TOI-260 (TIC 37749396) has been observed by \textit{TESS} three times: once in the prime mission (Cycle 1) in sector 3, using CCD 4 of camera 1; once in the first extended mission (Cycle 4) in sector 42, using CCD 4 of camera 2; and once in the second extended mission (Cycle 6) in sector 70, using CCD 1 of camera 1. All the observations were done in two-minute cadence.

TOI-286 (TIC 150030205) is in the \textit{TESS} continuous viewing zone. It was observed by \textit{TESS} in sectors 1-5 and 7-13 during the prime mission (Cycle 1); and 27-39 (Cycle 3) and 61-69 (Cycle 5) in the extended missions. It was observed in camera 4 throughout. CCD 1 was used for sectors 30-32; CCD 2 for sectors 3-5, 7-9, 33-36, and 61-63; CCD 3 for sectors 10-12, 37-39, and 64-66; and CCD 4 for sectors 1-2, 27-29, and 67-69. All the observations were done in two-minute cadence.  

TOI-134 (TIC 234994474) was observed by \textit{TESS} in the prime mission in sector 1 (Cycle 1), and in the extended missions in sector 28 (Cycle 3) and sector 68 (Cycle 5). It was observed in camera 2, in CCD 2 for sector 1 and CCD 1 for sectors 28 and 68. All the observations were done in two-minute cadence.

All the \textit{TESS} photometry was processed by the \textit{TESS} Science Processing Operation Center \citep[SPOC,][]{Jenkins2016} at NASA Ames Research Center. Potential transit signals -- corresponding to one planet candidate for TOI-260, two for TOI-286, and one for TOI-134 -- were identified in the SPOC transit search \citep{Jenkins2002, Jenkins2010, Jenkins2020} of the light curve, and designated as \textit{TESS} Objects of Interest (TOIs) by the \textit{TESS} Science Office \citep{Guerrero2021}. The SPOC fit an initial limb-darkened transit model to each of these planetary signatures \citep{Li2019} and subjected each to a suite of diagnostic tests \citep{Twicken2018} to help make or break their planetary nature. In particular, the difference image centroiding analysis for the multi-sector searches\footnote{The multi-sector search for TOI-260 was conducted over sectors 2-72, while those for TOI-286 and TOI-134 were conducted over sectors 1-69.} constrained the location of the transit source for TOI-260, TOI-286 and TOI-134 to within $2.6 \pm 6.7 \arcsec$, $2.8 \pm 5.7 \arcsec$, and $3.6 \pm 3.0 \arcsec$, respectively. For our analysis, we use the SPOC Presearch Data Conditioning Simple Aperture Photometry (PDC-SAP) light curves \citep{Stumpe2012,Stumpe2014,Smith2012}.

\begin{figure*}[htb]
    \centering
    \includegraphics[width=.45\textwidth]{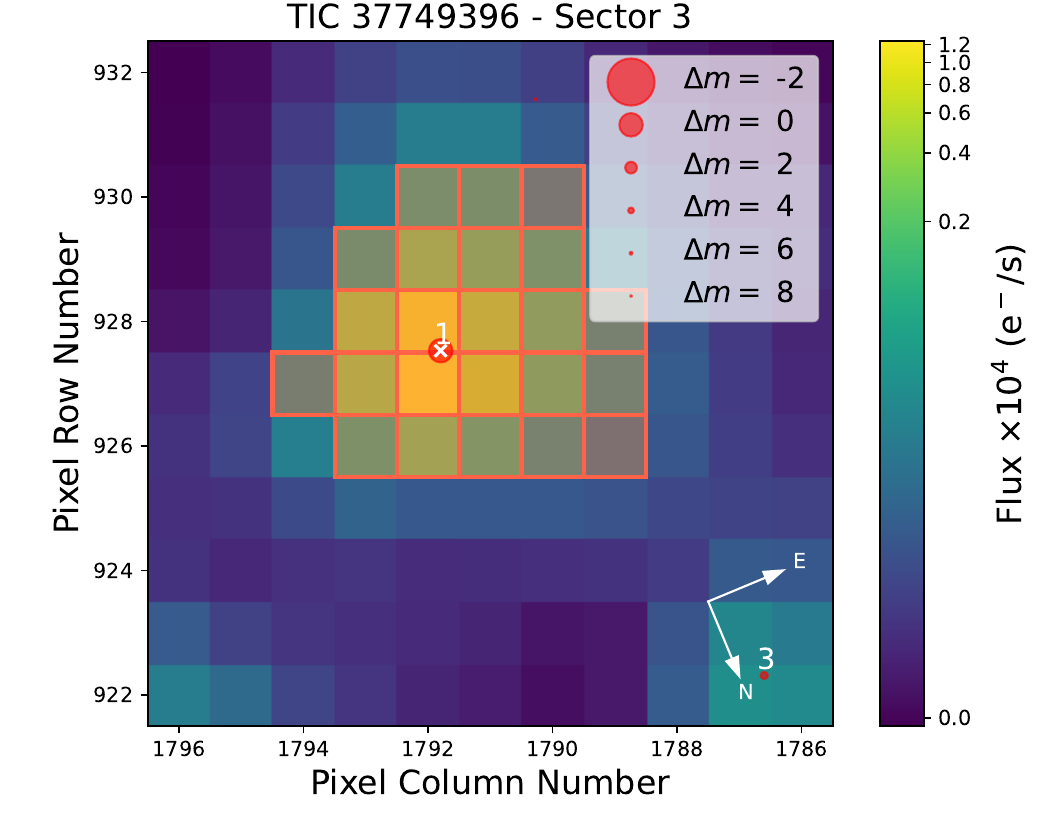}
    \includegraphics[width=.45\textwidth]{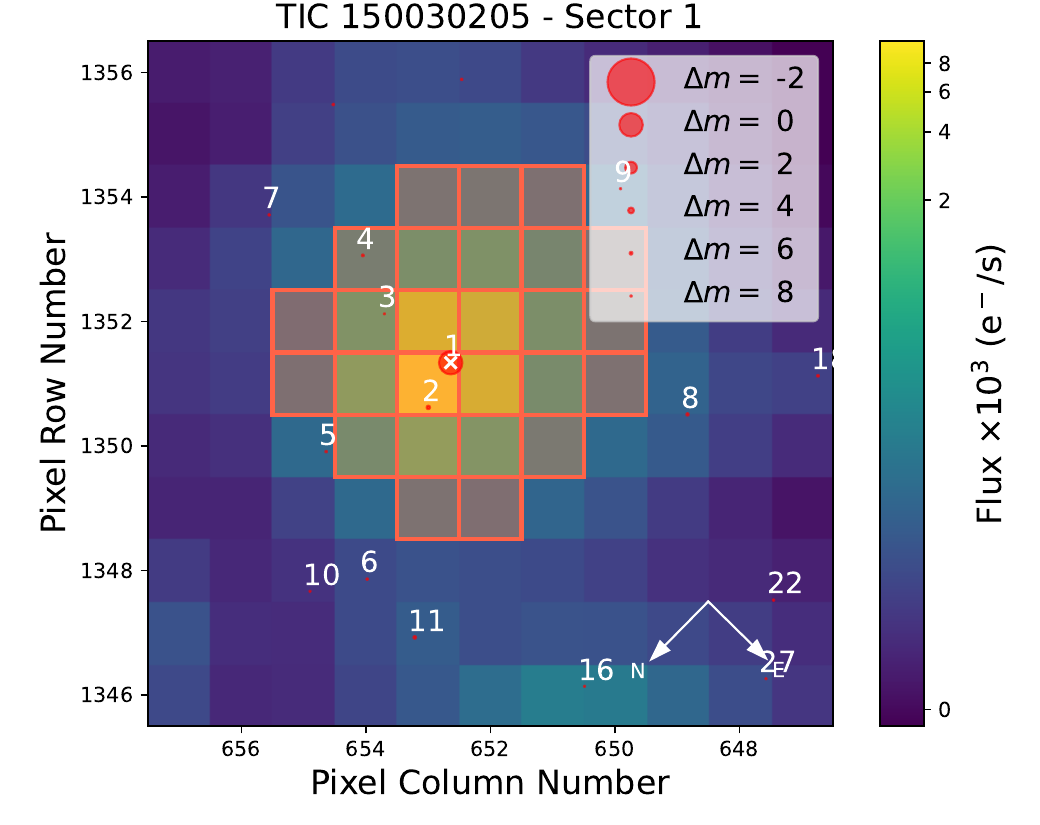}
    \caption{\textit{TESS} target pixel files for TOI-260 for sector 3 (left), and for TOI-286 for sector 1 ( right). The remaining sectors are shown in Appendix \ref{ap:tpfplotter}. The target star is labelled as 1 and marked by a white cross in each case. All sources from the Gaia DR3 catalogue down to a magnitude contrast of 8 are shown as red circles, with the size proportional to the contrast. The SPOC pipeline aperture is overplotted in shaded red squares.}
    \label{fig:tpfplotter}
\end{figure*}

Due to its large pixel size of $21\arcsec$ per pixel, the \textit{TESS} photometry can be contaminated by nearby companions. We used the \texttt{tpfplotter} package \citep{Aller2020} to search for potential contaminants in the \textit{TESS} apertures of our new planet hosts, TOI-260 and TOI-286. Since TOI-134~b is an already confirmed planet we do not apply this method to TOI-134.  Figure \ref{fig:tpfplotter} shows the resulting plots for sector 3 for TOI-260, and for sector 1 for TOI-286. The remaining sector plots are shown in Appendix \ref{ap:tpfplotter}. For TOI-260, we do not identify any contaminants within the \textit{TESS} aperture down to a magnitude contrast of 8, corresponding to the faintest blended binary that could mimic the transit depth \citep{LilloBox2014}. For TOI-286, there are three stars within the \textit{TESS} aperture, most notably the close companion Gaia DR3 5482316884791610880, with a Gaia magnitude of 15.08 (corresponding to a $\Delta m = 5.42$), and an angular separation of $16.7\arcsec$. This companion has a very different parallax and proper motion to TOI-286, and thus is likely not physically associated but is a chance alignment. We note that the SPOC difference image centroiding results strongly favour the target star as the source of the transit signal rather than the dimmer companion. Likewise, using a modified version of the \textit{TESS} positional probability code of \cite{Hadjigeorghiou2024}, we find probabilities of 34\% and 74\% for the transits of the inner and outer candidates respectively to come from the target star, versus 24\% probability for the transits of both candidates to come from the companion.

\subsection{Ground-based photometry}

Follow-up ground-based photometry is valuable to identify cases of false positives where the \textit{TESS} photometry is contaminated by nearby companions, or conversely to rule out the possibility that the transit-like signal is caused by a close companion that is an eclipsing binary. This is particularly important for TOI-286, which has a close bright companion. TOI-286 and TOI-260 were each observed three times; we describe the observations below. We used the {\tt TESS Transit Finder}, which is a customised version of the {\tt Tapir} software package \citep{Jensen2013}, to schedule our transit observations.

\subsubsection{LCO}
The Las Cumbres Observatory global telescope network \citep[LCO,][]{Brown2013} is a globally distributed network of telescopes. The 1\,m telescopes used for \textit{TESS} follow-up are equipped with $4096\times4096$ SINISTRO cameras which have an image scale of $0\farcs389$ per pixel, resulting in a $26\arcmin\times26\arcmin$ field of view. 

LCO observed TOI-260 three times from different sites: Once from the South African Astronomical Observatory (SAAO) site in Pan-STARRS Y-band ($\lambda_{\rm c} = 10040$\,\AA, ${\rm Width} =1120$\,\AA), on 19 August 2019; once from the Siding Springs Observatory (SSO) site in Pan-STARRS $z$-short band ($\lambda_{\rm c} = 8700$\,\AA, ${\rm Width} =1040$\,\AA), on 10 October 2020; and once from the Cerro Tololo Inter-American Observatory (CTIO) site in Pan-STARRS $z$-short band, on 7 July 2021. The images were calibrated by the standard LCOGT {\tt BANZAI} pipeline \citep{McCully2018}, and photometric data were extracted using {\tt AstroImageJ} \citep{Collins2017}. The target star light curves are shown in Fig. \ref{fig:toi-260_ground_phot}. Note that the target star in these and following image sequences was intentionally either strongly exposed or saturated (causing large scatter and/or systematics) for the purpose of searching for an eclipsing binary in fainter neighbouring stars that could be the cause of the TESS detection.

LCO also observed TOI-286 once, for a transit of the inner candidate TOI-286.01. The observation was done on 24th December 2018, in Sloan $r'$ filter, from the CTIO site. We highlight that the observation cleared the close companion Gaia DR3 5482316884791610880 of being a background eclipsing binary at this period. The target star light curve is shown in Fig. \ref{fig:toi-286_ground_phot}.

\subsubsection{MKO}

A full transit of TOI-286.01 was observed on 28 December 2018 in the Sloan $r^\prime$ (530=700 nm) band from the 0.61\,m University of Louisville Mt. Kent CDK700 (MKO CDK700), located at the University of Southern Queensland's Mt. Kent Observatory near Toowoomba, Australia. The telescope had a $4096\times4096$ SBIG STX-16803 camera providing an image scale of $0\farcs4$ per pixel, that resulted in a $27\arcmin\times27\arcmin$ field of view. The images were focused and seeing-limited to have typical stellar point-spread-functions with a full-width-half-maximum (FWHM) of $2\farcs3$. The images were calibrated for bias and dark corrections with WCS coordinates used to locate the target and potential nearby-stars that would affect TESS photometry. Astrometry was obtained with Astrometry.net\footnote{Avalaible online at \url{http://astrometry.net/}} \citep{Lang2010}. Photometric data were extracted with {\tt AstroImageJ} using a circular aperture with a $6\farcs0$ radius. The resulting light curve was detrended against airmass. While a transit in the target was anticipated to be too shallow to detect from the ground, potential nearby eclipsing binary stars or other contributors were inspected and eliminated as possible sources of a TESS signal. The target star light curve is shown in Fig. \ref{fig:toi-286_ground_phot}.

\subsubsection{PEST}
\begin{figure*}
    \centering
    \includegraphics[width=.95\textwidth]{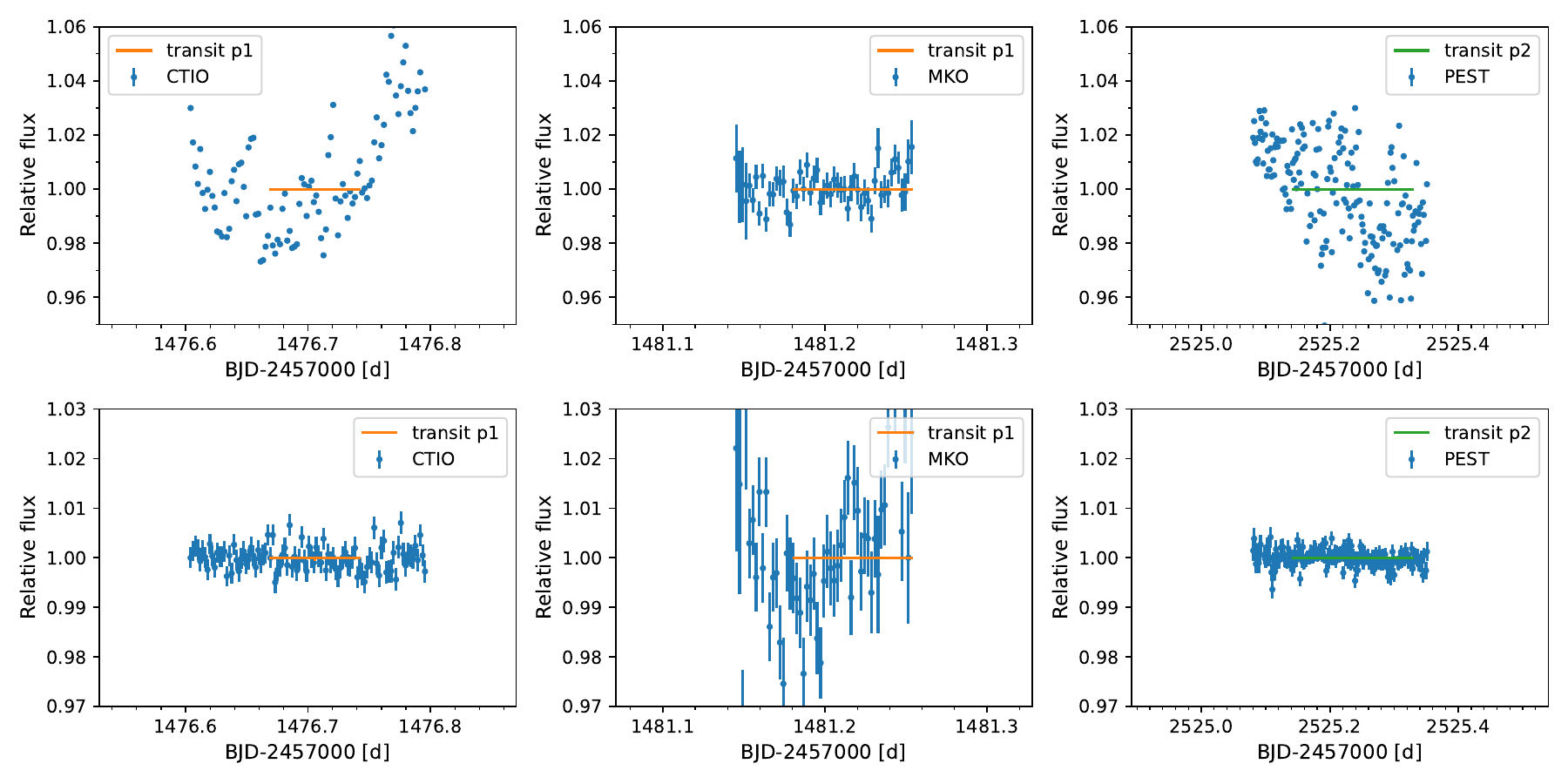}
    \caption{Top: Ground-based photometry for TOI-286, from LCO-CTIO (left), MKO (middle), and PEST (right). The orange (for planet 1) or green (for planet 2) horizontal bars indicate the timespan of the closest transit to each observation; the transit depth is too small to be seen. Note that the target star in these image sequences was intentionally either strongly exposed or saturated (causing large scatter and/or systematics) for the purpose of searching for an eclipsing binary in fainter neighbouring stars that could be the cause of the TESS detection. Bottom: ground based photometry for the close companion Gaia DR3 5482316884791610880, which clears it of being a background eclipsing binary.}
    \label{fig:toi-286_ground_phot}
\end{figure*}

The Perth Exoplanet Survey Telescope (PEST) is a backyard observatory near Perth, Australia, operated by Thiam-Guan Tan. The 0.3 m telescope is equipped with a $5544\times3694$ QHY183M camera.  Images are binned 2x2 in software giving an image scale of 0$\farcs$7 pixel$^{-1}$ resulting in a $32\arcmin\times21\arcmin$ field of view. PEST observed a transit of the outer candidate, TOI-286.02, on 6 November 2021, in r' band. A custom pipeline based on {\tt C-Munipack}\footnote{Available online at \url{http://c-munipack.sourceforge.net}} was used to calibrate the images and extract the differential photometry. We highlight that the observation cleared the close companion Gaia DR3 5482316884791610880 of being a background eclipsing binary at this period. The target star light curve is shown in Fig. \ref{fig:toi-286_ground_phot}.

\begin{figure*}
    \centering
    \includegraphics[width=.95\textwidth]{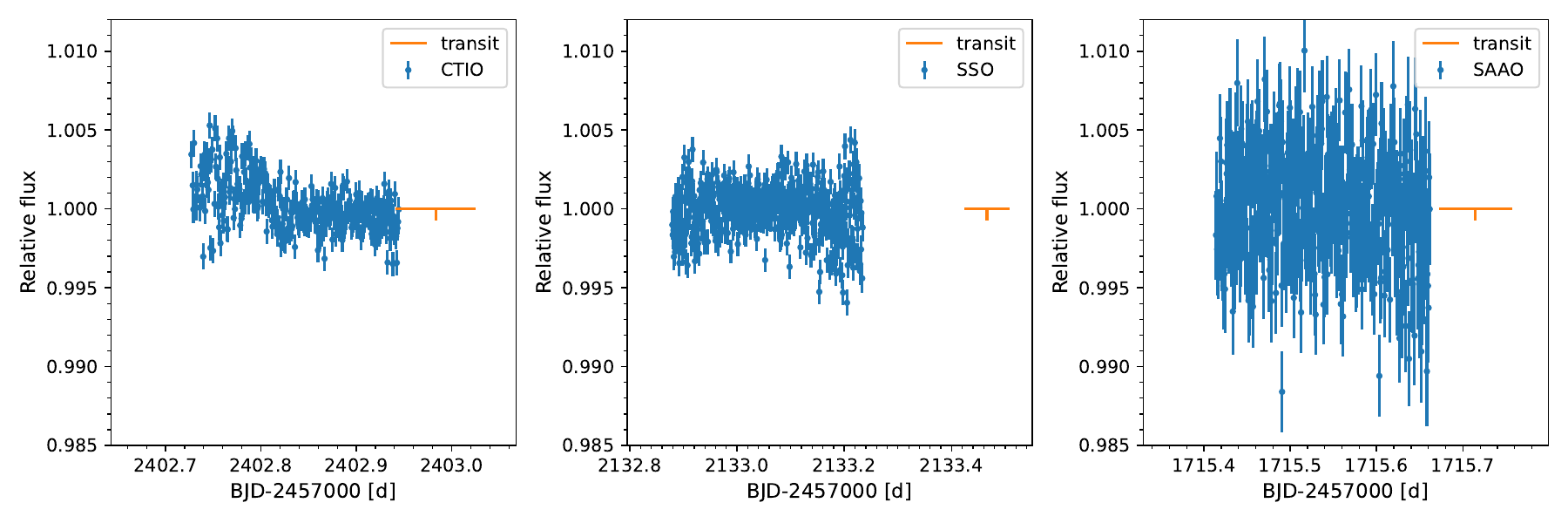}
    \caption{Ground-based photometry for TOI-260, from LCO-CTIO (left), LCO-SSO (middle), and LCO-SAAO (right). The orange horizontal bars indicate the timespan of the closest transit to each observation; the vertical bars indicate the transit midpoint and depth. The timing of the observations was planned from preliminary ephemerides; later additional \textit{TESS} data showed them to be out of transit. Note that the target star in these image sequences was intentionally either strongly exposed or saturated (causing large scatter and/or systematics) for the purpose of searching for an eclipsing binary in fainter neighbouring stars that could be the cause of the TESS detection.}
    \label{fig:toi-260_ground_phot}
\end{figure*}

\subsection{ESPRESSO radial velocities}

All three of our targets were observed with ESPRESSO in the context of the GTO. ESPRESSO underwent a major intervention in June-July 2019 when the fiber link was replaced \citep{Pepe2021}. Therefore, we split all ESPRESSO data into two sets pre- and post-upgrade, labelled ESPRESSO18 and ESPRESSO19 respectively.

TOI-260 was observed three times pre-upgrade, between 21 May 2019 and 5 June 2019, and 42 times post-upgrade, between 31 July 2019 and 25 January 2021, under program IDs 106.21M2, 1102.C-0744, and 1104.C-0350. All observations were performed with simultaneous Fabry-Perot calibration. An exposure time of $\mathrm{900\, s}$ was used for all observations except the first, which was taken with an exposure time of $\mathrm{1200\, s}$. The spectra have a mean signal to noise ratio (S/N) of 118 at $\mathrm{550\, nm}$, leading to a mean RV error of $\mathrm{\sigma_{RV} = 0.47 m \, s^{-1}}$. 

For TOI-286 we have 15 ESPRESSO18 observations, between 24 January 2019 and 5 May 2019, and 32 ESPRESSO19 observations, between 11 August 2019 and 22 March 2020, under program IDs 1102.C-0744, 1102.C-0958, and 1104.C-0350. 
All observations were performed with simultaneous Fabry-Perot calibration, with exposure times ranging from $\mathrm{900\, s}$ to $\mathrm{1800\, s}$. The spectra have a mean S/N of 116 at $\mathrm{550\, nm}$, leading to a mean RV error of $\mathrm{\sigma_{RV} = 0.43 m \, s^{-1}}$. 

Finally, TOI-134 was observed 15 times pre-upgrade, between 25 October 2018 and 13 June 2019, and 34 times post-upgrade, between 3 July 2019 and 10 January 2020, under program IDs 102.C-0744 and 1104.C-0350. 
This fainter star was observed with simultaneous sky observations, with exposure times ranging from $\mathrm{900\, s}$ to $\mathrm{1800\, s}$. One clear outlier on 13 June 2019 was removed from analysis, leaving a total of 14 ESPRESSO18 observations. The spectra have a mean S/N of 65 at $\mathrm{550\, nm}$, leading to a mean RV error of $\mathrm{\sigma_{RV} = 0.70 m \, s^{-1}}$.

All the observations were reduced with the data reduction software (DRS, \citealt{Pepe2021}) v.3.0.0, in which the radial velocities are obtained through the cross-correlation function (CCF) method \citep{Baranne1996}. For TOI-260, the K6 CCF mask was used; for TOI-286, the K2 mask; and for TOI-134, the M0 mask. The pipeline also computes several activity indicators: the Mount-Wilson S-index ($\mathrm{S_{MW}}$, \citealt{Vaughan1978}) and $\log R'_{\rm hk}$ \citep{Noyes1984}, which measure chromospheric emission in the cores of the Ca II H and K lines; the $\mathrm{H_\alpha}$ index \citep{Cincunegui2007,Bonfils2007}, which measures it for the $\mathrm{H_\alpha}$ line; the Na index \citep{Diaz2007}, which measures it for the Na I D1 and D2 lines; and the full width at half maximum (FWHM) and bisector inverse slope \citep[BIS,][]{Queloz2001} of the CCF. The RVs and activity indicators were obtained from the Data Analysis Center for Exoplanets (DACE) platform\footnote{Available online at \url{https://dace.unige.ch}}, and are listed in Tables \ref{tab:TOI-260_RV_ESPRESSO_data}, \ref{tab:TOI-286_RV_ESPRESSO_data}, and \ref{tab:TOI-134_RV_ESPRESSO_data} for TOI-260, TOI-286, and TOI-134 respectively. 

\begin{figure}[htb!]
    \centering
    \includegraphics[width=.4\textwidth]{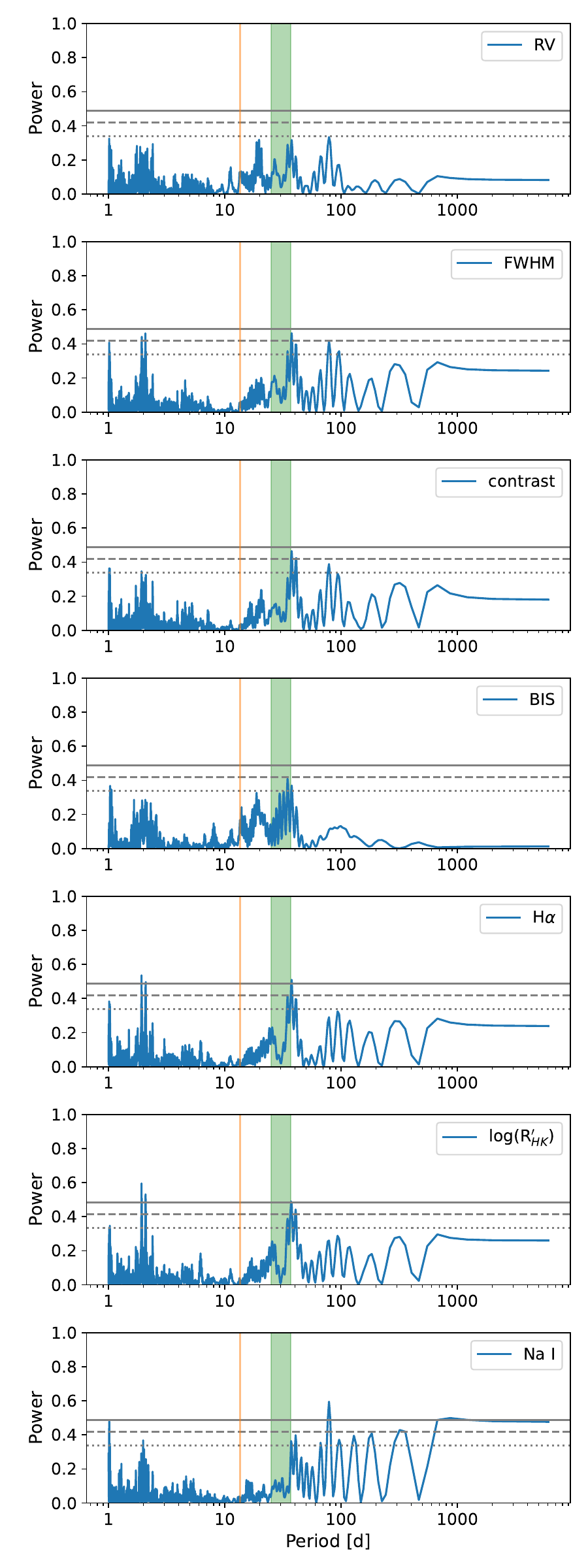}
    \caption{Periodograms of the ESPRESSO radial velocities (top) and activity indicators (second to bottom: CCF FWHM, CCF contrast, CCF bisector, $\mathrm{H_\alpha}$, $\mathrm{\log R'_{HK}}$, Na I) for TOI-260. The vertical orange line indicates the candidate planet period; the green shaded area, 1$\sigma$ around the stellar rotation period. The dotted, dashed, and solid horizontal grey lines indicate the 10\%, 1\%, and 0.1\% FAP levels respectively.}
    \label{fig:toi-260_periodograms}
\end{figure}

\begin{figure}[htb!]
    \centering
    \includegraphics[width=.4\textwidth]{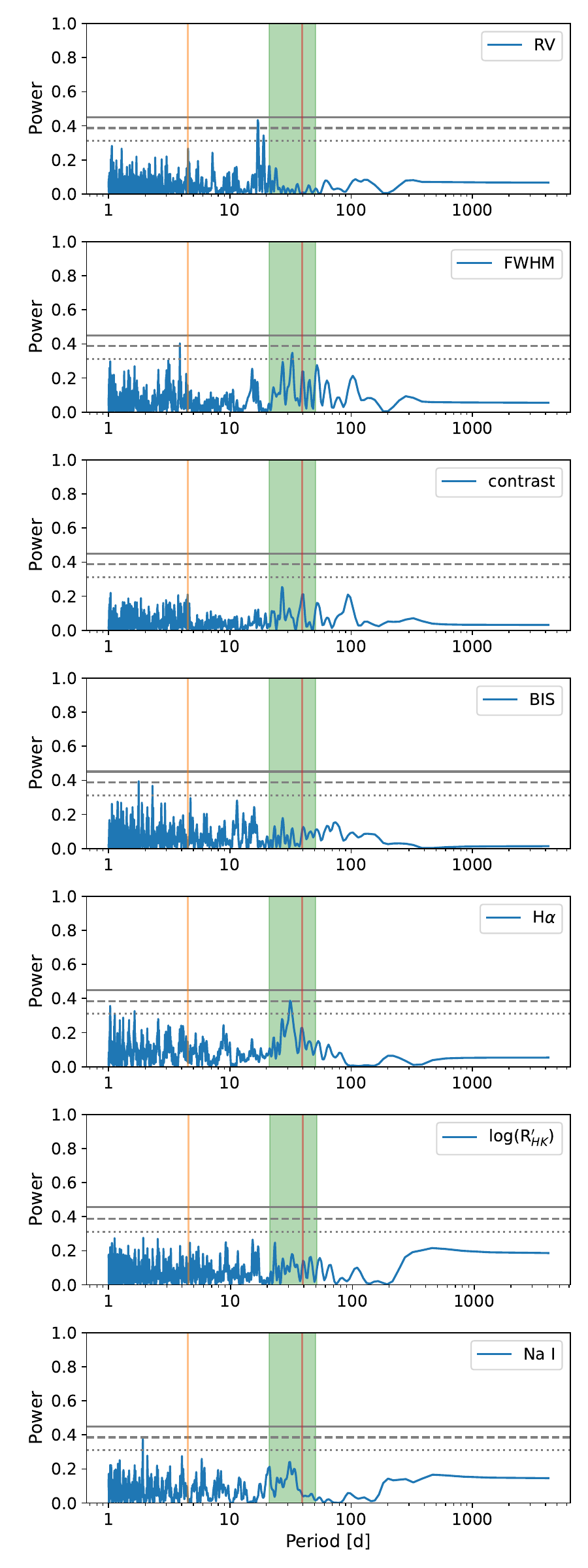}
    \caption{Periodograms of the ESPRESSO radial velocities (top) and activity indicators (second to bottom: CCF FWHM, CCF contrast, CCF bisector, $\mathrm{H_\alpha}$, $\mathrm{\log R'_{HK}}$, Na I) for TOI-286. The vertical orange and red lines indicate the candidate planet periods for TOI-286.01 and TOI-286.02. The green shaded area indicates 1$\sigma$ around the stellar rotation period. The dotted, dashed, and solid horizontal grey lines indicate the 10\%, 1\%, and 0.1\% FAP levels respectively.}
    \label{fig:toi-286_periodograms}
\end{figure}

\begin{figure}[htb!]
    \centering
    \includegraphics[width=.4\textwidth]{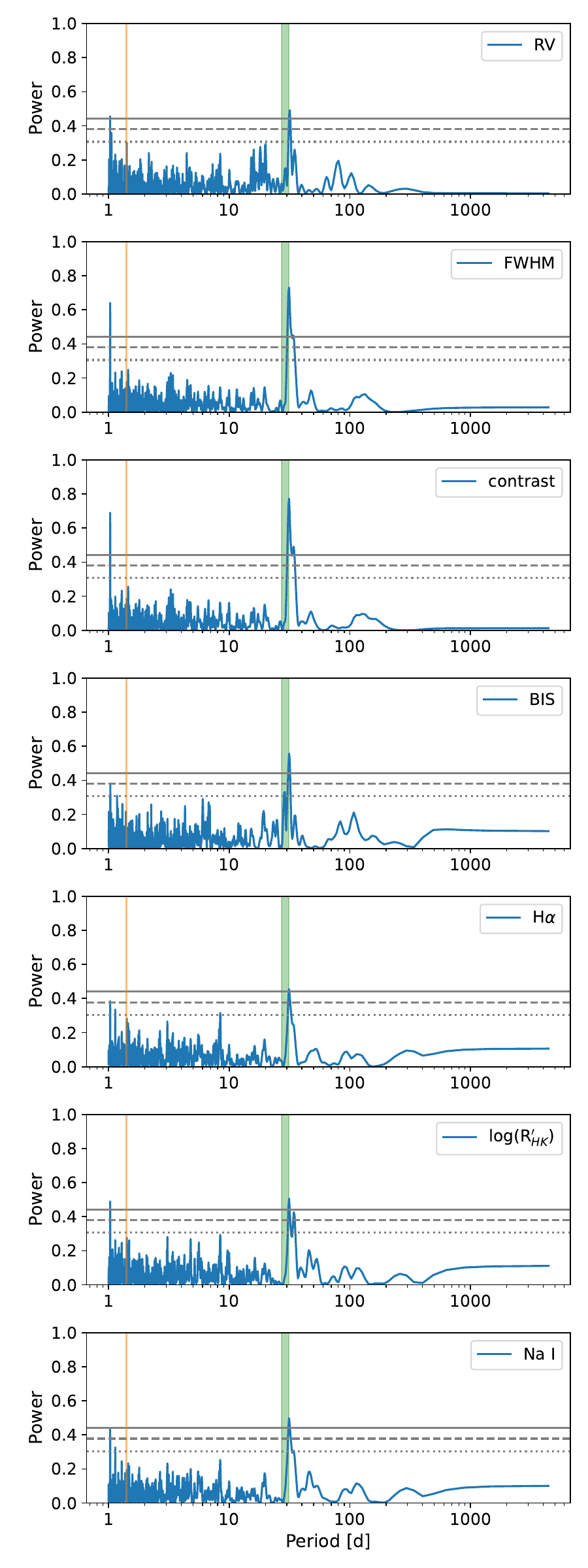}
    \caption{Periodograms of the ESPRESSO radial velocities (top) and activity indicators (second to bottom: CCF FWHM, CCF contrast, CCF bisector, $\mathrm{H_\alpha}$, $\mathrm{\log R'_{HK}}$, Na I) for TOI-134. The vertical orange line indicates the published planetary period. The green shaded area indicates 1$\sigma$ around the stellar rotation period. The dotted, dashed, and solid horizontal grey lines indicate the 10\%, 1\%, and 0.1\% FAP levels respectively.}
    \label{fig:toi-134_periodograms}
\end{figure}

We show the Generalized Lomb Scargle \citep[GLS,][]{Zechmeister2009} periodograms of the RVs and activity indicators in Fig. \ref{fig:toi-260_periodograms} for TOI-260, Fig. \ref{fig:toi-286_periodograms} for TOI-286, and Fig. \ref{fig:toi-134_periodograms} for TOI-134. In all cases, we do not see significant signals in the RV periodograms at the planet candidate periods. Instead, the main signals in both the RV and activity indicator periodograms are around the stellar rotation periods, indicating the presence of RV variability originated in stellar activity. This is particularly noteworthy for TOI-134, with highly significant signals in all periodograms at the stellar rotation period.

\subsection{Other radial velocities}

\subsubsection{HARPS}
TOI-286 was observed six times with the High Accuracy Radial velocity Planet Searcher spectrograph (HARPS, \citealt{Mayor2003}), which is mounted on the 3.6m telescope at La Silla Observatory. The observations were conducted between 18 January and 2 February 2019, under Program IDs 1102.C-0923(A), 1102.C-0249(A), and 60.A-9700(G). The exposure times ranged between $\mathrm{1200\, s}$ and $\mathrm{1800\, s}$, for a mean S/N of 81 at $\mathrm{550\, nm}$, leading to a mean RV error of $\mathrm{\sigma_{RV} = 0.62 m \, s^{-1}}$. The RVs were extracted with the HARPS-TERRA pipeline \citep{Anglada2012}. The RVs and activity indicators are listed in Table \ref{tab:TOI-286_RV_HARPS_data}.

\subsubsection{HIRES}
TOI-260 was observed 42 times with the Keck Observatory High Resolution Echelle Spectrometer (HIRES, \citealt{Vogt1994}) between 12 September 2008 and 19 January 2014, under the name HIP1532. The radial velocities obtained from these observations, carried out in the context of the Lick-Carnegie Exoplanet Survey, were taken from \cite{Butler2017}.

\subsection{High-resolution imaging}

\begin{figure*}[htb]
    \centering
    \includegraphics[width=.33\textwidth]{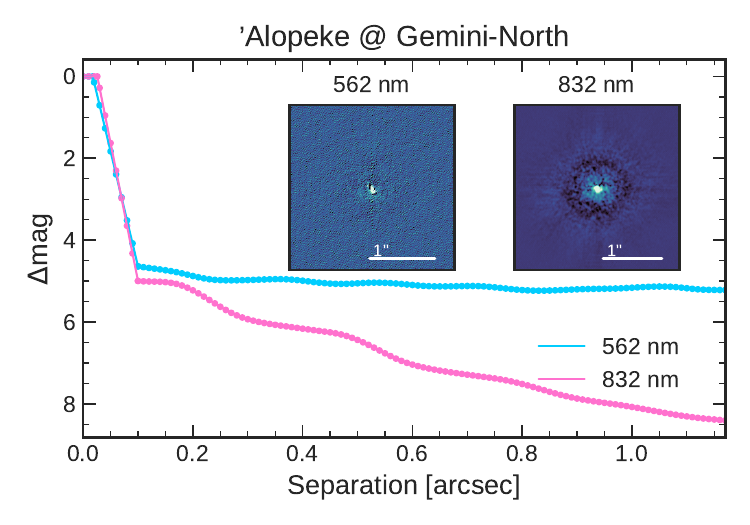}
    \includegraphics[width=.33\textwidth]{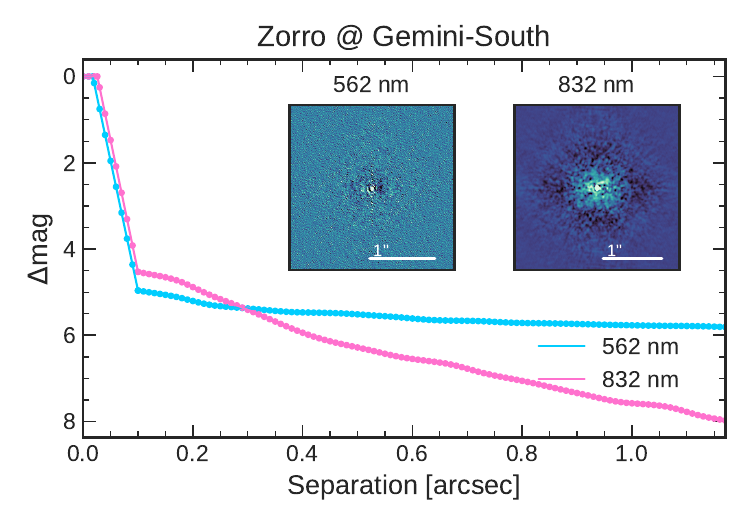}
    \includegraphics[width=.33\textwidth]{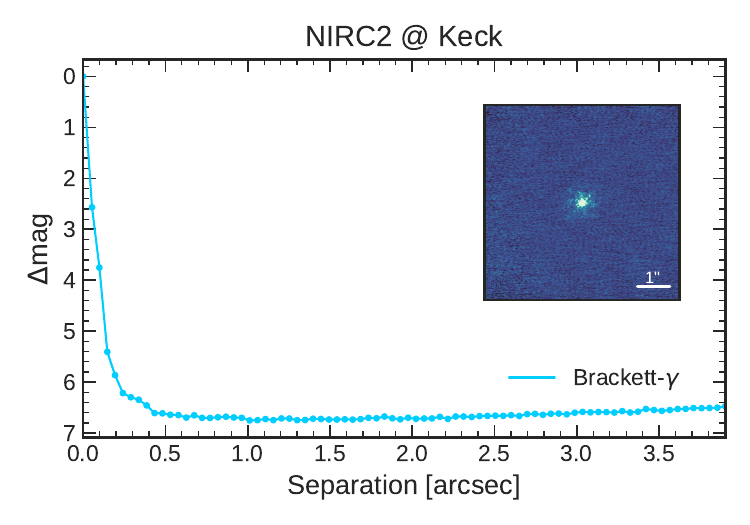}
    \includegraphics[width=.33\textwidth]{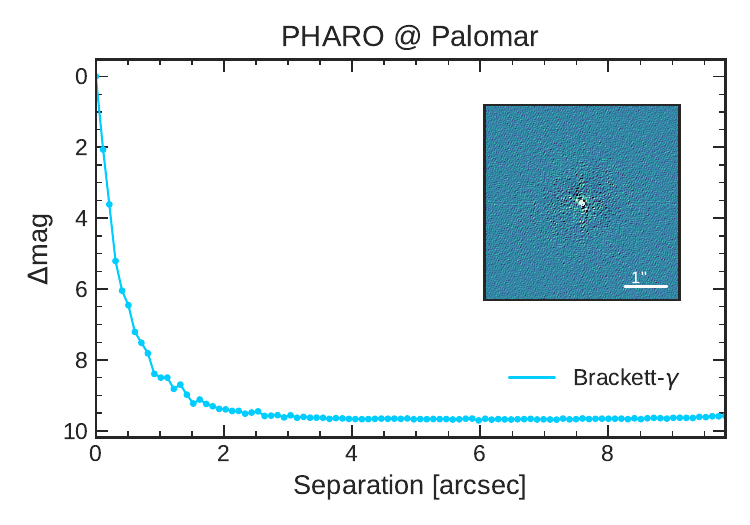}
    \includegraphics[width=.33\textwidth]{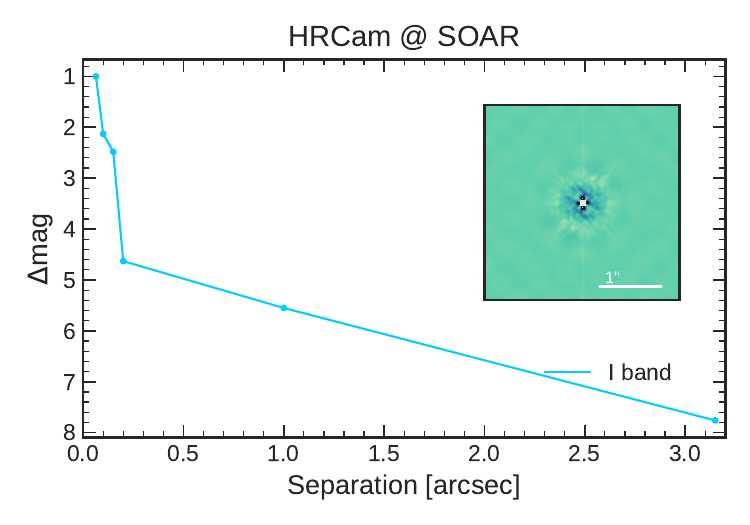}
    \includegraphics[width=.33\textwidth]{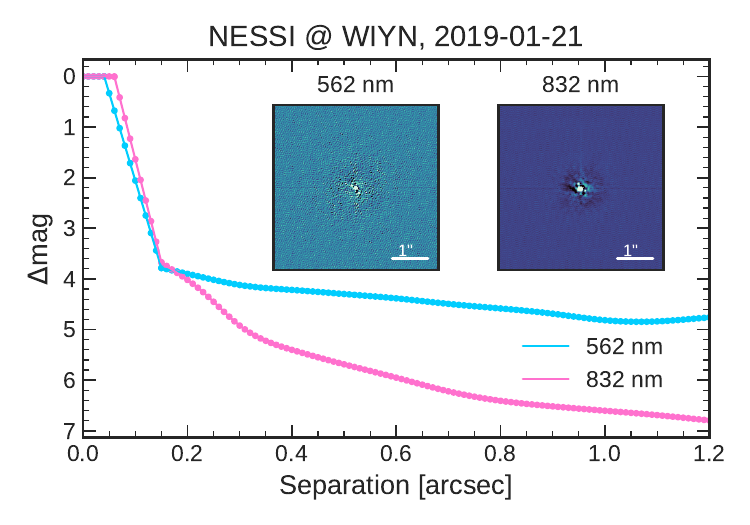}
    \includegraphics[width=.33\textwidth]{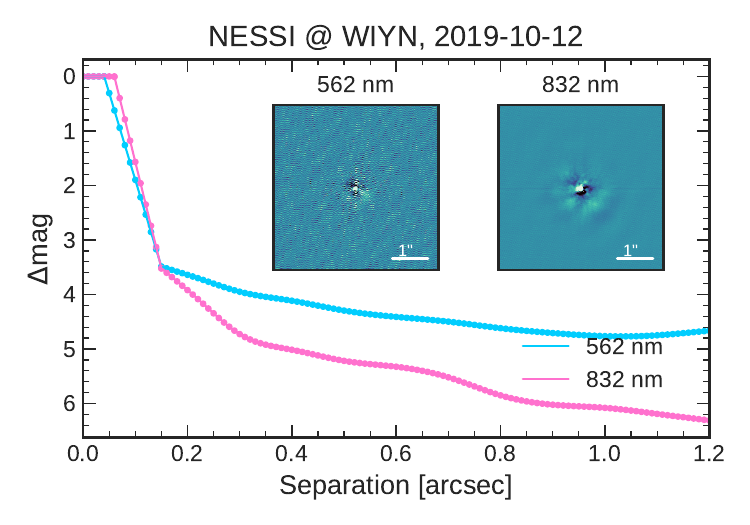}
    \caption{Companion sensitivity for the near-infrared and optical adaptive optics imaging for TOI-260. Top to bottom and left to right: 'Alopeke@Gemini-North, Zorro@Gemini-South, NIRC2@Keck, PHARO@Palomar, HRCam@SOAR, NESSI@WIYN (21 January 2019, 12 October 2019). 
    For all observations save SOAR, the inset image(s) is of the primary target showing no additional close-in companions. For SOAR the inset image shows the auto-correlation function.}
    \label{fig:speckle-imaging-toi260}
\end{figure*}

\begin{figure*}[htb]
    \centering
    \includegraphics[width=.45\textwidth]{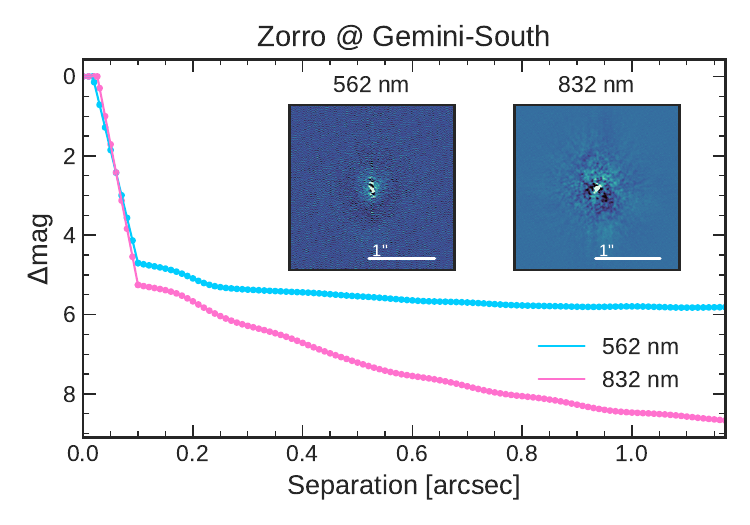}
    \includegraphics[width=.45\textwidth]{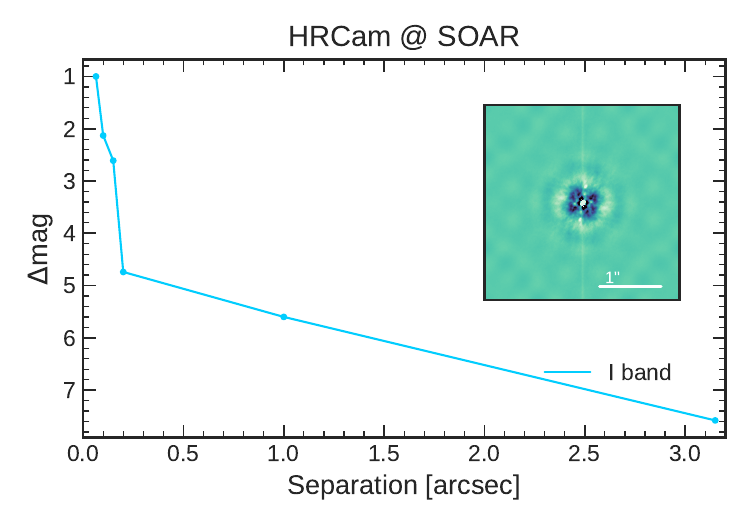}
    \caption{Companion sensitivity for the adaptive optics imaging for TOI-286. Left: Gemini Zorro@Gemini-South, where the inset image is of the primary target at 562 nm and 832 nm, showing no additional close-in companions. Right: HRCam@SOAR, where the inset image shows the auto-correlation function.}
    \label{fig:speckle-imaging-toi286}
\end{figure*}

If a star hosting a planet candidate has a close companion (bound or not), the companion can create a false-positive exoplanet detection if it is an eclipsing binary (EB) \citep{LilloBox2012}. Additionally, flux from the additional source(s) can lead to an underestimated planetary radius if not accounted for in the transit model \citep{Ciardi2015, Furlan2017, Matson2018, Castro2022}. As part of our standard process for validating transiting exoplanets to assess the the possible contamination of bound or unbound companions on the derived planetary radii, we therefore observed TOI-260 with high-resolution near-infrared adaptive optics (AO) imaging at Palomar and Keck Observatories and with optical speckle imaging at WIYN, SOAR, and Gemini-North/South, and TOI-286 with optical speckle imaging at SOAR and Gemini-South. The infrared observations provide the deepest sensitivities to faint companions while the optical speckle observations provide the highest resolution imaging, making the two techniques complementary.

\subsubsection{Gemini}

We obtained speckle imaging observations from both Gemini South’s Zorro instrument and Gemini North’s ‘Alopeke instrument for TOI-260, and from Gemini South's Zorro for TOI-286. Both instruments \citep{Scott2021} provide simultaneous speckle imaging in two bands (562nm and 832 nm) with output data products including a reconstructed image and robust contrast limits on companion detections \citep[see][]{Howell2011}.

TOI-260 was observed on 12 September 2019 with 'Alopeke at Gemini South and 12 October 2019 with Zorro at Gemini North. Both observations provided similar results, that TOI-260 has no close companions to within the 5$\sigma$ contrast limits obtained (5-8 magnitudes) and the angular limits sampled – 0.02 out to 1.2 arcsec (see Fig. \ref{fig:speckle-imaging-toi260}). At the distance to TOI-260 of $\mathrm{d=20.2\, pc}$, these angular limits correspond to 0.4 to 24 au. 

TOI-286 was observed on 26 December 2020 with Zorro on Gemini North. The high-resolution observation showed that TOI-286 has no close companions to within the 5$\sigma$ contrast limits obtained (5-9 magnitudes) and the angular limits sampled – 0.02 out to 1.2 arcsec (see Fig. \ref{fig:speckle-imaging-toi286}). At the distance to TOI-286 of $\mathrm{d=59.2\, pc}$, these angular limits correspond to 1.2 to 71 au.

\subsubsection{Palomar}

The Palomar Observatory observations of TOI-260 were made with the PHARO instrument \citep{Hayward2001} behind the natural guide star AO system P3K \citep{Dekany2013} on 22 December 2018 in a standard 5-point quincunx dither pattern with steps of 5\arcsec\ in the narrow-band $Br-\gamma$ filter $(\lambda_o = 2.1686; \Delta\lambda = 0.0326~\mu$m).  Each dither position was observed three times, offset in position from each other by 0.5\arcsec\ for a total of 15 frames; with an integration time of 1.4 seconds per frame, the total on-source time was 21 seconds. PHARO has a pixel scale of $0.025\arcsec$ per pixel for a total field of view of $\sim25\arcsec$. 

The science frames were flat-fielded and sky-subtracted.  The flat fields were generated from a median average of dark subtracted flats taken on-sky.  The flats were normalized such that the median value of the flats is unity.  The sky frames were generated from the median average of the 15 dithered science frames; each science image was then sky-subtracted and flat-fielded.  The reduced science frames were combined into a single combined image using a intra-pixel interpolation that conserves flux, shifts the individual dithered frames by the appropriate fractional pixels, and median-coadds the frames. The final resolution of the combined dithers was determined from the full-width half-maximum of the point spread function to be $0.105\arcsec$.

To within the limits of the AO observations, no stellar companions were detected. The sensitivities of the final combined AO image were determined by injecting simulated sources azimuthally around the primary target every $20^\circ $ at separations of integer multiples of the central source's FWHM \citep{Furlan2017, Clark2024}. The brightness of each injected source was scaled until standard aperture photometry detected it with $5\sigma $ significance. The resulting brightness of the injected sources relative to TOI-260 set the contrast limits at that injection location. The final $5\sigma $ limit at each separation was determined from the average of all of the determined limits at that separation and the uncertainty on the limit was set by the rms dispersion of the azimuthal slices at a given radial distance (see Fig. \ref{fig:speckle-imaging-toi260}).

\subsubsection{Keck}

The Keck Observatory observations of TOI-260 were made with the NIRC2 instrument on Keck-II behind the natural guide star AO system \citep{Wizinowich2000} on 9 September 2020 in the standard 3-point dither pattern that is used with NIRC2 to avoid the left lower quadrant of the detector which is typically noisier than the other three quadrants. The dither pattern step size was $3\arcsec$ and was repeated twice, with each dither offset from the previous dither by $0.5\arcsec$.  NIRC2 was used in the narrow-angle mode with a full field of view of $\sim10\arcsec$ and a pixel scale of approximately $0.0099442\arcsec$ per pixel.  The Keck observations were made in the narrow-band $Br-\gamma$ filter $(\lambda_o = 2.1686; \Delta\lambda = 0.0326~\mu$m) with an integration time in each filter of 0.18 second for a total of 1.6 seconds on target.

The frames were processed in the same way as described for the Palomar observations. The final resolution of the combined dithers was determined from the full-width half-maximum of the point spread function to be $0.050\arcsec$. To within the limits of the AO observations, no stellar companions were detected. The contrast curve is shown in Fig. \ref{fig:speckle-imaging-toi260}.

\subsubsection{SOAR}

The SOAR \textit{TESS} survey \citep{Ziegler2020} observes \textit{TESS} planet candidate hosts with speckle imaging using the high-resolution camera (HRCam) imager on the 4.1-m Southern Astrophysical Research (SOAR) telescope at Cerro Pachón, Chile \citep{Tokovinin2018}. TOI-260 was observed on 12 August 2019, while TOI-286 was observed on 18 February 2019. In both cases, no nearby sources were detected within $3\arcsec$. The contrast curves and auto-correlation functions are shown in Figs. \ref{fig:speckle-imaging-toi260} and \ref{fig:speckle-imaging-toi286} respectively.

\subsubsection{WIYN}

The NN-Explore Exoplanet Stellar Speckle Imager (NESSI) \citep{Scott2019}, which is mounted on the 3.5\,m WIYN telescope at Kitt Peak, observed TOI-260 twice, on 21 January 2019 and 12 October 2019. NESSI simultaneously acquires data in two bands centered at 562\,nm and 832\,nm using high speed electron-multiplying CCDs (EMCCDs). We collected and reduced the data following the procedures described in \citet{Howell2011}. The resulting reconstructed image achieved a contrast of $\Delta\mathrm{mag} \sim 6.5$ at a separation of 1\arcsec in the 832\,nm band for the observation of 21 January, and of $\Delta\mathrm{mag} \sim 6$ for the observation of 12 October  (see Fig.~\ref{fig:speckle-imaging-toi260}).

\subsection{Limits on outer companions from global astrometry}\label{sec:astrometry}

TOI-286 (CD-60 8051) and TOI-260 (BD-10 47) have been observed by both the Hipparcos and Gaia global astrometry missions. It is therefore possible to inspect the latest versions of the catalogues of astrometric accelerations constructed by \cite{Brandt2021} and \cite{Kervella2022} in order to probe for the existence of outer, massive companions. No statistically significant proper motion difference is found for either of the two stars. However, as both these stars are late-type and nearby, the limits that can be placed on the mass of an outer companion as a function of semi-major axis are rather interesting. Based on \cite{Kervella2022}, it is possible to rule out the presence of a giant planet in the approximate interval of orbital separations between 3 and 10 au with $\mathrm{M_p \gtrsim 0.6\, M_\mathrm{Jup}}$ and $\mathrm{M_p \gtrsim 0.3\, M_\mathrm{Jup}}$ around TOI-286 and TOI-260, respectively.

\section{Analysis}\label{sect:analysis}

\subsection{Stellar parameters}\label{sect:stellar-parameters}

The parameters of the three host stars are given in Table \ref{tab:starparams}. We obtained the stellar coordinates, proper motions, and parallaxes from the  \textit{\textit{Gaia}} Data Release 3 \citep{GAIA2016, GaiaDR3}. To compute the atmospheric parameters from the ESPRESSO spectra, we performed spectral synthesis with the {\sc SteParSyn} code\footnote{https://github.com/hmtabernero/SteParSyn/} \citep{Tabernero2022}. {\sc SteParSyn} provides the effective temperature $\mathrm{T_{eff}}$, metallicity [Fe/H], surface gravity $\log{g}$,  and broadening parameter $\mathrm{v_{broad}}$, which accounts for both the macroturbulence $\zeta$ and the projected rotational velocity $\mathrm{v \sin i}$. The first set of error bars corresponds to the internal errors alone. For a more realistic uncertainty on $\mathrm{T_{eff}}$, we adopt the 2\% error floor of \cite{Tayar2022}, reported in brackets in Table \ref{tab:starparams}.

As an independent comparison and validation, we also computed the atmospheric parameters through measurement of the equivalent widths of specific lines. For TOI-286, which is an early K dwarf, we used the ARES+MOOG approach described in \cite{Sousa2014}. TOI-134 and TOI-260 are cooler stars - an M dwarf and a late K dwarf - for which this approach is not suitable. Instead, we used the \texttt{ODUSSEAS} tool \citep{Antoniadis2020}, developed specifically for M dwarfs, although we note that TOI-260, which is on the K-M boundary, is on the edge of the temperature validity range. The parameters obtained for TOI-286 ($\mathrm{T_{eff} = 5059 \pm 78\, K}$, $\log{g} = 4.50 \pm 0.04\, \mathrm{dex}$, $\mathrm{[Fe/H] = -0.09 \pm 0.05\, dex}$) and TOI-134 ($\mathrm{T_{eff} = 3850 \pm 91\, K}$, $\log{g} = 4.44 \pm 0.08\, \mathrm{dex}$, $\mathrm{[Fe/H] = -0.01 \pm 0.11\, dex}$) are generally compatible with those obtained from {\sc SteParSyn}. For TOI-260, they differ more notably ($\mathrm{T_{eff} = 3851 \pm 101\, K}$, $\log{g} = 4.61 \pm 0.06\, \mathrm{dex}$, $\mathrm{[Fe/H] = 0.06 \pm 0.13\, dex}$); this is most likely due to the star being indeed too hot for \texttt{ODUSSEAS}. 

\begin{figure}
    \centering
    \includegraphics[width=\columnwidth]{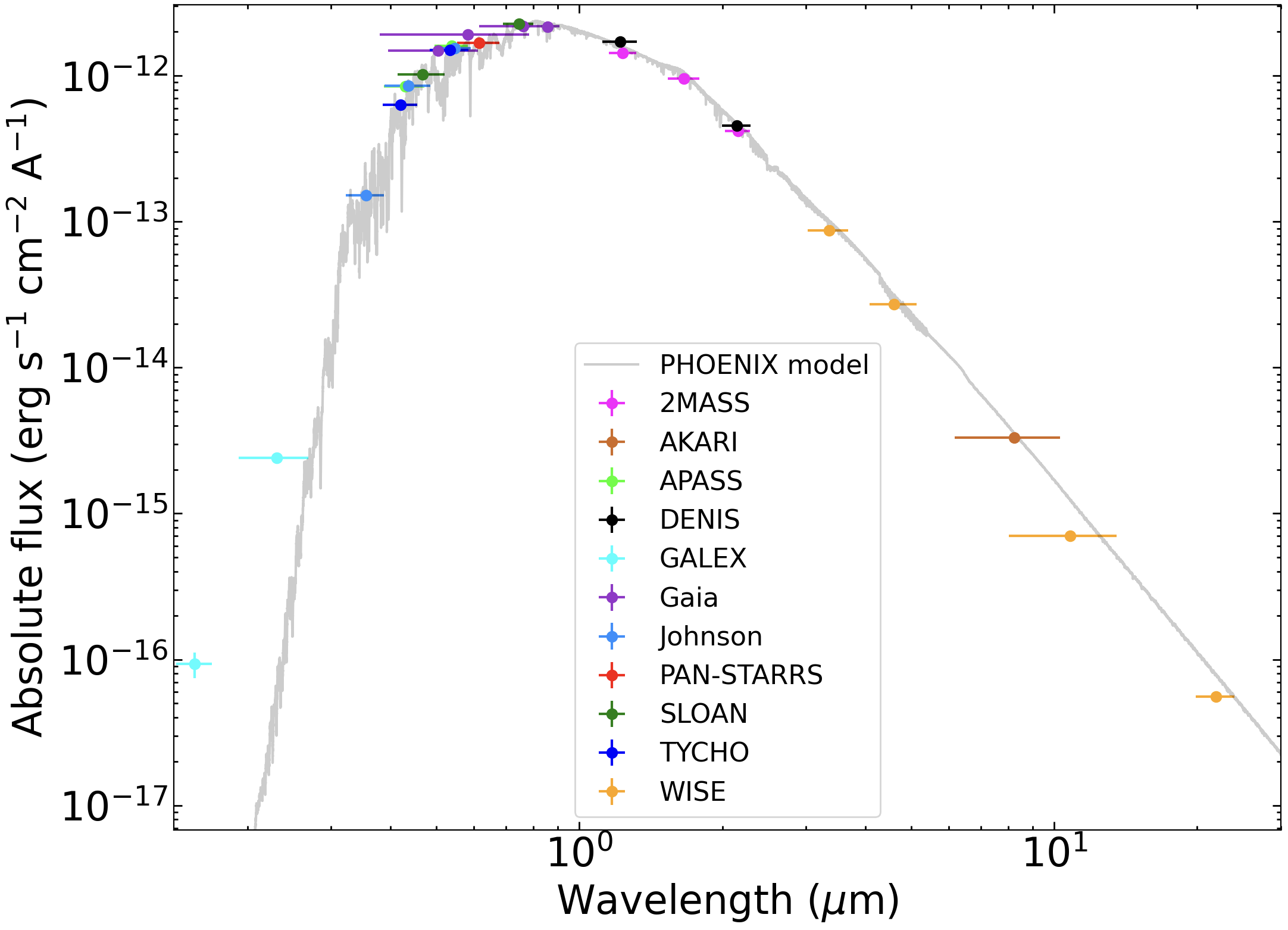}
    \includegraphics[width=\columnwidth]{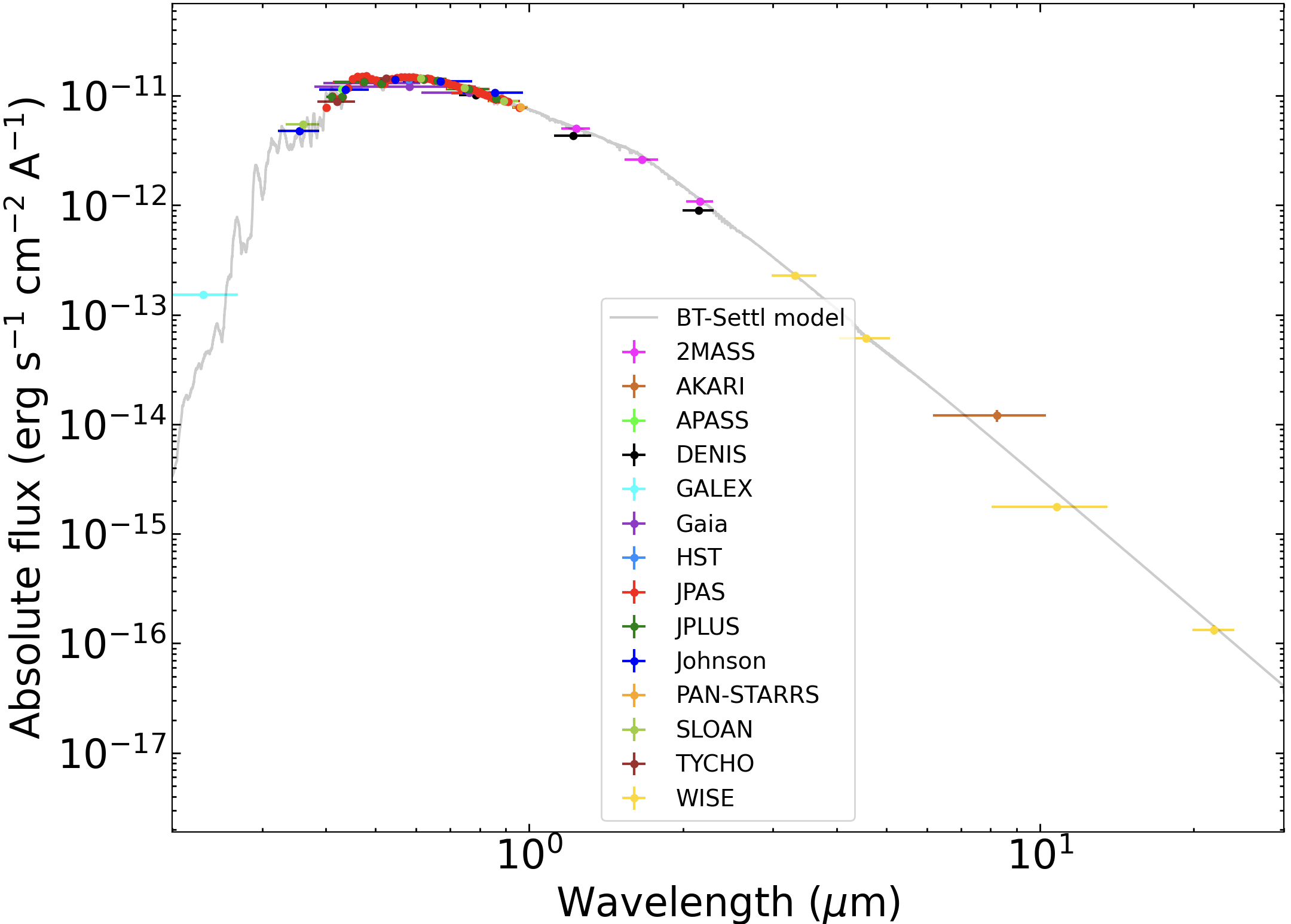}
    \includegraphics[width=\columnwidth]{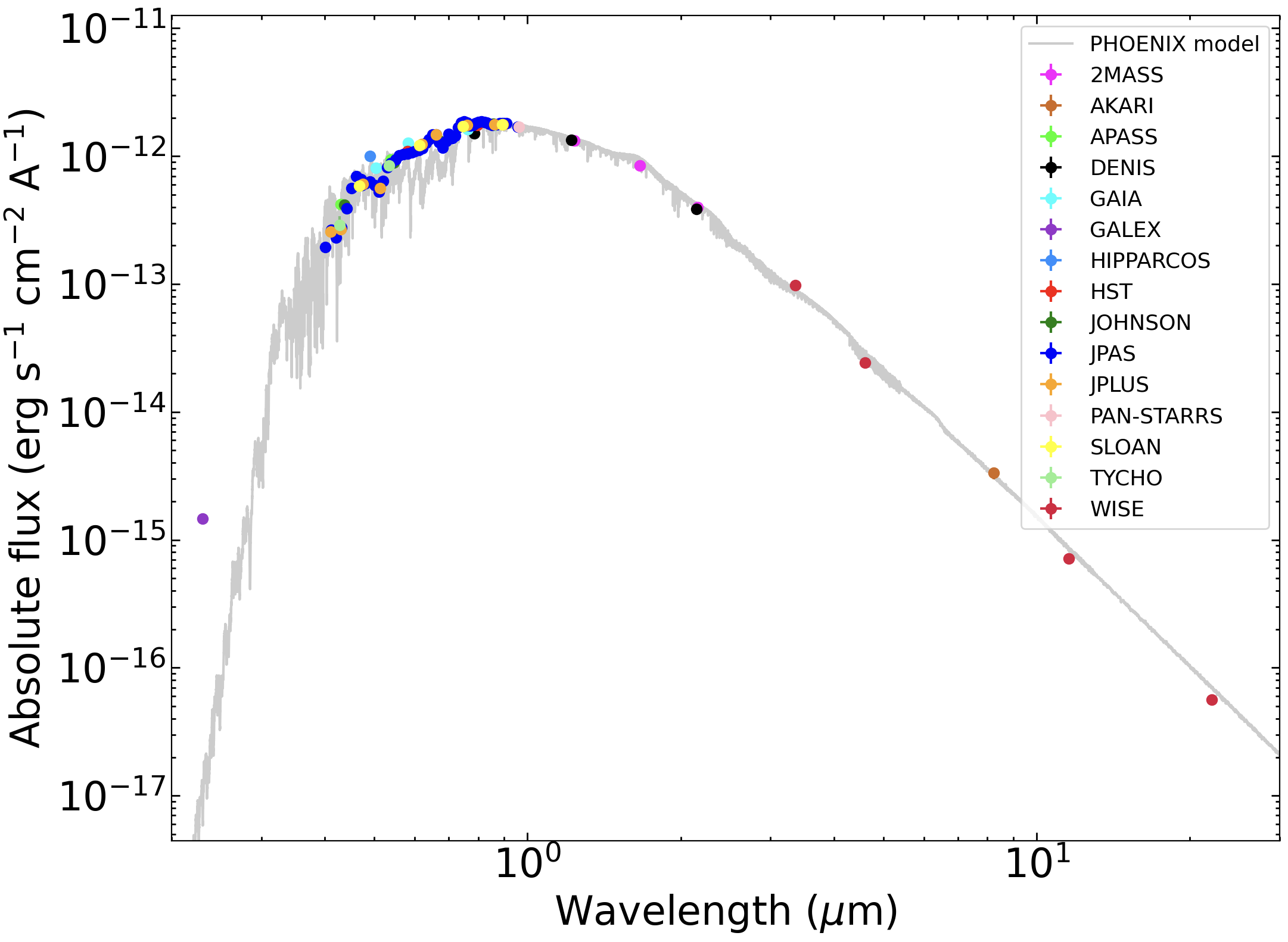}
    \caption{Top: SED for TOI-260. The grey line shows the BT-Settl model for $\mathrm{T_{eff} = 3800\, K}$, solar metallicity, and $\log{g} = 5.0\, \mathrm{dex}$. Middle: SED for TOI-286. The grey line shows the BT-Settl model for $\mathrm{T_{eff} = 5100\, K}$, solar metallicity, and $\log{g} = 4.5\, \mathrm{dex}$. Bottom: SED for TOI-134. The grey line shows the BT-Settl model for $\mathrm{T_{eff} = 3600\, K}$, solar metallicity, and $\log{g} =4.5\, \mathrm{dex}$.}
    \label{fig:sed}
\end{figure}

We also calculate the bolometric luminosities, masses, and radii. The bolometric luminosity $\mathrm{L_{bol}}$ is obtained by integrating the observed spectral energy distributions (SEDs), which are shown in Fig. \ref{fig:sed} together with the best-fit Bt-Settle models from \cite{Allard2013}. We note that the GALEX filters were removed from the SED integration, as only photospheric fluxes are relevant to the $\mathrm{L_{bol}}$ determination. The stellar radius R$_\star$ is obtained by using the Stefan-Boltzmann equation, which requires the $\mathrm{L_{bol}}$ and $\mathrm{T_{eff}}$. The stellar mass M$_\star$ is then obtained following the \cite{Schweitzer2019} mass-radius relationship for TOI-260 and TOI-134, and following the \cite{Eker2015} mass-luminosity relationship for TOI-286. The stellar density is calculated from M$_\star$ and R$_\star$. Our errors for these parameters are consistent with those expected by \cite{Tayar2022}. 

Additionally, we computed the Galactic space-velocity components U, V, and W using the \textit{Gaia} DR3 proper motion, parallax and radial velocity, following the prescription of \cite{Johnson1987}. U is in the direction of the Galactic centre, V in the direction of Galactic rotation, and W in the direction of the north Galactic pole. We note that the right-handed system is used, and that we do not subtract the solar motion. Kinematically, TOI-286 belongs to the young disk, but it does not appear to be a member of any known young moving group.

Finally, we report the median and standard deviation of the $\log R'_{\rm hk}$ activity indicator obtained from the ESPRESSO spectra for each target. From the $\log R'_{\rm hk}$, we can estimate the stellar rotation period $\mathrm{P_{rot}}$. For TOI-260 and TOI-286, we use the $\log R'_{\rm hk} - \mathrm{P_{rot}}$ relation of \cite{Suarez2015}. For the cooler M-dwarf TOI-134, we use the  $\log R'_{\rm hk} - \mathrm{P_{rot}}$ activity relation of \cite{Astudillo-Defru2017}.

\begin{table*}[htb]
\begin{center}
\caption{Stellar parameters of TOI-260, TOI-286, and TOI-134.}
\label{tab:starparams}
\centering
\begin{tabular}{lcccr}
\hline \hline
Parameter           & TOI-260 & TOI-286 & TOI-134 & Reference \\
\hline
Names               & BD-10 47 & CD-60 1348 & CD-60 8051 & Simbad \\
              & TIC 37749396 & TIC 150030205 & TIC 234994474 &\textit{TESS} \\
               & J00190556-0957530 & J06035606-6039587 & J23200751-6003545 & 2MASS \\
               & 2428162410789155328 & 5482316880495405056 & 6491962300492857472 & \textit{Gaia} DR3 \\
RA \dotfill (J2000) & $00^{\mathrm h}19^{\mathrm m}05^{\mathrm s}.56238842475$ & $06^{\mathrm h}03^{\mathrm m}56^{\mathrm s}.0652933046$ & $23^{\mathrm h}20^{\mathrm m}07^{\mathrm s}.5245143451$ & \textit{Gaia} DR3 \\
DEC \dotfill (J2000) & $-09{\degr}57{\arcmin}53{\arcsec}.468670700$ & $	-60{\degr}39{\arcmin}58{\arcsec}.752600686$ & $-60{\degr}03{\arcmin}54{\arcsec}.644648069$ & \textit{Gaia} DR3 \\
pm$^{\rm RA}$ \hfill [mas yr$^{-1}$] & 303.672 & 76.330 & -319.924 & \textit{Gaia} DR3 \\
pm$^{\rm DEC}$ \hfill [mas yr$^{-1}$] & -36.325 & -3.399 & -127.782 & \textit{Gaia} DR3 \\
$\pi$ \dotfill [mas] & 49.4757 & 16.8709 & 39.7113 & \textit{Gaia} DR3 \\
\hline
T \dotfill [mag] & 8.4991 & 9.1059 & 9.22978 & \textit{TESS} \\
B \dotfill [mag] & 11.246$^a$ & 10.80$^b$ & 12.45$^b$ & \multirow{2}{*}{$^a$ K10, $^b$ Tycho-2} \\
V \dotfill [mag] & 9.897$^a$ & 9.87$^b$ & 11.02$^b$ &           \\
J \dotfill [mag] & 7.376 & 8.357 & 7.941 & 2MASS \\
H \dotfill [mag] & 6.710 & 7.961 & 7.320 & 2MASS \\
K \dotfill [mag] & 6.549 & 7.851 & 7.082 & 2MASS \\
\hline
$\mathrm{T_{eff}}$ \dotfill [K] & $4026 \pm 14$ (81) & $5152 \pm 12$ (103) & $3842 \pm 32$ (77) & this work  \\
Spectral type  \dotfill & K8V & K3V & M0.5V & PM13 \\
Fe/H \dotfill [dex] & $-0.47 \pm 0.03$ & $-0.05 \pm 0.02$ & $-0.03 \pm 0.03$ & this work \\
$\log{g}$ \dotfill [dex] & $4.45 \pm 0.05$ & $4.59 \pm 0.02$ & $4.84 \pm 0.08$ & this work \\
$\mathrm{v_{broad}}$$^c$ \dotfill [$\mathrm{km\, s^{-1}}$] & $3.30 \pm 0.09$ & $2.58 \pm 0.03$ & $3.22 \pm 0.08$ & this work \\
R$_\star$ \dotfill [R$_\odot$] & $0.607 \pm 0.014$ & $0.780 \pm 0.036$ & $0.604 \pm 0.037$ & this work \\
M$_\star$ \dotfill [M$_\odot$] & $0.616 \pm 0.032$ & $0.832 \pm 0.049$ & $0.614 \pm 0.055$ & this work \\
L$_{bol,\star}$ \dotfill [L$_\odot$] & $0.0873 \pm 0.0027$ & $0.3858 \pm 0.0050$ & $0.0723 \pm 0.0018$ & this work \\
$\log R'_{\rm hk}$ \dotfill & $-4.979 \pm 0.069$ & $-5.14 \pm 0.21$ & $-4.768 \pm 0.032$ & this work \\
$\mathrm{P_{rot}}$$^d$ \dotfill [d] & $31 \pm 6$ & $36 \pm 15$ & $29 \pm 2$ & this work \\
\hline
U \dotfill [$\mathrm{km\, s^{-1}}$] &$ 17.863 \pm 0.004$ & $2.354 \pm 0.003$ & $49.39 \pm 0.12$ & this work \\
V \dotfill [$\mathrm{km\, s^{-1}}$] &$ -25.265 \pm 0.052$ & $-25.581 \pm 0.221$ & $-10.45 \pm 0.07$ & this work \\
W \dotfill [$\mathrm{km\, s^{-1}}$] & $2.015 \pm 0.184$ & $10.150 \pm 0.097$ & $-3.12 \pm 0.16$ & this work \\
\end{tabular}
\end{center}
    Simbad: Simbad astronomical database \citep{Wenger2000}; \textit{TESS}: \textit{TESS} Input Catalog \citep{Stassun2019}; 2MASS: Two-micron All Sky Survey \citep{2MASS}; \textit{Gaia} DR3: \textit{\textit{Gaia}} Data Release 3 \citep{GAIA2016, GaiaDR3}; K10: \cite{Koen2010}; Tycho-2: the Tycho-2 Catalogue \citep{Tycho-2}; PM13: using the tables of \cite{Pecaut2013}. Values in parentheses correspond to the fundamental uncertainty floors following \cite{Tayar2022}.\\
    $^c$: this parameter accounts for both the macroturbulence and the projected rotational velocity.\\
    $^d$: determined from $\log R'_{\rm hk} - \mathrm{P_{rot}}$ relations.
\end{table*}

\subsection{Data modelling}

We modelled the photometric and RV data simultaneously with the \texttt{juliet}\footnote{Available at \url{https://github.com/nespinoza/juliet}} software \citep{Espinoza2019juliet}. \texttt{juliet} enables the joint fitting of transit and RV data, through the \texttt{batman} package \citep{Kreidberg2015}) and the \texttt{radvel} package \citep{Fulton2018} respectively. It also incorporates Gaussian Process Regression (GPR) via the \texttt{celerite} package \citep{Foreman-Mackey2017}. Importance nested sampling, using the \texttt{dynesty} package \citep{Speagle2020}, is employed to explore the parameter space. By default, \texttt{juliet} adopts random walk sampling with 500 live points.

For each candidate transiting planet, we tested both circular and free-eccentricity models, fitting all the data simultaneously. Joint transit and RV fits in \texttt{juliet} require the stellar density $\rho$, and the following parameters for each planet $\mathrm{pi}$: period $\mathrm{P_{pi}}$, time of transit $\mathrm{t_{0,pi}}$, planet-to-star radius ratio $\mathrm{p_{pi}}$, impact parameter $\mathrm{b_{pi}}$, and either eccentricity $\mathrm{e_{pi}}$ and angle of periastron $\mathrm{\omega_{pi}}$ or a derived parametrization such as $\mathrm{\sqrt{e_{pi}} \sin \omega_{pi}, \sqrt{e_{pi}} \cos \omega_{pi}}$. There are also intrumental parameters: for each RV instrument, the systemic radial velocity $\mathrm{\mu_{instrument}}$ and the jitter $\mathrm{\sigma_{w,instrument}}$; for each photometric instrument, the dilution factor $\mathrm{m_{dilution,instrument}}$, the flux offset $\mathrm{m_{flux,instrument}}$, the jitter $\mathrm{\sigma_{w,instrument}}$, and the limb-darkening parameters $\mathrm{q_{1,instrument}}$ and $\mathrm{q_{2,instrument}}$ for the quadratic law parametrization of \cite{Kipping2013}. 
We also used GPs to model the effect of stellar activity on the radial velocities and instrumental effects on the photometry. For the RVs, we use the quasi-periodic kernel (called exp-sine-squared kernel
in \texttt{juliet}) of \cite{Haywood2014} . For the photometry, we employ the (approximate) Matern 3/2 kernel, as implemented in \texttt{celerite}. 

From the \texttt{juliet} fitted parameters and the stellar parameters we derived for each planet the semimajor axis $\mathrm{a_{pi}}$, the mass $\mathrm{M_{pi}}$, the radius $\mathrm{R_{pi}}$, the density $\mathrm{\rho_{pi}}$, and the equilibrium temperature $\mathrm{T_{eq,pi}}$. For this last we assume a Bond albedo of 0 and instant heat redistribution. The errors on the derived parameters are computed through error propagation, where for $\mathrm{T_{eff}}$ we take the error floor from \cite{Tayar2022}. The details of the fits for each star are discussed in the following subsections. 

\subsubsection{TOI-260}

\begin{figure}[bth]
    \centering
    \includegraphics[width=.5\textwidth]{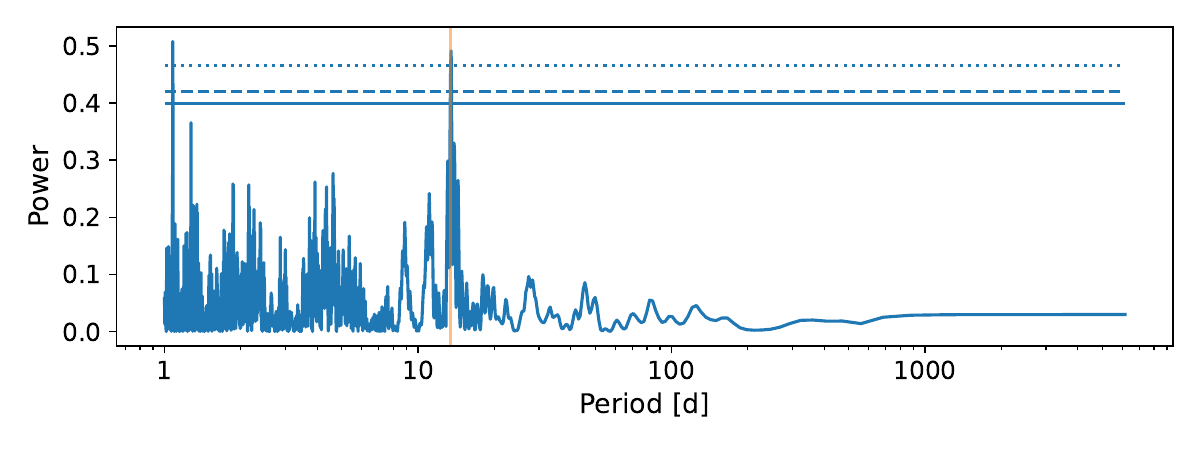}
    \caption{Periodogram of the TOI-260 ESPRESSO RVs with a GP component removed. The semitransparent orange line indicates the period of the transiting candidate, at which there is a clear signal. }
    \label{fig:toi-260_GPres_periodogram}
\end{figure}

We use ESPRESSO and HIRES RVs and TESS photometry to characterize the planetary candidate TOI-260.01. The ESPRESSO RVs do not immediately show evidence for a planet at the \textit{TESS} candidate period of $\mathrm{P \approx 13.48 \, d}$, as can be seen in the periodogram in Fig. \ref{fig:toi-260_periodograms}. However, most of the activity indicators show a strong signal at $\approx$ 37 d, which is also present in the RVs and which is close to the stellar rotation period, suggesting the stellar activity-induced variability may be overpowering the planet signal. To test this, we fitted a model consisting of a Keplerian signal and GP to the ESPRESSO RVs alone. In order to detrend the radial velocities from the stellar activity, we first fitted a GP to the FWHM activity indicator time series, which has previously shown to be a good tracer of stellar activity in ESPRESSO data \citep{LilloBox2020,Lavie2023,Castro2023}. We used broad log-uniform priors of $\mathcal{J}(0.001,1000)$ for $\mathrm{\sigma_{GP,FWHM}}$, $\mathrm{\alpha_{GP,FWHM}}$, and $\mathrm{\Gamma_{GP,FWHM}}$, and a normal prior of $\mathcal{N}(31,6)$ for the rotation period GP hyperparameter $\mathrm{P_{rot,GP, FWHM}}$, taking the value and uncertainty of the rotation period (see Sec. \ref{sect:stellar-parameters}) as the mean and standard deviation for this last. Subsequently, we used the resulting parameters as priors on the radial velocity GP. The periodogram of the RVs with the GP component removed is shown in Fig. \ref{fig:toi-260_GPres_periodogram}; a clear signal is evident at the planetary period. 

For the rest of the analysis, we jointly fitted the ESPRESSO RVs, the HIRES RVs, and the \textit{TESS} photometry. We tested two models: a circular transiting planet, and a transiting planet with free eccentricity. To detrend the RVs from stellar activity, we fit GPs to the HIRES data, and to the joint ESPRESSO data. We chose to use separate GPs both because these are different instruments, and because the $\mathrm{\approx 2000 \, d}$ gap between the observations means that we are likely tracing different parts of the star's activity cycle. However, we set the rotation period hyperparameter to be shared between the two GPs. We also tested a fit with a global GP on the RVs, with priors constrained by the $\mathrm{S_{MW}}$ indicators, but obtained no improvement in the Bayesian log-evidence and a much larger resulting jitter for ESPRESSO. We constrained the priors for these GPs using activity indicators; for the ESPRESSO data we used the FWHM, as previously described. For HIRES we have fewer indicators available, as it is an iodine cell spectrograph and therefore the RVs are not calculated via the CCF method. We selected the $\mathrm{S_{MW}}$ activity indicator and used the same procedure. The GP fits to these activity indicators are shown in Fig. \ref{fig:act-ind-gps}. 

For the \textit{TESS} photometry we employed a two-step process, where we first masked the transits and fitted GPs to the out-of-transit data only on a sector-by-sector basis, with broad log-uniform or Jeffreys priors of $\mathcal{J}(1\times10^{-6}, 1\times10^6)$ for $\mathrm{\sigma_{GP,sector}}$ and $\mathcal{J}(0.2, 1\times10^{3})$ for $\mathrm{\rho_{GP,sector}}$, where $\mathcal{J}(a,b)$ indicates a log-uniform distribution between $a$ and $b$. We then used the resulting parameters as priors for GPs on the joint full fit of the RVs and all photometric data. 

The ground-based photometry, taken after the \textit{TESS} sector 3 observations with preliminary ephemerides, is unfortunately not in transit according to the refined ephemerides obtained from the modelling of the full \textit{TESS} dataset and radial velocities (see Fig. \ref{fig:toi-260_ground_phot}). 
In any case, the observations were optimised for vetting faint nearby stars, so the scatter of the lightcurves for TOI-260 itself is too high to detect the transit. We therefore do not include the ground-based photometry in the modelling.

The priors and posteriors for the final models are given in Tables \ref{tab:TOI-260-circ} and \ref{tab:TOI-260-efree}, for the circular fit and the free-eccentricity fit respectively. The cornerplots of the posterior distributions are shown in Figs. \ref{fig:toi-260_cornerplot_circ_orbital}, \ref{fig:toi-260_cornerplot_circ_rv}, and \ref{fig:toi-260_cornerplot_circ_transit} for the circular fit; and in Figs. \ref{fig:toi-260_cornerplot_e07_orbital}, \ref{fig:toi-260_cornerplot_e07_rv}, and \ref{fig:toi-260_cornerplot_e07_transit} for the free-eccentricity fit. In the prior distributions, $\mathcal{U}(a,b)$ indicates a uniform distribution between $a$ and $b$; $\mathcal{N}(a, b)$ a normal distribution with mean $a$ and standard deviation $b$; $\mathcal{TN}(a, b,c,d)$ a truncated normal distribution with mean $a$, standard deviation $b$, and support in the $[c,d]$ interval; $\mathcal{J}(a,b)$ a Jeffreys or log-uniform distribution between $a$ and $b$. We used the values reported on ExoFOP as priors for the period $\mathrm{P_{p1}}$, time of transit $\mathrm{t_{0,p1}}$, and planet-to-star radius ratio $\mathrm{p_{p1}}$. 

Since we are using the PDC-SAP \textit{TESS} light curves, we can expect that most of the variability in the \textit{TESS} fluxes comes from instrumental systematics, which may vary between sectors. We therefore treat each \textit{TESS} sector as an individual instrument with separate priors for the flux offsets $m_\mathrm{{flux,sector}}$ and jitters $\sigma_{\mathrm{sector}}$. Likewise, the GP priors $\mathrm{\sigma_{GP,sector}}$ and $\mathrm{\rho_{GP,sector}}$ are set individually, with priors given by the previous separate fits on the out-of-transit data. \texttt{juliet} also requires priors for the dilution factor, which we have fixed to 1 for all \textit{TESS} sectors.The limb-darkening parameters q$_\mathrm{{1,TESS}}$ and q$_\mathrm{{2,TESS}}$, however, are common to all \textit{TESS} sectors. 

For the radial velocity offsets $\mathrm{\mu_{instrument}}$ we used broad uniform priors spanning the range of the obtained RVs for each instrument. For the RV GPs, we used the posteriors of the GP fits to the FWHM and $S_{MW}$ activity indicator as priors for ESPRESSO and HIRES respectively (in the case of $\mathrm{\sigma_{GP,rv}}$, scaled by a normalisation factor corresponding to the RV variation over the FWHM or $S_{MW}$ variation). For the free-eccentricity fit, we used the ($\mathrm{\sqrt{e} \sin \omega, \sqrt{e} \cos \omega}$) parametrization, and fixed an upper limit on the eccentricity of 0.7, to avoid convergence problems with \texttt{batman} at high eccentricities.  

The Bayesian log-evidence comparison of the two models strongly favours the eccentric model (henceforth model e) over the circular model (henceforth model c), with $\Delta \log Z_{c-e} \approx 6$. This model has a remarkably high eccentricity of $\mathrm{e_{p1} = \twosixtyeseveneccpone}$. However, it appears to be at least partly driven by a single ESPRESSO19 data point in an otherwise unsampled by ESPRESSO gap in phase coverage in the (-0.1,0) phase range, that is also poorly covered by HIRES. Additionally, the cornerplot (Fig. \ref{fig:toi-260_cornerplot_e07_orbital}) suggests the posterior for $\sqrt{e} \cos \omega$ parameter is at the edge of the prior space. Therefore, we ran other models for comparison: two models using the ($\mathrm{\sqrt{e} \sin \omega, \sqrt{e} \cos \omega}$) parametrization, one with a limiting eccentricity of 0.9 (model e1), and one with a limiting eccentricity of 0.5 (model e2), to explore the impact of this limit; one model using the ($\mathrm{e, \omega}$) parametrization (model e3), with a log-uniform prior on $\mathrm{e}$ between 0.001 and 0.9; and one model using the initial ($\mathrm{\sqrt{e} \sin \omega, \sqrt{e} \cos \omega}$) parametrization and 0.7 eccentricity limit, but with the ESPRESSO19 data point that appears to be driving the eccentricity removed (model e4). Model e1 returns an even higher eccentricity of $\mathrm{e_{p1} = 0.80 \pm 0.05}$, although with $\Delta \log Z_{e-e1} \approx 3$ in favour of model e, and a better-sampled posterior for $\sqrt{e} \cos \omega$. Model e2 returns a more poorly constrained eccentricity of $\mathrm{e_{p1} = 0.33^{+0.11}_{-0.14}}$, and is not favoured by the log-evidence comparison with the initial model ($\Delta \log Z_{e-e2} \approx 5$), though the posterior distributions do not appear visually poorly sampled. Meanwhile, model e3 returns a similarly high and more poorly constrained eccentricity of $\mathrm{e_{p1} = 0.75^{+0.06}_{-0.28}}$, with $\Delta \log Z_{e-e3} \approx 4$ in favour of the initial free-eccentricity model e. Finally, model e4 returns a slightly lower (though still compatible within error bars) and less constrained eccentricity to model e of $\mathrm{e_{p1} = 0.51^{+0.16}_{-0.29}}$. A $\Delta \log Z$ comparison is not appropriate here since the input data is different.

\begin{figure*}[thb!]
    \centering
    \includegraphics[width=\textwidth]{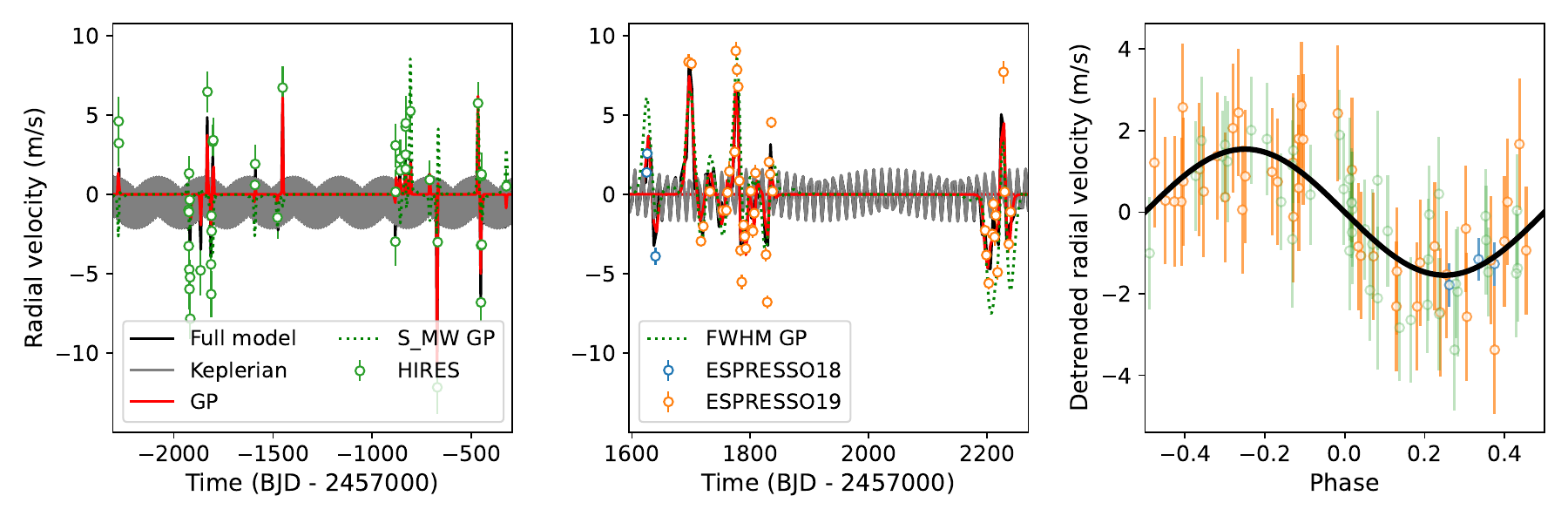}
    \caption{Left and centre: Radial velocities (HIRES: green dots, ESPRESSO18: blue dots, ESPRESSO19: orange dots), model components (GP: red, Keplerian: grey), and median circular model (black) for TOI-260. We have separated the RVs into two plots for ease of viewing, as there is a $\mathrm{\approx 2000 \, d}$ gap between the HIRES and ESPRESSO RVs. The error bars show the RV errors and jitter added in quadrature. The instrumental systemic velocity has been subtracted. The GP fits to the $S_{MW}$ and FWHM activity indicators are also shown (dotted green lines) for comparison. Right: phase-folded radial velocities plus median Keplerian, following the same colour scheme. }
    \label{fig:toi-260_RVs_circ}
\end{figure*}

\begin{figure}[thb!]
    \centering
    \includegraphics[width=.5\textwidth]{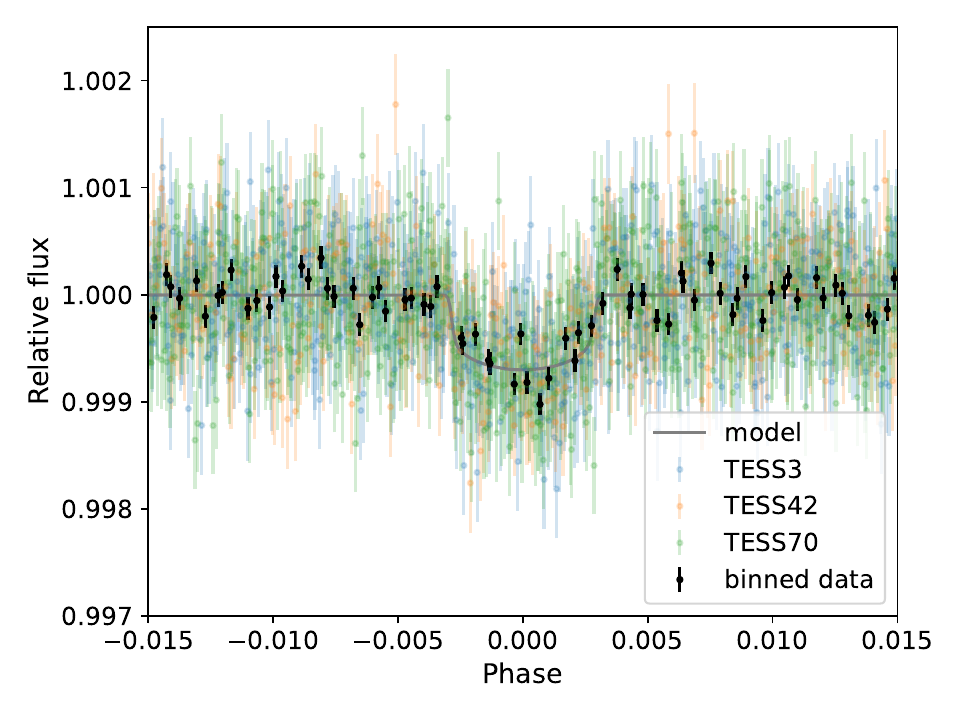}
    \caption{Stacked phase-folded \textit{TESS} data for TOI-260 with the median circular model (grey line). The points show the data from sectors 3 (blue), 42 (orange) and 70 (green), and binned data (black points). }
    \label{fig:toi-260_TESS_stacked_circ}
\end{figure}

We therefore conclude that the current ESPRESSO data do not allow us to properly constrain the eccentricity of the orbit, and thus prefer the simpler circular model. We also note that one mechanism for driving such a high eccentricity would be through interactions with other planets, but we see no evidence for outer companions in the RVs, and the astrometric data places limits on the existence of massive outer companions (see Sec. \ref{sec:astrometry}). We show the RVs and median circular model in Fig. \ref{fig:toi-260_RVs_circ}, and the phase-folded stacked \textit{TESS} data and model in Fig. \ref{fig:toi-260_TESS_stacked_circ}. The full \textit{TESS} light curves and corresponding median models are shown in Fig. \ref{fig:toi-260_TESS_sectors_circ}. With the circular model, we find an RV semi-amplitude of $\mathrm{K_{p1} = \twosixtycgpsepKpone \, m \, s^{-1}}$, corresponding to a detection at $3\sigma$, and leading to a planetary mass of $\mathrm{M_{p1} = 4.23 \pm 1.60 \, M_\oplus}$.

\begin{table}[pht] 
\begin{center} 
\caption{Prior and posterior planetary parameter distributions obtained with \texttt{juliet} for TOI-260, for a fit with eccentricity fixed to 0. \textit{Top}: Fitted parameters. \textit{Bottom}: derived orbital parameters and physical parameters.} 
\label{tab:TOI-260-circ} 
\centering 
\resizebox{\columnwidth}{!}{%
\begin{tabular}{lll} 
\hline  \hline 
Parameter & Prior & Posterior \\ 
\hline 
$\mathrm{P_{p1}}$ \dotfill [d] & $\mathcal{N}(13.5,0.1)$ & \twosixtycgpsepPpone \\
$\mathrm{t_{0,p1}}$ \dotfill [BJD] & $\mathcal{N}(2458392.3,0.1)$ & \twosixtycgpseptzeropone \\
$\mathrm{p_{p1}}$ \dotfill & $\mathcal{N}(0.02,0.1)$ & \twosixtycgpsepppone \\
$\mathrm{b_{p1}}$ \dotfill & $\mathcal{U}(0,1)$ & \twosixtycgpsepbpone \\
$\mathrm{e_{p1}}$ \dotfill & $\mathrm{fixed}$ & \twosixtycgpsepeccpone \\
$\mathrm{\omega_{p1}}$ \dotfill [$\degr$]& $\mathrm{fixed}$ & \twosixtycgpsepomegapone \\
$\mathrm{K_{p1}}$ \dotfill [$\mathrm{m \, s^{-1}}$] & $\mathcal{U}(0,10)$ [$\mathrm{m \, s^{-1}}$] & \twosixtycgpsepKpone \\
$\mathrm{\rho}$ \dotfill [$\mathrm{kg \, m^{-3}}$]& $\mathcal{J}(100,10000)$ & \twosixtycgpseprho \\
$\mathrm{q_{1,TESS}}$ \dotfill & $\mathcal{U}(0,1)$ & \twosixtycgpsepqoneTESS \\
$\mathrm{q_{2,TESS}}$ \dotfill & $\mathcal{U}(0,1)$ & \twosixtycgpsepqtwoTESS \\
$\mathrm{\mu_{ESPRESSO18}}$ \dotfill [$\mathrm{m \, s^{-1}}$] & $\mathcal{U}(-11000,-10900)$ & \twosixtycgpsepmuESPRESSOeighteen \\
$\mathrm{\sigma_{w,ESPRESSO18}}$ \dotfill [$\mathrm{m \, s^{-1}}$] & $\mathcal{J}(0.001,10)$ & \twosixtycgpsepsigmawESPRESSOeighteen \\
$\mathrm{\mu_{ESPRESSO19}}$ \dotfill [$\mathrm{m \, s^{-1}}$] & $\mathcal{U}(-11000,-10900)$ & \twosixtycgpsepmuESPRESSOnineteen \\
$\mathrm{\sigma_{w,ESPRESSO19}}$ \dotfill [$\mathrm{m \, s^{-1}}$] & $\mathcal{J}(0.001,10)$ & \twosixtycgpsepsigmawESPRESSOnineteen \\
$\mathrm{\mu_{HIRES}}$ \dotfill [$\mathrm{m \, s^{-1}}$] & $\mathcal{U}(-20,20)$ & \twosixtycgpsepmuHIRES \\
$\mathrm{\sigma_{w,HIRES}}$ \dotfill [$\mathrm{m \, s^{-1}}$] & $\mathcal{J}(0.001,100)$ & \twosixtycgpsepsigmawHIRES \\
$\mathrm{\sigma_{GP,ESPRESSO18,ESPRESSO19}}$ \dotfill & $\mathcal{TN}(3.8,6.4,0,1000)$ & \twosixtycgpsepGPsigmaESPRESSOeighteenESPRESSOnineteen \\
$\mathrm{\alpha_{GP,ESPRESSO18,ESPRESSO19}}$ \dotfill & $\mathcal{TN}(0.018,0.017,0,1000)$ & \twosixtycgpsepGPalphaESPRESSOeighteenESPRESSOnineteen \\
$\mathrm{\Gamma_{GP,ESPRESSO18,ESPRESSO19}}$ \dotfill & $\mathcal{TN}(0.1,1.5,0,1000)$ & \twosixtycgpsepGPGammaESPRESSOeighteenESPRESSOnineteen \\
$\mathrm{P_{rot,GP,ESPRESSO18,ESPRESSO19,HIRES}}$ \dotfill & $\mathcal{N}(31,6)$ & \twosixtycgpsepGPProtESPRESSOeighteenESPRESSOnineteenHIRES \\
$\mathrm{\sigma_{GP,HIRES}}$ \dotfill & $\mathcal{TN}(5.381,0.020,0,1000)$ & \twosixtycgpsepGPsigmaHIRES \\
$\mathrm{\alpha_{GP,HIRES}}$ \dotfill & $\mathcal{TN}(0.036,0.066,0,1000)$ & \twosixtycgpsepGPalphaHIRES \\
$\mathrm{\Gamma_{GP,HIRES}}$ \dotfill & $\mathcal{TN}(0.4,3.1,0,1000)$ & \twosixtycgpsepGPGammaHIRES \\
$\mathrm{m_{dilution,TESS3}}$ \dotfill & $\mathrm{fixed}$ & \twosixtycgpsepmdilutionTESSthree \\
$\mathrm{m_{flux,TESS3}}$ \dotfill & $\mathcal{N}(0,0.1)$ & \twosixtycgpsepmfluxTESSthree \\
$\mathrm{\sigma_{w,TESS3}}$ \dotfill & $\mathcal{J}(0.1,1000)$ & \twosixtycgpsepsigmawTESSthree \\
$\mathrm{\sigma_{GP,TESS3}}$ \dotfill & $\mathcal{N}(25.5 \times 10^{-5},2.5 \times 10^{-5})$ & \twosixtycgpsepGPsigmaTESSthree \\
$\mathrm{\rho_{GP,TESS3}}$ \dotfill & $\mathcal{N}(0.2031,0.0052)$ & \twosixtycgpsepGPrhoTESSthree \\
$\mathrm{m_{dilution,TESS42}}$ \dotfill & $\mathrm{fixed}$ & \twosixtycgpsepmdilutionTESSfortytwo \\
$\mathrm{m_{flux,TESS42}}$ \dotfill & $\mathcal{N}(0,0.1)$ & \twosixtycgpsepmfluxTESSfortytwo \\
$\mathrm{\sigma_{w,TESS42}}$ \dotfill & $\mathcal{J}(0.1,1000)$ & \twosixtycgpsepsigmawTESSfortytwo \\
$\mathrm{\sigma_{GP,TESS42}}$ \dotfill & $\mathcal{N}(9.0 \times 10^{-5},1.2 \times 10^{-5})$ & \twosixtycgpsepGPsigmaTESSfortytwo \\
$\mathrm{\rho_{GP,TESS42}}$ \dotfill & $\mathcal{N}(0.23,0.04)$ & \twosixtycgpsepGPrhoTESSfortytwo \\
$\mathrm{m_{dilution,TESS70}}$ \dotfill & $\mathrm{fixed}$ & \twosixtycgpsepmdilutionTESSseventy \\
$\mathrm{m_{flux,TESS70}}$ \dotfill & $\mathcal{N}(0,0.1)$ & \twosixtycgpsepmfluxTESSseventy \\
$\mathrm{\sigma_{w,TESS70}}$ \dotfill & $\mathcal{J}(0.1,1000)$ & \twosixtycgpsepsigmawTESSseventy \\
$\mathrm{\sigma_{GP,TESS70}}$ \dotfill & $\mathcal{N}(24.6 \times 10^{-5},5.1 \times 10^{-5})$ & \twosixtycgpsepGPsigmaTESSseventy \\
$\mathrm{\rho_{GP,TESS70}}$ \dotfill & $\mathcal{N}(0.81,0.21)$ & \twosixtycgpsepGPrhoTESSseventy \\
\hline 
$\mathrm{a_{p1}}$ \dotfill [au] & $-$ & $0.0943 \pm 0.0016$ \\
$\mathrm{i_{p1}}$ \dotfill [$\degr$] & $-$ & $88.84 \pm 0.17$ \\
$\mathrm{T_{14,p1}}$ \dotfill [h] & $-$ & $2.37 \pm 0.27$ \\
$\mathrm{M_{p1}}$ \dotfill [$\mathrm{M_{e}}$] & $-$ & $4.23 \pm 1.60$ \\
$\mathrm{R_{p1}}$ \dotfill [$\mathrm{R_{e}}$] & $-$ & $1.71 \pm 0.08$ \\
$\mathrm{\rho_{p1}}$ \dotfill [$\mathrm{g \, cm{-3}}$] & $-$ & $4.66 \pm 1.89$ \\
$\mathrm{T_{eq,p1}}$ \dotfill [K] & $-$ & $493.0 \pm 12.0$ \\
$\mathrm{\log Z}$ \dotfill  & $-$ & $240842.8 \pm 0.5$ \\
\hline
\end{tabular} 
} 
\end{center} 
\end{table} 

\begin{table}[pht] 
\begin{center} 
\caption{Prior and posterior planetary parameter distributions obtained with \texttt{juliet} for TOI-260, for the free-eccentricity model using ($\mathrm{\sqrt{e} \sin \omega, \sqrt{e} \cos \omega}$) parametrization, with a limiting eccentricity of 0.7. \textit{Top}: Fitted parameters. \textit{Bottom}: derived orbital parameters and physical parameters.} 
\label{tab:TOI-260-efree} 
\centering 
\resizebox{\columnwidth}{!}{%
\begin{tabular}{lll} 
\hline  \hline 
Parameter & Prior & Posterior \\ 
\hline 
$\mathrm{P_{p1}}$ \dotfill [d] & $\mathcal{N}(13.5,0.1)$ & \twosixtyesevenPpone \\
$\mathrm{t_{0,p1}}$ \dotfill [BJD] & $\mathcal{N}(2458392.3,0.1)$ & \twosixtyeseventzeropone \\
$\mathrm{p_{p1}}$ \dotfill & $\mathcal{N}(0.02,0.1)$ & \twosixtyesevenppone \\
$\mathrm{b_{p1}}$ \dotfill & $\mathcal{U}(0,1)$ & \twosixtyesevenbpone \\
$\mathrm{\sqrt{e}\sin\omega_{p1}}$ \dotfill & $\mathcal{U}(0,1)$ & \twosixtyesevensesinomegapone \\
$\mathrm{\sqrt{e}\cos\omega_{p1}}$ \dotfill & $\mathcal{U}(0,1)$ & \twosixtyesevensecosomegapone \\
$\mathrm{K_{p1}}$ \dotfill [$\mathrm{m \, s^{-1}}$] & $\mathcal{U}(0,10)$ [$\mathrm{m \, s^{-1}}$] & \twosixtyesevenKpone \\
$\mathrm{\rho_{}}$ \dotfill [$\mathrm{kg \, m^{-3}}$]& $\mathcal{J}(100,10000)$ & \twosixtyesevenrho \\
$\mathrm{q_{1,TESS}}$ \dotfill & $\mathcal{U}(0,1)$ & \twosixtyesevenqoneTESS \\
$\mathrm{q_{2,TESS}}$ \dotfill & $\mathcal{U}(0,1)$ & \twosixtyesevenqtwoTESS \\
$\mathrm{\mu_{ESPRESSO18}}$ \dotfill [$\mathrm{m \, s^{-1}}$] & $\mathcal{U}(-11000,-10900)$ & \twosixtyesevenmuESPRESSOeighteen \\
$\mathrm{\sigma_{w,ESPRESSO18}}$ \dotfill [$\mathrm{m \, s^{-1}}$] & $\mathcal{J}(0.001,10)$ & \twosixtyesevensigmawESPRESSOeighteen \\
$\mathrm{\mu_{ESPRESSO19}}$ \dotfill [$\mathrm{m \, s^{-1}}$] & $\mathcal{U}(-11000,-10900)$ & \twosixtyesevenmuESPRESSOnineteen \\
$\mathrm{\sigma_{w,ESPRESSO19}}$ \dotfill [$\mathrm{m \, s^{-1}}$] & $\mathcal{J}(0.001,10)$ & \twosixtyesevensigmawESPRESSOnineteen \\
$\mathrm{\mu_{HIRES}}$ \dotfill [$\mathrm{m \, s^{-1}}$] & $\mathcal{U}(-20,20)$ & \twosixtyesevenmuHIRES \\
$\mathrm{\sigma_{w,HIRES}}$ \dotfill [$\mathrm{m \, s^{-1}}$] & $\mathcal{J}(0.001,100)$ & \twosixtyesevensigmawHIRES \\
$\mathrm{\sigma_{GP,ESPRESSO18,ESPRESSO19}}$ \dotfill & $\mathcal{TN}(3.80,0.81,0,1000)$ & \twosixtyesevenGPsigmaESPRESSOeighteenESPRESSOnineteen \\
$\mathrm{\alpha_{GP,ESPRESSO18,ESPRESSO19}}$ \dotfill & $\mathcal{TN}(0.018,0.017,0,1000)$ & \twosixtyesevenGPalphaESPRESSOeighteenESPRESSOnineteen \\
$\mathrm{\Gamma_{GP,ESPRESSO18,ESPRESSO19}}$ \dotfill & $\mathcal{TN}(0.1,1.5,0,1000)$ & \twosixtyesevenGPGammaESPRESSOeighteenESPRESSOnineteen \\
$\mathrm{P_{rot,GP,ESPRESSO18,ESPRESSO19,HIRES}}$ \dotfill & $\mathcal{N}(31,6)$ & \twosixtyesevenGPProtESPRESSOeighteenESPRESSOnineteenHIRES \\
$\mathrm{\sigma_{GP,HIRES}}$ \dotfill & $\mathcal{TN}(5.4,1.1,0,1000)$ & \twosixtyesevenGPsigmaHIRES \\
$\mathrm{\alpha_{GP,HIRES}}$ \dotfill & $\mathcal{TN}(0.036,0.066,0,1000)$ & \twosixtyesevenGPalphaHIRES \\
$\mathrm{\Gamma_{GP,HIRES}}$ \dotfill & $\mathcal{TN}(0.4,3.1,0,1000)$ & \twosixtyesevenGPGammaHIRES \\
$\mathrm{m_{dilution,TESS3}}$ \dotfill & $\mathrm{fixed}$ & \twosixtyesevenmdilutionTESSthree \\
$\mathrm{m_{flux,TESS3}}$ \dotfill & $\mathcal{N}(0,0.1)$ & \twosixtyesevenmfluxTESSthree \\
$\mathrm{\sigma_{w,TESS3}}$ \dotfill & $\mathcal{J}(0.1,1000)$ & \twosixtyesevensigmawTESSthree \\
$\mathrm{\sigma_{GP,TESS3}}$ \dotfill & $\mathcal{N}(25.5\times 10^{-5},2.5\times 10^{-5})$ & \twosixtyesevenGPsigmaTESSthree \\
$\mathrm{\rho_{GP,TESS3}}$ \dotfill & $\mathcal{N}(0.2031,0.0052)$ & \twosixtyesevenGPrhoTESSthree \\
$\mathrm{m_{dilution,TESS42}}$ \dotfill & $\mathrm{fixed}$ & \twosixtyesevenmdilutionTESSfortytwo \\
$\mathrm{m_{flux,TESS42}}$ \dotfill & $\mathcal{N}(0,0.1)$ & \twosixtyesevenmfluxTESSfortytwo \\
$\mathrm{\sigma_{w,TESS42}}$ \dotfill & $\mathcal{J}(0.1,1000)$ & \twosixtyesevensigmawTESSfortytwo \\
$\mathrm{\sigma_{GP,TESS42}}$ \dotfill & $\mathcal{N}(9.0\times 10^{-5},1.2\times 10^{-5})$ & \twosixtyesevenGPsigmaTESSfortytwo \\
$\mathrm{\rho_{GP,TESS42}}$ \dotfill & $\mathcal{N}(0.23,0.04)$ & \twosixtyesevenGPrhoTESSfortytwo \\
$\mathrm{m_{dilution,TESS70}}$ \dotfill & $\mathrm{fixed}$ & \twosixtyesevenmdilutionTESSseventy \\
$\mathrm{m_{flux,TESS70}}$ \dotfill & $\mathcal{N}(0,0.1)$ & \twosixtyesevenmfluxTESSseventy \\
$\mathrm{\sigma_{w,TESS70}}$ \dotfill & $\mathcal{J}(0.1,1000)$ & \twosixtyesevensigmawTESSseventy \\
$\mathrm{\sigma_{GP,TESS70}}$ \dotfill & $\mathcal{N}(24.6\times 10^{-5},5.1\times 10^{-5})$ & \twosixtyesevenGPsigmaTESSseventy \\
$\mathrm{\rho_{GP,TESS70}}$ \dotfill & $\mathcal{N}(0.81,0.21)$ & \twosixtyesevenGPrhoTESSseventy \\
\hline
$\mathrm{e_{p1}}$ \dotfill & $-$ & \twosixtyeseveneccpone \\
$\mathrm{\omega_{p1}}$ \dotfill [$\degr$]& $-$ & \twosixtyesevenomegapone \\
$\mathrm{a_{p1}}$ \dotfill [au] & $-$ & $0.0943 \pm 0.0016$ \\
$\mathrm{i_{p1}}$ \dotfill [$\degr$] & $-$ & $88.62 \pm 0.72$ \\
$\mathrm{T_{14,p1}}$ \dotfill [h] & $-$ & $2.99 \pm 0.19$ \\
$\mathrm{M_{p1}}$ \dotfill [$\mathrm{M_{e}}$] & $-$ & $9.41 \pm 2.96$ \\
$\mathrm{R_{p1}}$ \dotfill [$\mathrm{R_{e}}$] & $-$ & $1.63 \pm 0.07$ \\
$\mathrm{\rho_{p1}}$ \dotfill [$\mathrm{g \, cm{-3}}$] & $-$ & $11.99 \pm 4.04$ \\
$\mathrm{T_{eq,p1}}$ \dotfill [K] & $-$ & $532.0 \pm 15.0$ \\
$\mathrm{\log Z}$ \dotfill  & $-$ & $240848.9 \pm 0.6$ \\
\hline
\end{tabular} 
} 
\end{center} 
\end{table}

\subsubsection{TOI-286}

\begin{figure}
    \centering
    \includegraphics[width=.5\textwidth]{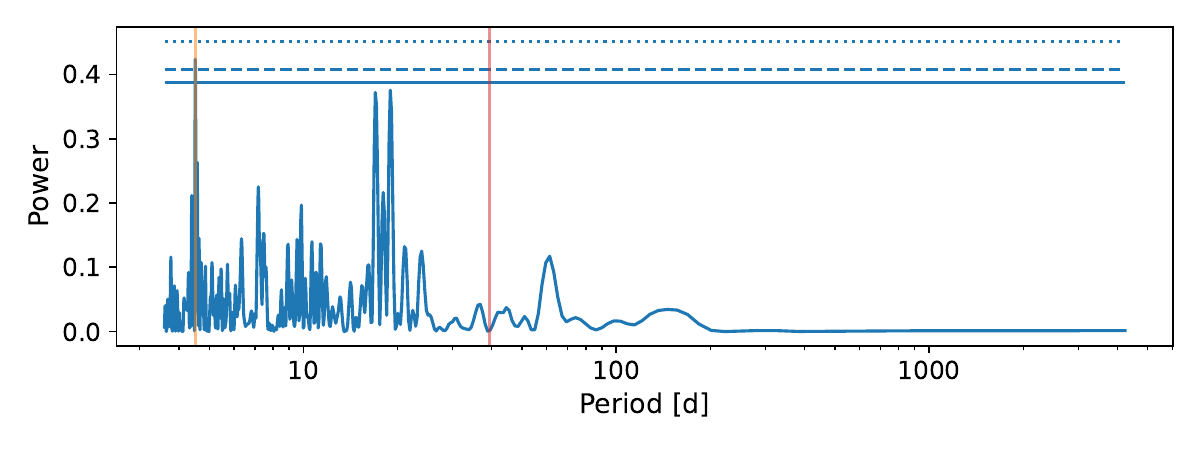}
    \caption{Periodogram of the TOI-286 ESPRESSO RVs with a GP component removed. The orange and red semitransparent lines indicate the periods of the two transiting candidates. The strongest signal in the periodogram is located at the period of the inner candidate, but there is none at the period of the outer candidate.}
    \label{fig:toi-286_GPres_periodogram}
\end{figure}

We use ESPRESSO and HARPS RVs and TESS photometry to characterise the planetary candidates TOI-286.01 and TOI-286.02. As with TOI-260, the ESPRESSO RVs for TOI-286 do not immediately show evidence for periodic signals at either of the \textit{TESS} candidate periods of $\mathrm{P \approx 4.5 \, d}$ and $\mathrm{P \approx 39.4 \, d}$, as can be seen in the periodogram in Fig. \ref{fig:toi-286_periodograms}. The stellar activity indicators, however, show evidence for both a long-term trend and some signal at $\mathrm{\approx 32 \, d}$, close to the stellar rotation period determined from the $\log R'_{\rm hk} - \mathrm{P_{rot}}$ relations (see Sec. \ref{sect:stellar-parameters}). 
To test whether the stellar activity is also masking the planetary signals for this target, we fitted a model consisting of two Keplerian signals  and a GP to the ESPRESSO RVs alone. In order to detrend the radial velocities from the stellar activity, we first fitted a GP to the FWHM activity indicator time series. We used broad log-uniform priors of $\mathcal{J}(0.001,1000)$ for $\mathrm{\sigma_{GP,FWHM}}$, $\mathrm{\alpha_{GP,FWHM}}$, and $\mathrm{\Gamma_{GP,FWHM}}$, and a normal prior of $\mathcal{N}(36,15)$ for the rotation period GP hyperparameter $\mathrm{P_{rot,GP, FWHM}}$, taking the value and uncertainty of the rotation period (see Sec. \ref{sect:stellar-parameters}) as the mean and standard deviation. Subsequently, we used the resulting parameters as priors on the radial velocity GP. The periodogram of the RVs with the GP component removed is shown in Fig. \ref{fig:toi-286_GPres_periodogram}; a signal can be seen at the period of the inner candidate, but there is none at the period of the outer candidate. This is not unexpected, as it is close to the estimated stellar rotation period and is thus likely to be partly absorbed by the stellar activity GP unless both the activity and the planet are fitted simultaneously.

We therefore tested three models for the full fit of all the data acquired for TOI-286: a single circular transiting planet corresponding to the inner signal, two circular transiting planets, and two transiting planets with free eccentricity. Since the HARPS RVs are few and coeval with the start of the ESPRESSO18 data, we detrended the RVs using a global GP constrained by the GP on the ESPRESSO FWHM activity indicator. The GP fit to the FWHM is shown in Fig. \ref{fig:act-ind-gps}. 

As it is in the \textit{TESS} continuous viewing zone, TOI-286 has a wealth of \textit{TESS} data that makes simultaneous modelling of the photometric instrumental parameters and GPs for all sectors together with the planet computationally prohibitive. Therefore, we employed a variation of the two-step process used for TOI-260, where we first masked the transits (with a margin of 5 hours around each transit window) and fitted GPs and instrumental parameters to the out-of-transit data only on a sector-by-sector basis, then fixed the resulting parameters for the joint full fit in which we employed only the in-transit \textit{TESS} data. We report the priors and posteriors of the GP fits to the out-of-transit data in Appendix \ref{ap:TOI286_TESSGP}.

While the ground-based photometry is in transit, its main purpose was to vet nearby stars, particularly the companion at 17$\arcsec$ separation, and verify they are not background eclipsing binaries. The variation of the photometry is too high for the small transits measured by \textit{TESS} to be detectable, so we do not include them in the full model. 

The priors and posteriors of the three models are given in Table \ref{tab:TOI-286_1planet} for the one-planet model, Table \ref{tab:TOI-286-circ} for the model with two circular planets, and Table \ref{tab:TOI-286-efree} for the model with two planets with free eccentricities, respectively. The prior distributions were chosen in an analogous way to those for TOI-260, with the following differences: For the parameters of the outer candidate, period $\mathrm{P_{p2}}$, time of transit $\mathrm{t_{0,p2}}$, and planet-to-star radius ratio $\mathrm{p_{p2}}$, we also use the ExoFOP values as priors. The \textit{TESS} instrumental parameters for each sector are fixed in this case, other than the global limb-darkening parameters q$_\mathrm{{1,TESS}}$ and q$_\mathrm{{2,TESS}}$. For the global RV GP, we used the posteriors of the GP fit to the FWHM activity indicator as priors (in the case of $\mathrm{\sigma_{GP,rv}}$, scaled by a normalisation factor corresponding to the RV variation over the FWHM variation). For the free-eccentricity fit, we used the ($\mathrm{\sqrt{e} \sin \omega, \sqrt{e} \cos \omega}$) parametrization for both planets, and fixed an upper limit on the eccentricities of 0.7.  The cornerplots of the posterior distributions are shown in Figs. \ref{fig:toi-286_cornerplot_1pl_p1} and \ref{fig:toi-286_cornerplot_1pl_instr} for the one-planet fit; in Figs. \ref{fig:toi-286_cornerplot_circ_p1}, \ref{fig:toi-286_cornerplot_circ_p2}, and \ref{fig:toi-286_cornerplot_circ_instr} for the two-planet circular fit; and in Figs. \ref{fig:toi-286_cornerplot_e07_p1}, \ref{fig:toi-286_cornerplot_e07_p2}, and \ref{fig:toi-286_cornerplot_e07_instr} for the free-eccentricity fit. 

\begin{figure*}
    \centering
    \includegraphics[width=.95\textwidth]{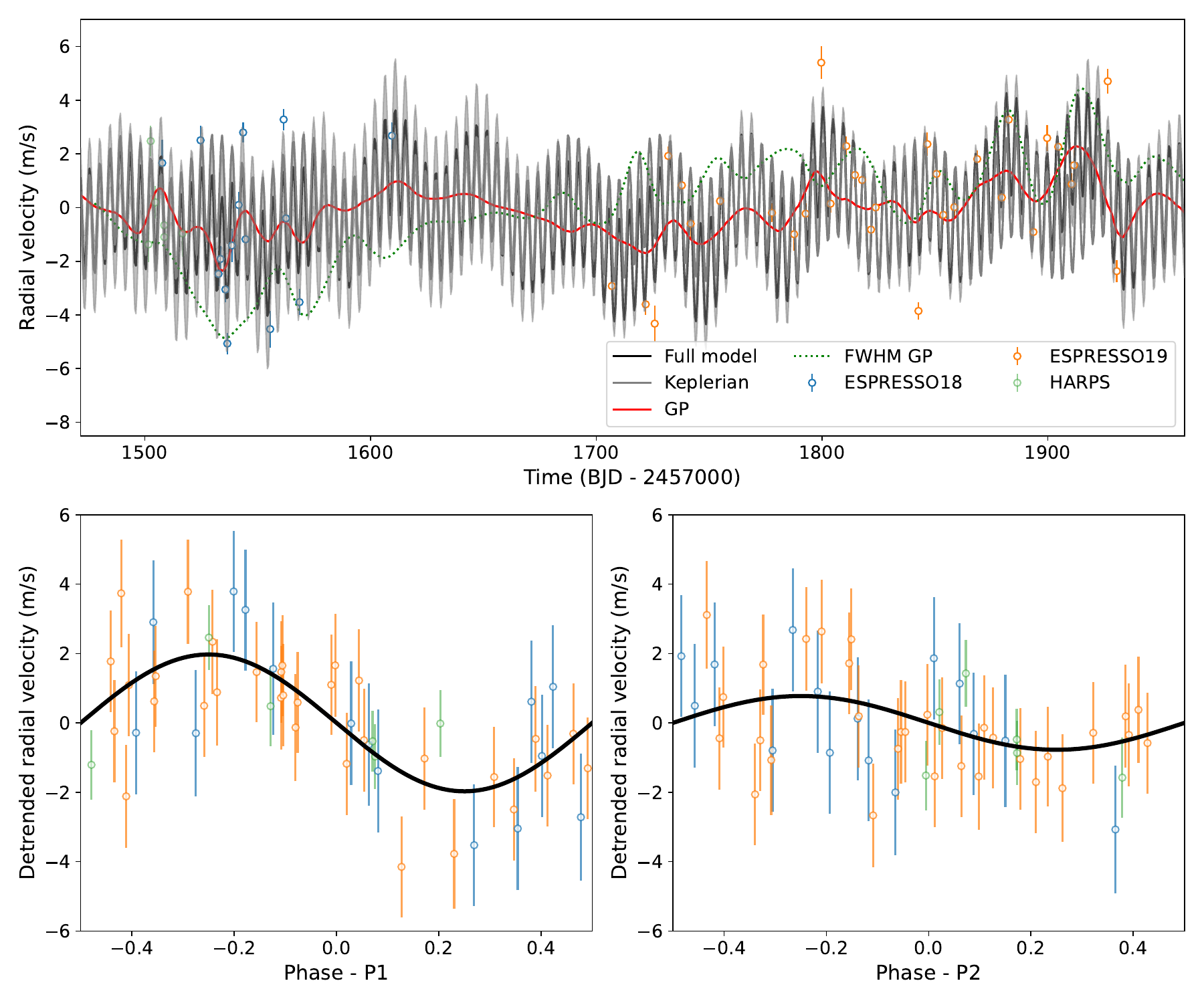}
    \caption{Top: Radial velocities (ESPRESSO18: blue dots, ESPRESSO19: orange dots, HARPS: green dots), model components (GP: red, Keplerian: grey), and median circular model (black) for TOI-286. The error bars show the RV errors and jitter added in quadrature. The instrumental systemic velocity has been subtracted. The GP fit to the FWHM activity indicator is also shown (dotted green line) for comparison. Bottom: phase-folded radial velocities plus median Keplerian for the inner (left) and outer (right) candidates, following the same colour scheme. }
    \label{fig:toi-286_RVs_circ}
\end{figure*}

\begin{figure*}
    \centering
    \includegraphics[width=.95\textwidth]{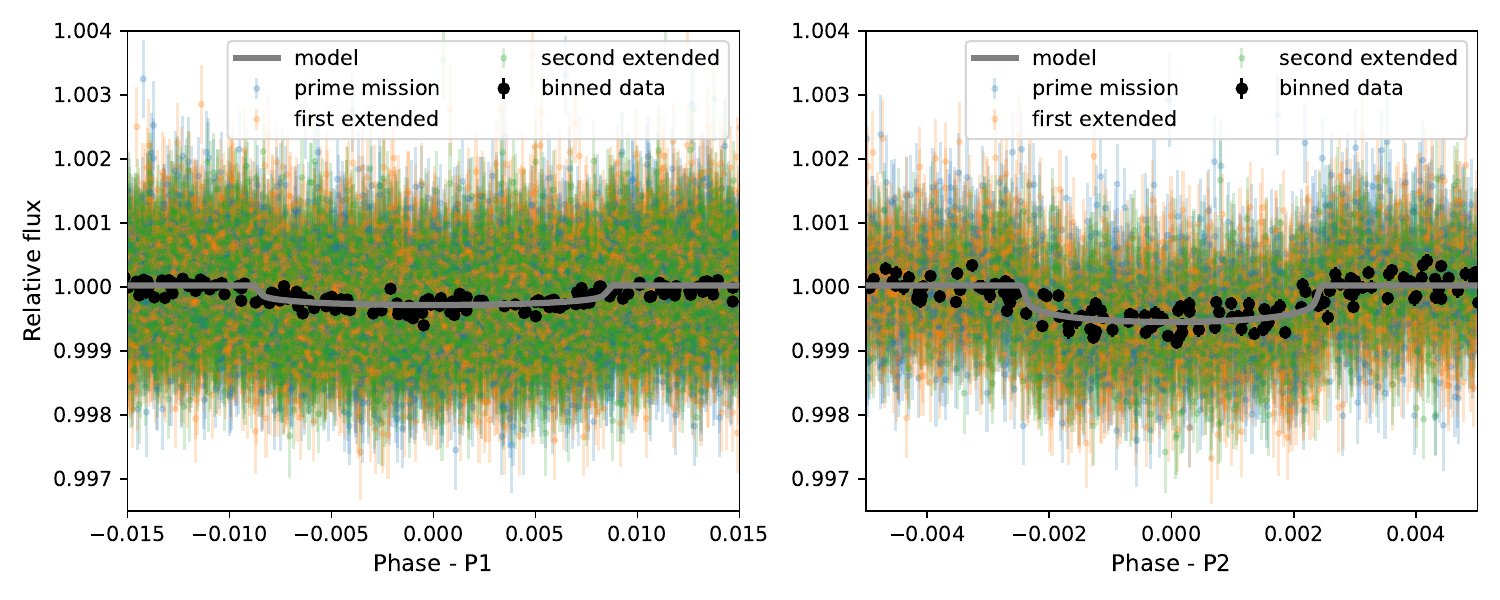}
    \caption{Stacked phase-folded \textit{TESS} data for TOI-286 with the median circular model (grey line), for the inner candidate (left) and outer candidate (right). The points show the data from the \textit{TESS} primary mission (blue), first extended mission (orange), and second extended mission (green), and binned data (black points). }
    \label{fig:toi-286_TESS_stacked_circ}
\end{figure*}

The log-evidence comparison of the three models very strongly favours the two-planet models over the one-planet model, with $\Delta \log Z \approx 769$ for both two-planet models versus the one-planet circular model. Comparison between the two 2-planet models slightly favours the model with circular orbits, with $\Delta \log Z \approx 2.9$, although the difference in log-evidence is only marginally significant; therefore, we prefer the simpler circular model. We also note that the eccentricities in the free-eccentricity model are low, with both being compatible with 0 at the $2\sigma$ level. We show the RVs and median circular model in Fig. \ref{fig:toi-286_RVs_circ}, and the phase-folded stacked \textit{TESS} data and model in Fig. \ref{fig:toi-286_TESS_stacked_circ}. The full \textit{TESS} light curves and corresponding median models are shown in Fig.\ref{fig:toi-286_TESS_sectors_circ} for sector 1 as an illustration. With the circular model, we find for TOI-268~b an RV semi-amplitude of $\mathrm{K_{p1} = \twoeightysixtwopcKpone \, m \, s^{-1}}$, corresponding to a detection at $3\sigma$, and leading to a planetary mass of $\mathrm{M_{p1} = 4.67 \pm 0.75 \, M_\oplus}$. The RV semi-amplitude found for TOI-286~c of $\mathrm{K_{p1} = \twoeightysixtwopcKptwo \, m \, s^{-1}}$ is only marginally significant; it leads to a planetary mass of $\mathrm{M_{p1} = 3.72 \pm 2.22 \, M_\oplus}$.

\begin{table}[pht] 
\begin{center} 
\caption{Prior and posterior planetary parameter distributions obtained with \texttt{juliet} for TOI-286, for a one-planet circular model. \textit{Top}: Fitted parameters. \textit{Bottom}: derived orbital parameters and physical parameters.} 
\label{tab:TOI-286_1planet} 
\centering 
\resizebox{\columnwidth}{!}{%
\begin{tabular}{lll} 
\hline  \hline 
Parameter & Prior & Posterior \\ 
\hline 
$\mathrm{P_{p1}}$ \dotfill [d] & $\mathcal{N}(4.5,0.1)$ & \twoeightysixoneplPpone \\
$\mathrm{t_{0,p1}}$ \dotfill [BJD] & $\mathcal{N}(2460186.6,0.1)$ & \twoeightysixonepltzeropone \\
$\mathrm{p_{p1}}$ \dotfill & $\mathcal{N}(0.02,0.1)$ & \twoeightysixoneplppone \\
$\mathrm{b_{p1}}$ \dotfill & $\mathcal{U}(0,1)$ & \twoeightysixoneplbpone \\
$\mathrm{e_{p1}}$ \dotfill & $\mathrm{fixed}$ & \twoeightysixonepleccpone \\
$\mathrm{\omega_{p1}}$ \dotfill [$\degr$] & $\mathrm{fixed}$ & \twoeightysixoneplomegapone \\
$\mathrm{K_{p1}}$ \dotfill [$\mathrm{m \, s^{-1}}$] & $\mathcal{U}(0,10)$ & \twoeightysixoneplKpone \\
$\mathrm{\rho_{}}$ \dotfill [$\mathrm{kg \, m^{-3}}$] & $\mathcal{J}(100,10000)$ & \twoeightysixoneplrho \\
$\mathrm{q_{1,TESS}}$ \dotfill & $\mathcal{U}(0,1)$ & \twoeightysixoneplqoneTESS \\
$\mathrm{q_{2,TESS}}$ \dotfill & $\mathcal{U}(0,1)$ & \twoeightysixoneplqtwoTESS \\
$\mathrm{\mu_{ESPRESSO18}}$ \dotfill [$\mathrm{m \, s^{-1}}$] & $\mathcal{U}(17774.8,17785.3)$ & \twoeightysixoneplmuESPRESSOeighteen \\
$\mathrm{\sigma_{w,ESPRESSO18}}$ \dotfill [$\mathrm{m \, s^{-1}}$] & $\mathcal{J}(0.001,10)$ & \twoeightysixoneplsigmawESPRESSOeighteen \\
$\mathrm{\mu_{ESPRESSO19}}$ \dotfill [$\mathrm{m \, s^{-1}}$] & $\mathcal{U}(17774.8,17785.3)$ & \twoeightysixoneplmuESPRESSOnineteen \\
$\mathrm{\sigma_{w,ESPRESSO19}}$ \dotfill [$\mathrm{m \, s^{-1}}$] & $\mathcal{J}(0.001,10)$ & \twoeightysixoneplsigmawESPRESSOnineteen \\
$\mathrm{\mu_{HARPS}}$ \dotfill [$\mathrm{m \, s^{-1}}$] & $\mathcal{U}(-0.71,3.13)$ & \twoeightysixoneplmuHARPS \\
$\mathrm{\sigma_{w,HARPS}}$ \dotfill [$\mathrm{m \, s^{-1}}$] & $\mathcal{J}(0.001,10)$ & \twoeightysixoneplsigmawHARPS \\
$\mathrm{\sigma_{GP,rv}}$ \dotfill & $\mathcal{TN}(3.89,0.85,0,1000)$ & \twoeightysixoneplGPsigmarv \\
$\mathrm{\alpha_{GP,rv}}$ \dotfill & $\mathcal{TN}(0.00124,0.00056,0,1000)$ & \twoeightysixoneplGPalpharv \\
$\mathrm{\Gamma_{GP,rv}}$ \dotfill & $\mathcal{TN}(0.54,0.43,0,1000)$ & \twoeightysixoneplGPGammarv \\
$\mathrm{P_{rot,GP,rv}}$ \dotfill & $\mathcal{N}(35.9,4.2)$ & \twoeightysixoneplGPProtrv \\
\hline
$\mathrm{a_{p1}}$ \dotfill [au] & $-$ & $0.0503 \pm 0.001$ \\
$\mathrm{i_{p1}}$ \dotfill [$\degr$] & $-$ & $87.44 \pm 1.37$ \\
$\mathrm{T_{14,p1}}$ \dotfill [h] & $-$ & $2.01 \pm 0.64$ \\
$\mathrm{M_{p1}}$ \dotfill [$\mathrm{M_{e}}$] & $-$ & $4.32 \pm 1.04$ \\
$\mathrm{R_{p1}}$ \dotfill [$\mathrm{R_{e}}$] & $-$ & $1.45 \pm 0.12$ \\
$\mathrm{\rho_{p1}}$ \dotfill [$\mathrm{g \, cm{-3}}$] & $-$ & $7.85 \pm 2.75$ \\
$\mathrm{T_{eq,p1}}$ \dotfill [K] & $-$ & $979.0 \pm 31.0$ \\
$\mathrm{\log Z}$ \dotfill  & $-$ & $214793.8 \pm 0.4$ \\
\hline 
\end{tabular} 
} 
\end{center} 
\end{table} 

\begin{table}[pht] 
\begin{center} 
\caption{Prior and posterior planetary parameter distributions obtained with \texttt{juliet} for TOI-286, for two circular planets. \textit{Top}: Fitted parameters. \textit{Bottom}: derived orbital parameters and physical parameters.} 
\label{tab:TOI-286-circ} 
\centering 
\resizebox{\columnwidth}{!}{%
\begin{tabular}{lll} 
\hline  \hline 
Parameter & Prior & Posterior \\ 
\hline 
$\mathrm{P_{p1}}$ \dotfill [d] & $\mathcal{N}(4.5,0.1)$ & \twoeightysixtwopcPpone \\
$\mathrm{t_{0,p1}}$ \dotfill [BJD] & $\mathcal{N}(2460186.6,0.1)$ & \twoeightysixtwopctzeropone \\
$\mathrm{p_{p1}}$ \dotfill & $\mathcal{N}(0.02,0.1)$ & \twoeightysixtwopcppone \\
$\mathrm{b_{p1}}$ \dotfill & $\mathcal{U}(0,1)$ & \twoeightysixtwopcbpone \\
$\mathrm{e_{p1}}$ \dotfill & $\mathrm{fixed}$ & \twoeightysixtwopceccpone \\
$\mathrm{\omega_{p1}}$ \dotfill [$\degr$]& $\mathrm{fixed}$ & \twoeightysixtwopcomegapone \\
$\mathrm{K_{p1}}$ \dotfill [$\mathrm{m \, s^{-1}}$] & $\mathcal{U}(0,10)$ & \twoeightysixtwopcKpone \\
$\mathrm{P_{p2}}$ \dotfill [d] & $\mathcal{N}(39.4,0.1)$ & \twoeightysixtwopcPptwo \\
$\mathrm{t_{0,p2}}$ \dotfill [BJD] & $\mathcal{N}(2460155.0,0.1)$ & \twoeightysixtwopctzeroptwo \\
$\mathrm{p_{p2}}$ \dotfill & $\mathcal{N}(0.02,0.1)$ & \twoeightysixtwopcpptwo \\
$\mathrm{b_{p2}}$ \dotfill & $\mathcal{U}(0,1)$ & \twoeightysixtwopcbptwo \\
$\mathrm{e_{p2}}$ \dotfill & $\mathrm{fixed}$ & \twoeightysixtwopceccptwo \\
$\mathrm{\omega_{p2}}$ \dotfill [$\degr$]& $\mathrm{fixed}$ & \twoeightysixtwopcomegaptwo \\
$\mathrm{K_{p2}}$ \dotfill [$\mathrm{m \, s^{-1}}$] & $\mathcal{U}(0,10)$ & \twoeightysixtwopcKptwo \\
$\mathrm{\rho_{}}$ \dotfill [$\mathrm{kg \, m^{-3}}$] & $\mathcal{J}(100,10000)$ & \twoeightysixtwopcrho \\
$\mathrm{q_{1,TESS}}$ \dotfill & $\mathcal{U}(0,1)$ & \twoeightysixtwopcqoneTESS \\
$\mathrm{q_{2,TESS}}$ \dotfill & $\mathcal{U}(0,1)$ & \twoeightysixtwopcqtwoTESS \\
$\mathrm{\mu_{ESPRESSO18}}$ \dotfill [$\mathrm{m \, s^{-1}}$] & $\mathcal{U}(17774.8,17785.3)$ & \twoeightysixtwopcmuESPRESSOeighteen \\
$\mathrm{\sigma_{w,ESPRESSO18}}$ \dotfill [$\mathrm{m \, s^{-1}}$] & $\mathcal{J}(0.001,10)$ & \twoeightysixtwopcsigmawESPRESSOeighteen \\
$\mathrm{\mu_{ESPRESSO19}}$ \dotfill [$\mathrm{m \, s^{-1}}$] & $\mathcal{U}(17774.8,17785.3)$ & \twoeightysixtwopcmuESPRESSOnineteen \\
$\mathrm{\sigma_{w,ESPRESSO19}}$ \dotfill [$\mathrm{m \, s^{-1}}$] & $\mathcal{J}(0.001,10)$ & \twoeightysixtwopcsigmawESPRESSOnineteen \\
$\mathrm{\mu_{HARPS}}$ \dotfill [$\mathrm{m \, s^{-1}}$] & $\mathcal{U}(-0.71,3.13)$ & \twoeightysixtwopcmuHARPS \\
$\mathrm{\sigma_{w,HARPS}}$ \dotfill [$\mathrm{m \, s^{-1}}$] & $\mathcal{J}(0.001,10)$ & \twoeightysixtwopcsigmawHARPS \\
$\mathrm{\sigma_{GP,rv}}$ \dotfill & $\mathcal{TN}(3.9,4.5,0,1000)$ & \twoeightysixtwopcGPsigmarv \\
$\mathrm{\alpha_{GP,rv}}$ \dotfill & $\mathcal{TN}(0.00124,0.00056,0,1000)$ & \twoeightysixtwopcGPalpharv \\
$\mathrm{\Gamma_{GP,rv}}$ \dotfill & $\mathcal{TN}(0.54,0.43,0,1000)$ & \twoeightysixtwopcGPGammarv \\
$\mathrm{P_{rot,GP,rv}}$ \dotfill & $\mathcal{N}(35.9,4.2)$ & \twoeightysixtwopcGPProtrv \\
\hline
$\mathrm{a_{p1}}$ \dotfill [au] & $-$ & $0.0503 \pm 0.001$ \\
$\mathrm{i_{p1}}$ \dotfill [$\degr$] & $-$ & $87.52 \pm 0.59$ \\
$\mathrm{T_{14,p1}}$ \dotfill [h] & $-$ & $2.04 \pm 0.27$ \\
$\mathrm{M_{p1}}$ \dotfill [$\mathrm{M_{e}}$] & $-$ & $4.53 \pm 0.78$ \\
$\mathrm{R_{p1}}$ \dotfill [$\mathrm{R_{e}}$] & $-$ & $1.42 \pm 0.10$ \\
$\mathrm{\rho_{p1}}$ \dotfill [$\mathrm{g \, cm{-3}}$] & $-$ & $8.63 \pm 2.32$ \\
$\mathrm{T_{eq,p1}}$ \dotfill [K] & $-$ & $979.0 \pm 31.0$ \\
$\mathrm{a_{p1}}$ \dotfill [au] & $-$ & $0.213 \pm 0.0042$ \\
$\mathrm{i_{p1}}$ \dotfill [$\degr$] & $-$ & $89.69 \pm 0.28$ \\
$\mathrm{T_{14,p1}}$ \dotfill [h] & $-$ & $4.97 \pm 0.55$ \\
$\mathrm{M_{p1}}$ \dotfill [$\mathrm{M_{e}}$] & $-$ & $3.72 \pm 2.22$ \\
$\mathrm{R_{p1}}$ \dotfill [$\mathrm{R_{e}}$] & $-$ & $1.88 \pm 0.12$ \\
$\mathrm{\rho_{p1}}$ \dotfill [$\mathrm{g \, cm{-3}}$] & $-$ & $3.09 \pm 1.93$ \\
$\mathrm{T_{eq,p1}}$ \dotfill [K] & $-$ & $475.0 \pm 15.0$ \\
$\mathrm{\log Z}$ \dotfill  & $-$ & $215562.9 \pm 0.5$ \\
\hline 
\end{tabular} 
} 
\end{center} 
\end{table} 

\begin{table}[pht] 
\begin{center} 
\caption{Prior and posterior planetary parameter distributions obtained with \texttt{juliet} for TOI-286, for two planets with free eccentricity. \textit{Top}: Fitted parameters. \textit{Bottom}: derived orbital parameters and physical parameters.} 
\label{tab:TOI-286-efree} 
\centering 
\resizebox{\columnwidth}{!}{%
\begin{tabular}{lll} 
\hline  \hline 
Parameter & Prior & Posterior \\ 
\hline 
$\mathrm{P_{p1}}$ \dotfill [d] & $\mathcal{N}(4.5,0.1)$ & \twoeightysixtwopePpone \\
$\mathrm{t_{0,p1}}$ \dotfill [BJD] & $\mathcal{N}(2460186.63,0.1)$ & \twoeightysixtwopetzeropone \\
$\mathrm{p_{p1}}$ \dotfill & $\mathcal{N}(0.02,0.1)$ & \twoeightysixtwopeppone \\
$\mathrm{b_{p1}}$ \dotfill & $\mathcal{U}(0,1)$ & \twoeightysixtwopebpone \\
$\mathrm{\sqrt{e}\sin\omega_{p1}}$ \dotfill & $\mathcal{U}(0,1)$ & \twoeightysixtwopesesinomegapone \\
$\mathrm{\sqrt{e} \cos\omega_{p1}}$ \dotfill & $\mathcal{U}(0,1)$ & \twoeightysixtwopesecosomegapone \\
$\mathrm{K_{p1}}$ \dotfill [$\mathrm{m \, s^{-1}}$] & $\mathcal{U}(0,10)$ & \twoeightysixtwopeKpone \\
$\mathrm{P_{p2}}$ \dotfill [d] & $\mathcal{N}(39.4,0.1)$ & \twoeightysixtwopePptwo \\
$\mathrm{t_{0,p2}}$ \dotfill [BJD] & $\mathcal{N}(2460155.0,0.1)$ & \twoeightysixtwopetzeroptwo \\
$\mathrm{p_{p2}}$ \dotfill & $\mathcal{N}(0.02,0.1)$ & \twoeightysixtwopepptwo \\
$\mathrm{b_{p2}}$ \dotfill & $\mathcal{U}(0,1)$ & \twoeightysixtwopebptwo \\
$\mathrm{\sqrt{e}\sin\omega_{p2}}$ \dotfill & $\mathcal{U}(0,1)$ & \twoeightysixtwopesesinomegaptwo \\
$\mathrm{\sqrt{e}\cos\omega_{p2}}$ \dotfill & $\mathcal{U}(0,1)$ & \twoeightysixtwopesecosomegaptwo \\
$\mathrm{K_{p2}}$ \dotfill [$\mathrm{m \, s^{-1}}$] & $\mathcal{U}(0,10)$ & \twoeightysixtwopeKptwo \\
$\mathrm{\rho_{}}$ \dotfill [$\mathrm{kg \, m^{-3}}$] & $\mathcal{J}(100,10000)$ & \twoeightysixtwoperho \\
$\mathrm{q_{1,TESS}}$ \dotfill & $\mathcal{U}(0,1)$ & \twoeightysixtwopeqoneTESS \\
$\mathrm{q_{2,TESS}}$ \dotfill & $\mathcal{U}(0,1)$ & \twoeightysixtwopeqtwoTESS \\
$\mathrm{\mu_{ESPRESSO18}}$ \dotfill [$\mathrm{m \, s^{-1}}$] & $\mathcal{U}(17774.8,17785.3)$ & \twoeightysixtwopemuESPRESSOeighteen \\
$\mathrm{\sigma_{w,ESPRESSO18}}$ \dotfill [$\mathrm{m \, s^{-1}}$] & $\mathcal{J}(0.001,10.0)$ & \twoeightysixtwopesigmawESPRESSOeighteen \\
$\mathrm{\mu_{ESPRESSO19}}$ \dotfill [$\mathrm{m \, s^{-1}}$] & $\mathcal{U}(17774.8,17785.3)$ & \twoeightysixtwopemuESPRESSOnineteen \\
$\mathrm{\sigma_{w,ESPRESSO19}}$ \dotfill [$\mathrm{m \, s^{-1}}$] & $\mathcal{J}(0.001,10)$ & \twoeightysixtwopesigmawESPRESSOnineteen \\
$\mathrm{\mu_{HARPS}}$ \dotfill & $\mathcal{U}(-0.71,3.13)$ & \twoeightysixtwopemuHARPS \\
$\mathrm{\sigma_{w,HARPS}}$ \dotfill & $\mathcal{J}(0.001,10)$ & \twoeightysixtwopesigmawHARPS \\
$\mathrm{\sigma_{GP,rv}}$ \dotfill & $\mathcal{TN}(3.87,0.85,0,1000)$ & \twoeightysixtwopeGPsigmarv \\
$\mathrm{\alpha_{GP,rv}}$ \dotfill & $\mathcal{TN}(0.00124,0.00056,0,1000)$ & \twoeightysixtwopeGPalpharv \\
$\mathrm{\Gamma_{GP,rv}}$ \dotfill & $\mathcal{TN}(0.54,0.435,0,1000)$ & \twoeightysixtwopeGPGammarv \\
$\mathrm{P_{rot,GP,rv}}$ \dotfill & $\mathcal{N}(35.9,4.2)$ & \twoeightysixtwopeGPProtrv \\
\hline
$\mathrm{e_{p1}}$ \dotfill & $-$ & \twoeightysixtwopeeccpone \\
$\mathrm{\omega_{p1}}$ \dotfill [$\degr$]& $-$ & \twoeightysixtwopeomegapone \\
$\mathrm{a_{p1}}$ \dotfill [au] & $-$ & $0.0503 \pm 0.001$ \\
$\mathrm{i_{p1}}$ \dotfill [$\degr$] & $-$ & $87.39 \pm 0.76$ \\
$\mathrm{T_{14,p1}}$ \dotfill [h] & $-$ & $2.11 \pm 0.27$ \\
$\mathrm{M_{p1}}$ \dotfill [$\mathrm{M_{e}}$] & $-$ & $4.48 \pm 0.78$ \\
$\mathrm{R_{p1}}$ \dotfill [$\mathrm{R_{e}}$] & $-$ & $1.41 \pm 0.08$ \\
$\mathrm{\rho_{p1}}$ \dotfill [$\mathrm{g \, cm{-3}}$] & $-$ & $8.81 \pm 2.12$ \\
$\mathrm{T_{eq,p1}}$ \dotfill [K] & $-$ & $981.0 \pm 32.0$ \\
$\mathrm{e_{p2}}$ \dotfill & $-$ & \twoeightysixtwopeeccptwo \\
$\mathrm{\omega_{p2}}$ \dotfill [$\degr$]& $-$ & \twoeightysixtwopeomegaptwo \\
$\mathrm{a_{p2}}$ \dotfill [au] & $-$ & $0.213 \pm 0.0042$ \\
$\mathrm{i_{p2}}$ \dotfill [$\degr$] & $-$ & $89.7 \pm 0.19$ \\
$\mathrm{T_{14,p2}}$ \dotfill [h] & $-$ & $5.02 \pm 0.37$ \\
$\mathrm{M_{p2}}$ \dotfill [$\mathrm{M_{e}}$] & $-$ & $4.75 \pm 3.02$ \\
$\mathrm{R_{p2}}$ \dotfill [$\mathrm{R_{e}}$] & $-$ & $1.87 \pm 0.09$ \\
$\mathrm{\rho_{p2}}$ \dotfill [$\mathrm{g \, cm{-3}}$] & $-$ & $4.0 \pm 2.61$ \\
$\mathrm{T_{eq,p2}}$ \dotfill [K] & $-$ & $476.0 \pm 15.0$ \\
$\mathrm{\log Z}$ \dotfill  & $-$ & $215560.0 \pm 0.5$ \\
\hline 
\end{tabular} 
} 
\end{center} 
\end{table} 


\subsubsection{TOI-134}

The transiting super-Earth TOI-134~b was first confirmed by \cite{Astudillo-Defru2020}, using \textit{TESS} photometry and HARPS and Magellan/PFS RVs. The orbital parameters were subsequently updated by \cite{Patel2022}. In this work, we present refined values of the planetary parameters based on ESPRESSO observations in addition to the published HARPS and PFS RVs, and additional TESS data. We tested two models: a circular transiting planet, and a transiting planet with free eccentricity. 

The periodograms of the ESPRESSO data in Fig. \ref{fig:toi-134_periodograms} show a very strong signal at  $\mathrm{\approx 31 \, d}$ in the activity indicators and RV alike. The RVs will therefore require detrending from the stellar activity. Unfortunately, only two activity indicators are available for all four instruments, $\mathrm{H_\alpha}$, and $\mathrm{S_{MW}}$, and their values are not consistent from one instrument to another. However, the HARPS, PFS, and ESPRESSO18 RVs are roughly coeval, and the ESPRESSO19 RVs are close in time to them. We therefore chose to use a global RV GP, since it should be modelling the same underlying stellar activity. To constrain the parameters, we first fitted a GP to the FWHM activity indicator time series of the ESPRESSO data, and used the resulting parameters as priors on the RV GP. This GP fit is shown in Fig. \ref{fig:act-ind-gps}.  For the \textit{TESS} photometry we employed the same two-step process as for TOI-260. 

The priors and posteriors are given in Tables \ref{tab:TOI-134-circ} and \ref{tab:TOI-134-efree} for the fixed-eccentricity and free-eccentricity fits, respectively. The prior distributions were chosen similarly to those for TOI-260, except that we used the published orbital parameters as priors.  The cornerplots of the posterior distributions are shown in Figs. \ref{fig:toi-134_cornerplot_circ_orbital}, \ref{fig:toi-134_cornerplot_circ_rv}, and \ref{fig:toi-134_cornerplot_circ_transit} for the circular fit; and in Figs. \ref{fig:toi-134_cornerplot_efree_orbital}, \ref{fig:toi-134_cornerplot_efree_rv}, and \ref{fig:toi-134_cornerplot_efree_transit} for the free-eccentricity fit. 

\begin{figure*}
    \centering
    \includegraphics[width=\textwidth]{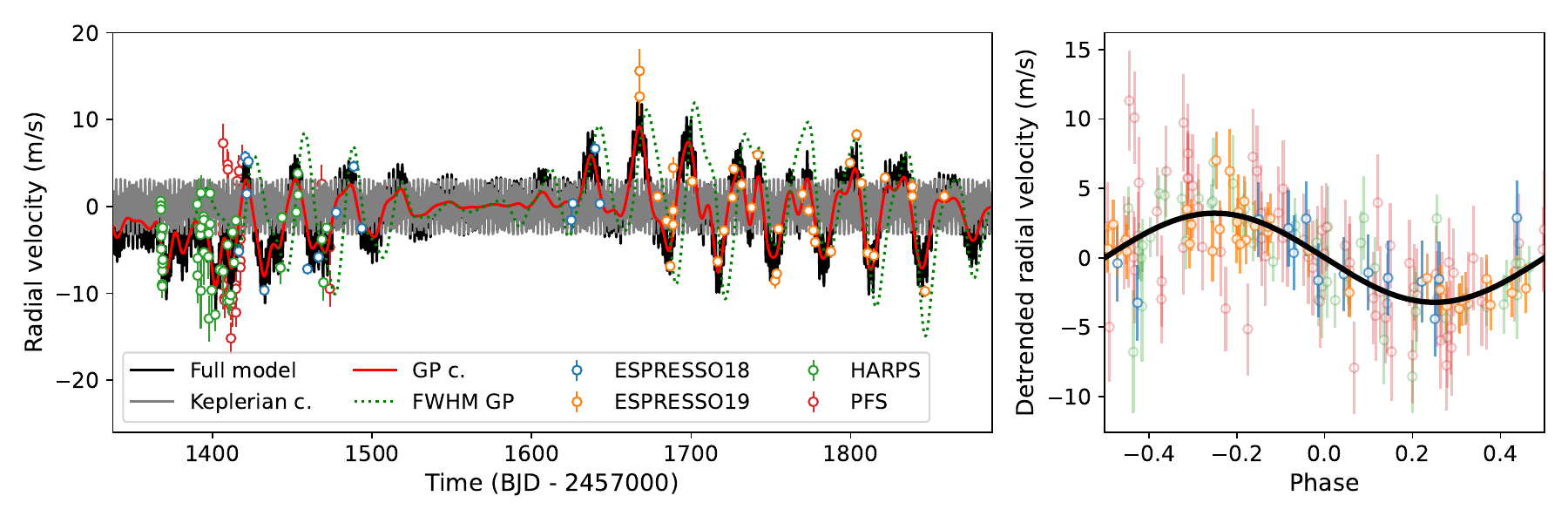}
    \caption{Left: Radial velocities (ESPRESSO18: blue dots, ESPRESSO19: orange dots, HARPS: green dots, PFS: red dots), model components (GP: red, Keplerian: grey), and median circular model (black) for TOI-134. The error bars show the RV errors and jitter added in quadrature. The instrumental systemic velocity has been subtracted. The GP fit to the FWHM activity indicator for ESPRESSO is also shown (dotted green line) for comparison. Right: phase-folded radial velocities plus median Keplerian, following the same colour scheme. }
    \label{fig:toi-134_RVs_circ}
\end{figure*}

\begin{figure}
    \centering
    \includegraphics[width=\columnwidth]{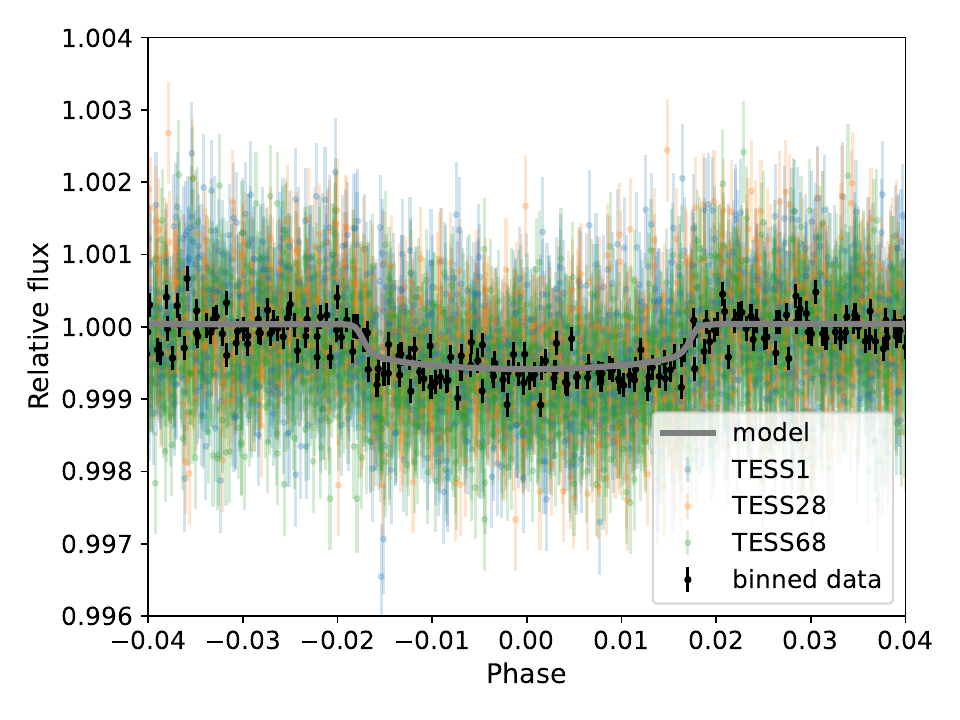}
    \caption{Stacked phase-folded \textit{TESS} data for TOI-134 with the median circular model (grey line). The points show the data from sectors 1 (blue), 28 (orange) and 68 (green), and binned data (black points).}
    \label{fig:toi-134_TESS_stacked_circ}
\end{figure}

The log-evidence comparison of the two models slightly favours the eccentric model, with $\Delta \log Z \approx 2.6$. However, the difference is only marginally significant, and the eccentricity of the free-eccentricity fit is very small and compatible with zero at the $2\sigma$ level. Therefore, we adopt the circular model. We show the RVs and median circular model in Fig. \ref{fig:toi-134_RVs_circ}, and the phase-folded stacked \textit{TESS} data and model in Fig. \ref{fig:toi-134_TESS_stacked_circ}. The full \textit{TESS} light curves and corresponding median models are shown in \ref{fig:toi-134_TESS_sectors_circ}.

\begin{table}[pht] 
\begin{center} 
\caption{Prior and posterior planetary parameter distributions obtained with \texttt{juliet} for TOI-134, for a fit with eccentricity fixed to 0. \textit{Top}: Fitted parameters. \textit{Bottom}: derived orbital parameters and physical parameters.} 
\label{tab:TOI-134-circ} 
\centering 
\resizebox{\columnwidth}{!}{%
\begin{tabular}{lll} 
\hline  \hline 
Parameter & Prior & Posterior \\ 
\hline 
$\mathrm{P_{p1}}$ \dotfill [d] & $\mathcal{N}(1.4,0.1)$ & \onethirtyfourcPpone \\
$\mathrm{t_{0,p1}}$ \dotfill [BJD] & $\mathcal{N}(2459082.9,0.1)$ & \onethirtyfourctzeropone \\
$\mathrm{p_{p1}}$ \dotfill & $\mathcal{N}(0.02,0.1)$ & \onethirtyfourcppone \\
$\mathrm{b_{p1}}$ \dotfill & $\mathcal{U}(0,1)$ & \onethirtyfourcbpone \\
$\mathrm{e_{p1}}$ \dotfill & $\mathrm{fixed}$ & \onethirtyfourceccpone \\
$\mathrm{\omega_{p1}}$ \dotfill [$\degr$]& $\mathrm{fixed}$ & \onethirtyfourcomegapone \\
$\mathrm{K_{p1}}$ \dotfill [$\mathrm{m \, s^{-1}}$] & $\mathcal{U}(0,10)$ & \onethirtyfourcKpone \\
$\mathrm{\rho_{}}$ \dotfill [$\mathrm{kg \, m^{-3}}$] & $\mathcal{J}(100,10000)$ & \onethirtyfourcrho \\
$\mathrm{q_{1,TESS}}$ \dotfill & $\mathcal{U}(0,1)$ & \onethirtyfourcqoneTESS \\
$\mathrm{q_{2,TESS}}$ \dotfill & $\mathcal{U}(0,1)$ & \onethirtyfourcqtwoTESS \\
$\mathrm{\mu_{ESPRESSO18}}$ \dotfill [$\mathrm{m \, s^{-1}}$] & $\mathcal{U}(29700,29750)$ & \onethirtyfourcmuESPRESSOeighteen \\
$\mathrm{\sigma_{w,ESPRESSO18}}$ \dotfill [$\mathrm{m \, s^{-1}}$] & $\mathcal{J}(0.001,10)$ & \onethirtyfourcsigmawESPRESSOeighteen \\
$\mathrm{\mu_{ESPRESSO19}}$ \dotfill [$\mathrm{m \, s^{-1}}$] & $\mathcal{U}(29700,29750)$ & \onethirtyfourcmuESPRESSOnineteen \\
$\mathrm{\sigma_{w,ESPRESSO19}}$ \dotfill [$\mathrm{m \, s^{-1}}$] & $\mathcal{J}(0.001,10)$ & \onethirtyfourcsigmawESPRESSOnineteen \\
$\mathrm{\mu_{HARPS}}$ \dotfill [$\mathrm{m \, s^{-1}}$] & $\mathcal{U}(29750,29790)$ & \onethirtyfourcmuHARPS \\
$\mathrm{\sigma_{w,HARPS}}$ \dotfill [$\mathrm{m \, s^{-1}}$] & $\mathcal{J}(0.001,10)$ & \onethirtyfourcsigmawHARPS \\
$\mathrm{\mu_{PFS}}$ \dotfill [$\mathrm{m \, s^{-1}}$] & $\mathcal{U}(-12,12)$ & \onethirtyfourcmuPFS \\
$\mathrm{\sigma_{w,PFS}}$ \dotfill [$\mathrm{m \, s^{-1}}$] & $\mathcal{J}(0.001,10)$ & \onethirtyfourcsigmawPFS \\
$\mathrm{\sigma_{GP,rv}}$ \dotfill & $\mathcal{TN}(4.87,0.93,0,1000)$ & \onethirtyfourcGPsigmarv \\
$\mathrm{\alpha_{GP,rv}}$ \dotfill & $\mathcal{TN}(0.00108,0.00015,0,1000)$ & \onethirtyfourcGPalpharv \\
$\mathrm{\Gamma_{GP,rv}}$ \dotfill & $\mathcal{TN}(2.21,0.99,0,1000)$ & \onethirtyfourcGPGammarv \\
$\mathrm{P_{rot,GP,rv}}$ \dotfill & $\mathcal{N}(30.79,0.86)$ & \onethirtyfourcGPProtrv \\
$\mathrm{m_{dilution,TESS1}}$ \dotfill & $\mathrm{fixed}$ & \onethirtyfourcmdilutionTESSone \\
$\mathrm{m_{flux,TESS1}}$ \dotfill & $\mathcal{N}(0,0.1)$ & \onethirtyfourcmfluxTESSone \\
$\mathrm{\sigma_{w,TESS1}}$ \dotfill & $\mathcal{J}(0.1,1000)$ & \onethirtyfourcsigmawTESSone \\
$\mathrm{\sigma_{GP,TESS1}}$ \dotfill & $\mathcal{N}(25.0\times 10^{-5},2.7 \times 10^{-5})$ & \onethirtyfourcGPsigmaTESSone \\
$\mathrm{\rho_{GP,TESS1}}$ \dotfill & $\mathcal{N}(0.311,0.098)$ & \onethirtyfourcGPrhoTESSone \\
$\mathrm{m_{dilution,TESS28}}$ \dotfill & $\mathrm{fixed}$ & \onethirtyfourcmdilutionTESStwentyeight \\
$\mathrm{m_{flux,TESS28}}$ \dotfill & $\mathcal{N}(0,0.1)$ & \onethirtyfourcmfluxTESStwentyeight \\
$\mathrm{\sigma_{w,TESS28}}$ \dotfill & $\mathcal{J}(0.1,1000)$ & \onethirtyfourcsigmawTESStwentyeight \\
$\mathrm{\sigma_{GP,TESS28}}$ \dotfill & $\mathcal{N}(54.3\times 10^{-5},6.6\times 10^{-5})$ & \onethirtyfourcGPsigmaTESStwentyeight \\
$\mathrm{\rho_{GP,TESS28}}$ \dotfill & $\mathcal{N}(0.324,0.066)$ & \onethirtyfourcGPrhoTESStwentyeight \\
$\mathrm{m_{dilution,TESS68}}$ \dotfill & $\mathrm{fixed}$ & \onethirtyfourcmdilutionTESSsixtyeight \\
$\mathrm{m_{flux,TESS68}}$ \dotfill & $\mathcal{N}(0,0.1)$ & \onethirtyfourcmfluxTESSsixtyeight \\
$\mathrm{\sigma_{w,TESS68}}$ \dotfill & $\mathcal{J}(0.1,1000)$ & \onethirtyfourcsigmawTESSsixtyeight \\
$\mathrm{\sigma_{GP,TESS68}}$ \dotfill & $\mathcal{N}(61.6\times 10^{-5},54.0\times 10^{-5})$ & \onethirtyfourcGPsigmaTESSsixtyeight \\
$\mathrm{\rho_{GP,TESS68}}$ \dotfill & $\mathcal{N}(2.7,2.0)$ & \onethirtyfourcGPrhoTESSsixtyeight \\
\hline 
$\mathrm{a_{p1}}$ \dotfill [au] & $-$ & $0.0208 \pm 0.0006$ \\
$\mathrm{i_{p1}}$ \dotfill [$\degr$] & $-$ & $84.27 \pm 1.01$ \\
$\mathrm{T_{14,p1}}$ \dotfill [h] & $-$ & ${1.02 \pm 0.19}$ \\
$\mathrm{M_{p1}}$ \dotfill [$\mathrm{M_{e}}$] & $-$ & $4.07 \pm 0.45$ \\
$\mathrm{R_{p1}}$ \dotfill [$\mathrm{R_{e}}$] & $-$ & $1.63 \pm 0.14$ \\
$\mathrm{\rho_{p1}}$ \dotfill [$\mathrm{g \, cm{-3}}$] & $-$ & $5.18 \pm 1.46$ \\
$\mathrm{T_{eq,p1}}$ \dotfill [K] & $-$ & $998.0 \pm 39.0$ \\
$\mathrm{\log Z}$ \dotfill  & $-$ & $262430.1 \pm 0.6$ \\
\hline
\end{tabular} 
} 
\end{center} 
\end{table} 

\begin{table}[pht] 
\begin{center} 
\caption{Prior and posterior planetary parameter distributions obtained with \texttt{juliet} for TOI-134, for a fit with free eccentricity. \textit{Top}: Fitted parameters. \textit{Bottom}: derived orbital parameters and physical parameters.} 
\label{tab:TOI-134-efree} 
\centering 
\resizebox{\columnwidth}{!}{%
\begin{tabular}{lll} 
\hline  \hline 
Parameter & Prior & Posterior \\ 
\hline 
$\mathrm{P_{p1}}$ \dotfill [d] & $\mathcal{N}(1.4,0.1)$ & \onethirtyfourePpone \\
$\mathrm{t_{0,p1}}$ \dotfill [BJD] & $\mathcal{N}(2459082.9,0.1)$ & \onethirtyfouretzeropone \\
$\mathrm{p_{p1}}$ \dotfill & $\mathcal{N}(0.02,0.1)$ & \onethirtyfoureppone \\
$\mathrm{b_{p1}}$ \dotfill & $\mathcal{U}(0,1)$ & \onethirtyfourebpone \\
$\mathrm{\sqrt{e}\sin\omega_{p1}}$ \dotfill & $\mathcal{U}(0,1)$ & \onethirtyfouresesinomegapone \\
$\mathrm{\sqrt{e}\cos\omega_{p1}}$ \dotfill & $\mathcal{U}(0,1)$ & \onethirtyfouresecosomegapone \\
$\mathrm{K_{p1}}$ \dotfill [$\mathrm{m \, s^{-1}}$] & $\mathcal{U}(0,10)$ & \onethirtyfoureKpone \\
$\mathrm{\rho_{}}$ \dotfill [$\mathrm{kg \, m^{-3}}$] & $\mathcal{J}(100,10000)$ & \onethirtyfourerho \\
$\mathrm{q_{1,TESS}}$ \dotfill & $\mathcal{U}(0,1)$ & \onethirtyfoureqoneTESS \\
$\mathrm{q_{2,TESS}}$ \dotfill & $\mathcal{U}(0,1)$ & \onethirtyfoureqtwoTESS \\
$\mathrm{\mu_{ESPRESSO18}}$ \dotfill [$\mathrm{m \, s^{-1}}$] & $\mathcal{U}(29700,29750)$ & \onethirtyfouremuESPRESSOeighteen \\
$\mathrm{\sigma_{w,ESPRESSO18}}$ \dotfill [$\mathrm{m \, s^{-1}}$] & $\mathcal{J}(0.001,10)$ & \onethirtyfouresigmawESPRESSOeighteen \\
$\mathrm{\mu_{ESPRESSO19}}$ \dotfill [$\mathrm{m \, s^{-1}}$] & $\mathcal{U}(29700,29750)$ & \onethirtyfouremuESPRESSOnineteen \\
$\mathrm{\sigma_{w,ESPRESSO19}}$ \dotfill [$\mathrm{m \, s^{-1}}$] & $\mathcal{J}(0.001,10)$ & \onethirtyfouresigmawESPRESSOnineteen \\
$\mathrm{\mu_{HARPS}}$ \dotfill [$\mathrm{m \, s^{-1}}$] & $\mathcal{U}(29750,29790)$ & \onethirtyfouremuHARPS \\
$\mathrm{\sigma_{w,HARPS}}$ \dotfill [$\mathrm{m \, s^{-1}}$] & $\mathcal{J}(0.001,10)$ & \onethirtyfouresigmawHARPS \\
$\mathrm{\mu_{PFS}}$ \dotfill &[$\mathrm{m \, s^{-1}}$]  $\mathcal{U}(-12,12)$ & \onethirtyfouremuPFS \\
$\mathrm{\sigma_{w,PFS}}$ \dotfill [$\mathrm{m \, s^{-1}}$] & $\mathcal{J}(0.001,10)$ & \onethirtyfouresigmawPFS \\
$\mathrm{\sigma_{GP,rv}}$ \dotfill & $\mathcal{TN}(4.87,0.93,0,1000)$ & \onethirtyfoureGPsigmarv \\
$\mathrm{\alpha_{GP,rv}}$ \dotfill & $\mathcal{TN}(0.00108,0.00015,0,1000)$ & \onethirtyfoureGPalpharv \\
$\mathrm{\Gamma_{GP,rv}}$ \dotfill & $\mathcal{TN}(2.21,0.99,0,1000)$ & \onethirtyfoureGPGammarv \\
$\mathrm{P_{rot,GP,rv}}$ \dotfill & $\mathcal{N}(30.79,0.86)$ & \onethirtyfoureGPProtrv \\
$\mathrm{m_{dilution,TESS1}}$ \dotfill & $\mathrm{fixed}$ & \onethirtyfouremdilutionTESSone \\
$\mathrm{m_{flux,TESS1}}$ \dotfill & $\mathcal{N}(0,0.1)$ & \onethirtyfouremfluxTESSone \\
$\mathrm{\sigma_{w,TESS1}}$ \dotfill & $\mathcal{J}(0.1,1000)$ & \onethirtyfouresigmawTESSone \\
$\mathrm{\sigma_{GP,TESS1}}$ \dotfill & $\mathcal{N}(25.0\times 10^{-5},2.7 \times 10^{-5})$ & \onethirtyfoureGPsigmaTESSone \\
$\mathrm{\rho_{GP,TESS1}}$ \dotfill & $\mathcal{N}(0.311,0.098)$ & \onethirtyfoureGPrhoTESSone \\
$\mathrm{m_{dilution,TESS28}}$ \dotfill & $\mathrm{fixed}$ & \onethirtyfouremdilutionTESStwentyeight \\
$\mathrm{m_{flux,TESS28}}$ \dotfill & $\mathcal{N}(0,0.1)$ & \onethirtyfouremfluxTESStwentyeight \\
$\mathrm{\sigma_{w,TESS28}}$ \dotfill & $\mathcal{J}(0.1,1000)$ & \onethirtyfouresigmawTESStwentyeight \\
$\mathrm{\sigma_{GP,TESS28}}$ \dotfill & $\mathcal{N}(54.3\times 10^{-5},6.6\times 10^{-5})$ & \onethirtyfoureGPsigmaTESStwentyeight \\
$\mathrm{\rho_{GP,TESS28}}$ \dotfill & $\mathcal{N}(0.324,0.066)$ & \onethirtyfoureGPrhoTESStwentyeight \\
$\mathrm{m_{dilution,TESS68}}$ \dotfill & $\mathrm{fixed}$ & \onethirtyfouremdilutionTESSsixtyeight \\
$\mathrm{m_{flux,TESS68}}$ \dotfill & $\mathcal{N}(0,0.1)$ & \onethirtyfouremfluxTESSsixtyeight \\
$\mathrm{\sigma_{w,TESS68}}$ \dotfill & $\mathcal{J}(0.1,1000)$ & \onethirtyfouresigmawTESSsixtyeight \\
$\mathrm{\sigma_{GP,TESS68}}$ \dotfill & $\mathcal{N}(61.6\times 10^{-5},54.0\times 10^{-5})$ & \onethirtyfoureGPsigmaTESSsixtyeight \\
$\mathrm{\rho_{GP,TESS68}}$ \dotfill & $\mathcal{N}(2.7,2.0)$ & \onethirtyfoureGPrhoTESSsixtyeight \\
\hline 
$\mathrm{e_{p1}}$ \dotfill & $-$ & \onethirtyfoureeccpone \\
$\mathrm{\omega_{p1}}$ \dotfill [$\degr$] & $-$ & \onethirtyfoureomegapone \\
$\mathrm{a_{p1}}$ \dotfill [au] & $-$ & $0.0208 \pm 0.0006$ \\
$\mathrm{i_{p1}}$ \dotfill [$\degr$] & $-$ & $85.75 \pm 1.92$ \\
$\mathrm{T_{14,p1}}$ \dotfill [h] & $-$ & $1.26 \pm 0.23$ \\
$\mathrm{M_{p1}}$ \dotfill [$\mathrm{M_{e}}$] & $-$ & $4.02 \pm 0.44$ \\
$\mathrm{R_{p1}}$ \dotfill [$\mathrm{R_{e}}$] & $-$ & $1.56 \pm 0.11$ \\
$\mathrm{\rho_{p1}}$ \dotfill [$\mathrm{g \, cm{-3}}$] & $-$ & $5.8 \pm 1.39$ \\
$\mathrm{T_{eq,p1}}$ \dotfill [K] & $-$ & $998.0 \pm 39.0$ \\
$\mathrm{\log Z}$ \dotfill  & $-$ & $262432.7 \pm 0.6$ \\
\hline
\end{tabular} 
} 
\end{center} 
\end{table}

\section{Discussion}\label{sect:discussion}

\subsection{Update to the parameters of TOI-134}
The goal of our analysis of TOI-134 was to update and refine the planetary parameters from \cite{Astudillo-Defru2020} (henceforward AD20) and \cite{Patel2022} (henceforward PE22) with the more accurate ESPRESSO RVs, as well as additional \textit{TESS} sectors (two compared to AD20, one compared to PE22). Table \ref{tab:TOI-134-params-comp} lists the main orbital and physical parameters we compare. While all the reported orbital periods are equivalent at the $1\sigma$ level, we improve the precision on the period by a factor of $\sim$200 compared to AD20. We also constrain the orbit to circularity. Regarding the physical parameters of the planet, we find a slightly lower mass, with improved precision, though it remains compatible with the value of AD20 at the $2\sigma$ level. We also find a slightly larger radius ratio than both AD20 and PE22. The larger radius and lower mass combined lead to a notably lower density for TOI-134~b than that reported by AD20. We note that this is robust to the stellar parameter determination; our stellar mass and radius are compatible with those of AD20 within error bars, and computing the planetary mass and radius with radius and mass with their stellar parameters leads to an even lower planetary density.

\begin{table}[htb]
\begin{center}
\caption{Parameter comparison for TOI-134.} 
\label{tab:TOI-134-params-comp} 
\centering 
\resizebox{\columnwidth}{!}{%
\begin{tabular}{llll} 
\hline  \hline 
Parameter & AD20 & PE22 & this work \\ 
\hline 
P \dotfill [d] & $1.40150\pm 0.00018$ & $1.4015272^{+0.0000011}_{-0.0000012}$ & \onethirtyfourcPpone \\
$\mathrm{e_{p1}}$ \dotfill & $< 0.21$ & $-$ & \onethirtyfourceccpone \\
$\mathrm{M_{p1}}$ \dotfill [$\mathrm{M_\oplus}$] & $4.60\pm 0.56$ & $-$ & $4.07 \pm 0.45$ \\
$\mathrm{p_{p1}}$ \dotfill & $0.0212\pm 0.0010$ & $0.0233\pm0.0007$ & \onethirtyfourcppone \\
$\mathrm{R_{p1}}$ \dotfill [$\mathrm{R_\oplus}$] & $1.39\pm 0.09$ & $-$ & $1.63 \pm 0.14$ \\
$\mathrm{\rho_{p1}}$ \dotfill [$\mathrm{g \, cm^{-3}}$] & $9.60^{+2.40}_{-1.80}$ & $-$ & $5.18 \pm 1.46$ \\
\hline
\end{tabular}
}
\end{center} 
\end{table}

\subsection{Planet composition}

\begin{figure*}[htb]
    \centering
    \includegraphics[width=\textwidth]{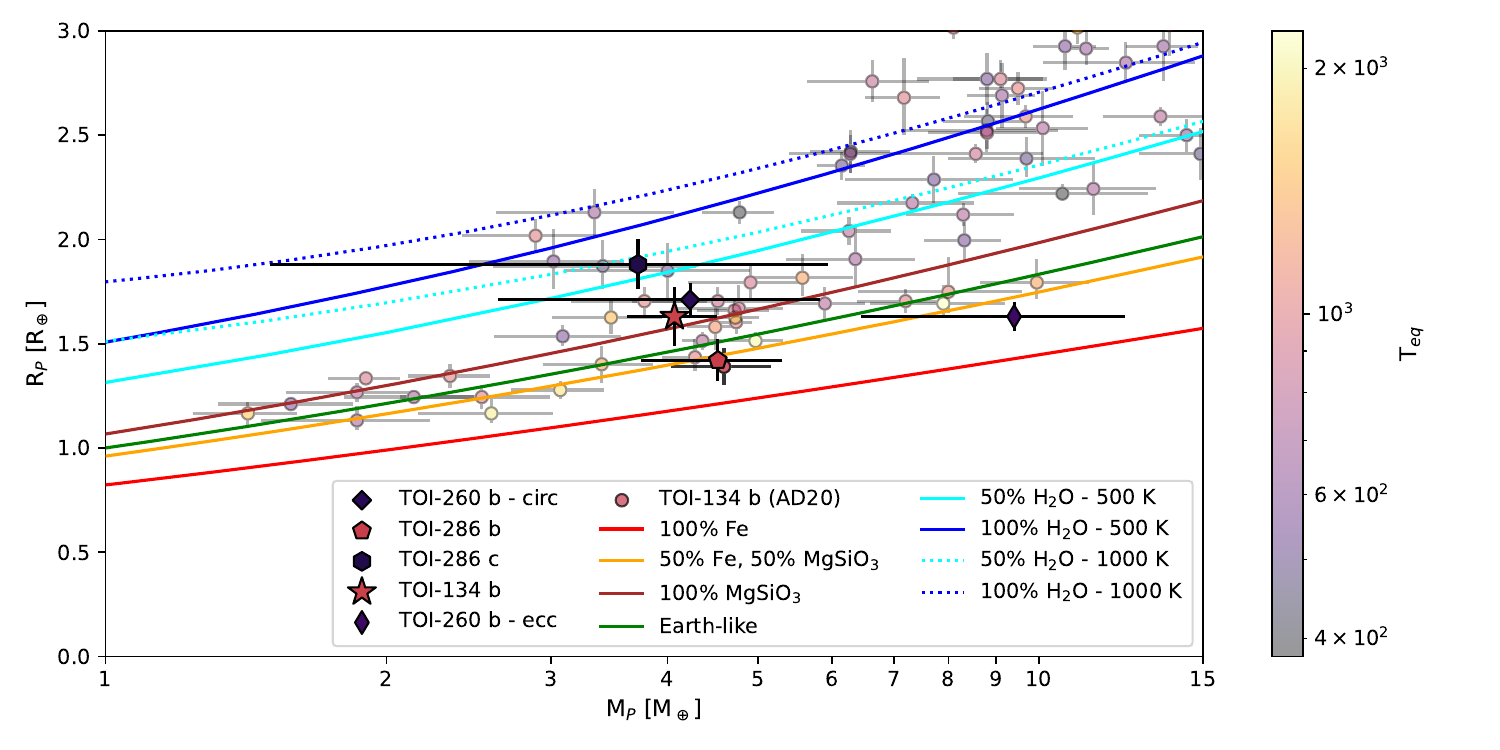}
    \caption{Mass-radius diagram for all planets in the PlanetS catalog with $\mathrm{R_p < 3\, R_\oplus}$ and $\mathrm{M_p < 15\, M_\oplus}$, well-constrained masses and radii (5\% error on radius, 25\% error on mass), hosted by early M and K stars ($\mathrm{0.35\, M_\odot < M_s < 0.9\, M_\odot}$). The points are coloured by $\mathrm{T_{eq}}$. The four planets presented in this paper are highlighted. We also show the eccentric solution for TOI-260~b, and the previous solution from AD20 for TOI-134~b (overlapped by TOI-286~b). The coloured curves show the composition models of \cite{Zeng2019}.}
    \label{fig:mass-radius-teq}
\end{figure*}

Knowing both the mass and radius of a planet allows us to compute its density, and compare it to composition models. Fig. \ref{fig:mass-radius-teq} shows the four planets presented in this paper, in the context of both compositional models and the population of well-characterized small, low-mass planets around similar stellar hosts. The planets are taken from the PlanetS catalog.
The compositional models are from \cite{Zeng2019}\footnote{Available at \url{https://lweb.cfa.harvard.edu/~lzeng/planetmodels.html}, accessed 14 February 2024.}.

For TOI-260~b, considering its $4.23 \pm 1.60 \, \mathrm{M_\oplus}$ mass and $1.71 \pm 0.08 \, \mathrm{R_\oplus}$ radius from the circular model, the most likely composition is between a pure rock planet, and a rocky core with a 50\% condensed water layer. However, the large uncertainty on the mass (primarily due to the large relative uncertainty on the RV semi-amplitude) makes a clear determination challenging. The uncertainty on the eccentricity is also important here; if the high-eccentricity model is actually correct, we obtain a much higher mass. In this case, the most likely composition would be between a pure iron and a half-iron, half-silicates model.

The two planets orbiting TOI-286 have similar masses, but notably different radii, leading to different probable compositions. The inner, hotter, smaller planet TOI-286~b, with a $4.53 \pm 0.78 \, \mathrm{M_\oplus}$ mass and $1.42 \pm 0.10\, \mathrm{R_\oplus}$ radius, is likely close to a 50\% iron core and 50\% silicate mantle, or an Earth-like composition. The cooler, outer, larger planet TOI-286~c should have a higher volatile content considering its $3.72 \pm 2.22 \, \mathrm{M_\oplus}$ mass and $1.88 \pm 0.12\, \mathrm{R_\oplus}$ radius, and most likely has a significant water layer. Considering their bulk densities (normalized by the scaled Earth bulk density), these planets likewise sit on either side of the density gap proposed by \cite{Luque2022}, with TOI-286~b being consistent with the rocky planet population, and TOI-286~c with the water worlds. This in turn suggests that TOI-286~b likely formed within the snow line, while TOI-286~c would have formed beyond it and migrated inwards \citep[][and references therein]{Luque2022}. 

Our updated parameters for TOI-134~b suggest a pure rock composition. This contrasts with the results of AD20, who obtained a density more compatible with a 50\% iron core and 50\% silicate mantle.

Two of our planets - TOI-260~b and TOI-286~c - fall within the Fulton radius gap or radius valley \citep{Fulton2017}, with radii between $\mathrm{1.5 \, R_\oplus}$ and $\mathrm{2 \, R_\oplus}$. The precise location of the radius valley has been shown to depend on orbital period \citep{Martinez2019} and stellar type \citep{Cloutier2020, Venturini2024}. Using the period-radius relation of \cite{Cloutier2020}, we find the centre of the valley at $\mathrm{1.64 \, R_\oplus}$ at the period of TOI-260~b, and at $\mathrm{1.70 \, R_\oplus}$ at the period of TOI-286~c. TOI-260~b, with a radius of $\mathrm{1.71 \pm 0.08 \, R_\oplus}$, is thus close to the centre of the valley, while TOI-286~c, with a radius of $1.88 \pm 0.12\, \mathrm{R_\oplus}$, is close to the upper border. Likewise, using the stellar mass-planet radius relation of \cite{Venturini2024}, we find the centre of the valley at $\mathrm{1.70 \, R_\oplus}$ at the stellar mass of TOI-260, and at $\mathrm{1.77 \, R_\oplus}$ at the stellar mass of TOI-286. Again, this places TOI-260~b in the centre of the valley, and TOI-286~c towards the upper part. The normalised bulk density of TOI-260~b suggests it is most likely a rocky world formed within the snow line, potentially in situ \citep{Luque2022, Burn2024}. However, given the error bars on the mass and the uncertainty regarding the eccentricity, we cannot exclude a water-rich composition for this planet, and thus a formation beyond the snow line. Meanwhile, given both its location in the upper part of the radius valley and its bulk density, TOI-286~c is likely water-rich. The TOI-286 system thus has one rocky planet below the radius gap and one water-rich planet close to the upper border orbiting the same star. This is a similar configuration to the TOI-238 system, also consisting of two planets - an inner rocky planet and an outer water-rich planet - orbiting a K dwarf \citep{Suarez2024}. Systems such as these are highly valuable probes of the radius gap.

\begin{figure*}
    \centering
    \includegraphics[width=.4\textwidth]{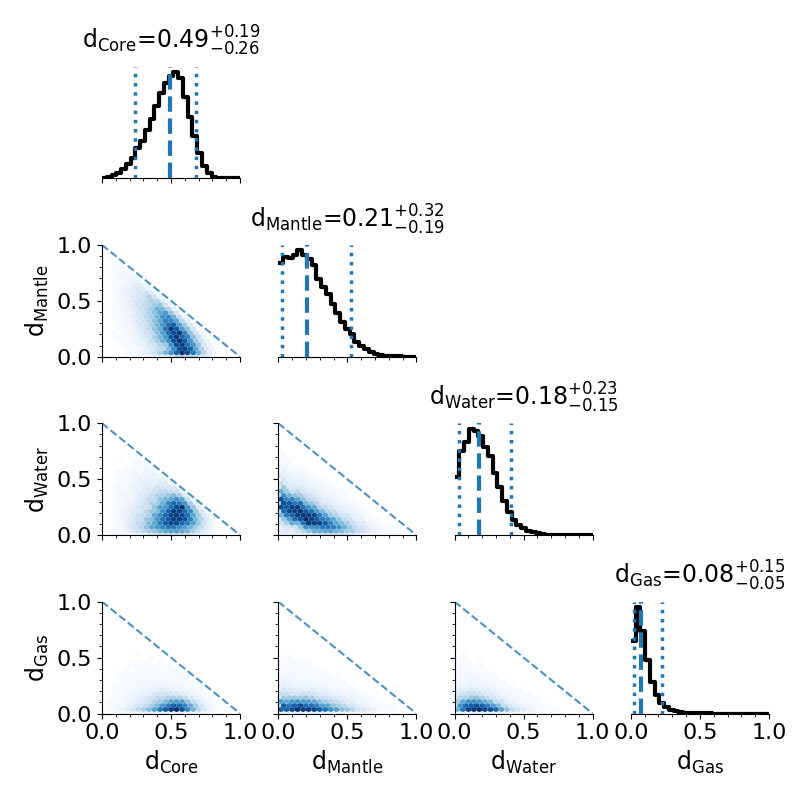}
    \includegraphics[width=.4\textwidth]{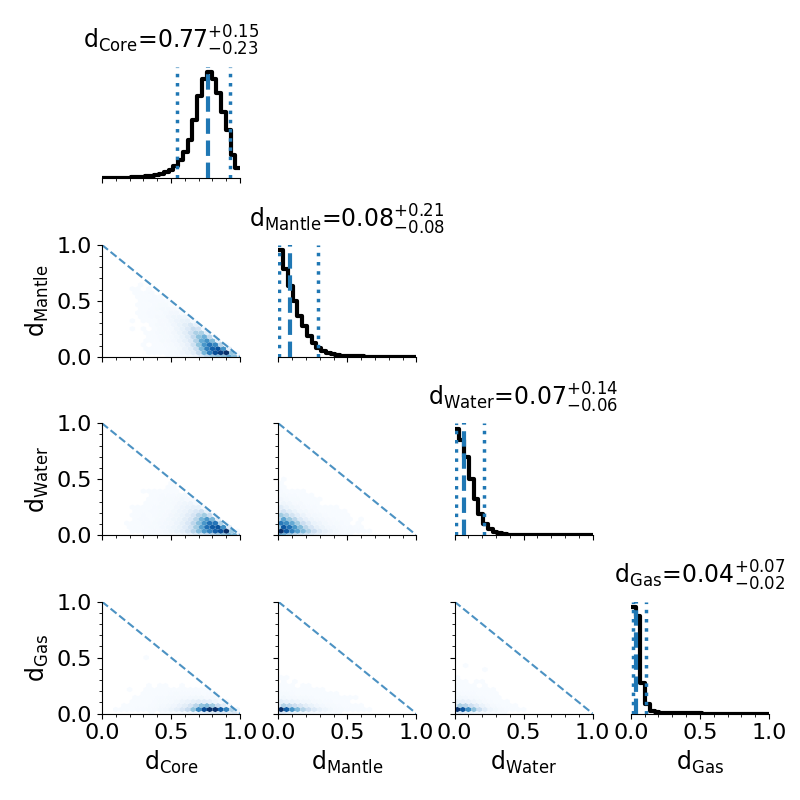}
    \includegraphics[width=.4\textwidth]{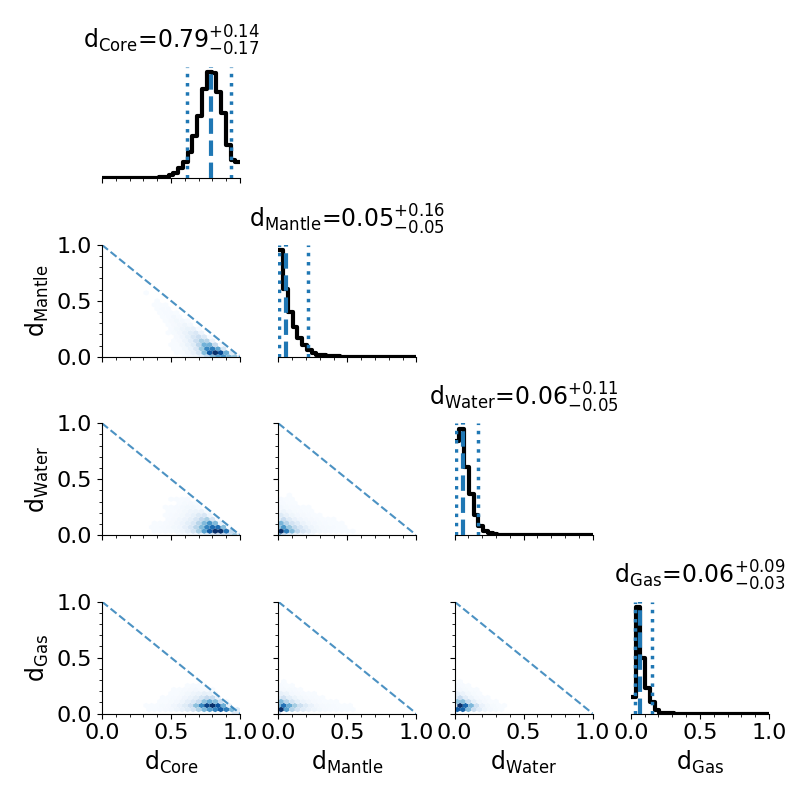}
    \includegraphics[width=.4\textwidth]{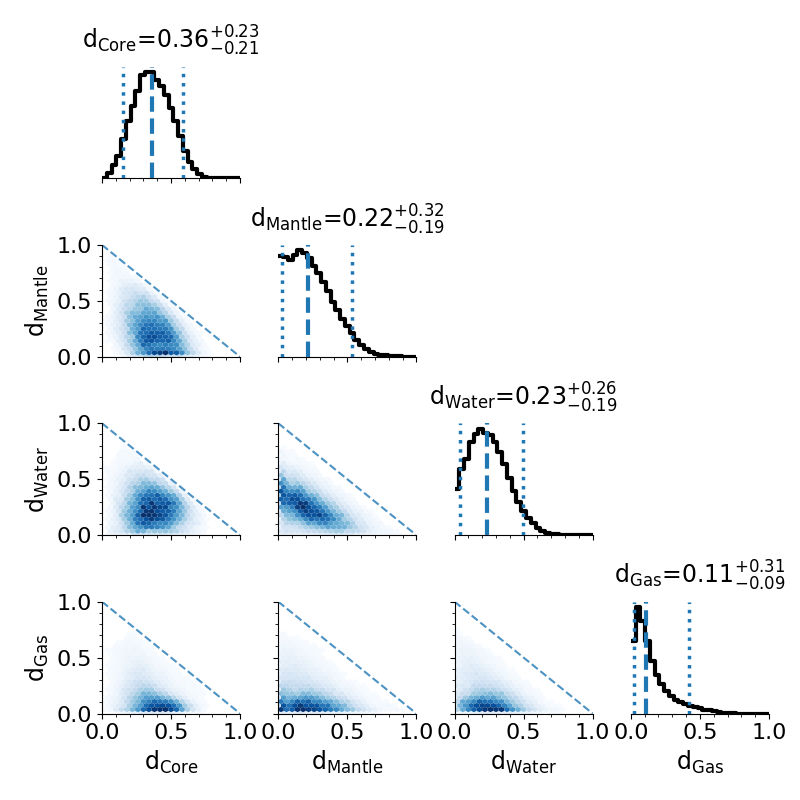}
    \includegraphics[width=.4\textwidth]{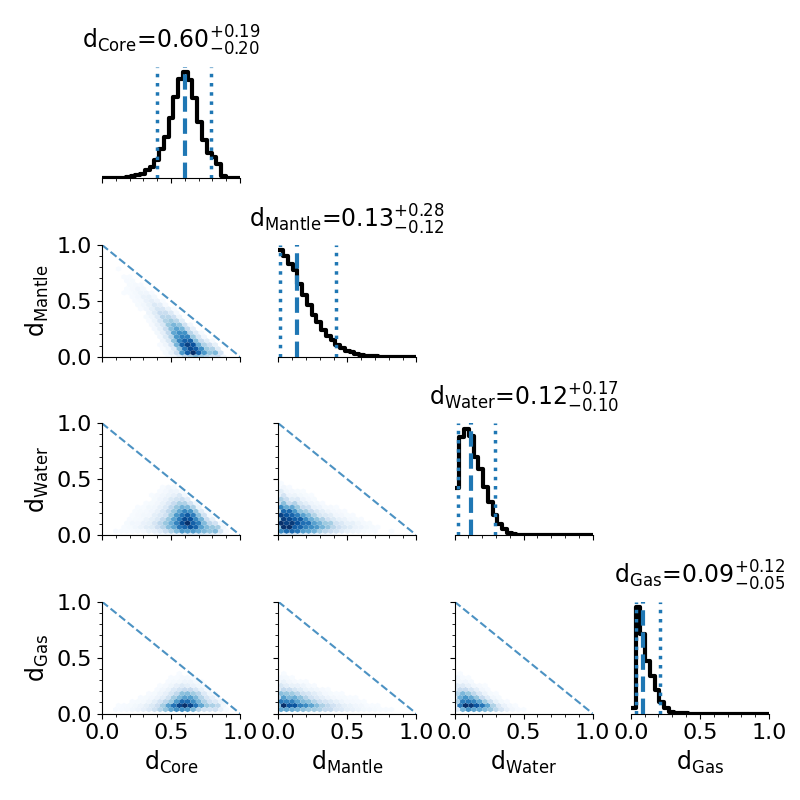}
    \includegraphics[width=.4\textwidth]{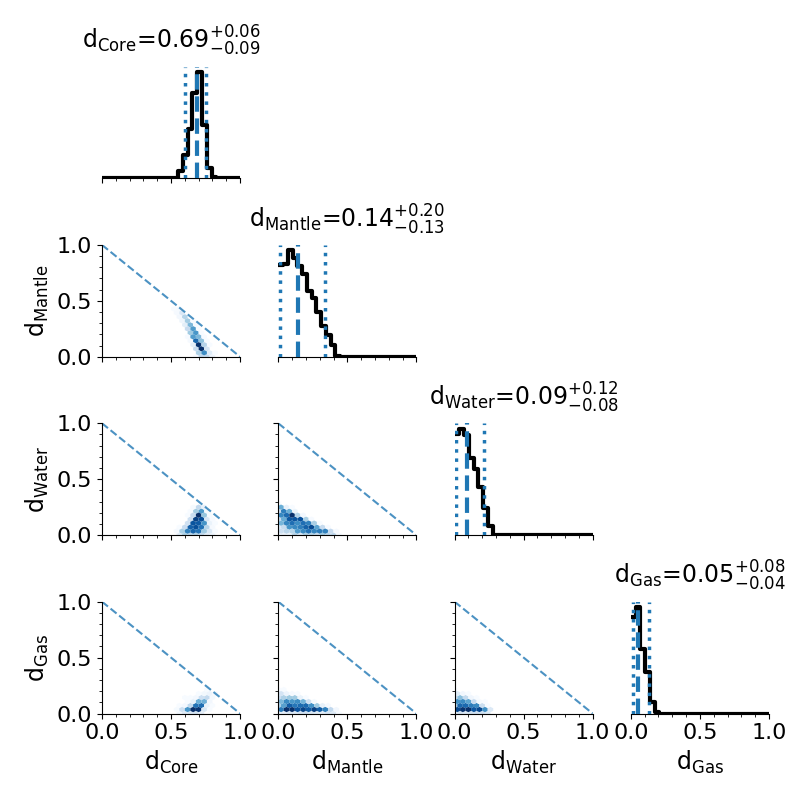}
    \caption{Cornerplots of the radius fractions of the four-layer interior structure models for (top to bottom, left to right): the circular model for TOI-260 b, the eccentric model for TOI-260 b, TOI-286 b, TOI-286 c,  TOI-134 b, and Earth for comparison.}
    \label{fig:interior-structure-models}
\end{figure*}

In addition to comparing with theoretical composition curves, we can also use interior inference models to characterize the interior structure of our planets. We employ the ExoMDN code \citep{Baumeister2023}\footnote{Available at \url{https://github.com/philippbaumeister/ExoMDN}} to model our planets using a four-layer model consisting of an iron core, a silicate mantle, a water layer, and a H/He atmosphere. The inputs to ExoMDN are the planet's mass, radius, and equilibrium temperature. We include both the circular and eccentric models for TOI-260~b. The resulting models are shown in Fig. \ref{fig:interior-structure-models}, together with a model of the Earth for comparison. The circular model for TOI-260~b has higher mantle and water radius fractions compared to the eccentric model, which is more similar to the Earth model proportions, though with a larger core fraction and lower mantle fraction. Regarding the TOI-286 system, TOI-286~b is strongly core-dominated, while TOI-286~c has a significant water radius fraction. Finally, the proportions of the model for TOI-134~b are similar to that of the Earth model.   

\subsection{Orbital parameters}

\begin{figure}[htb]
    \centering
    \includegraphics[width=\columnwidth]{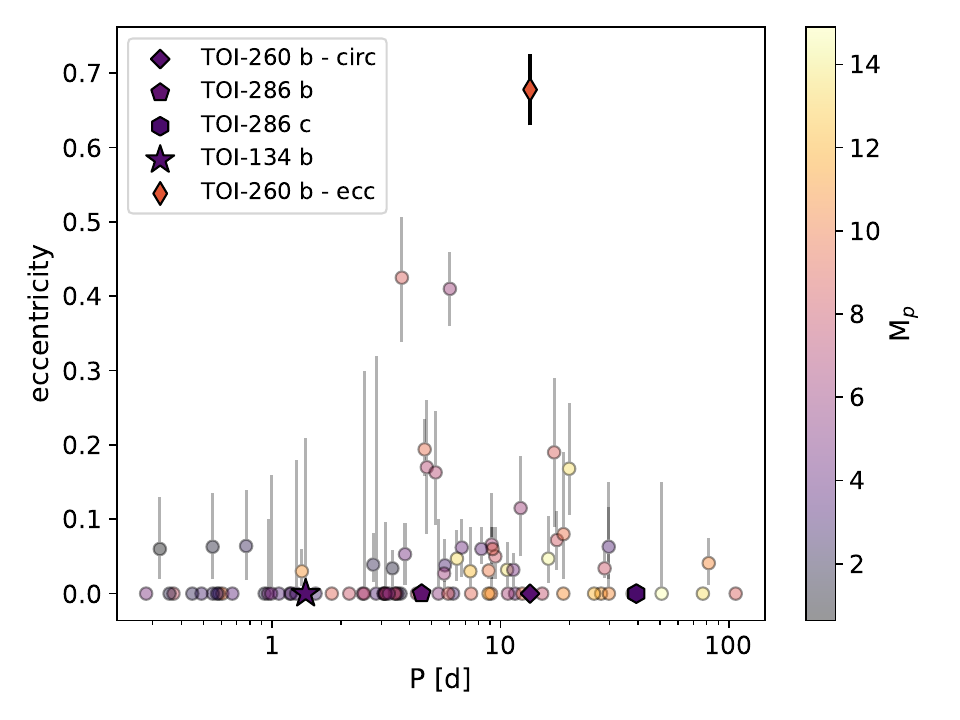}
    \caption{Period-eccentricity diagram for the same population of planets as Fig. \ref{fig:mass-radius-teq}. The points are coloured by planet mass. The four planets presented in this paper are highlighted. We also show the eccentric solution for TOI-260~b.}
    \label{fig:per-ecc-pmass}
\end{figure}

To place our planets in a population context in regard to orbital parameters, we plot in Fig. \ref{fig:per-ecc-pmass} the period-eccentricity distribution of the small, well-characterized planets around low-mass stellar hosts. Of these planets, slightly under half (45\%) have tabulated eccentricities of 0, as is the case for the planets presented here. However, the majority have low eccentricities, with 86\% being below 0.1. The relatively few planets with higher eccentricities are preferentially found at longer periods above $\mathrm{\approx 3\, d}$.  If the eccentric model for TOI-260~b should prove to be correct, it would be by far the most eccentric such planet known. 

The prevailing low eccentricities of this population contrast with the nonzero eccentricities found for warm Neptunes by \cite{Correia2020}. These raised eccentricities are suggested to be a product of the same combination of high-eccentricity migration and atmospheric evaporation that sculpts the hot Neptune desert \citep{Owen2018}, where atmospheric evaporation could excite the eccentricities through thermal tidal torque. Regarding the high-eccentricity model for TOI-260 b, since both mechanisms produce similar lower boundaries for the desert, we cannot exclude either of them as a potential originator.

All our planets are among the lower-mass part of the population. TOI-286~c, in particular, is the lowest-mass planet of this population at its orbital period. 

\subsection{Atmospheric characterization potential}

To assess the potential of these planets for atmospheric characterization, we computed the Transmission Spectroscopy Metric (TSM) and Emission Spectroscopy Metric (ESM) as defined by \cite{Kempton2018}. For TOI-260 b we obtain a TSM of $\sim 70.5$ and an ESM of $\sim 3.1$. For TOI-286 b and c we find TSM values of $\sim 4.1$ and $\sim 37.4$ respectively, and ESM values of $\sim 2.9$ and $\sim 0.7$ respectively. Finally, for TOI-134 b we obtain a TSM of $\sim 94.7$ and an ESM of $\sim 13.8$. Regarding transmission spectroscopy, none of the newly confirmed planets reach the thresholds proposed by \cite{Kempton2018} of 10 for small planets ($\mathrm{R_p \leq 1.5 R_\oplus}$) or 90 for larger planets. TOI-134 b, however, surpasses the threshold, making it a promising candidate for transmission spectroscopy. This contrasts with the low TSM computed by \cite{Astudillo-Defru2020}; the difference is primarily due to the larger radius we find. Regarding emission spectroscopy, once again the newly confirmed planets are below the 7.5 threshold of \cite{Kempton2018}. On the other hand, TOI-134 b surpasses it and is thus also a promising candidate for characterization in emission, as was previously noted by \cite{Astudillo-Defru2020}.
 
\section{Conclusions}\label{sect:conclusions}

We have presented the confirmation and characterization of three new super-Earths: 
\begin{itemize}
    \item TOI-260~b, the only known planet around TOI-260, with an orbital period of \twosixtycgpsepPpone d. It has a $4.23 \pm 1.60 \, \mathrm{M_\oplus}$ mass and a $1.71 \pm 0.08 \, \mathrm{R_\oplus}$ radius, suggesting a rocky composition with a water layer. The eccentricity of this planet is poorly constrained; further observations would be needed to better evaluate it.
    \item TOI-286~b, a \twoeightysixtwopcPpone d planet with a $4.53 \pm 0.78 \, \mathrm{M_\oplus}$ mass and $1.42 \pm 0.10\, \mathrm{R_\oplus}$ radius, suggesting an earth-like composition and thus a formation within the snowline.
    \item TOI-286~c, a \twoeightysixtwopcPptwo d planet with a $3.72 \pm 2.22 \, \mathrm{M_\oplus}$ mass and $1.88 \pm 0.12\, \mathrm{R_\oplus}$ radius, suggesting a significant water layer and therefore a formation outside the snowline. 
\end{itemize}

We also updated and refined the parameters of TOI-134~b, finding a lower mass and larger radius compared to the literature, that point to a rocky composition.

\begin{acknowledgements}
We thank the referee for their helpful comments and careful reading that improved this manuscript. 
We thank the Swiss National Science Foundation (SNSF) and the Geneva University for their continuous support to our planet low-mass companion search programs. This work has been carried out within the framework of the National Centre of Competence in Research PlanetS supported by the Swiss National Science Foundation. 
This publication makes use of The Data \& Analysis Center for Exoplanets (DACE), which is a facility based at the University of Geneva (CH) dedicated to extrasolar planets data visualisation, exchange and analysis. DACE is a platform of the Swiss National Centre of Competence in Research (NCCR) PlanetS, federating the Swiss expertise in Exoplanet research. The DACE platform is available at \url{https://dace.unige.ch}.
This work made use of \texttt{tpfplotter} by J. Lillo-Box (publicly available in www.github.com/jlillo/tpfplotter), which also made use of the python packages \texttt{astropy}, \texttt{lightkurve}, \texttt{matplotlib} and \texttt{numpy}.
This research has made use of the Exoplanet Follow-up Observation Program (ExoFOP; DOI: 10.26134/ExoFOP5) website, which is operated by the California Institute of Technology, under contract with the National Aeronautics and Space Administration under the Exoplanet Exploration Program.
Based on observations carried out at the European Southern Observatory (ESO; La Silla, Chile) using the 3.6m telescope, under ESO programmes 1102.C-0923(A), 1102.C-0249(A), and 60.A-9700(G).
The observations in the paper made use of the NN-EXPLORE Exoplanet and Stellar Speckle Imager (NESSI). NESSI was funded by the NASA Exoplanet Exploration Program and the NASA Ames Research Center. NESSI was built at the Ames Research Center by Steve B. Howell, Nic Scott, Elliott P. Horch, and Emmett Quigley. The authors are honored to be permitted to conduct observations on Iolkam Du'ag (Kitt Peak), a mountain within the Tohono O'odham Nation with particular significance to the Tohono O'odham people. 
Some of the observations in this paper made use of the High-Resolution Imaging instruments ‘Alopeke and Zorro and were obtained under Gemini LLP Proposal Number: GN/S-2021A-LP-105. ‘Alopeke and Zorro were funded by the NASA Exoplanet Exploration Program and built at the NASA Ames Research Center by Steve B. Howell, Nic Scott, Elliott P. Horch, and Emmett Quigley. Both instruments were mounted on the Gemini North/South telescopes of the international Gemini Observatory, a program of NSF’s OIR Lab, which is managed by the Association of Universities for Research in Astronomy (AURA) under a cooperative agreement with the National Science Foundation. on behalf of the Gemini partnership: the National Science Foundation (United States), National Research Council (Canada), Agencia Nacional de Investigación y Desarrollo (Chile), Ministerio de Ciencia, Tecnología e Innovación (Argentina), Ministério da Ciência, Tecnologia, Inovações e Comunicações (Brazil), and Korea Astronomy and Space Science Institute (Republic of Korea).

Funding for the TESS mission is provided by NASA's Science Mission Directorate. KAC acknowledges support from the TESS mission via subaward s3449 from MIT.
This paper made use of data collected by the TESS mission and are publicly available from the Mikulski Archive for Space Telescopes (MAST) operated by the Space Telescope Science Institute (STScI). 
We acknowledge the use of public TESS data from pipelines at the TESS Science Office and at the TESS Science Processing Operations Center. 
Resources supporting this work were provided by the NASA High-End Computing (HEC) Program through the NASA Advanced Supercomputing (NAS) Division at Ames Research Center for the production of the SPOC data products.

This work makes use of observations from the LCOGT network. Part of the LCOGT telescope time was granted by NOIRLab through the Mid-Scale Innovations Program (MSIP). MSIP is funded by NSF.

JIGH, RR, CAP and ASM acknowledge financial support from the Spanish Ministry of Science and Innovation (MICINN) project PID2020-117493GB-I00.
This work was financed by Portuguese funds through FCT (Funda\c c\~ao para a Ci\^encia e a Tecnologia) in the framework of the project 2022.04048.PTDC (Phi in the Sky, DOI 10.54499/2022.04048.PTDC). CJM also acknowledges FCT and POCH/FSE (EC) support through Investigador FCT Contract 2021.01214.CEECIND/CP1658/CT0001.
We acknowledge financial support from the Agencia Estatal de Investigaci\'on of the Ministerio de Ciencia e Innovaci\'on MCIN/AEI/10.13039/501100011033 and the ERDF “A way of making Europe” through project PID2021-125627OB-C32, and from the Centre of Excellence “Severo Ochoa” award to the Instituto de Astrofisica de Canarias.
 FPE and CLO would like to acknowledge the Swiss National Science Foundation (SNSF) for supporting research with ESPRESSO through the SNSF grants nr. 140649, 152721, 166227, 184618 and 215190. The ESPRESSO Instrument Project was partially funded through SNSF’s FLARE Programme for large infrastructures.
 Funded/Co-funded by the European Union (ERC, FIERCE, 101052347). Views and opinions expressed are however those of the author(s) only and do not necessarily reflect those of the European Union or the European Research Council. Neither the European Union nor the granting authority can be held responsible for them. This work was supported by FCT - Fundação para a Ciência e a Tecnologia through national funds and by FEDER through COMPETE2020 - Programa Operacional Competitividade e Internacionalização by these grants: UIDB/04434/2020; UIDP/04434/2020.
 S.G.S acknowledges the support from FCT through Investigador FCT contract nr. CEECIND/00826/2018 and  POPH/FSE (EC).
This work has made use of data from the European Space Agency (ESA) mission {\it Gaia} (\url{https://www.cosmos.esa.int/gaia}), processed by the {\it Gaia} Data Processing and Analysis Consortium (DPAC, \url{https://www.cosmos.esa.int/web/gaia/dpac/consortium}). Funding for the DPAC has been provided by national institutions, in particular the institutions participating in the {\it Gaia} Multilateral Agreement.
A.S.M. acknowledges financial support from the Spanish Ministry of Science and Innovation (MICINN) project PID2020-117493GB-I00 and from the Government of the Canary Islands project ProID2020010129.
A.C.-G. and J.L.-B. are funded by the Spanish Ministry of Science through MCIN/AEI/10.13039/501100011033 grants PID2019-107061GB-C61 and CNS2023-144309.
This research was funded in part by the UKRI (Grants ST/X001121/1, EP/X027562/1).
This work has been carried out within the framework of the NCCR PlanetS supported by the Swiss National Science Foundation under grants 51NF40\_182901 and 51NF40\_205606. X.D acknowledges the support from the European Research Council (ERC) under the European Union’s Horizon 2020 research and innovation programme (grant agreement SCORE No 851555) and from the Swiss National Science Foundation under the grant SPECTRE (No 200021\_215200).
\end{acknowledgements}

\bibliographystyle{aa} 
\bibliography{biblio} 

\appendix

\section{\textit{TESS} pixel file plots}\label{ap:tpfplotter}

In this appendix we show the \textit{TESS} pixel file plots for TOI-260 and TOI-286, for all observed sectors save the first sector for each target, which are shown in Figure \ref{fig:tpfplotter}.

\begin{figure*}
    \centering
    \includegraphics[width=.45\textwidth]{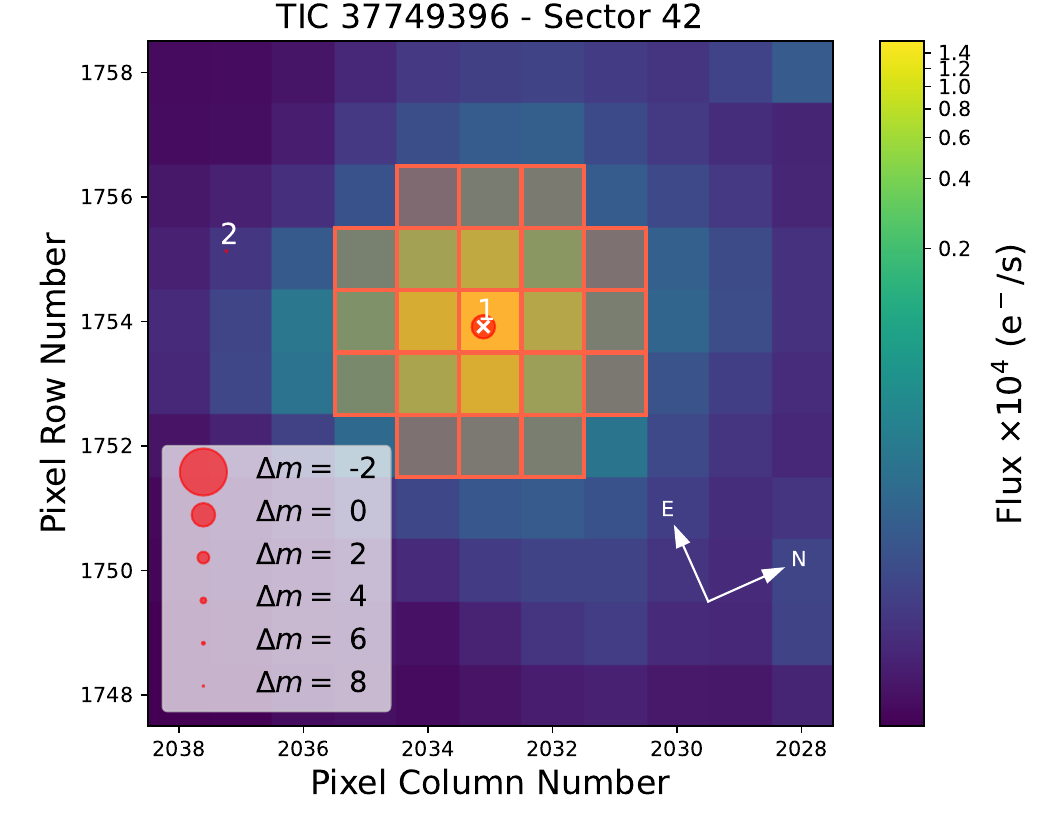}
    \includegraphics[width=.45\textwidth]{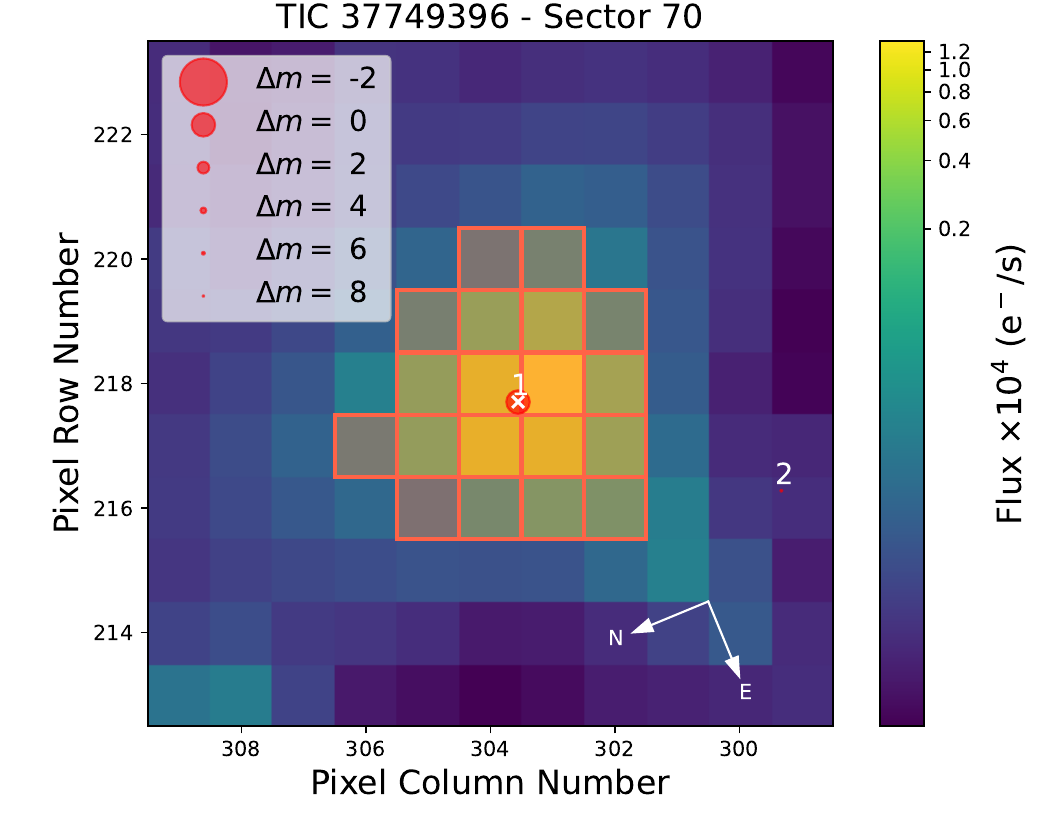}
    \caption{\textit{TESS} target pixel files for TOI-260 for sector 42 (left) and 70 (right). The target star is labelled as 1 and marked by a white cross in each case. All sources from the Gaia DR3 catalogue down to a magnitude contrast of 8 are shown as red circles, with the size proportional to the contrast. The SPOC pipeline aperture is overplotted in shaded red squares.}
    \label{fig:tpfplotter-toi-260-ap}
\end{figure*}
\begin{figure*}
    \centering
    \includegraphics[width=.33\textwidth]{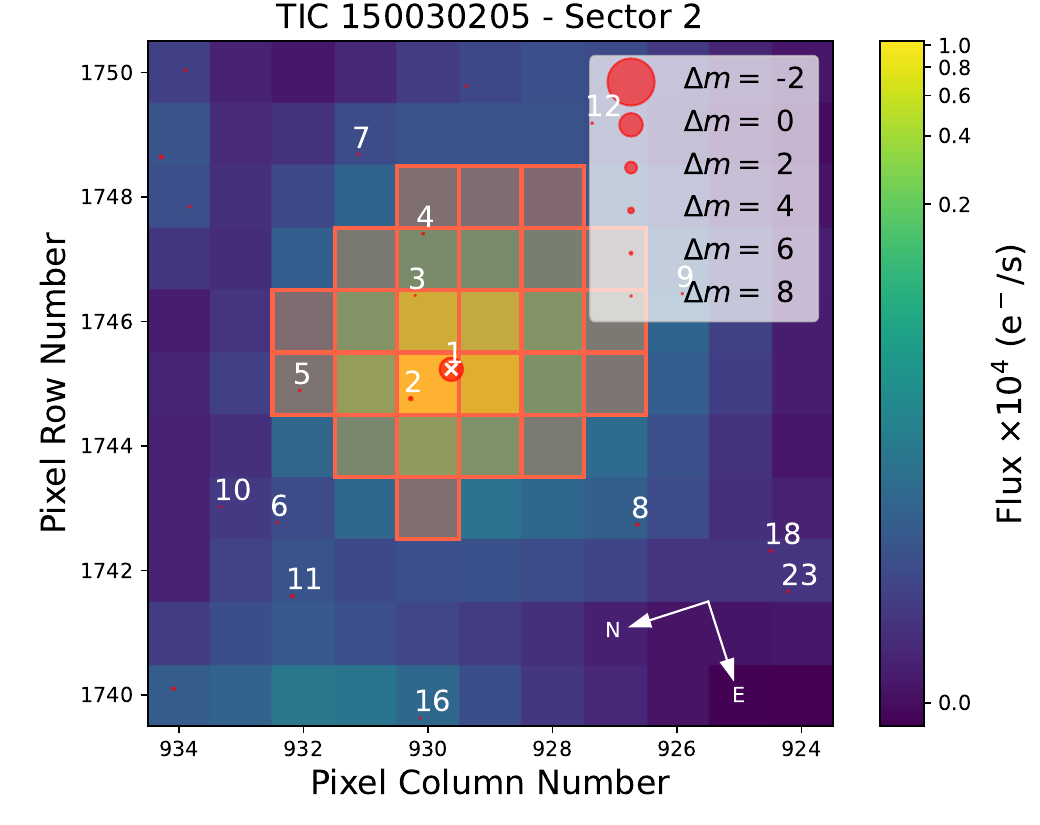}
    \includegraphics[width=.33\textwidth]{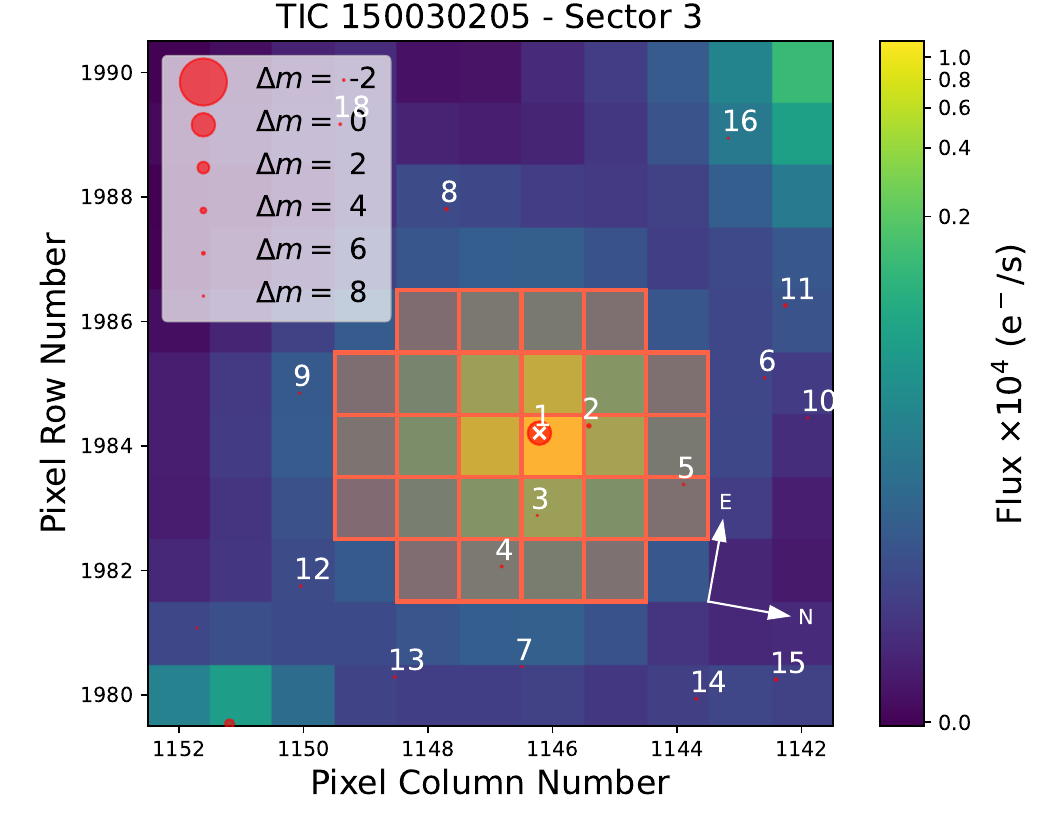}
    \includegraphics[width=.33\textwidth]{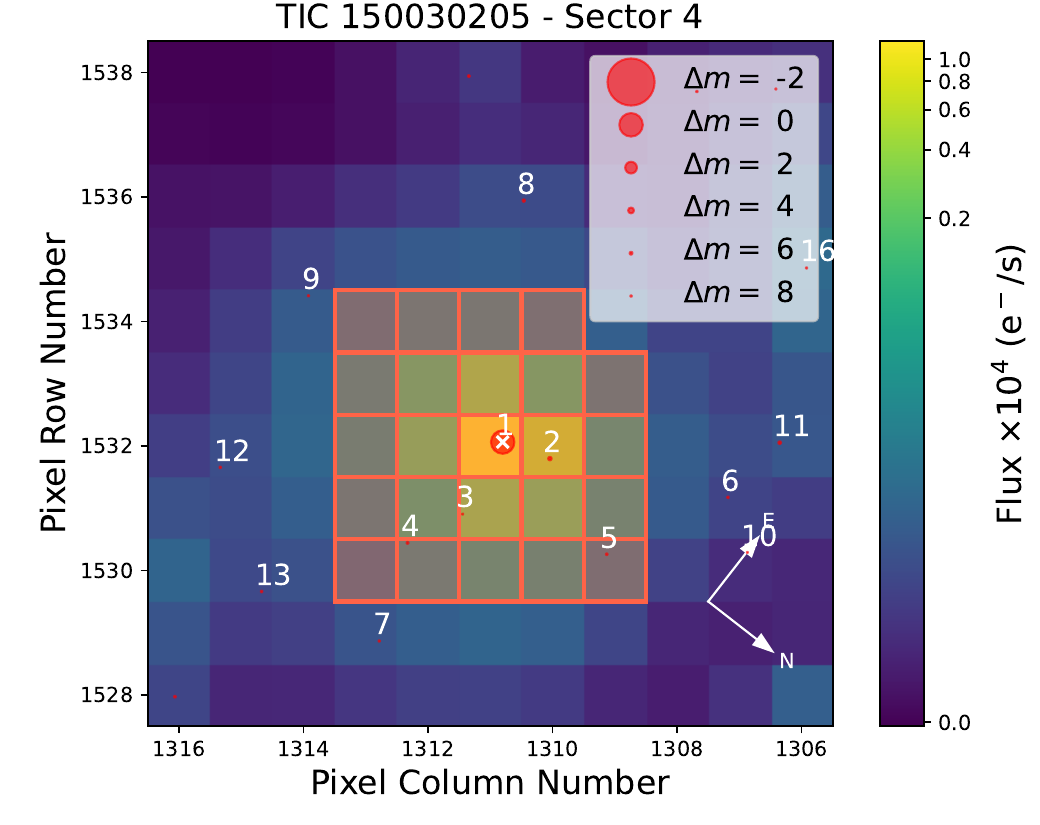}
    \includegraphics[width=.33\textwidth]{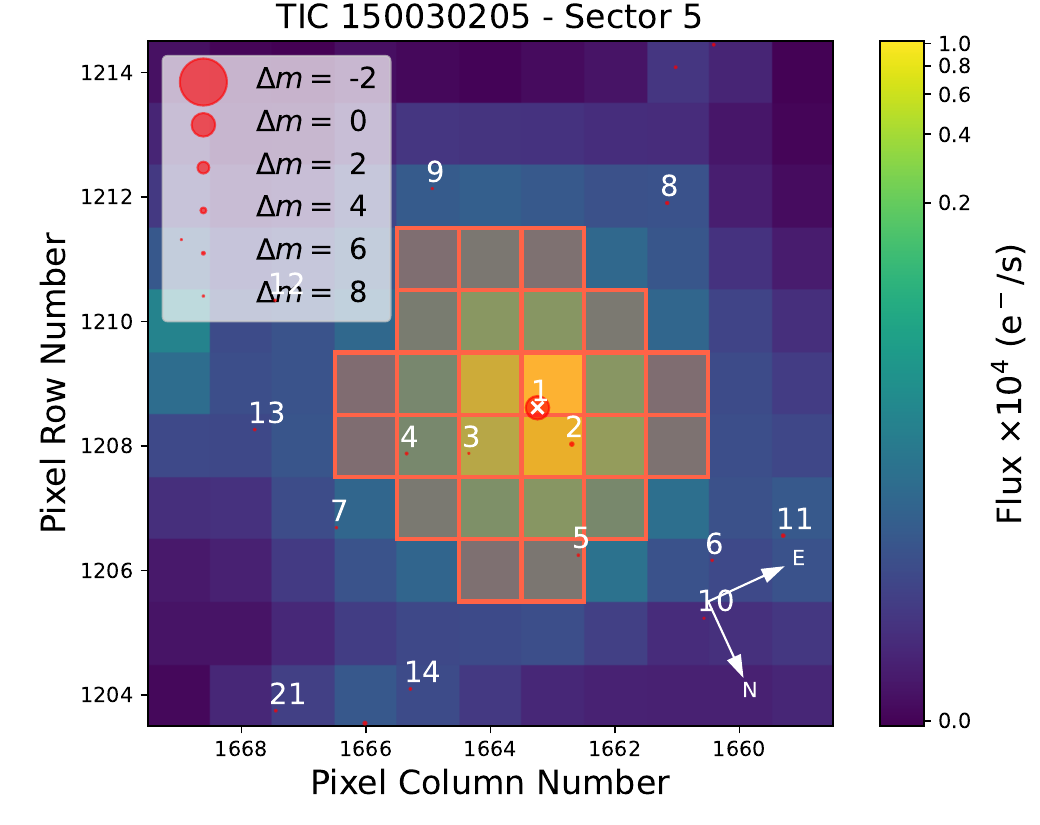}
    \includegraphics[width=.33\textwidth]{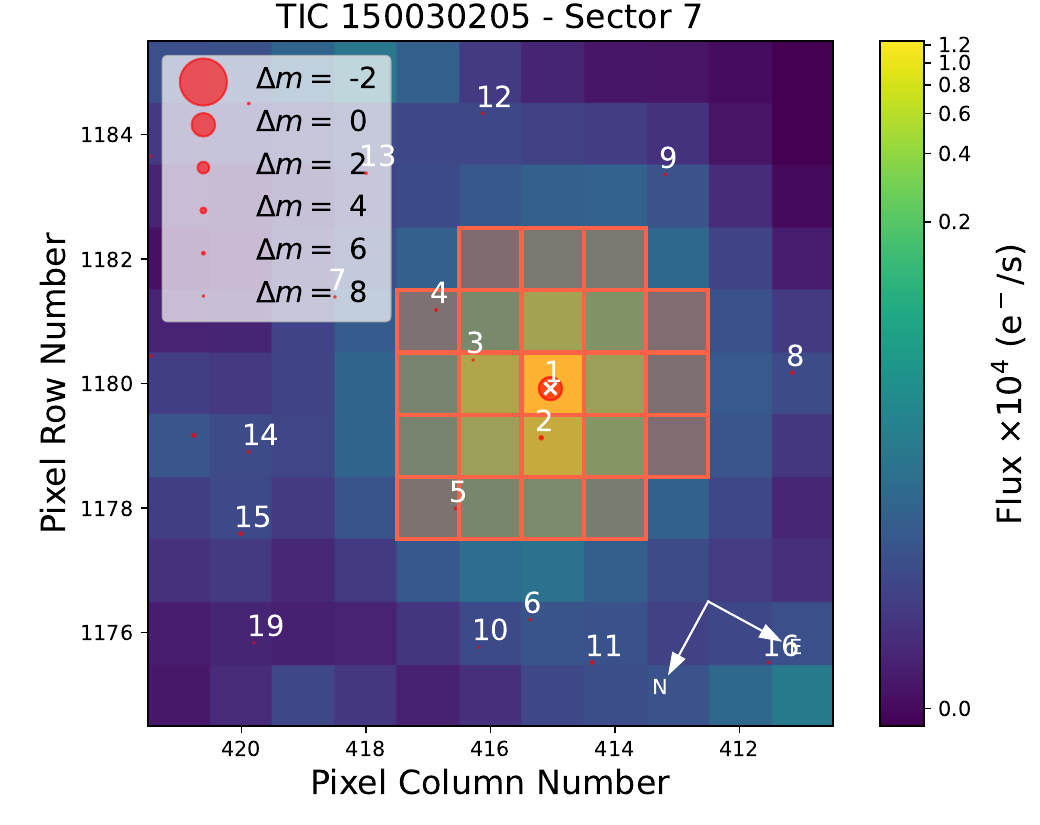}
    \includegraphics[width=.33\textwidth]{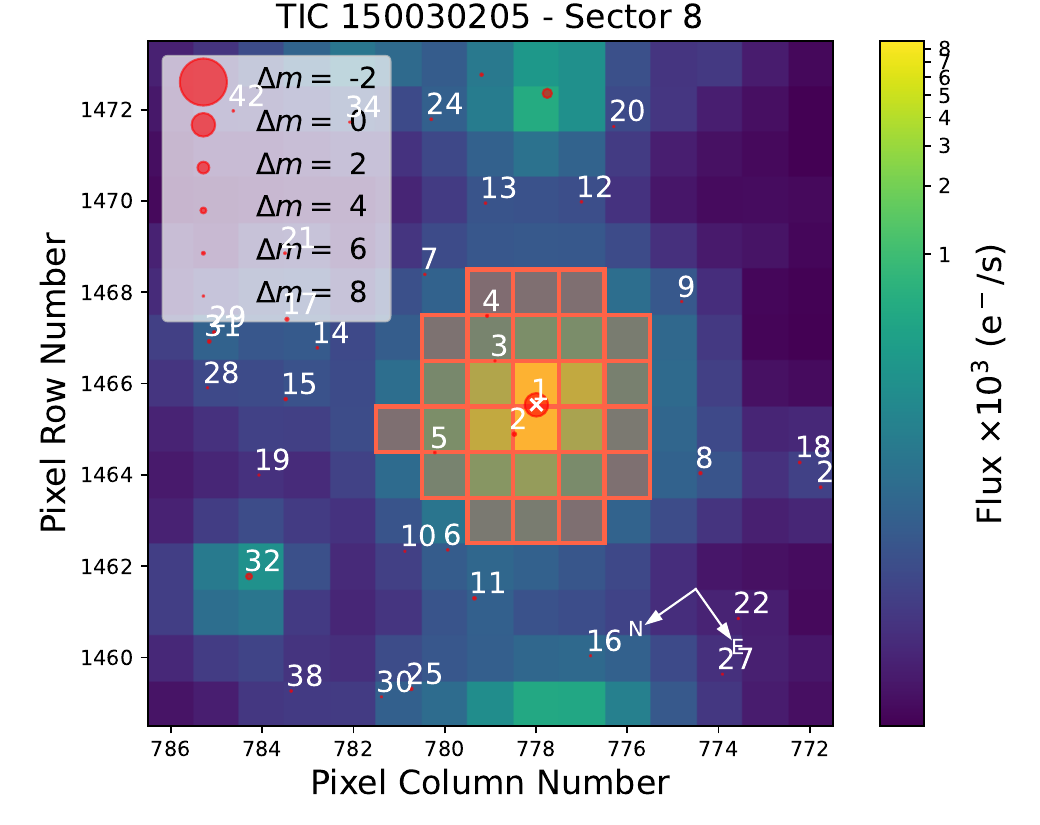}
    \includegraphics[width=.33\textwidth]{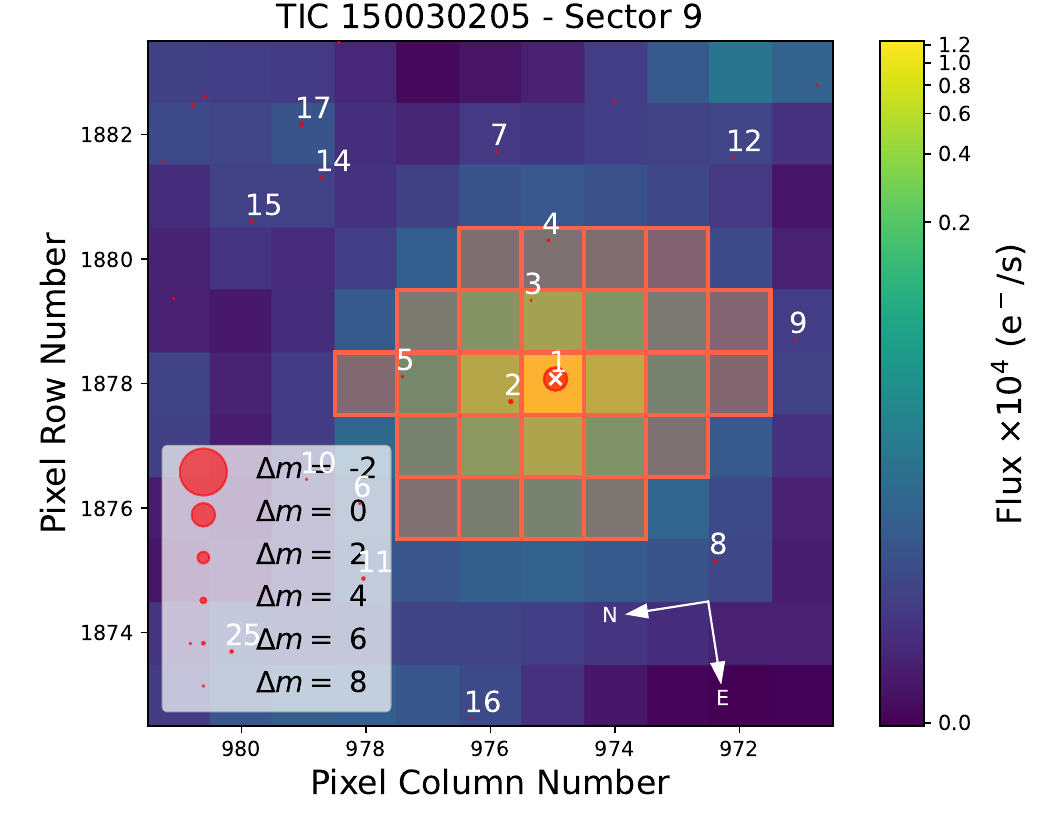}
    \includegraphics[width=.33\textwidth]{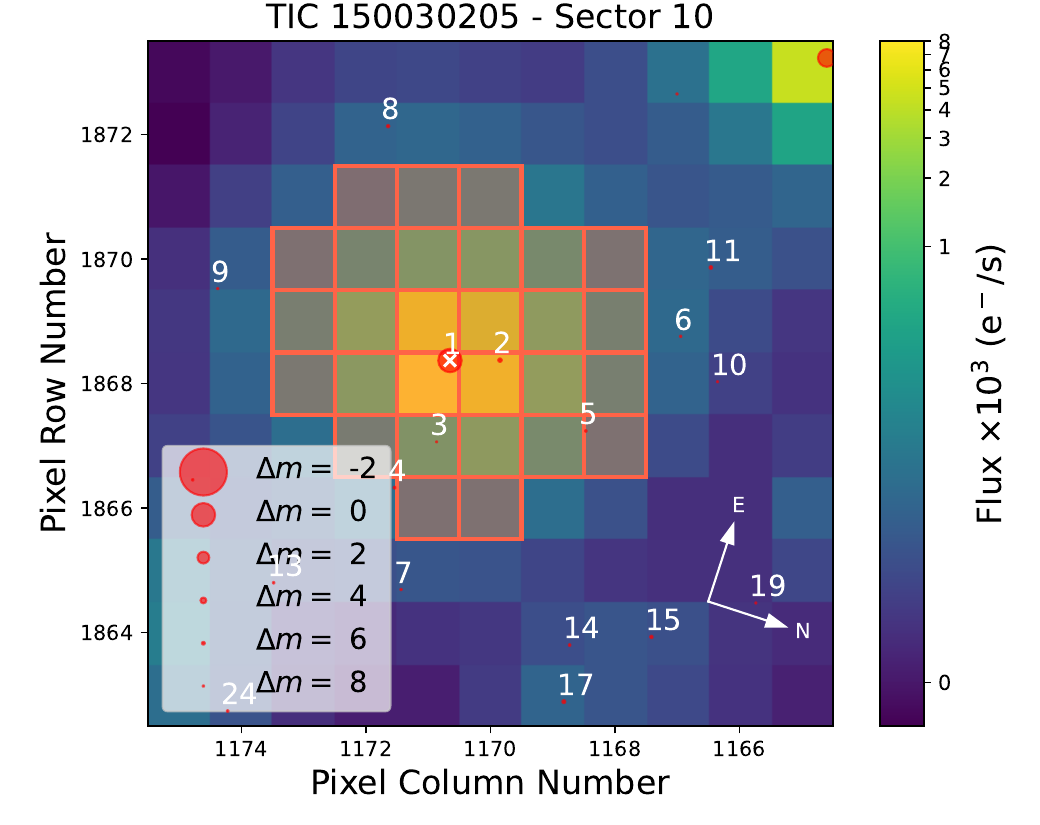}
    \includegraphics[width=.33\textwidth]{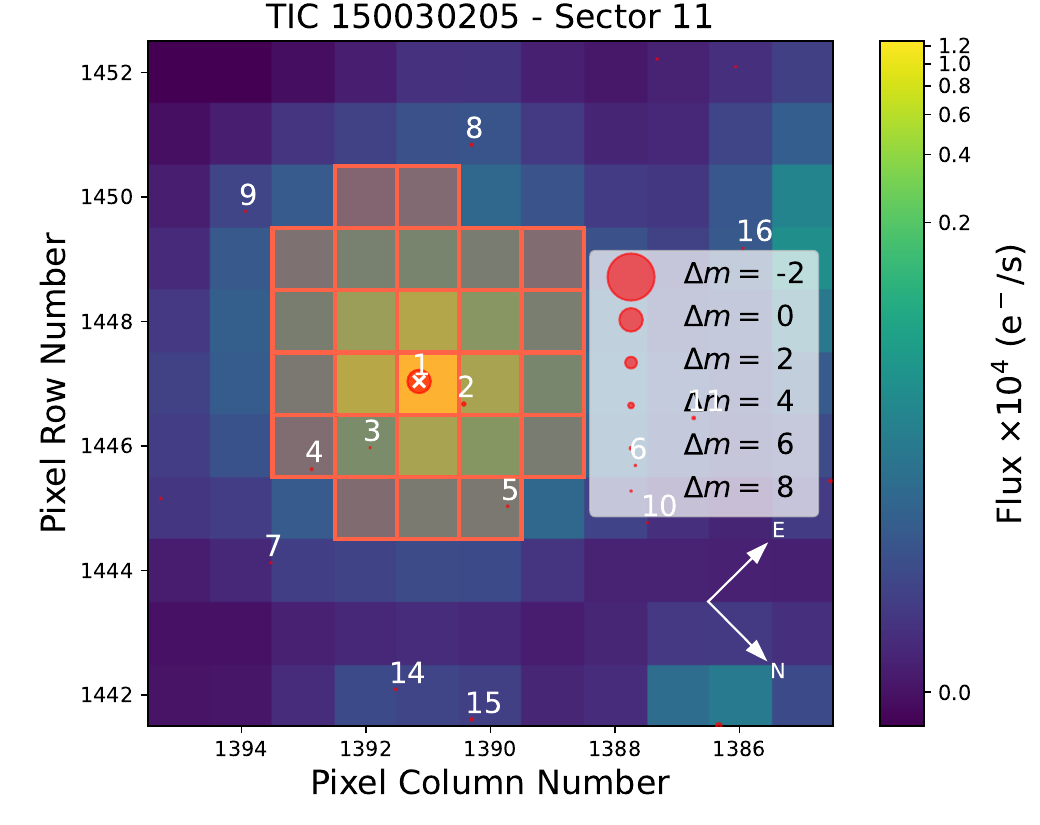}
    \includegraphics[width=.33\textwidth]{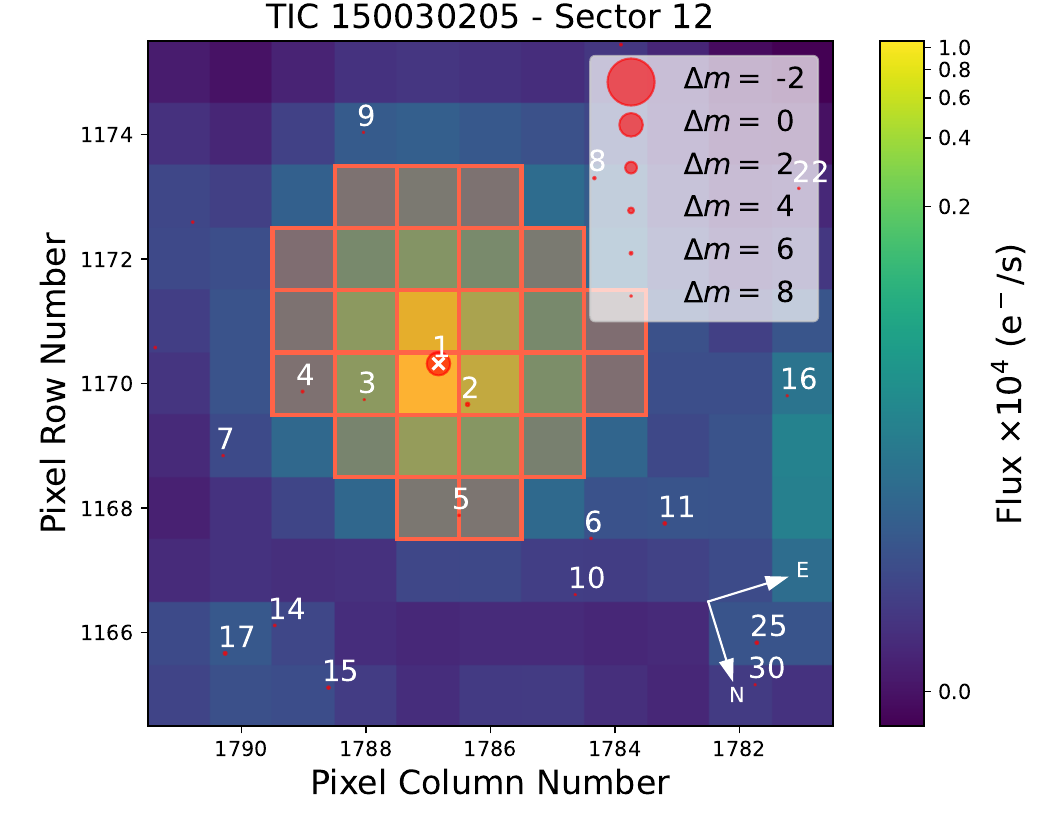}
    \includegraphics[width=.33\textwidth]{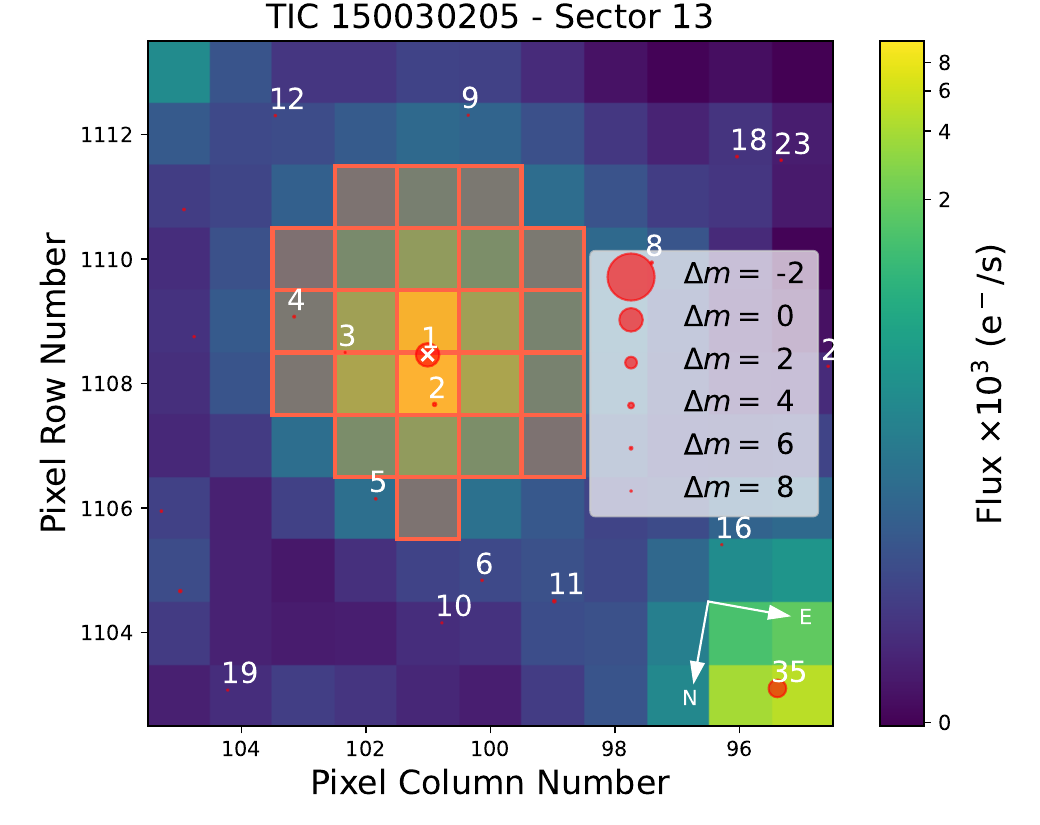}
    \caption{\textit{TESS} target pixel files for TOI-286 for sectors 2-5 and 7-13, observed during the prime mission. The target star is labelled as 1 and marked by a white cross in each case. All sources from the Gaia DR3 catalogue down to a magnitude contrast of 8 are shown as red circles, with the size proportional to the contrast. The SPOC pipeline aperture is overplotted in shaded red squares.}
    \label{fig:tpfplotter-toi286-1}
\end{figure*}

\begin{figure*}
    \centering
    \includegraphics[width=.24\textwidth]{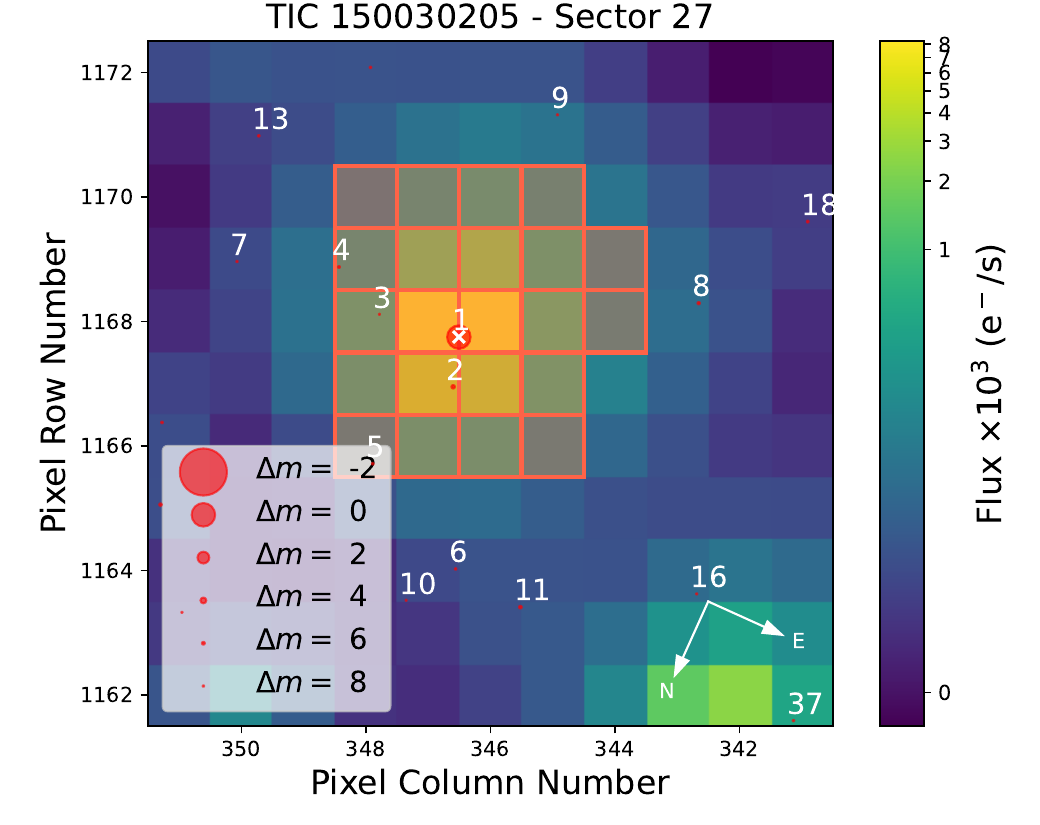}
    \includegraphics[width=.24\textwidth]{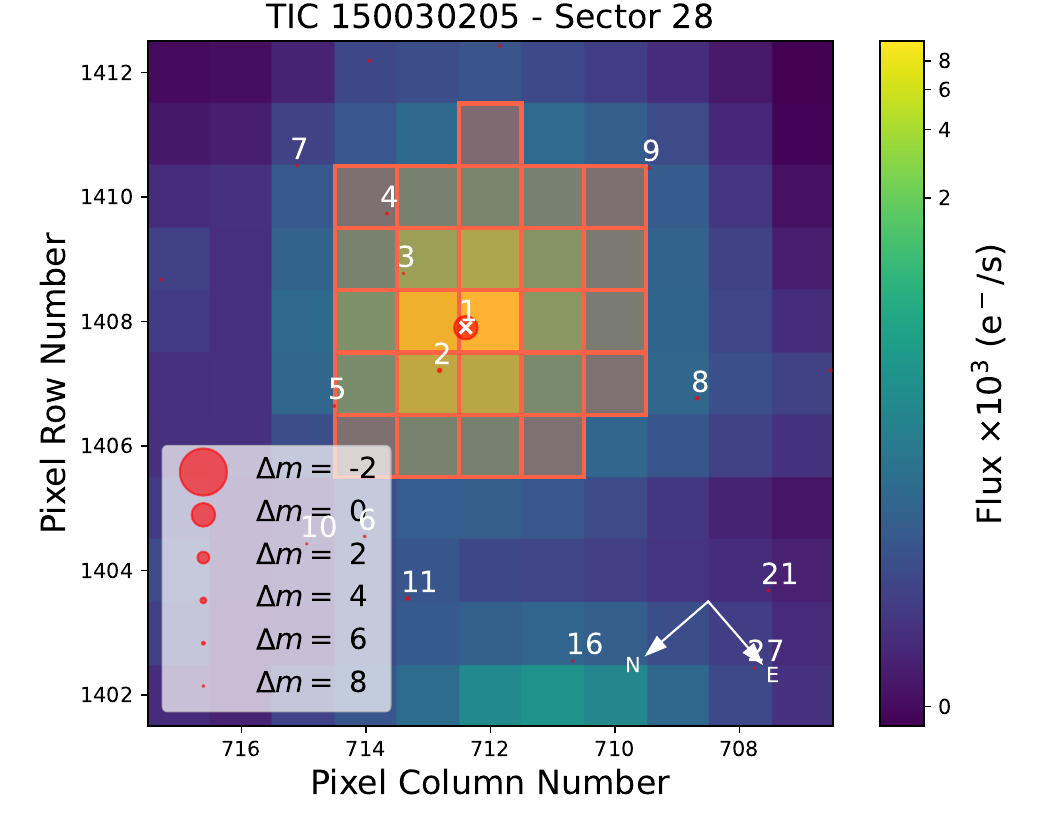}
    \includegraphics[width=.24\textwidth]{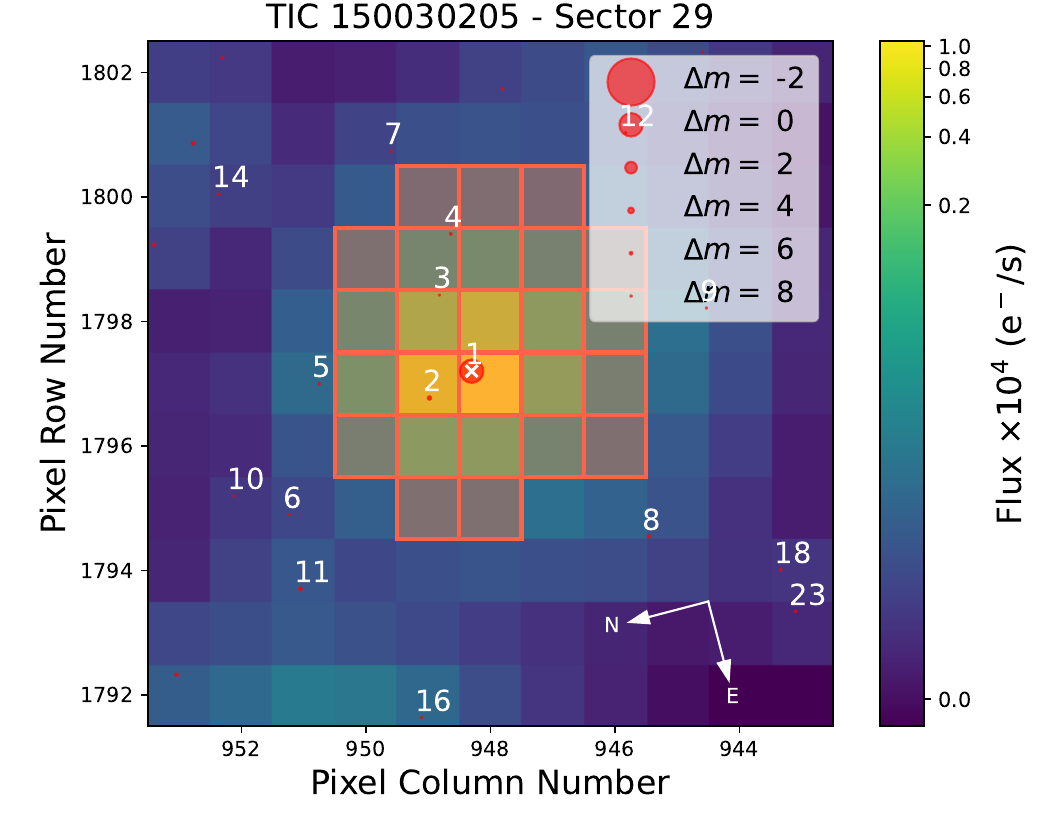}
    \includegraphics[width=.24\textwidth]{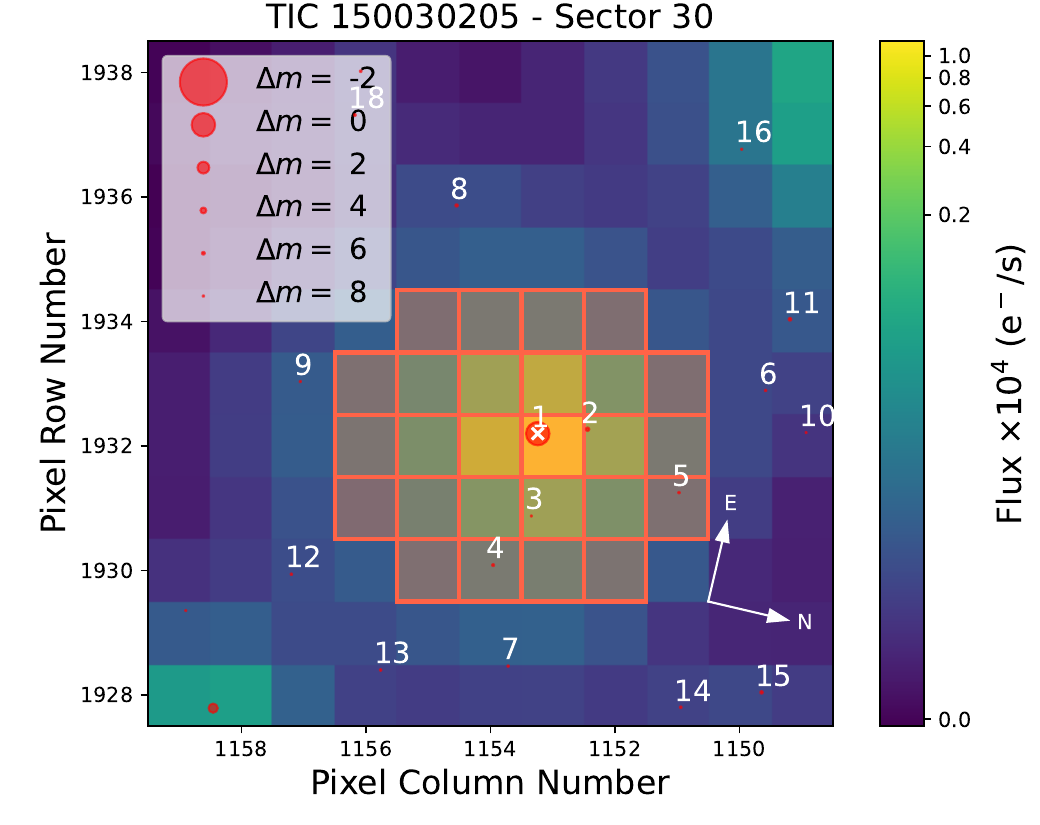}
    \includegraphics[width=.24\textwidth]{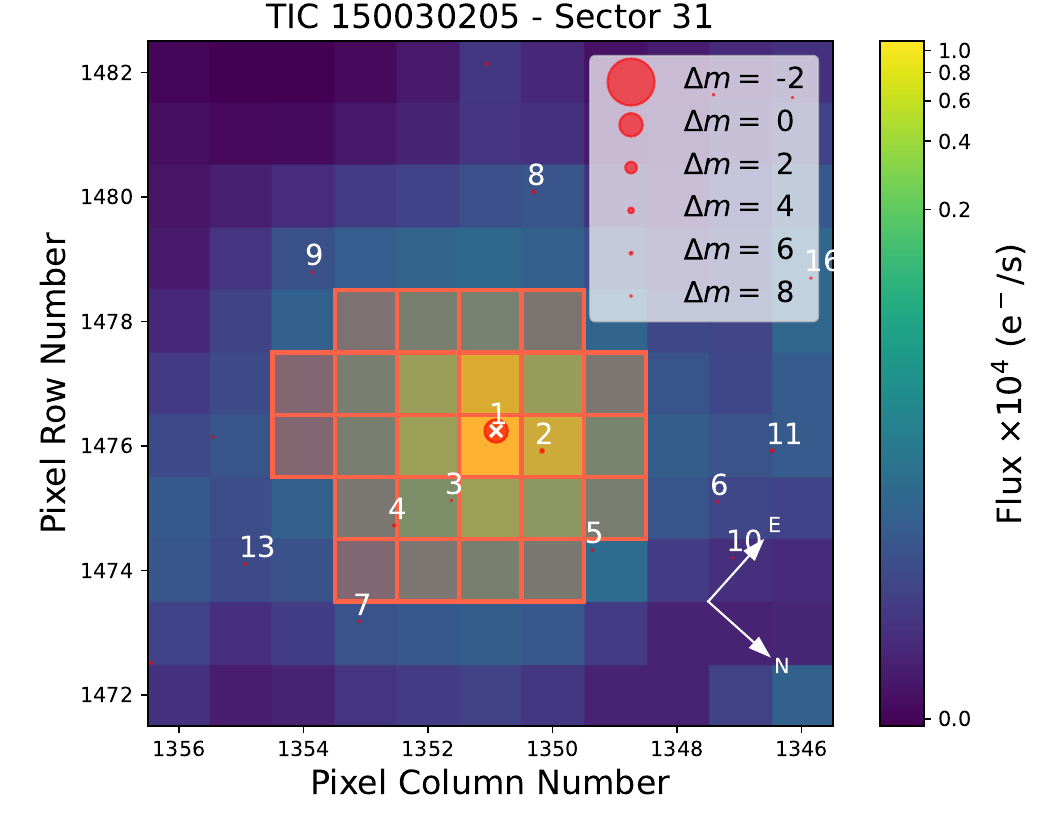}
    \includegraphics[width=.24\textwidth]{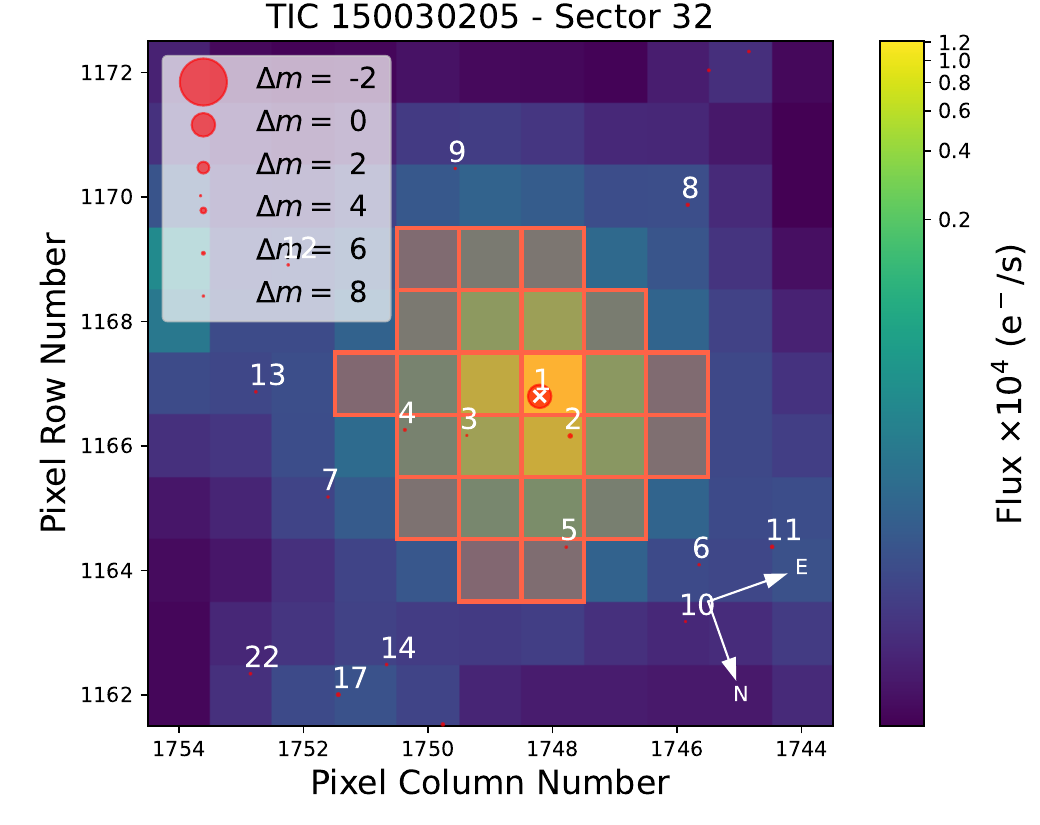}
    \includegraphics[width=.24\textwidth]{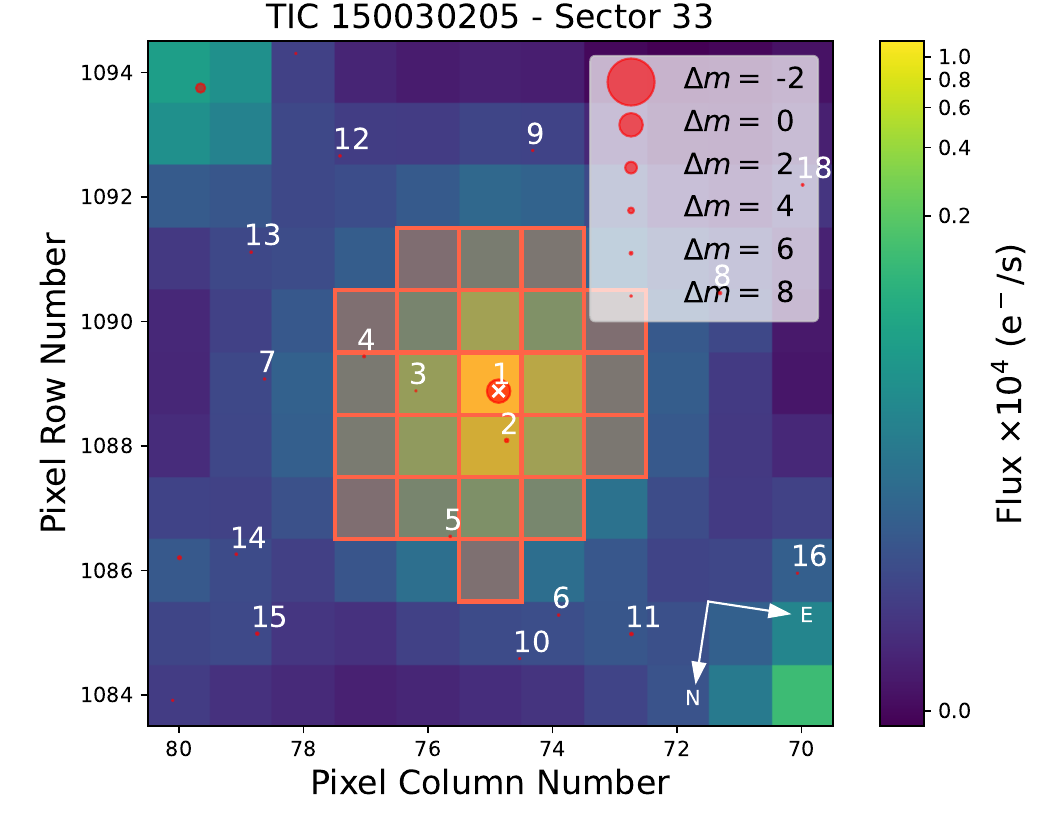}
    \includegraphics[width=.24\textwidth]{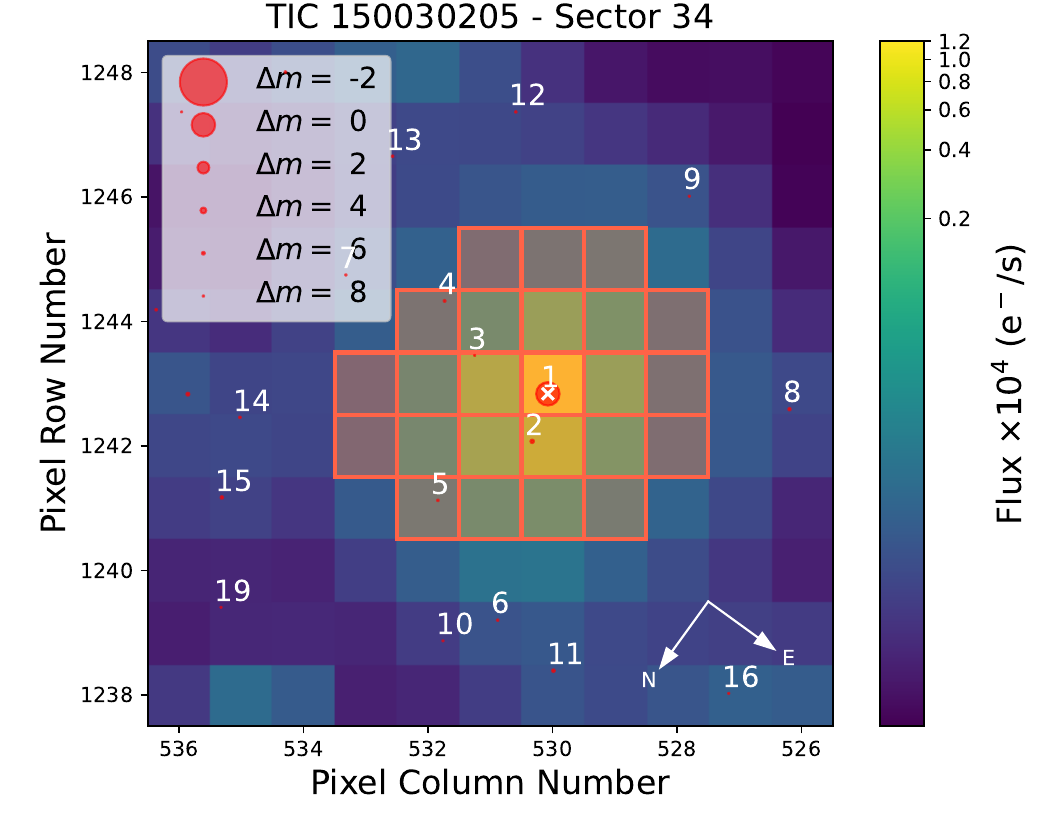}
    \includegraphics[width=.24\textwidth]{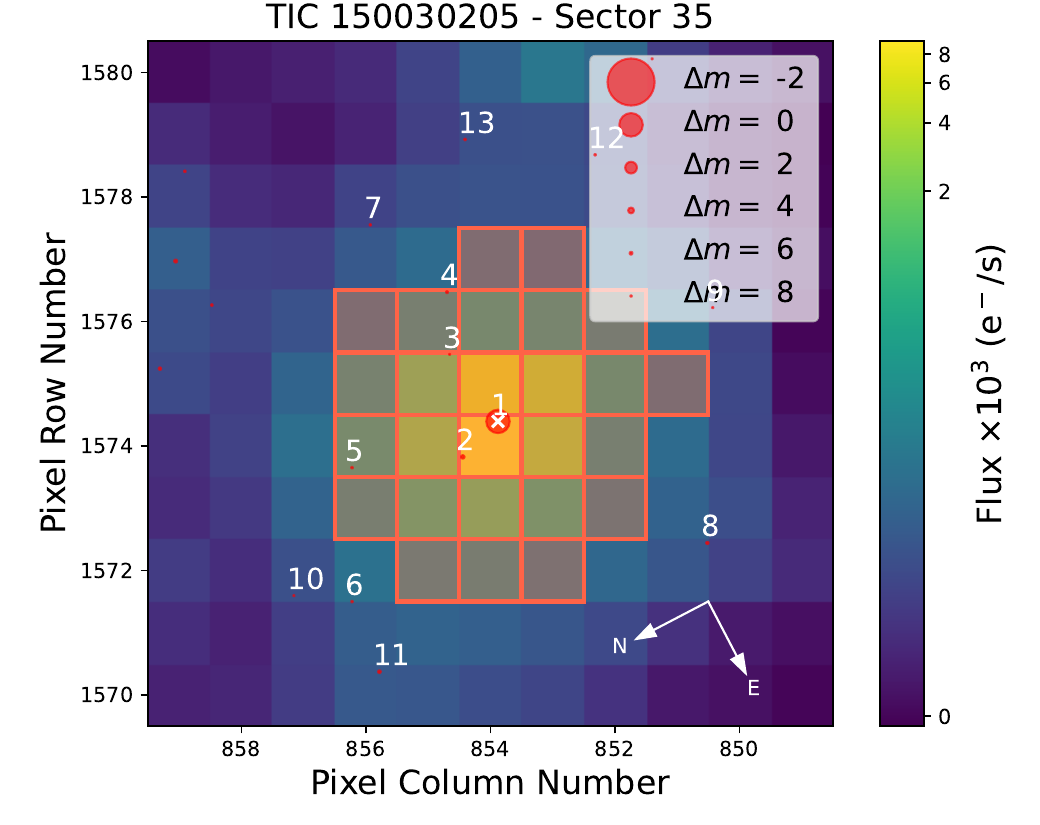}
    \includegraphics[width=.24\textwidth]{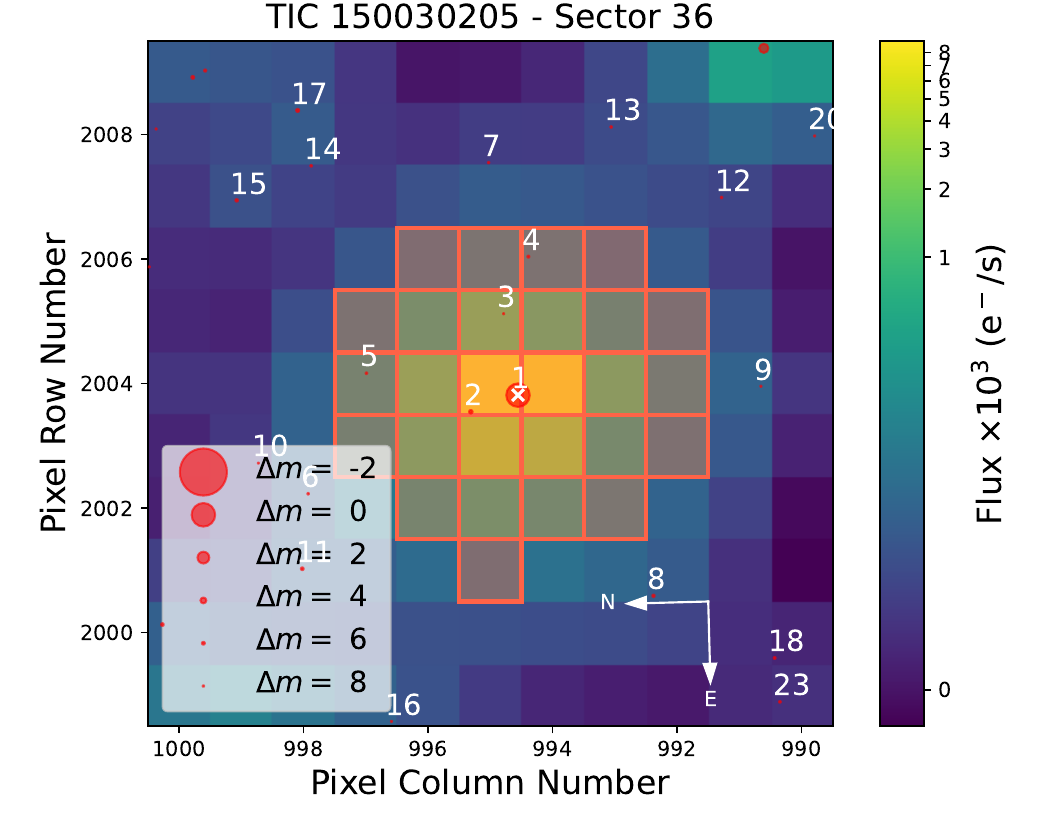}
    \includegraphics[width=.24\textwidth]{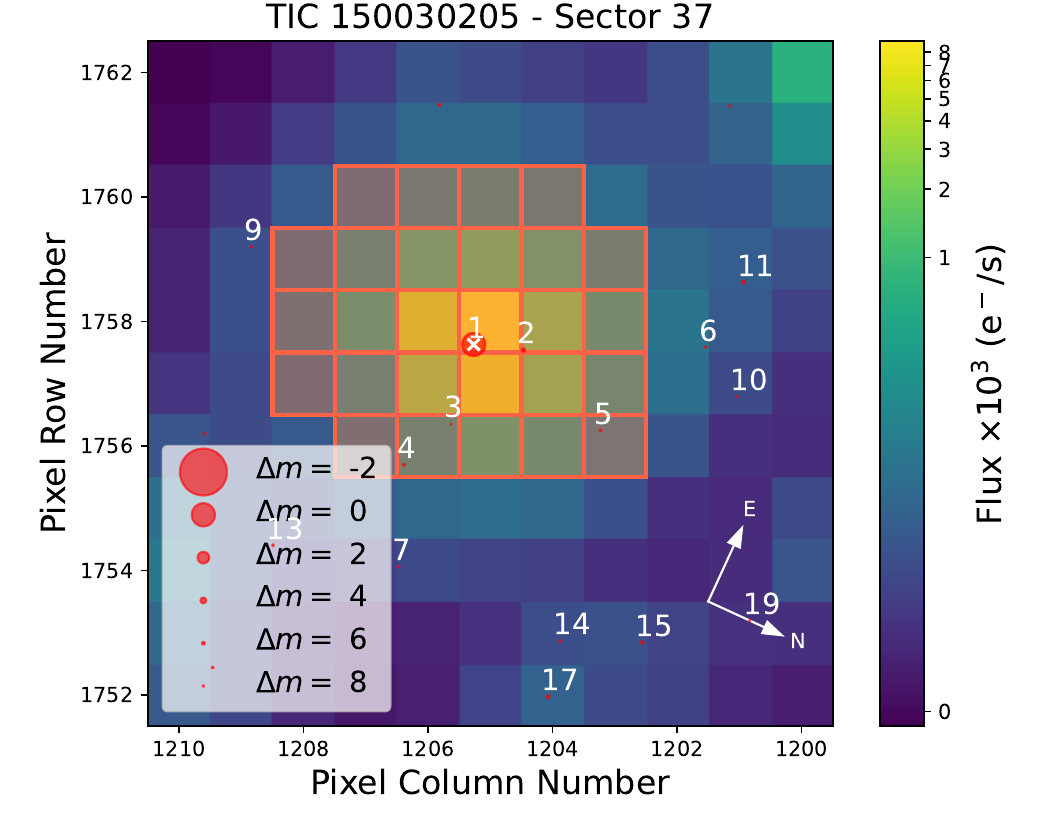}    
    \includegraphics[width=.24\textwidth]{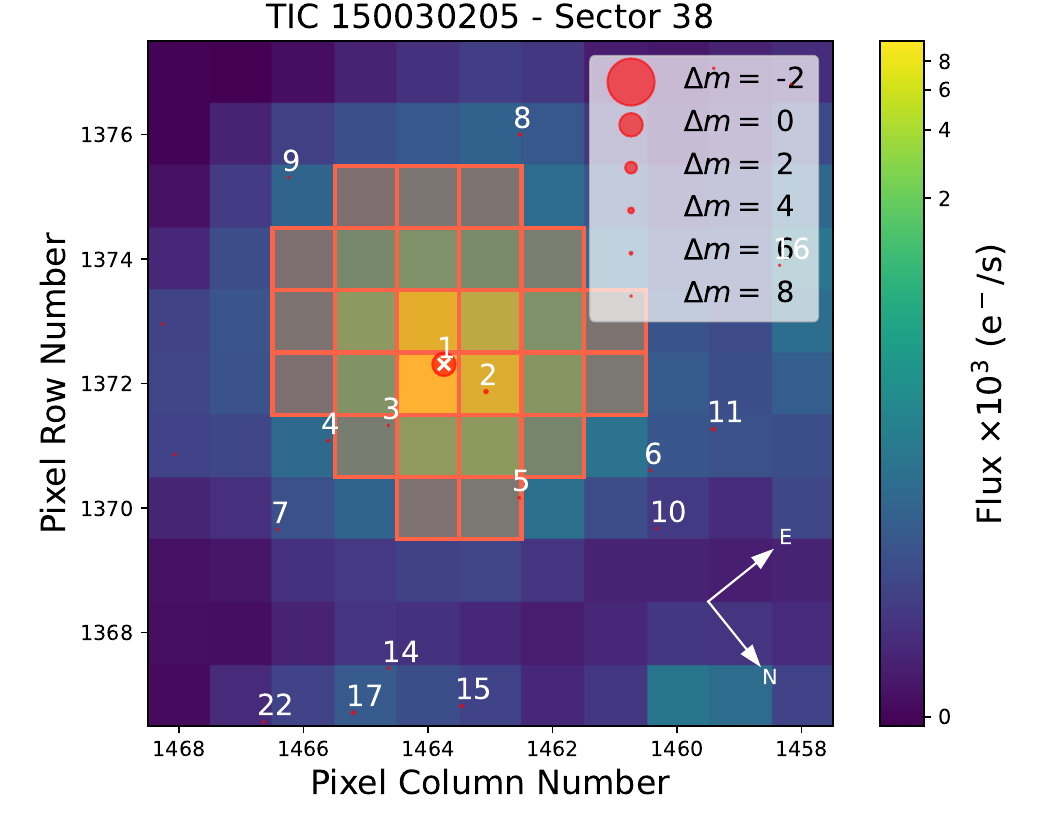}    
    \includegraphics[width=.24\textwidth]{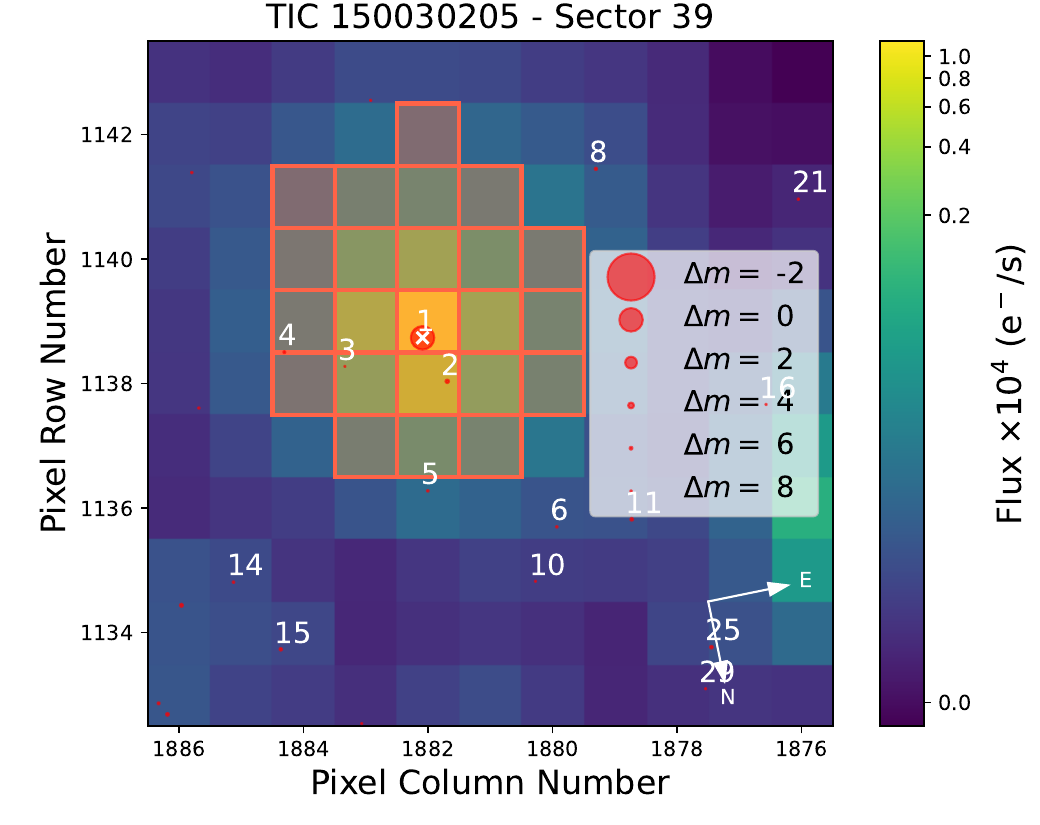}
    \caption{\textit{TESS} target pixel files for TOI-286 for sectors 27-39, observed during the first extended mission. The target star is labelled as 1 and marked by a white cross in each case. All sources from the Gaia DR3 catalogue down to a magnitude contrast of 8 are shown as red circles, with the size proportional to the contrast. The SPOC pipeline aperture is overplotted in shaded red squares.}
    \label{fig:tpfplotter-toi286-2}
\end{figure*}

\begin{figure*}
    \centering
    \includegraphics[width=.33\textwidth]{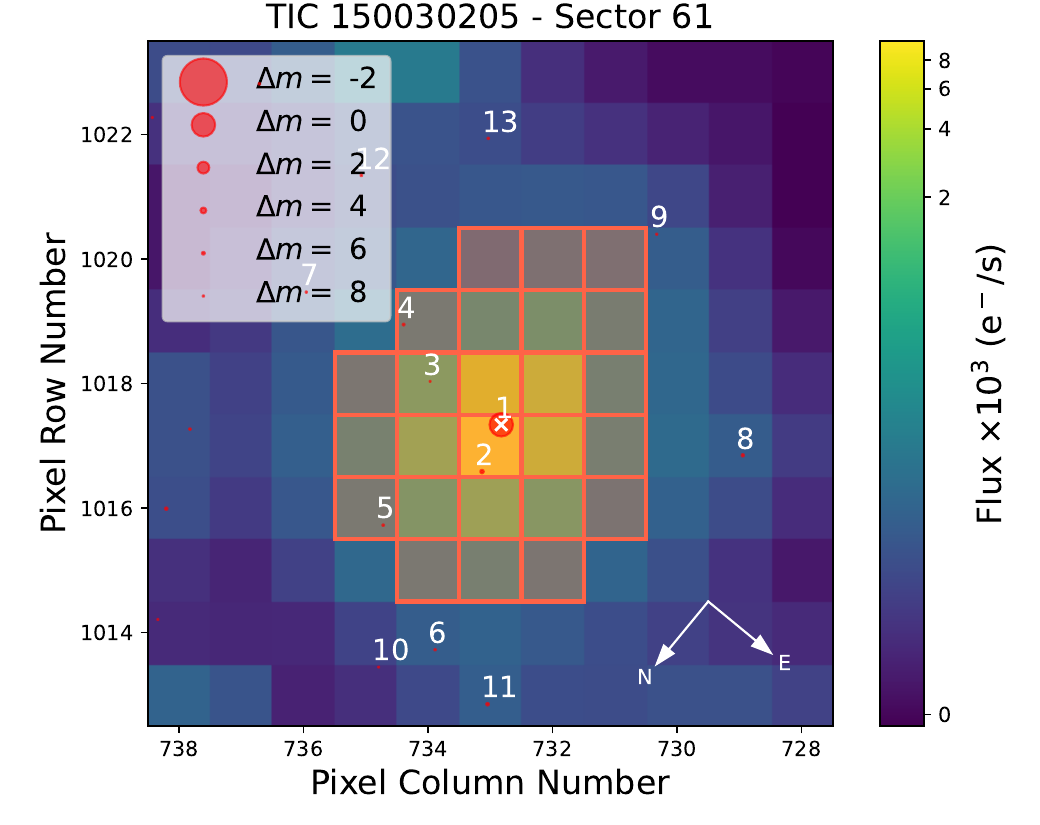}
    \includegraphics[width=.33\textwidth]{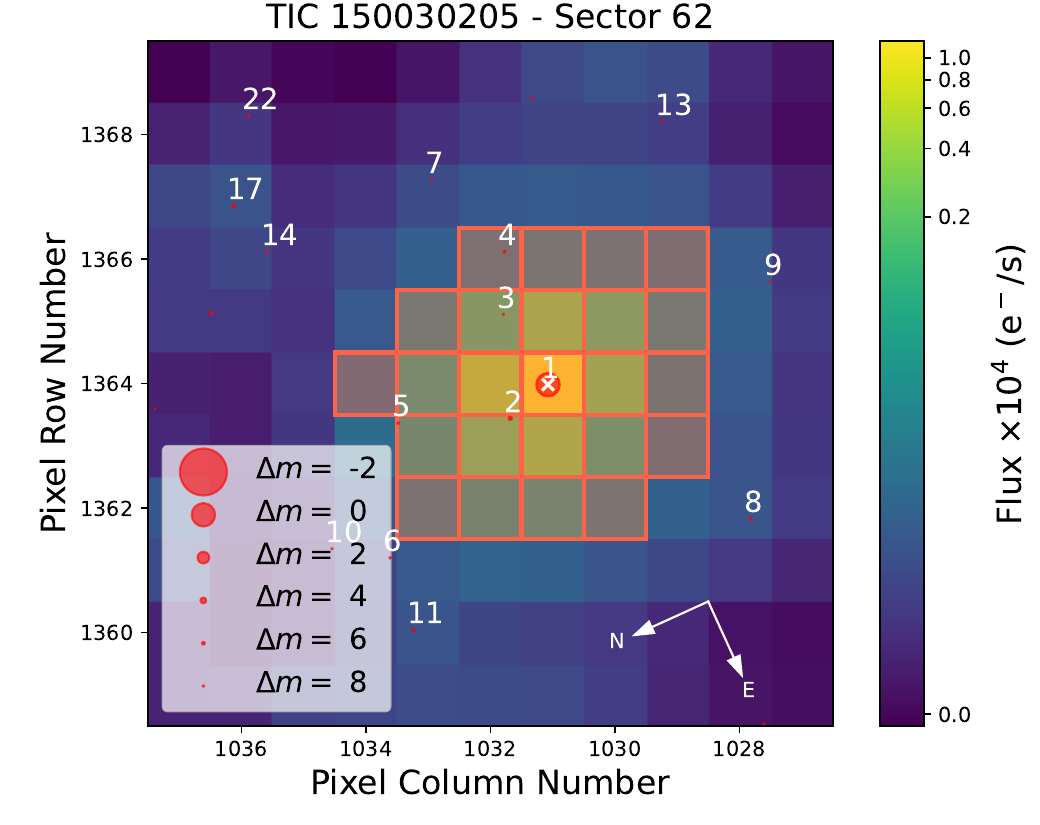}
    \includegraphics[width=.33\textwidth]{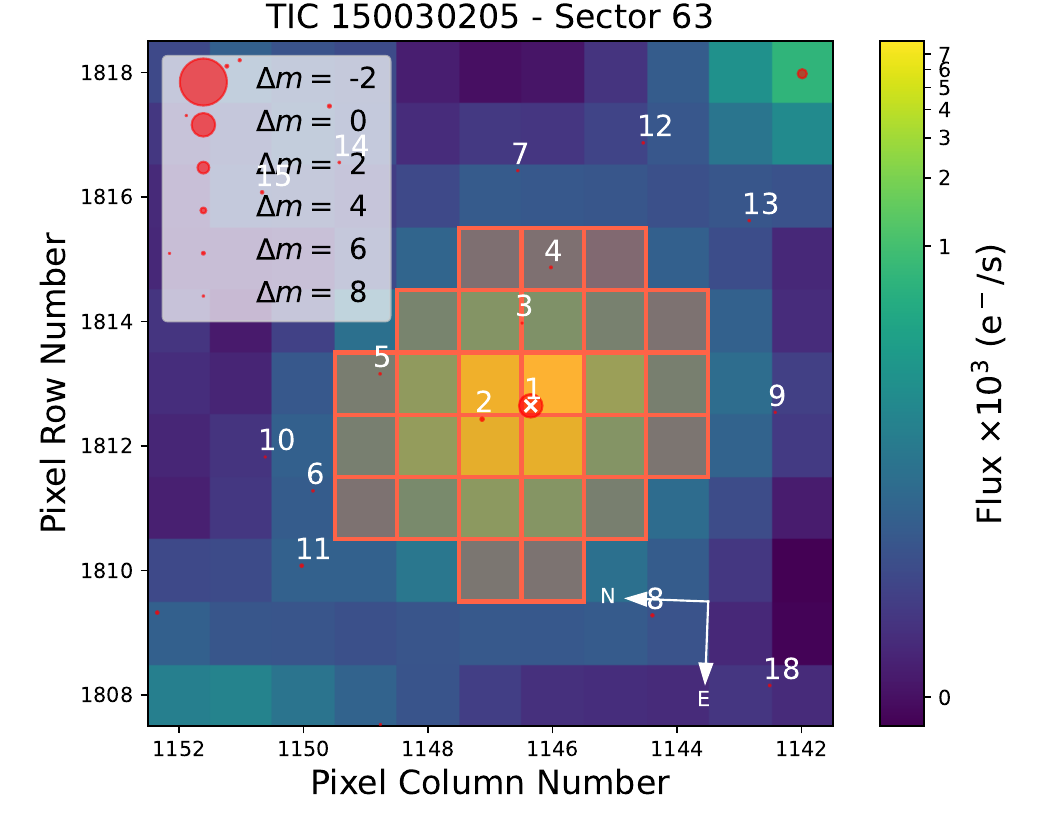}
    \includegraphics[width=.33\textwidth]{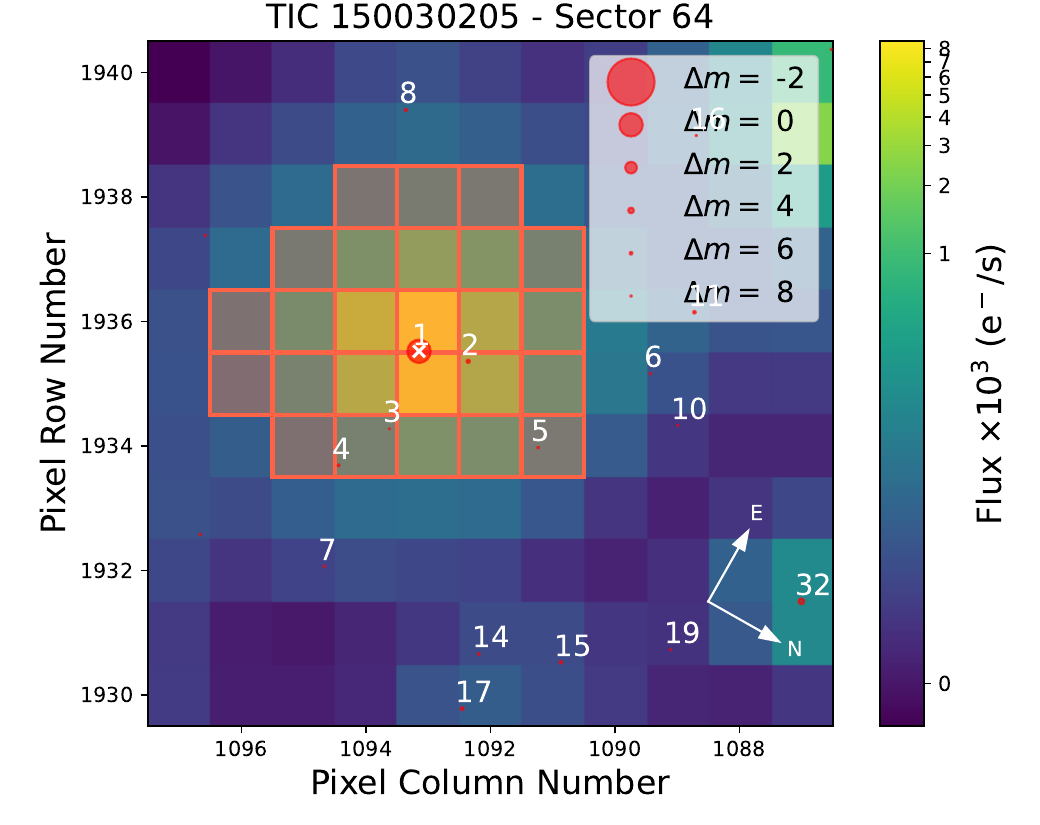}
    \includegraphics[width=.33\textwidth]{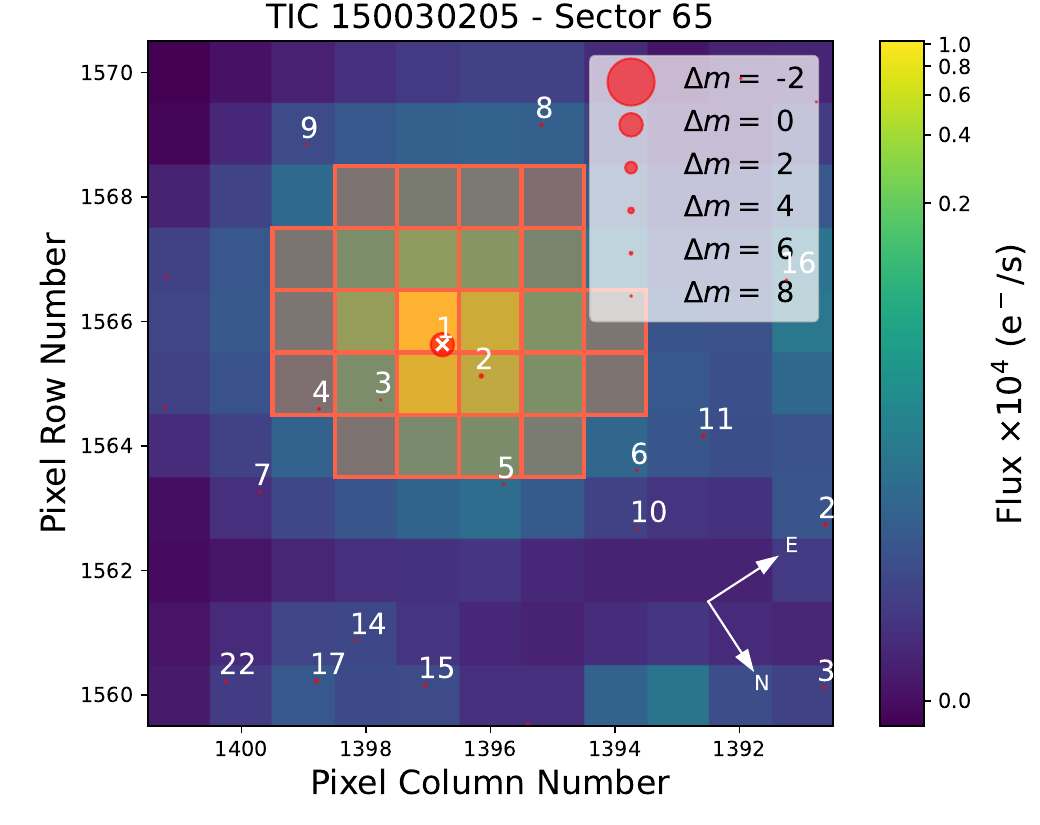}
    \includegraphics[width=.33\textwidth]{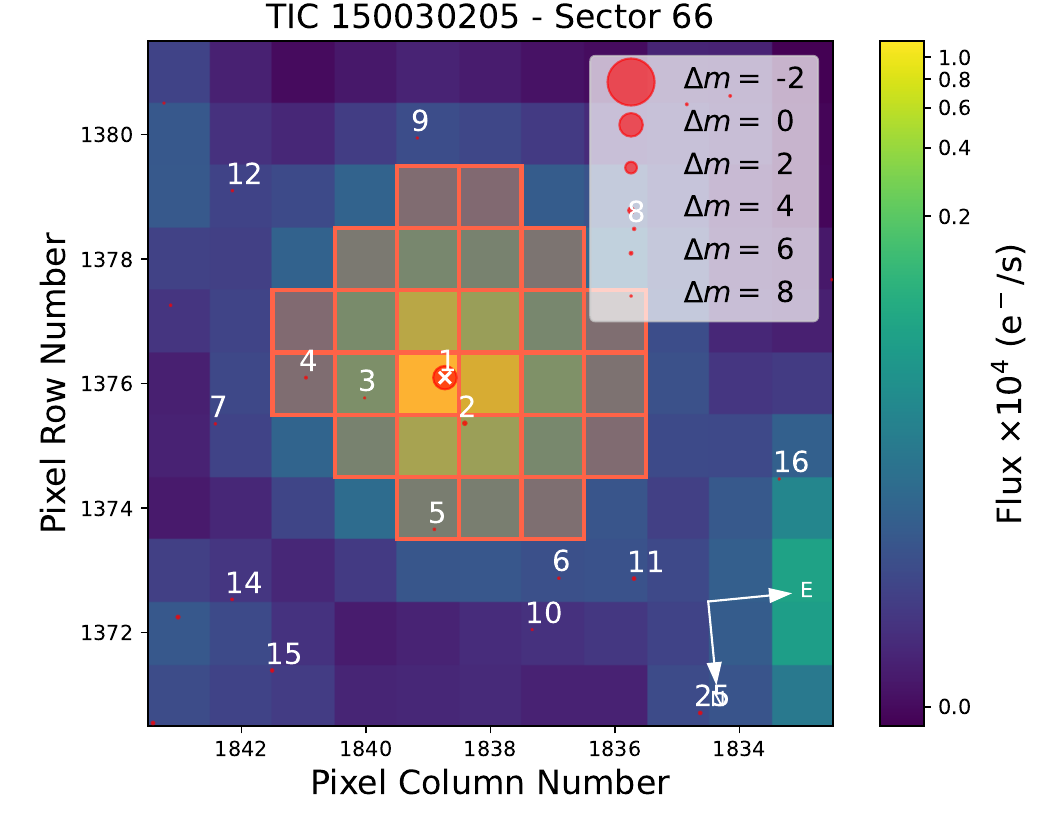}
    \includegraphics[width=.33\textwidth]{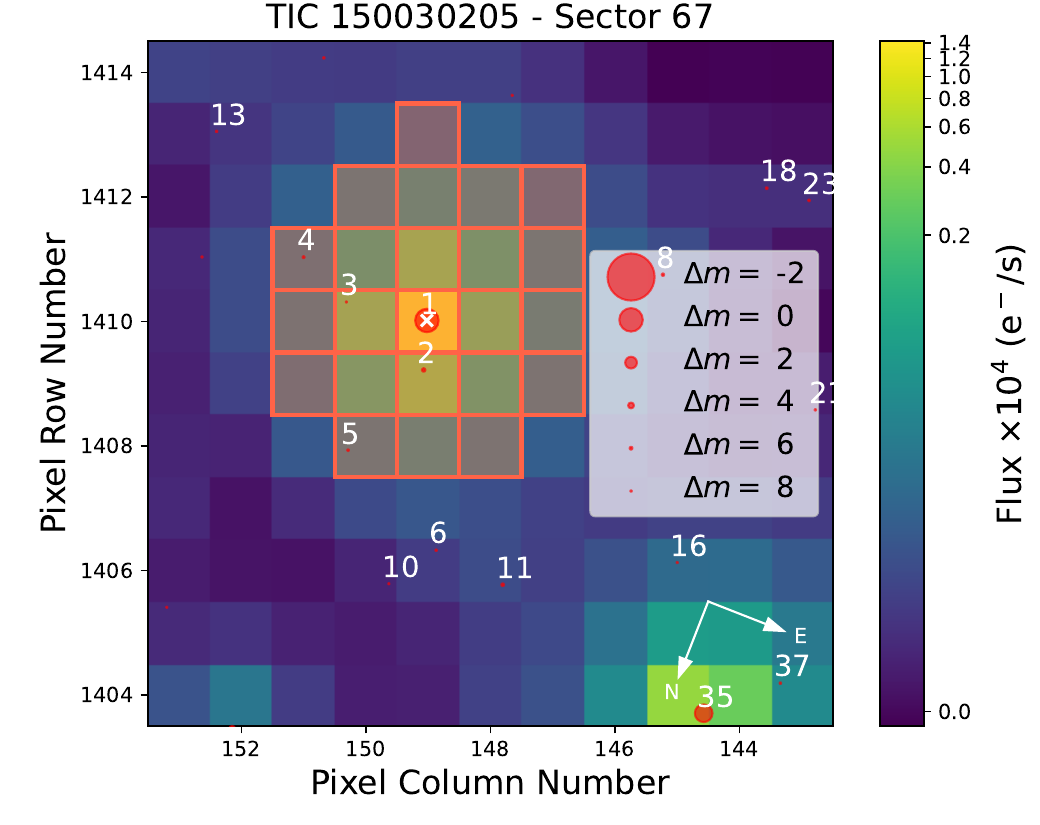}
    \includegraphics[width=.33\textwidth]{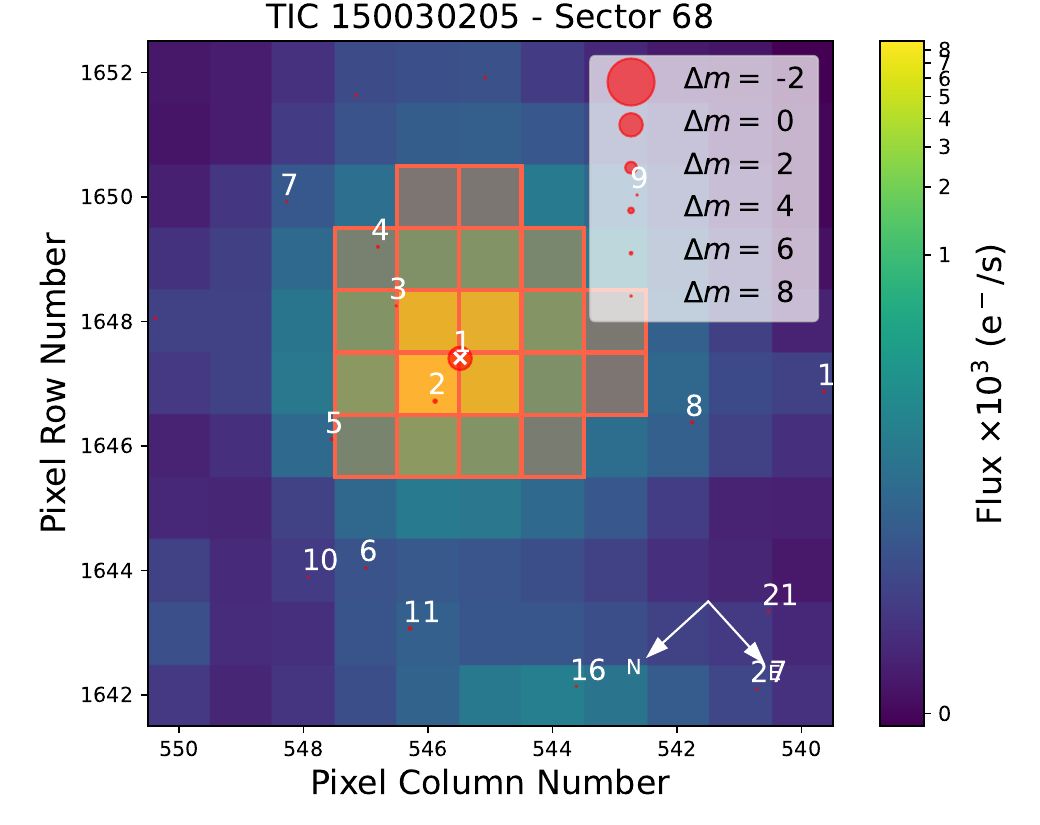}
    \includegraphics[width=.33\textwidth]{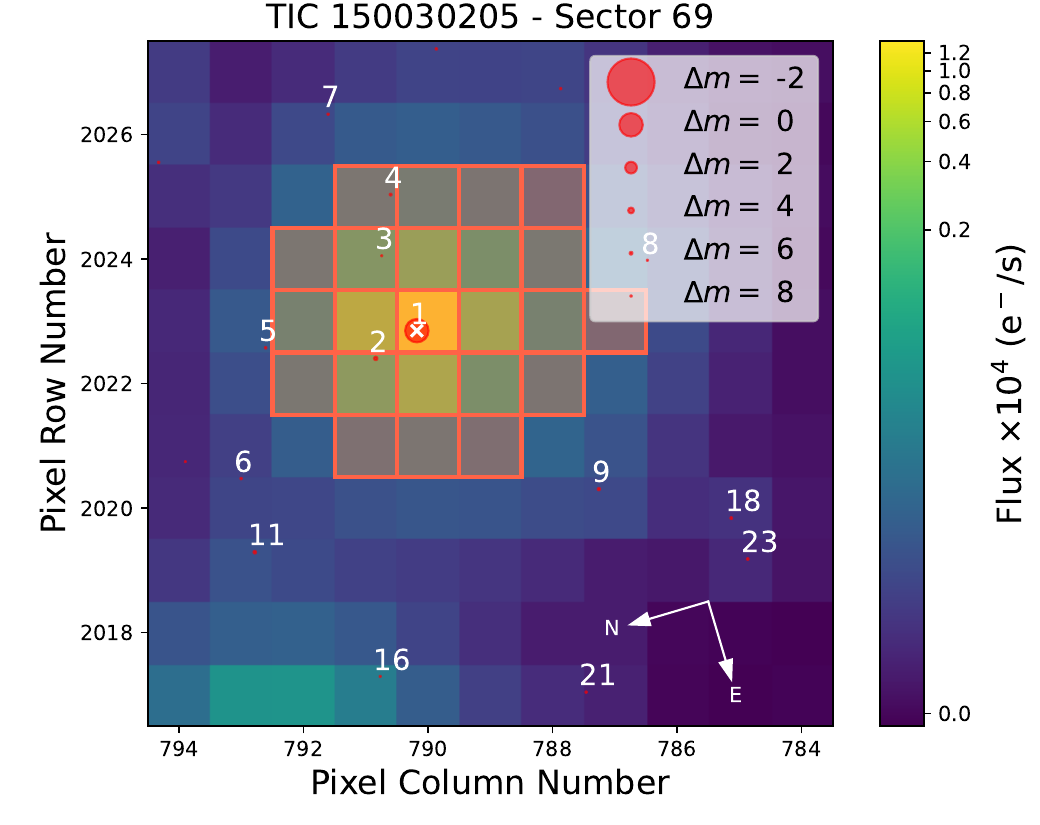}
    \caption{\textit{TESS} target pixel files for TOI-286 for sectors 61-69, observed during the first extended mission. The target star is labelled as 1 and marked by a white cross in each case. All sources from the Gaia DR3 catalogue down to a magnitude contrast of 8 are shown as red circles, with the size proportional to the contrast. The SPOC pipeline aperture is overplotted in shaded red squares.}
    \label{fig:tpfplotter-toi286-3}
\end{figure*}

\clearpage

\section{Radial velocity and activity indices}

In this appendix, we present the radial velocities and activity indices. Table \ref{tab:TOI-260_RV_ESPRESSO_data} shows the ESPRESSO data for TOI-260, Table \ref{tab:TOI-286_RV_ESPRESSO_data} shows the ESPRESSO data for TOI-286, Table \ref{tab:TOI-134_RV_ESPRESSO_data} shows the ESPRESSO data for TOI-134, and Table \ref{tab:TOI-286_RV_HARPS_data} shows the HARPS data for TOI-286.

\begin{table*}[pht]
\begin{center}
\caption{RVs and activity indices obtained from the ESPRESSO spectra for TOI-260. Values above the horizontal line correspond to ESPRESSO18 data, values below the horizontal line to ESPRESSO19 data.}
\label{tab:TOI-260_RV_ESPRESSO_data}
\centering
\resizebox{\textwidth}{!}{%
\begin{tabular}{lllllllll}
\hline  \hline
BJD - 2457000 [d] & RV [$\mathrm{km \, s^{-1}}$] & Bisector & FWHM & Contrast & H$_\alpha$ & $S_{MW}$ & log($R^\prime_{HK}$) & Na~I \\
\hline
1624.9155 & -10941.77617 $\pm$ 0.00053 & 39.28 $\pm$ 1.05 & 5420.31 $\pm$ 1.05 & 48.6536 $\pm$ 0.0095 & 0.385423 $\pm$ 0.000045 & 1.03784 $\pm$ 0.00077 & -4.88636 $\pm$ 0.00037 & 0.126590 $\pm$ 0.000028 \\ 
1625.8988 & -10940.59085 $\pm$ 0.00053 & 40.23 $\pm$ 1.06 & 5421.60 $\pm$ 1.06 & 48.6542 $\pm$ 0.0095 & 0.391076 $\pm$ 0.000045 & 1.05219 $\pm$ 0.00078 & -4.87957 $\pm$ 0.00036 & 0.127015 $\pm$ 0.000029 \\ 
1639.9110 & -10947.05923 $\pm$ 0.00054 & 36.42 $\pm$ 1.07 & 5367.48 $\pm$ 1.07 & 49.1624 $\pm$ 0.0098 & 0.371145 $\pm$ 0.000047 & 0.90568 $\pm$ 0.00081 & -4.95443 $\pm$ 0.00045 & 0.119632 $\pm$ 0.000030 \\ 
\hline
1695.8184 & -10936.99909 $\pm$ 0.00051 & 37.75 $\pm$ 1.02 & 5421.49 $\pm$ 1.02 & 48.5764 $\pm$ 0.0091 & 0.385472 $\pm$ 0.000041 & 0.98585 $\pm$ 0.00076 & -4.91188 $\pm$ 0.00038 & 0.126291 $\pm$ 0.000027 \\ 
1700.6865 & -10937.10936 $\pm$ 0.00041 & 37.83 $\pm$ 0.82 & 5422.11 $\pm$ 0.82 & 48.5608 $\pm$ 0.0073 & 0.390577 $\pm$ 0.000030 & 1.01640 $\pm$ 0.00056 & -4.89670 $\pm$ 0.00027 & 0.126073 $\pm$ 0.000020 \\ 
1716.6857 & -10948.29164 $\pm$ 0.00035 & 36.77 $\pm$ 0.69 & 5355.40 $\pm$ 0.69 & 49.1781 $\pm$ 0.0064 & 0.360964 $\pm$ 0.000025 & 0.81831 $\pm$ 0.00039 & -5.00610 $\pm$ 0.00025 & 0.121070 $\pm$ 0.000016 \\ 
1720.6411 & -10947.35240 $\pm$ 0.00044 & 30.64 $\pm$ 0.88 & 5360.30 $\pm$ 0.88 & 49.1085 $\pm$ 0.0081 & 0.358732 $\pm$ 0.000033 & 0.79238 $\pm$ 0.00062 & -5.02270 $\pm$ 0.00041 & 0.120673 $\pm$ 0.000022 \\ 
1731.7332 & -10945.15490 $\pm$ 0.00032 & 34.49 $\pm$ 0.63 & 5394.83 $\pm$ 0.63 & 48.9025 $\pm$ 0.0057 & 0.380737 $\pm$ 0.000024 & 0.94832 $\pm$ 0.00032 & -4.93128 $\pm$ 0.00017 & 0.125054 $\pm$ 0.000015 \\ 
1753.6880 & -10946.35812 $\pm$ 0.00038 & 34.71 $\pm$ 0.76 & 5357.19 $\pm$ 0.76 & 49.3298 $\pm$ 0.0070 & 0.357409 $\pm$ 0.000030 & 0.81142 $\pm$ 0.00045 & -5.01045 $\pm$ 0.00028 & 0.122561 $\pm$ 0.000019 \\ 
1759.5795 & -10946.34677 $\pm$ 0.00036 & 32.06 $\pm$ 0.73 & 5371.06 $\pm$ 0.73 & 49.1836 $\pm$ 0.0067 & 0.363955 $\pm$ 0.000028 & 0.82325 $\pm$ 0.00043 & -5.00301 $\pm$ 0.00027 & 0.122094 $\pm$ 0.000018 \\ 
1761.5278 & -10945.22359 $\pm$ 0.00052 & 32.19 $\pm$ 1.04 & 5374.76 $\pm$ 1.04 & 49.2190 $\pm$ 0.0096 & 0.369657 $\pm$ 0.000042 & 0.80601 $\pm$ 0.00085 & -5.01389 $\pm$ 0.00054 & 0.121363 $\pm$ 0.000028 \\ 
1764.7147 & -10943.89008 $\pm$ 0.00064 & 38.66 $\pm$ 1.27 & 5390.68 $\pm$ 1.27 & 49.0911 $\pm$ 0.0116 & 0.376129 $\pm$ 0.000055 & 0.82926 $\pm$ 0.00109 & -4.99927 $\pm$ 0.00067 & 0.122222 $\pm$ 0.000036 \\ 
1773.7389 & -10942.66162 $\pm$ 0.00045 & 31.92 $\pm$ 0.90 & 5420.93 $\pm$ 0.90 & 48.7332 $\pm$ 0.0080 & 0.388480 $\pm$ 0.000036 & 0.96552 $\pm$ 0.00060 & -4.92228 $\pm$ 0.00031 & 0.125928 $\pm$ 0.000023 \\ 
1775.5104 & -10936.29858 $\pm$ 0.00055 & 31.72 $\pm$ 1.10 & 5432.45 $\pm$ 1.10 & 48.6620 $\pm$ 0.0099 & 0.417495 $\pm$ 0.000047 & 1.12249 $\pm$ 0.00088 & -4.84777 $\pm$ 0.00038 & 0.129466 $\pm$ 0.000031 \\ 
1777.6520 & -10937.49530 $\pm$ 0.00043 & 37.83 $\pm$ 0.85 & 5441.77 $\pm$ 0.85 & 48.5984 $\pm$ 0.0076 & 0.393579 $\pm$ 0.000034 & 1.00005 $\pm$ 0.00055 & -4.90476 $\pm$ 0.00027 & 0.126913 $\pm$ 0.000022 \\ 
1779.5144 & -10938.55668 $\pm$ 0.00037 & 42.28 $\pm$ 0.73 & 5434.31 $\pm$ 0.73 & 48.5931 $\pm$ 0.0065 & 0.394364 $\pm$ 0.000028 & 0.99512 $\pm$ 0.00043 & -4.90722 $\pm$ 0.00022 & 0.126986 $\pm$ 0.000018 \\ 
1781.5462 & -10944.48169 $\pm$ 0.00037 & 46.67 $\pm$ 0.74 & 5417.39 $\pm$ 0.74 & 48.7452 $\pm$ 0.0067 & 0.380686 $\pm$ 0.000029 & 0.97410 $\pm$ 0.00043 & -4.91786 $\pm$ 0.00022 & 0.127938 $\pm$ 0.000018 \\ 
1783.5516 & -10948.86283 $\pm$ 0.00040 & 45.16 $\pm$ 0.80 & 5395.09 $\pm$ 0.80 & 48.9484 $\pm$ 0.0072 & 0.370263 $\pm$ 0.000032 & 0.92299 $\pm$ 0.00048 & -4.94488 $\pm$ 0.00026 & 0.124847 $\pm$ 0.000020 \\ 
1785.5385 & -10950.85422 $\pm$ 0.00045 & 39.26 $\pm$ 0.90 & 5375.24 $\pm$ 0.90 & 49.1396 $\pm$ 0.0082 & 0.363986 $\pm$ 0.000036 & 0.88002 $\pm$ 0.00064 & -4.96898 $\pm$ 0.00037 & 0.124126 $\pm$ 0.000024 \\ 
1788.5874 & -10947.27113 $\pm$ 0.00032 & 34.40 $\pm$ 0.65 & 5357.39 $\pm$ 0.65 & 49.2567 $\pm$ 0.0060 & 0.368010 $\pm$ 0.000025 & 0.88642 $\pm$ 0.00034 & -4.96530 $\pm$ 0.00020 & 0.122228 $\pm$ 0.000016 \\ 
1790.5234 & -10947.98729 $\pm$ 0.00041 & 33.84 $\pm$ 0.81 & 5356.53 $\pm$ 0.81 & 49.2901 $\pm$ 0.0075 & 0.357164 $\pm$ 0.000033 & 0.81452 $\pm$ 0.00051 & -5.00849 $\pm$ 0.00032 & 0.120587 $\pm$ 0.000021 \\ 
1792.5378 & -10948.75186 $\pm$ 0.00050 & 24.92 $\pm$ 1.00 & 5366.99 $\pm$ 1.00 & 49.3602 $\pm$ 0.0092 & 0.360088 $\pm$ 0.000043 & 0.85841 $\pm$ 0.00074 & -4.98162 $\pm$ 0.00044 & 0.123089 $\pm$ 0.000028 \\ 
1800.6578 & -10945.12884 $\pm$ 0.00036 & 33.88 $\pm$ 0.71 & 5386.59 $\pm$ 0.71 & 49.0094 $\pm$ 0.0065 & 0.387253 $\pm$ 0.000028 & 0.93504 $\pm$ 0.00041 & -4.93836 $\pm$ 0.00022 & 0.118903 $\pm$ 0.000017 \\ 
1802.6901 & -10947.54836 $\pm$ 0.00041 & 39.67 $\pm$ 0.82 & 5381.04 $\pm$ 0.82 & 49.0912 $\pm$ 0.0075 & 0.372320 $\pm$ 0.000032 & 0.90878 $\pm$ 0.00054 & -4.95270 $\pm$ 0.00030 & 0.121021 $\pm$ 0.000021 \\ 
1804.5346 & -10947.65735 $\pm$ 0.00041 & 39.67 $\pm$ 0.82 & 5384.89 $\pm$ 0.82 & 49.1513 $\pm$ 0.0075 & 0.380442 $\pm$ 0.000033 & 0.92558 $\pm$ 0.00053 & -4.94347 $\pm$ 0.00029 & 0.118453 $\pm$ 0.000021 \\ 
1806.6944 & -10946.52736 $\pm$ 0.00062 & 39.69 $\pm$ 1.24 & 5371.67 $\pm$ 1.24 & 49.2142 $\pm$ 0.0114 & 0.376005 $\pm$ 0.000052 & 0.82942 $\pm$ 0.00116 & -4.99918 $\pm$ 0.00072 & 0.119776 $\pm$ 0.000035 \\ 
1808.6438 & -10943.99286 $\pm$ 0.00053 & 34.20 $\pm$ 1.05 & 5387.26 $\pm$ 1.05 & 49.1196 $\pm$ 0.0096 & 0.375296 $\pm$ 0.000042 & 0.90325 $\pm$ 0.00088 & -4.95579 $\pm$ 0.00050 & 0.118521 $\pm$ 0.000028 \\ 
1826.5563 & -10949.13710 $\pm$ 0.00044 & 35.44 $\pm$ 0.89 & 5349.96 $\pm$ 0.89 & 49.3771 $\pm$ 0.0082 & 0.357788 $\pm$ 0.000036 & 0.80390 $\pm$ 0.00060 & -5.01524 $\pm$ 0.00038 & 0.119712 $\pm$ 0.000024 \\ 
1828.5824 & -10952.12639 $\pm$ 0.00042 & 31.18 $\pm$ 0.84 & 5354.82 $\pm$ 0.84 & 49.4118 $\pm$ 0.0077 & 0.359587 $\pm$ 0.000033 & 0.80012 $\pm$ 0.00056 & -5.01768 $\pm$ 0.00036 & 0.119511 $\pm$ 0.000022 \\ 
1831.5372 & -10943.30031 $\pm$ 0.00038 & 33.27 $\pm$ 0.76 & 5368.22 $\pm$ 0.76 & 49.2581 $\pm$ 0.0069 & 0.365068 $\pm$ 0.000030 & 0.86618 $\pm$ 0.00046 & -4.97703 $\pm$ 0.00027 & 0.120023 $\pm$ 0.000019 \\ 
1833.5446 & -10944.05862 $\pm$ 0.00040 & 30.03 $\pm$ 0.81 & 5378.18 $\pm$ 0.81 & 49.1901 $\pm$ 0.0074 & 0.366078 $\pm$ 0.000032 & 0.86786 $\pm$ 0.00050 & -4.97604 $\pm$ 0.00030 & 0.120272 $\pm$ 0.000021 \\ 
1835.5369 & -10940.80274 $\pm$ 0.00034 & 34.44 $\pm$ 0.69 & 5384.98 $\pm$ 0.69 & 49.0986 $\pm$ 0.0063 & 0.367572 $\pm$ 0.000026 & 0.89148 $\pm$ 0.00038 & -4.96242 $\pm$ 0.00022 & 0.121087 $\pm$ 0.000017 \\ 
1837.5436 & -10945.14442 $\pm$ 0.00037 & 37.09 $\pm$ 0.73 & 5391.08 $\pm$ 0.73 & 49.0579 $\pm$ 0.0066 & 0.369649 $\pm$ 0.000028 & 0.88513 $\pm$ 0.00043 & -4.96604 $\pm$ 0.00024 & 0.120767 $\pm$ 0.000018 \\ 
2196.5451 & -10947.60784 $\pm$ 0.00035 & 41.56 $\pm$ 0.70 & 5357.39 $\pm$ 0.70 & 49.2298 $\pm$ 0.0065 & 0.358309 $\pm$ 0.000027 & 0.83258 $\pm$ 0.00040 & -4.99723 $\pm$ 0.00024 & 0.119176 $\pm$ 0.000018 \\ 
2198.5843 & -10949.16287 $\pm$ 0.00037 & 37.47 $\pm$ 0.74 & 5342.94 $\pm$ 0.74 & 49.3860 $\pm$ 0.0069 & 0.355255 $\pm$ 0.000029 & 0.79223 $\pm$ 0.00044 & -5.02280 $\pm$ 0.00029 & 0.118246 $\pm$ 0.000019 \\ 
2202.6067 & -10950.93693 $\pm$ 0.00038 & 34.78 $\pm$ 0.76 & 5328.65 $\pm$ 0.76 & 49.5718 $\pm$ 0.0071 & 0.352270 $\pm$ 0.000029 & 0.75243 $\pm$ 0.00049 & -5.04958 $\pm$ 0.00034 & 0.114442 $\pm$ 0.000020 \\ 
2208.5754 & -10947.83603 $\pm$ 0.00052 & 29.65 $\pm$ 1.05 & 5325.52 $\pm$ 1.05 & 49.5396 $\pm$ 0.0097 & 0.349274 $\pm$ 0.000043 & 0.72902 $\pm$ 0.00088 & -5.06615 $\pm$ 0.00063 & 0.112671 $\pm$ 0.000029 \\ 
2210.5504 & -10945.93734 $\pm$ 0.00044 & 36.31 $\pm$ 0.88 & 5325.87 $\pm$ 0.88 & 49.5534 $\pm$ 0.0082 & 0.352338 $\pm$ 0.000036 & 0.77935 $\pm$ 0.00061 & -5.03129 $\pm$ 0.00041 & 0.113303 $\pm$ 0.000024 \\ 
2212.5915 & -10948.02165 $\pm$ 0.00053 & 32.84 $\pm$ 1.06 & 5326.05 $\pm$ 1.06 & 49.5374 $\pm$ 0.0099 & 0.351217 $\pm$ 0.000044 & 0.74023 $\pm$ 0.00087 & -5.05813 $\pm$ 0.00061 & 0.112939 $\pm$ 0.000030 \\ 
2214.5649 & -10946.66233 $\pm$ 0.00092 & 41.12 $\pm$ 1.85 & 5361.35 $\pm$ 1.85 & 48.8468 $\pm$ 0.0168 & 0.348083 $\pm$ 0.000081 & 0.50984 $\pm$ 0.00276 & -5.26266 $\pm$ 0.00314 & 0.115749 $\pm$ 0.000059 \\ 
2217.5431 & -10950.24556 $\pm$ 0.00045 & 33.40 $\pm$ 0.90 & 5329.28 $\pm$ 0.90 & 49.6074 $\pm$ 0.0084 & 0.352633 $\pm$ 0.000036 & 0.70613 $\pm$ 0.00067 & -5.08297 $\pm$ 0.00050 & 0.113612 $\pm$ 0.000024 \\ 
2227.5646 & -10937.62718 $\pm$ 0.00071 & 45.37 $\pm$ 1.43 & 5417.06 $\pm$ 1.43 & 48.5572 $\pm$ 0.0128 & 0.361443 $\pm$ 0.000060 & 0.77760 $\pm$ 0.00165 & -5.03245 $\pm$ 0.00110 & 0.120304 $\pm$ 0.000043 \\ 
2229.5401 & -10945.18112 $\pm$ 0.00051 & 41.02 $\pm$ 1.01 & 5374.36 $\pm$ 1.01 & 49.2166 $\pm$ 0.0093 & 0.370459 $\pm$ 0.000041 & 0.86451 $\pm$ 0.00080 & -4.97801 $\pm$ 0.00047 & 0.119580 $\pm$ 0.000028 \\ 
2236.5182 & -10948.46748 $\pm$ 0.00052 & 37.67 $\pm$ 1.05 & 5340.13 $\pm$ 1.05 & 49.4102 $\pm$ 0.0097 & 0.360876 $\pm$ 0.000043 & 0.75023 $\pm$ 0.00083 & -5.05111 $\pm$ 0.00058 & 0.114238 $\pm$ 0.000029 \\ 
2239.5297 & -10946.44744 $\pm$ 0.00078 & 41.89 $\pm$ 1.56 & 5331.51 $\pm$ 1.56 & 49.6497 $\pm$ 0.0145 & 0.354254 $\pm$ 0.000068 & 0.76698 $\pm$ 0.00176 & -5.03960 $\pm$ 0.00119 & 0.113457 $\pm$ 0.000048 \\ 
\hline
\end{tabular}
}
\end{center}
\end{table*}

\begin{table*}[pht]
\begin{center}
\caption{RVs and activity indices obtained from the ESPRESSO spectra for TOI-286. Values above the horizontal line correspond to ESPRESSO18 data, values below the horizontal line to ESPRESSO19 data.}
\label{tab:TOI-286_RV_ESPRESSO_data}
\centering
\resizebox{\textwidth}{!}{%
\begin{tabular}{lllllllll}
\hline  \hline
BJD - 2457000 [d] & RV [$\mathrm{km \, s^{-1}}$] & Bisector & FWHM & Contrast & H$_\alpha$ & $S_{MW}$ & log($R^\prime_{HK}$) & Na~I \\
\hline
1507.7295 & 17781.57345 $\pm$ 0.00086 & -45.13 $\pm$ 1.72 & 6270.60 $\pm$ 1.72 & 64.0916 $\pm$ 0.0176 & 0.232877 $\pm$ 0.000104 & 0.02724 $\pm$ 0.00065 & $-$ & 0.178377 $\pm$ 0.000074 \\ 
1524.7193 & 17782.41829 $\pm$ 0.00055 & -51.39 $\pm$ 1.09 & 6246.36 $\pm$ 1.09 & 64.8458 $\pm$ 0.0113 & 0.230858 $\pm$ 0.000061 & 0.13113 $\pm$ 0.00033 & -5.54457 $\pm$ 0.00338 & 0.176863 $\pm$ 0.000042 \\ 
1532.6583 & 17777.44273 $\pm$ 0.00045 & -42.06 $\pm$ 0.91 & 6237.78 $\pm$ 0.91 & 64.8323 $\pm$ 0.0094 & 0.232868 $\pm$ 0.000050 & 0.14201 $\pm$ 0.00024 & -5.44634 $\pm$ 0.00196 & 0.182475 $\pm$ 0.000034 \\ 
1533.5905 & 17777.99974 $\pm$ 0.00047 & -46.29 $\pm$ 0.94 & 6237.44 $\pm$ 0.94 & 64.8208 $\pm$ 0.0098 & 0.235509 $\pm$ 0.000052 & 0.13485 $\pm$ 0.00025 & -5.50839 $\pm$ 0.00234 & 0.182294 $\pm$ 0.000036 \\ 
1535.7242 & 17776.86197 $\pm$ 0.00048 & -43.85 $\pm$ 0.97 & 6235.87 $\pm$ 0.97 & 64.8668 $\pm$ 0.0101 & 0.232384 $\pm$ 0.000053 & 0.08207 $\pm$ 0.00029 & $-$ & 0.178278 $\pm$ 0.000036 \\ 
1536.5728 & 17774.84898 $\pm$ 0.00040 & -47.14 $\pm$ 0.80 & 6230.00 $\pm$ 0.80 & 64.9237 $\pm$ 0.0083 & 0.233528 $\pm$ 0.000044 & 0.15796 $\pm$ 0.00019 & -5.33354 $\pm$ 0.00119 & 0.182538 $\pm$ 0.000030 \\ 
1538.6301 & 17778.50210 $\pm$ 0.00063 & -48.75 $\pm$ 1.25 & 6240.48 $\pm$ 1.25 & 64.7635 $\pm$ 0.0130 & 0.235126 $\pm$ 0.000071 & 0.10258 $\pm$ 0.00041 & -6.02031 $\pm$ 0.01254 & 0.179989 $\pm$ 0.000050 \\ 
1541.5888 & 17780.00988 $\pm$ 0.00048 & -47.03 $\pm$ 0.97 & 6242.11 $\pm$ 0.97 & 64.7832 $\pm$ 0.0100 & 0.233058 $\pm$ 0.000054 & 0.15720 $\pm$ 0.00026 & -5.33832 $\pm$ 0.00162 & 0.181832 $\pm$ 0.000037 \\ 
1543.5788 & 17782.70897 $\pm$ 0.00038 & -49.82 $\pm$ 0.76 & 6242.24 $\pm$ 0.76 & 64.8624 $\pm$ 0.0079 & 0.233393 $\pm$ 0.000041 & 0.16324 $\pm$ 0.00017 & -5.30181 $\pm$ 0.00101 & 0.183252 $\pm$ 0.000028 \\ 
1544.6667 & 17778.73201 $\pm$ 0.00043 & -46.28 $\pm$ 0.86 & 6239.18 $\pm$ 0.86 & 64.8488 $\pm$ 0.0090 & 0.233570 $\pm$ 0.000047 & 0.14524 $\pm$ 0.00023 & -5.42100 $\pm$ 0.00174 & 0.182981 $\pm$ 0.000032 \\ 
1555.5608 & 17775.38652 $\pm$ 0.00069 & -43.03 $\pm$ 1.37 & 6258.51 $\pm$ 1.37 & 64.6856 $\pm$ 0.0142 & 0.238006 $\pm$ 0.000080 & 0.15029 $\pm$ 0.00045 & -5.38415 $\pm$ 0.00318 & 0.183600 $\pm$ 0.000056 \\ 
1561.5216 & 17783.19208 $\pm$ 0.00040 & -44.25 $\pm$ 0.81 & 6260.98 $\pm$ 0.81 & 64.7383 $\pm$ 0.0083 & 0.236252 $\pm$ 0.000044 & 0.20195 $\pm$ 0.00019 & -5.12109 $\pm$ 0.00074 & 0.183931 $\pm$ 0.000030 \\ 
1562.5582 & 17779.51079 $\pm$ 0.00052 & -43.83 $\pm$ 1.05 & 6258.01 $\pm$ 1.05 & 64.8468 $\pm$ 0.0109 & 0.238159 $\pm$ 0.000061 & 0.20341 $\pm$ 0.00031 & -5.11552 $\pm$ 0.00117 & 0.180552 $\pm$ 0.000041 \\ 
1568.5391 & 17776.39057 $\pm$ 0.00048 & -48.47 $\pm$ 0.97 & 6243.89 $\pm$ 0.97 & 64.8417 $\pm$ 0.0101 & 0.233227 $\pm$ 0.000055 & 0.15050 $\pm$ 0.00026 & -5.38268 $\pm$ 0.00184 & 0.182286 $\pm$ 0.000037 \\ 
1609.4563 & 17782.58879 $\pm$ 0.00050 & -46.96 $\pm$ 0.99 & 6265.95 $\pm$ 0.99 & 64.5660 $\pm$ 0.0102 & 0.240338 $\pm$ 0.000056 & 0.21216 $\pm$ 0.00028 & -5.08374 $\pm$ 0.00100 & 0.183978 $\pm$ 0.000038 \\ 
\hline
1706.8932 & 17776.96139 $\pm$ 0.00040 & -44.57 $\pm$ 0.79 & 6279.66 $\pm$ 0.79 & 64.5536 $\pm$ 0.0081 & 0.243018 $\pm$ 0.000039 & 0.19412 $\pm$ 0.00022 & -5.15206 $\pm$ 0.00091 & 0.183810 $\pm$ 0.000028 \\ 
1721.9036 & 17776.27705 $\pm$ 0.00039 & -37.17 $\pm$ 0.78 & 6304.30 $\pm$ 0.78 & 64.4522 $\pm$ 0.0080 & 0.248586 $\pm$ 0.000039 & 0.22886 $\pm$ 0.00020 & -5.02883 $\pm$ 0.00063 & 0.183840 $\pm$ 0.000027 \\ 
1725.8885 & 17775.55942 $\pm$ 0.00068 & -43.54 $\pm$ 1.36 & 6282.45 $\pm$ 1.36 & 64.4067 $\pm$ 0.0139 & 0.245452 $\pm$ 0.000072 & 0.17312 $\pm$ 0.00050 & -5.24805 $\pm$ 0.00255 & 0.184551 $\pm$ 0.000053 \\ 
1731.8831 & 17781.80461 $\pm$ 0.00035 & -41.79 $\pm$ 0.69 & 6287.82 $\pm$ 0.69 & 64.6882 $\pm$ 0.0071 & 0.246291 $\pm$ 0.000035 & 0.20756 $\pm$ 0.00017 & -5.10015 $\pm$ 0.00062 & 0.184109 $\pm$ 0.000024 \\ 
1737.8990 & 17780.70991 $\pm$ 0.00035 & -39.77 $\pm$ 0.71 & 6285.31 $\pm$ 0.71 & 64.7143 $\pm$ 0.0073 & 0.243864 $\pm$ 0.000036 & 0.21786 $\pm$ 0.00017 & -5.06420 $\pm$ 0.00058 & 0.185215 $\pm$ 0.000025 \\ 
1741.8442 & 17779.28006 $\pm$ 0.00056 & -41.27 $\pm$ 1.13 & 6285.83 $\pm$ 1.13 & 64.7897 $\pm$ 0.0116 & 0.245002 $\pm$ 0.000062 & 0.20847 $\pm$ 0.00039 & -5.09687 $\pm$ 0.00140 & 0.183821 $\pm$ 0.000044 \\ 
1754.8388 & 17780.12503 $\pm$ 0.00031 & -39.33 $\pm$ 0.63 & 6302.09 $\pm$ 0.63 & 64.6023 $\pm$ 0.0064 & 0.246549 $\pm$ 0.000031 & 0.22772 $\pm$ 0.00014 & -5.03237 $\pm$ 0.00045 & 0.186835 $\pm$ 0.000021 \\ 
1777.8239 & 17779.68478 $\pm$ 0.00035 & -40.56 $\pm$ 0.70 & 6302.94 $\pm$ 0.70 & 64.6602 $\pm$ 0.0072 & 0.244818 $\pm$ 0.000035 & 0.22674 $\pm$ 0.00018 & -5.03544 $\pm$ 0.00055 & 0.184643 $\pm$ 0.000024 \\ 
1787.6540 & 17778.88927 $\pm$ 0.00061 & -45.52 $\pm$ 1.23 & 6303.60 $\pm$ 1.23 & 64.7707 $\pm$ 0.0126 & 0.247897 $\pm$ 0.000067 & 0.16332 $\pm$ 0.00052 & -5.30135 $\pm$ 0.00303 & 0.180124 $\pm$ 0.000048 \\ 
1792.7690 & 17779.65316 $\pm$ 0.00040 & -37.50 $\pm$ 0.80 & 6301.48 $\pm$ 0.80 & 64.5317 $\pm$ 0.0082 & 0.249930 $\pm$ 0.000042 & 0.24296 $\pm$ 0.00022 & -4.98732 $\pm$ 0.00061 & 0.185291 $\pm$ 0.000030 \\ 
1799.6513 & 17785.27818 $\pm$ 0.00062 & -46.24 $\pm$ 1.24 & 6287.15 $\pm$ 1.24 & 64.5196 $\pm$ 0.0127 & 0.246690 $\pm$ 0.000069 & 0.18211 $\pm$ 0.00046 & -5.20435 $\pm$ 0.00212 & 0.178080 $\pm$ 0.000050 \\ 
1803.7665 & 17780.02632 $\pm$ 0.00032 & -38.71 $\pm$ 0.64 & 6297.88 $\pm$ 0.64 & 64.5964 $\pm$ 0.0066 & 0.246543 $\pm$ 0.000032 & 0.22973 $\pm$ 0.00015 & -5.02614 $\pm$ 0.00046 & 0.182195 $\pm$ 0.000022 \\ 
1810.7723 & 17782.17134 $\pm$ 0.00036 & -44.18 $\pm$ 0.72 & 6302.03 $\pm$ 0.72 & 64.6639 $\pm$ 0.0073 & 0.245792 $\pm$ 0.000037 & 0.22010 $\pm$ 0.00018 & -5.05676 $\pm$ 0.00060 & 0.181593 $\pm$ 0.000026 \\ 
1814.5958 & 17781.09472 $\pm$ 0.00046 & -40.73 $\pm$ 0.92 & 6305.16 $\pm$ 0.92 & 64.5965 $\pm$ 0.0095 & 0.246674 $\pm$ 0.000049 & 0.20803 $\pm$ 0.00030 & -5.09845 $\pm$ 0.00109 & 0.184152 $\pm$ 0.000035 \\ 
1817.7661 & 17780.90714 $\pm$ 0.00036 & -39.35 $\pm$ 0.72 & 6306.09 $\pm$ 0.72 & 64.5991 $\pm$ 0.0074 & 0.247541 $\pm$ 0.000038 & 0.26236 $\pm$ 0.00018 & -4.93602 $\pm$ 0.00045 & 0.184455 $\pm$ 0.000026 \\ 
1821.6856 & 17779.05916 $\pm$ 0.00031 & -42.38 $\pm$ 0.61 & 6298.12 $\pm$ 0.61 & 64.5847 $\pm$ 0.0063 & 0.246400 $\pm$ 0.000031 & 0.23237 $\pm$ 0.00014 & -5.01812 $\pm$ 0.00042 & 0.183621 $\pm$ 0.000021 \\ 
1823.7659 & 17779.88513 $\pm$ 0.00031 & -43.10 $\pm$ 0.62 & 6296.88 $\pm$ 0.62 & 64.5942 $\pm$ 0.0063 & 0.246982 $\pm$ 0.000031 & 0.23562 $\pm$ 0.00014 & -5.00844 $\pm$ 0.00041 & 0.182904 $\pm$ 0.000021 \\ 
1842.7292 & 17776.03053 $\pm$ 0.00032 & -44.87 $\pm$ 0.64 & 6279.77 $\pm$ 0.64 & 64.7319 $\pm$ 0.0066 & 0.243919 $\pm$ 0.000032 & 0.22245 $\pm$ 0.00015 & -5.04909 $\pm$ 0.00049 & 0.182919 $\pm$ 0.000022 \\ 
1846.6557 & 17782.24436 $\pm$ 0.00043 & -45.84 $\pm$ 0.87 & 6288.63 $\pm$ 0.87 & 64.6609 $\pm$ 0.0089 & 0.241227 $\pm$ 0.000046 & 0.24736 $\pm$ 0.00026 & -4.97516 $\pm$ 0.00070 & 0.178662 $\pm$ 0.000033 \\ 
1850.7092 & 17781.13340 $\pm$ 0.00038 & -40.46 $\pm$ 0.77 & 6297.82 $\pm$ 0.77 & 64.6492 $\pm$ 0.0079 & 0.245555 $\pm$ 0.000040 & 0.24473 $\pm$ 0.00020 & -4.98240 $\pm$ 0.00057 & 0.183237 $\pm$ 0.000028 \\ 
1853.7321 & 17779.60771 $\pm$ 0.00036 & -42.03 $\pm$ 0.73 & 6296.67 $\pm$ 0.73 & 64.6300 $\pm$ 0.0075 & 0.246085 $\pm$ 0.000037 & 0.22342 $\pm$ 0.00019 & -5.04598 $\pm$ 0.00062 & 0.182590 $\pm$ 0.000026 \\ 
1858.5947 & 17779.89539 $\pm$ 0.00035 & -41.30 $\pm$ 0.70 & 6288.11 $\pm$ 0.70 & 64.6435 $\pm$ 0.0072 & 0.243830 $\pm$ 0.000036 & 0.21323 $\pm$ 0.00018 & -5.08000 $\pm$ 0.00061 & 0.184315 $\pm$ 0.000025 \\ 
1868.7485 & 17781.68737 $\pm$ 0.00032 & -40.30 $\pm$ 0.65 & 6292.29 $\pm$ 0.65 & 64.6058 $\pm$ 0.0066 & 0.238746 $\pm$ 0.000031 & 0.22111 $\pm$ 0.00017 & -5.05346 $\pm$ 0.00054 & 0.169666 $\pm$ 0.000021 \\ 
1879.6403 & 17780.25646 $\pm$ 0.00025 & -41.25 $\pm$ 0.51 & 6315.65 $\pm$ 0.51 & 64.4895 $\pm$ 0.0052 & 0.247694 $\pm$ 0.000023 & 0.24000 $\pm$ 0.00011 & -4.99571 $\pm$ 0.00031 & 0.184474 $\pm$ 0.000016 \\ 
1882.7150 & 17783.15726 $\pm$ 0.00023 & -34.91 $\pm$ 0.46 & 6321.59 $\pm$ 0.46 & 64.4297 $\pm$ 0.0046 & 0.250763 $\pm$ 0.000020 & 0.26306 $\pm$ 0.00009 & -4.93428 $\pm$ 0.00022 & 0.184157 $\pm$ 0.000014 \\ 
1893.6461 & 17778.97123 $\pm$ 0.00032 & -40.07 $\pm$ 0.64 & 6287.76 $\pm$ 0.64 & 64.7047 $\pm$ 0.0066 & 0.238640 $\pm$ 0.000030 & 0.23122 $\pm$ 0.00016 & -5.02162 $\pm$ 0.00050 & 0.174204 $\pm$ 0.000021 \\ 
1899.7150 & 17782.46986 $\pm$ 0.00048 & -43.31 $\pm$ 0.97 & 6286.66 $\pm$ 0.97 & 64.5597 $\pm$ 0.0099 & 0.241957 $\pm$ 0.000050 & 0.19157 $\pm$ 0.00033 & -5.16266 $\pm$ 0.00137 & 0.178992 $\pm$ 0.000036 \\ 
1904.6154 & 17782.14442 $\pm$ 0.00023 & -43.02 $\pm$ 0.45 & 6294.25 $\pm$ 0.45 & 64.5740 $\pm$ 0.0046 & 0.242972 $\pm$ 0.000019 & 0.23179 $\pm$ 0.00009 & -5.01987 $\pm$ 0.00027 & 0.180343 $\pm$ 0.000013 \\ 
1910.6080 & 17780.75300 $\pm$ 0.00040 & -41.45 $\pm$ 0.81 & 6323.80 $\pm$ 0.81 & 64.4885 $\pm$ 0.0082 & 0.248201 $\pm$ 0.000040 & 0.26005 $\pm$ 0.00023 & -4.94184 $\pm$ 0.00059 & 0.182810 $\pm$ 0.000029 \\ 
1911.5885 & 17781.45598 $\pm$ 0.00054 & -41.52 $\pm$ 1.08 & 6323.04 $\pm$ 1.08 & 64.4913 $\pm$ 0.0111 & 0.249049 $\pm$ 0.000057 & 0.27227 $\pm$ 0.00037 & -4.91198 $\pm$ 0.00087 & 0.184110 $\pm$ 0.000041 \\ 
1926.5684 & 17784.58613 $\pm$ 0.00046 & -37.46 $\pm$ 0.93 & 6303.53 $\pm$ 0.93 & 64.7154 $\pm$ 0.0095 & 0.246158 $\pm$ 0.000048 & 0.25393 $\pm$ 0.00030 & -4.95758 $\pm$ 0.00078 & 0.182886 $\pm$ 0.000034 \\ 
1930.5353 & 17777.51337 $\pm$ 0.00042 & -38.91 $\pm$ 0.85 & 6295.68 $\pm$ 0.85 & 64.6550 $\pm$ 0.0087 & 0.242927 $\pm$ 0.000043 & 0.24155 $\pm$ 0.00026 & -4.99132 $\pm$ 0.00072 & 0.182278 $\pm$ 0.000030 \\ 
\hline
\end{tabular}
}
\end{center}
\end{table*}

\begin{table*}[pht]
\begin{center}
\caption{RVs and activity indices obtained from the ESPRESSO spectra for TOI-134. Values above the horizontal line correspond to ESPRESSO18 data, values below the horizontal line to ESPRESSO19 data.}
\label{tab:TOI-134_RV_ESPRESSO_data}
\centering
\resizebox{\textwidth}{!}{%
\begin{tabular}{lllllllll}
\hline  \hline
BJD - 2457000 [d] & RV [$\mathrm{km \, s^{-1}}$] & Bisector & FWHM & Contrast & H$_\alpha$ & $S_{MW}$ & log($R^\prime_{HK}$) & Na~I \\
\hline
1416.5357 & 29705.31080 $\pm$ 0.00079 & 33.78 $\pm$ 1.58 & 4667.99 $\pm$ 1.58 & 39.5613 $\pm$ 0.0134 & 0.460253 $\pm$ 0.000104 & 1.70354 $\pm$ 0.00210 & -4.80109 $\pm$ 0.00058 & 0.123551 $\pm$ 0.000069 \\ 
1420.5481 & 29716.18159 $\pm$ 0.00071 & 35.49 $\pm$ 1.41 & 4679.62 $\pm$ 1.41 & 39.4005 $\pm$ 0.0119 & 0.524765 $\pm$ 0.000093 & 2.08297 $\pm$ 0.00183 & -4.70649 $\pm$ 0.00041 & 0.132104 $\pm$ 0.000061 \\ 
1421.5384 & 29711.93089 $\pm$ 0.00075 & 36.63 $\pm$ 1.50 & 4685.98 $\pm$ 1.50 & 39.3738 $\pm$ 0.0126 & 0.515943 $\pm$ 0.000100 & 1.99090 $\pm$ 0.00201 & -4.72763 $\pm$ 0.00047 & 0.129836 $\pm$ 0.000065 \\ 
1422.5275 & 29715.64434 $\pm$ 0.00065 & 37.41 $\pm$ 1.29 & 4690.88 $\pm$ 1.29 & 39.3338 $\pm$ 0.0108 & 0.494207 $\pm$ 0.000082 & 1.90507 $\pm$ 0.00152 & -4.74830 $\pm$ 0.00038 & 0.128854 $\pm$ 0.000053 \\ 
1432.5284 & 29700.82250 $\pm$ 0.00053 & 41.13 $\pm$ 1.06 & 4683.71 $\pm$ 1.06 & 39.4199 $\pm$ 0.0089 & 0.489339 $\pm$ 0.000064 & 1.87938 $\pm$ 0.00104 & -4.75468 $\pm$ 0.00026 & 0.126868 $\pm$ 0.000041 \\ 
1459.5293 & 29703.27457 $\pm$ 0.00064 & 41.90 $\pm$ 1.28 & 4704.89 $\pm$ 1.28 & 39.2714 $\pm$ 0.0107 & 0.485370 $\pm$ 0.000081 & 1.87648 $\pm$ 0.00141 & -4.75541 $\pm$ 0.00035 & 0.128312 $\pm$ 0.000052 \\ 
1466.5502 & 29704.65694 $\pm$ 0.00074 & 40.16 $\pm$ 1.48 & 4672.88 $\pm$ 1.48 & 39.4351 $\pm$ 0.0125 & 0.485205 $\pm$ 0.000095 & 1.92654 $\pm$ 0.00192 & -4.74303 $\pm$ 0.00047 & 0.126752 $\pm$ 0.000062 \\ 
1477.5369 & 29709.79069 $\pm$ 0.00079 & 31.48 $\pm$ 1.59 & 4652.29 $\pm$ 1.59 & 39.7315 $\pm$ 0.0136 & 0.458693 $\pm$ 0.000105 & 1.71189 $\pm$ 0.00207 & -4.79877 $\pm$ 0.00058 & 0.122675 $\pm$ 0.000069 \\ 
1488.5509 & 29715.07490 $\pm$ 0.00072 & 40.01 $\pm$ 1.43 & 4703.17 $\pm$ 1.43 & 39.2046 $\pm$ 0.0119 & 0.509888 $\pm$ 0.000089 & 1.95166 $\pm$ 0.00183 & -4.73696 $\pm$ 0.00044 & 0.127515 $\pm$ 0.000058 \\ 
1493.5556 & 29707.95949 $\pm$ 0.00085 & 39.11 $\pm$ 1.70 & 4683.82 $\pm$ 1.70 & 39.4317 $\pm$ 0.0143 & 0.501803 $\pm$ 0.000110 & 1.94788 $\pm$ 0.00250 & -4.73786 $\pm$ 0.00060 & 0.127836 $\pm$ 0.000073 \\ 
1624.8674 & 29708.88793 $\pm$ 0.00143 & 38.36 $\pm$ 2.86 & 4671.85 $\pm$ 2.86 & 39.5836 $\pm$ 0.0243 & 0.476315 $\pm$ 0.000208 & 1.77610 $\pm$ 0.00564 & -4.78134 $\pm$ 0.00150 & 0.124653 $\pm$ 0.000146 \\ 
1625.8608 & 29710.82636 $\pm$ 0.00059 & 38.81 $\pm$ 1.19 & 4675.68 $\pm$ 1.19 & 39.5442 $\pm$ 0.0100 & 0.475657 $\pm$ 0.000071 & 1.78141 $\pm$ 0.00134 & -4.77993 $\pm$ 0.00036 & 0.124574 $\pm$ 0.000046 \\ 
1639.8591 & 29717.12347 $\pm$ 0.00088 & 40.23 $\pm$ 1.76 & 4706.90 $\pm$ 1.76 & 39.2025 $\pm$ 0.0146 & 0.541451 $\pm$ 0.000119 & 2.00316 $\pm$ 0.00273 & -4.72475 $\pm$ 0.00064 & 0.132112 $\pm$ 0.000076 \\ 
1642.8386 & 29710.79052 $\pm$ 0.00052 & 36.98 $\pm$ 1.03 & 4710.26 $\pm$ 1.03 & 39.2293 $\pm$ 0.0086 & 0.504908 $\pm$ 0.000060 & 1.88191 $\pm$ 0.00104 & -4.75405 $\pm$ 0.00026 & 0.129372 $\pm$ 0.000038 \\ 
\hline
1667.7026 & 29730.92913 $\pm$ 0.00253 & 27.87 $\pm$ 5.06 & 4707.48 $\pm$ 5.06 & 39.3096 $\pm$ 0.0423 & 0.517128 $\pm$ 0.000393 & 1.90430 $\pm$ 0.01277 & -4.74849 $\pm$ 0.00315 & 0.129429 $\pm$ 0.000289 \\ 
1667.7305 & 29727.97595 $\pm$ 0.00217 & 30.45 $\pm$ 4.35 & 4699.72 $\pm$ 4.35 & 39.3135 $\pm$ 0.0364 & 0.518110 $\pm$ 0.000343 & 1.96666 $\pm$ 0.00995 & -4.73336 $\pm$ 0.00237 & 0.131012 $\pm$ 0.000247 \\ 
1678.9386 & 29716.46567 $\pm$ 0.00048 & 38.96 $\pm$ 0.96 & 4699.41 $\pm$ 0.96 & 39.3177 $\pm$ 0.0080 & 0.503518 $\pm$ 0.000053 & 1.86557 $\pm$ 0.00094 & -4.75815 $\pm$ 0.00024 & 0.128625 $\pm$ 0.000035 \\ 
1684.6502 & 29713.67675 $\pm$ 0.00059 & 37.46 $\pm$ 1.18 & 4681.46 $\pm$ 1.18 & 39.4775 $\pm$ 0.0099 & 0.526411 $\pm$ 0.000066 & 1.95665 $\pm$ 0.00151 & -4.73576 $\pm$ 0.00036 & 0.131566 $\pm$ 0.000044 \\ 
1686.8663 & 29708.46950 $\pm$ 0.00056 & 38.55 $\pm$ 1.12 & 4670.30 $\pm$ 1.12 & 39.5840 $\pm$ 0.0095 & 0.464455 $\pm$ 0.000065 & 1.69978 $\pm$ 0.00122 & -4.80214 $\pm$ 0.00034 & 0.123148 $\pm$ 0.000043 \\ 
1688.6708 & 29713.24388 $\pm$ 0.00139 & 29.95 $\pm$ 2.78 & 4657.79 $\pm$ 2.78 & 39.6895 $\pm$ 0.0237 & 0.471515 $\pm$ 0.000202 & 1.75014 $\pm$ 0.00574 & -4.78830 $\pm$ 0.00155 & 0.122910 $\pm$ 0.000143 \\ 
1688.6822 & 29719.79901 $\pm$ 0.00126 & 32.27 $\pm$ 2.52 & 4663.83 $\pm$ 2.52 & 39.7025 $\pm$ 0.0214 & 0.477099 $\pm$ 0.000180 & 1.73454 $\pm$ 0.00484 & -4.79254 $\pm$ 0.00132 & 0.125128 $\pm$ 0.000126 \\ 
1688.6933 & 29714.85951 $\pm$ 0.00116 & 37.13 $\pm$ 2.33 & 4657.59 $\pm$ 2.33 & 39.7316 $\pm$ 0.0198 & 0.475391 $\pm$ 0.000164 & 1.75140 $\pm$ 0.00429 & -4.78796 $\pm$ 0.00116 & 0.124232 $\pm$ 0.000114 \\ 
1700.6294 & 29718.21560 $\pm$ 0.00054 & 33.72 $\pm$ 1.08 & 4715.37 $\pm$ 1.08 & 39.1533 $\pm$ 0.0090 & 0.557015 $\pm$ 0.000060 & 2.13521 $\pm$ 0.00130 & -4.69494 $\pm$ 0.00028 & 0.138393 $\pm$ 0.000040 \\ 
1716.6285 & 29708.99734 $\pm$ 0.00053 & 41.92 $\pm$ 1.07 & 4678.14 $\pm$ 1.07 & 39.4930 $\pm$ 0.0090 & 0.484520 $\pm$ 0.000059 & 1.77027 $\pm$ 0.00119 & -4.78289 $\pm$ 0.00032 & 0.124982 $\pm$ 0.000039 \\ 
1720.5718 & 29712.56456 $\pm$ 0.00054 & 34.82 $\pm$ 1.07 & 4650.90 $\pm$ 1.07 & 39.7366 $\pm$ 0.0092 & 0.462058 $\pm$ 0.000058 & 1.63242 $\pm$ 0.00125 & -4.82137 $\pm$ 0.00036 & 0.120357 $\pm$ 0.000038 \\ 
1725.7941 & 29716.37457 $\pm$ 0.00063 & 28.84 $\pm$ 1.26 & 4666.89 $\pm$ 1.26 & 39.6539 $\pm$ 0.0107 & 0.469030 $\pm$ 0.000075 & 1.68233 $\pm$ 0.00155 & -4.80704 $\pm$ 0.00044 & 0.122391 $\pm$ 0.000050 \\ 
1726.6188 & 29719.65897 $\pm$ 0.00053 & 28.20 $\pm$ 1.06 & 4669.96 $\pm$ 1.06 & 39.6021 $\pm$ 0.0090 & 0.481925 $\pm$ 0.000059 & 1.74331 $\pm$ 0.00117 & -4.79016 $\pm$ 0.00032 & 0.124685 $\pm$ 0.000039 \\ 
1731.5901 & 29717.85743 $\pm$ 0.00063 & 32.27 $\pm$ 1.27 & 4703.09 $\pm$ 1.27 & 39.2599 $\pm$ 0.0106 & 0.504754 $\pm$ 0.000074 & 1.88192 $\pm$ 0.00158 & -4.75405 $\pm$ 0.00040 & 0.127374 $\pm$ 0.000049 \\ 
1737.8008 & 29715.22313 $\pm$ 0.00057 & 35.97 $\pm$ 1.14 & 4701.90 $\pm$ 1.14 & 39.2976 $\pm$ 0.0095 & 0.503282 $\pm$ 0.000065 & 1.90524 $\pm$ 0.00132 & -4.74826 $\pm$ 0.00032 & 0.130165 $\pm$ 0.000043 \\ 
1741.6123 & 29721.28545 $\pm$ 0.00055 & 38.13 $\pm$ 1.11 & 4701.95 $\pm$ 1.11 & 39.2684 $\pm$ 0.0092 & 0.511297 $\pm$ 0.000062 & 1.95838 $\pm$ 0.00126 & -4.73534 $\pm$ 0.00030 & 0.132073 $\pm$ 0.000041 \\ 
1752.5454 & 29706.80017 $\pm$ 0.00092 & 38.88 $\pm$ 1.85 & 4667.45 $\pm$ 1.85 & 39.5896 $\pm$ 0.0157 & 0.485970 $\pm$ 0.000120 & 1.78191 $\pm$ 0.00302 & -4.77979 $\pm$ 0.00080 & 0.125874 $\pm$ 0.000082 \\ 
1753.5769 & 29707.59852 $\pm$ 0.00052 & 35.02 $\pm$ 1.03 & 4670.48 $\pm$ 1.03 & 39.5875 $\pm$ 0.0088 & 0.482761 $\pm$ 0.000057 & 1.76984 $\pm$ 0.00112 & -4.78301 $\pm$ 0.00030 & 0.125283 $\pm$ 0.000037 \\ 
1754.7193 & 29712.75430 $\pm$ 0.00046 & 37.02 $\pm$ 0.92 & 4668.00 $\pm$ 0.92 & 39.5745 $\pm$ 0.0078 & 0.480980 $\pm$ 0.000050 & 1.75916 $\pm$ 0.00090 & -4.78587 $\pm$ 0.00024 & 0.124547 $\pm$ 0.000032 \\ 
1769.6861 & 29716.74328 $\pm$ 0.00106 & 33.33 $\pm$ 2.13 & 4707.57 $\pm$ 2.13 & 39.2408 $\pm$ 0.0177 & 0.500313 $\pm$ 0.000141 & 1.86356 $\pm$ 0.00359 & -4.75866 $\pm$ 0.00091 & 0.128678 $\pm$ 0.000097 \\ 
1773.5490 & 29714.84821 $\pm$ 0.00073 & 36.25 $\pm$ 1.47 & 4711.28 $\pm$ 1.47 & 39.1645 $\pm$ 0.0122 & 0.509285 $\pm$ 0.000090 & 1.94779 $\pm$ 0.00199 & -4.73789 $\pm$ 0.00048 & 0.130218 $\pm$ 0.000060 \\ 
1776.5199 & 29712.58645 $\pm$ 0.00057 & 39.92 $\pm$ 1.14 & 4702.67 $\pm$ 1.14 & 39.2789 $\pm$ 0.0096 & 0.501934 $\pm$ 0.000065 & 1.89209 $\pm$ 0.00132 & -4.75151 $\pm$ 0.00033 & 0.128994 $\pm$ 0.000043 \\ 
1777.6919 & 29711.22043 $\pm$ 0.00059 & 41.38 $\pm$ 1.17 & 4692.57 $\pm$ 1.17 & 39.3719 $\pm$ 0.0098 & 0.497104 $\pm$ 0.000067 & 1.87383 $\pm$ 0.00138 & -4.75607 $\pm$ 0.00035 & 0.129282 $\pm$ 0.000045 \\ 
1787.5724 & 29710.14763 $\pm$ 0.00071 & 33.99 $\pm$ 1.43 & 4665.82 $\pm$ 1.43 & 39.5871 $\pm$ 0.0121 & 0.452101 $\pm$ 0.000083 & 1.69261 $\pm$ 0.00199 & -4.80415 $\pm$ 0.00056 & 0.121037 $\pm$ 0.000057 \\ 
1799.5679 & 29720.35507 $\pm$ 0.00059 & 34.24 $\pm$ 1.19 & 4701.24 $\pm$ 1.19 & 39.2351 $\pm$ 0.0099 & 0.522633 $\pm$ 0.000067 & 2.00830 $\pm$ 0.00150 & -4.72355 $\pm$ 0.00035 & 0.130735 $\pm$ 0.000044 \\ 
1803.6971 & 29723.60978 $\pm$ 0.00069 & 36.21 $\pm$ 1.38 & 4706.29 $\pm$ 1.38 & 39.2106 $\pm$ 0.0115 & 0.500291 $\pm$ 0.000077 & 1.89962 $\pm$ 0.00203 & -4.74964 $\pm$ 0.00050 & 0.129240 $\pm$ 0.000053 \\ 
1806.5335 & 29718.00498 $\pm$ 0.00070 & 37.16 $\pm$ 1.41 & 4700.31 $\pm$ 1.41 & 39.2757 $\pm$ 0.0118 & 0.495717 $\pm$ 0.000083 & 1.82852 $\pm$ 0.00195 & -4.76760 $\pm$ 0.00050 & 0.126518 $\pm$ 0.000056 \\ 
1810.6921 & 29709.99996 $\pm$ 0.00137 & 35.11 $\pm$ 2.74 & 4671.23 $\pm$ 2.74 & 39.5611 $\pm$ 0.0232 & 0.466239 $\pm$ 0.000185 & 1.67344 $\pm$ 0.00594 & -4.80956 $\pm$ 0.00169 & 0.122401 $\pm$ 0.000132 \\ 
1814.5040 & 29709.66793 $\pm$ 0.00079 & 33.18 $\pm$ 1.59 & 4649.60 $\pm$ 1.59 & 39.8121 $\pm$ 0.0136 & 0.441038 $\pm$ 0.000098 & 1.59682 $\pm$ 0.00227 & -4.83189 $\pm$ 0.00068 & 0.118887 $\pm$ 0.000067 \\ 
1821.5392 & 29718.65271 $\pm$ 0.00050 & 28.76 $\pm$ 1.00 & 4668.72 $\pm$ 1.00 & 39.5613 $\pm$ 0.0085 & 0.459229 $\pm$ 0.000054 & 1.65604 $\pm$ 0.00105 & -4.81453 $\pm$ 0.00030 & 0.122176 $\pm$ 0.000036 \\ 
1838.5526 & 29717.63847 $\pm$ 0.00090 & 38.46 $\pm$ 1.81 & 4694.63 $\pm$ 1.81 & 39.3131 $\pm$ 0.0151 & 0.475470 $\pm$ 0.000115 & 1.79821 $\pm$ 0.00288 & -4.77549 $\pm$ 0.00076 & 0.123840 $\pm$ 0.000078 \\ 
1838.5664 & 29716.50644 $\pm$ 0.00085 & 37.87 $\pm$ 1.70 & 4693.28 $\pm$ 1.70 & 39.3116 $\pm$ 0.0142 & 0.477970 $\pm$ 0.000106 & 1.78083 $\pm$ 0.00262 & -4.78008 $\pm$ 0.00070 & 0.125544 $\pm$ 0.000072 \\ 
1846.5277 & 29705.58639 $\pm$ 0.00068 & 29.93 $\pm$ 1.36 & 4637.40 $\pm$ 1.36 & 39.8114 $\pm$ 0.0117 & 0.440417 $\pm$ 0.000078 & 1.59287 $\pm$ 0.00188 & -4.83307 $\pm$ 0.00056 & 0.116660 $\pm$ 0.000053 \\ 
1858.5617 & 29716.52385 $\pm$ 0.00094 & 41.54 $\pm$ 1.89 & 4684.26 $\pm$ 1.89 & 39.3992 $\pm$ 0.0159 & 0.498129 $\pm$ 0.000118 & 1.80793 $\pm$ 0.00328 & -4.77295 $\pm$ 0.00086 & 0.125457 $\pm$ 0.000082 \\ 

\hline
\end{tabular}
}
\end{center}
\end{table*}

\begin{table*}[pht]
\begin{center}
\caption{RVs and activity indices obtained from the HARPS spectra for TOI-286.}
\label{tab:TOI-286_RV_HARPS_data}
\centering
\resizebox{\textwidth}{!}{%
\begin{tabular}{lllllllll}
\hline  \hline
BJD - 2457000 [d] & RV [km/s] & Bisector & FWHM & Contrast & H$_\alpha$ & $S_{MW}$ & Na~D1 & Na~D2 \\
\hline
1501.6148 & -0.00071 $\pm$ 0.00067 & 0.017 & 6.115  & 44.00 & 0.8794  & 0.1912 & 1.0430 & 0.7758  \\ 
1502.6504 & 0.00313 $\pm$ 0.00057 & 0.015 & 6.114  & 43.99 & 0.8745  & 0.1938 & 1.0444 & 0.7718  \\ 
1504.6919 & 0.00085 $\pm$ 0.00060 & 0.018 & 6.126  & 43.98 & 0.8752  & 0.1903 & 1.0456 & 0.7709  \\ 
1508.6066 & 0.00000\tablefootmark{a} $\pm$ 0.00046 & 0.022 & 6.123  & 43.94 & 0.8956  & 0.1984 & 1.0448 & 0.7529  \\ 
1508.6288 & -0.00044 $\pm$ 0.00055 & 0.020 & 6.125  & 43.92 & 0.8973  & 0.2055 & 1.0470 & 0.7547  \\ 
1516.7304 & -0.00030 $\pm$ 0.00088 & 0.018 & 6.119  & 44.01 & 0.8877  & 0.1846 & 1.0464 & 0.7577  \\ 
\hline
\end{tabular}
}
\tablefoottext{a}{HARPS-TERRA RVs are relative; this is the zero-point RV.}
\end{center}
\end{table*}

\clearpage

\section{GP fits to activity indicators}

In this appendix, we show the GP fits to the selected activity indicators that were used to determine priors for the GP fits to the RVs.

\begin{figure*}
    \centering
    \includegraphics[width=.95\textwidth]{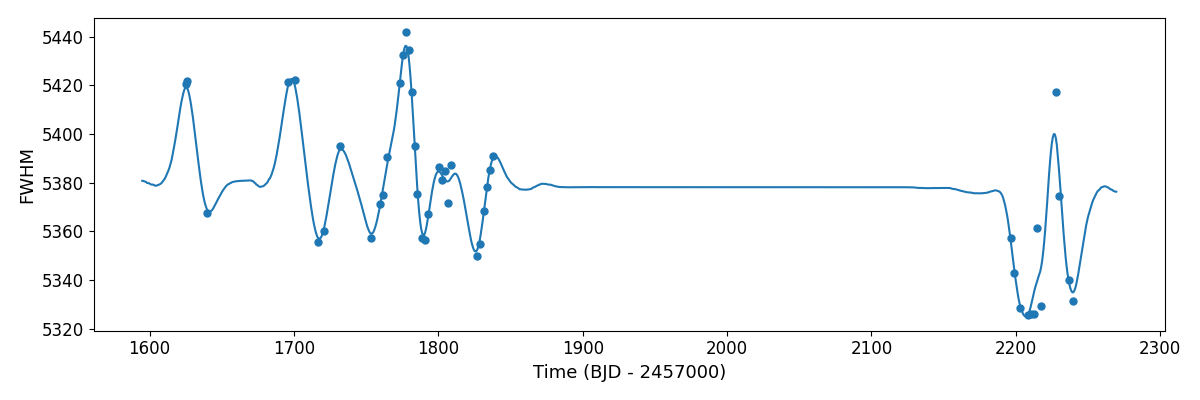}
    \includegraphics[width=.95\textwidth]{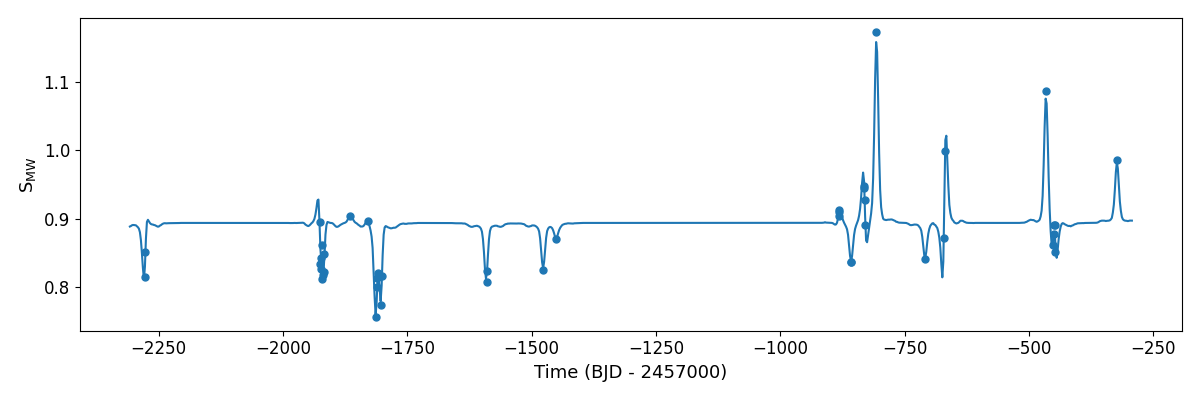}
    \includegraphics[width=.95\textwidth]{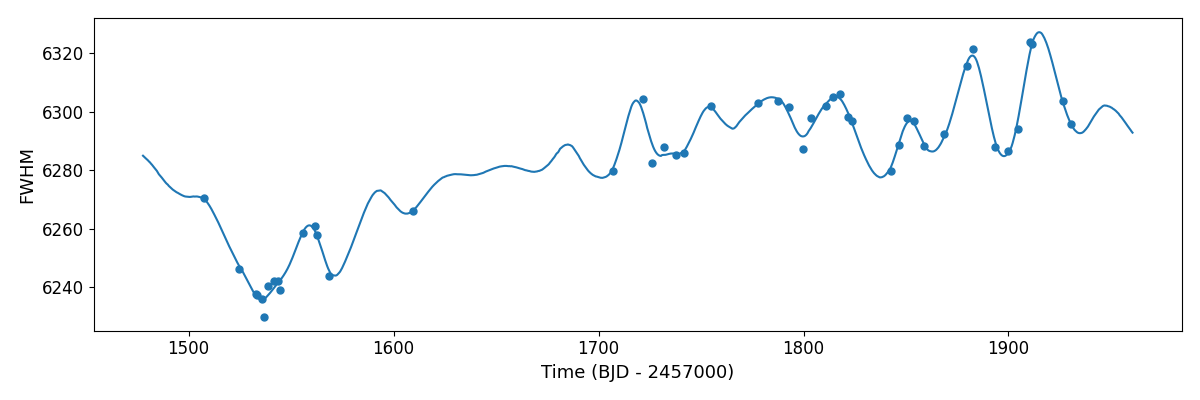}
    \includegraphics[width=.95\textwidth]{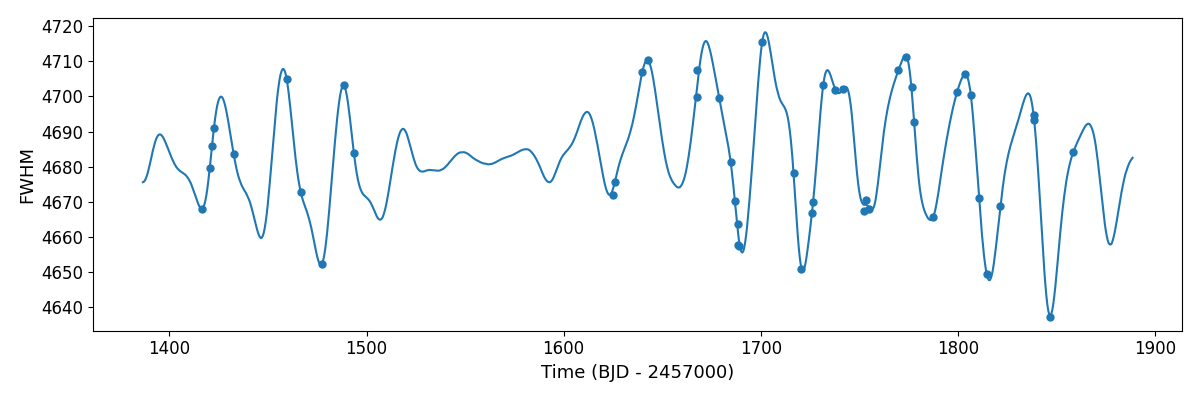}
    \caption{Quasiperiodic GP fits to selected activity indicators for: (top) the FWHM from ESPRESSO data for TOI-260; (second) the $\mathrm{S_{MW}}$ from HIRES data for TOI-260; (third) the FWHM from ESPRESSO data for TOI-286; (bottom) the FWHM from ESPRESSO data for TOI-134.}
    \label{fig:act-ind-gps}
\end{figure*}

\clearpage

\section{\textit{TESS} sector-by-sector lightcurves}

In this appendix, we show the full \textit{TESS} light curves, fitted models, and phase-folded data for each sector for TOI-260 and TOI-134. For TOI-286, we show the fitted models to the out-of-transit and in-transit data for sector 1 as an example.

\begin{figure*}
    \centering
    \includegraphics[width=.95\textwidth]{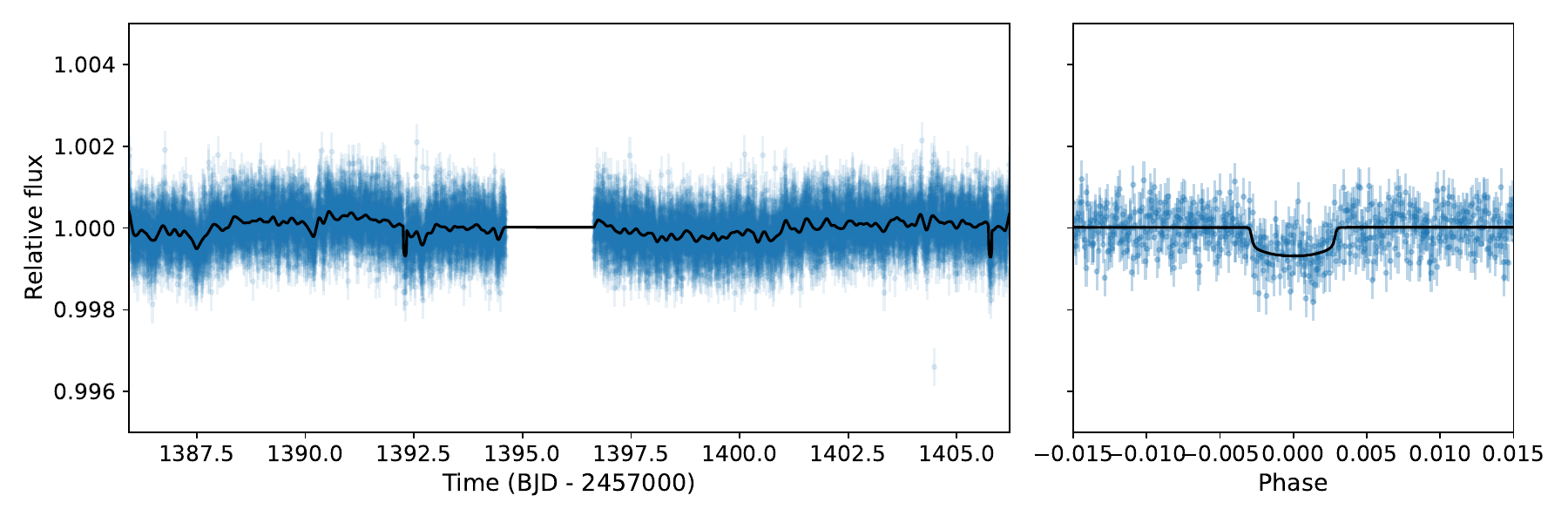}
    \includegraphics[width=.95\textwidth]{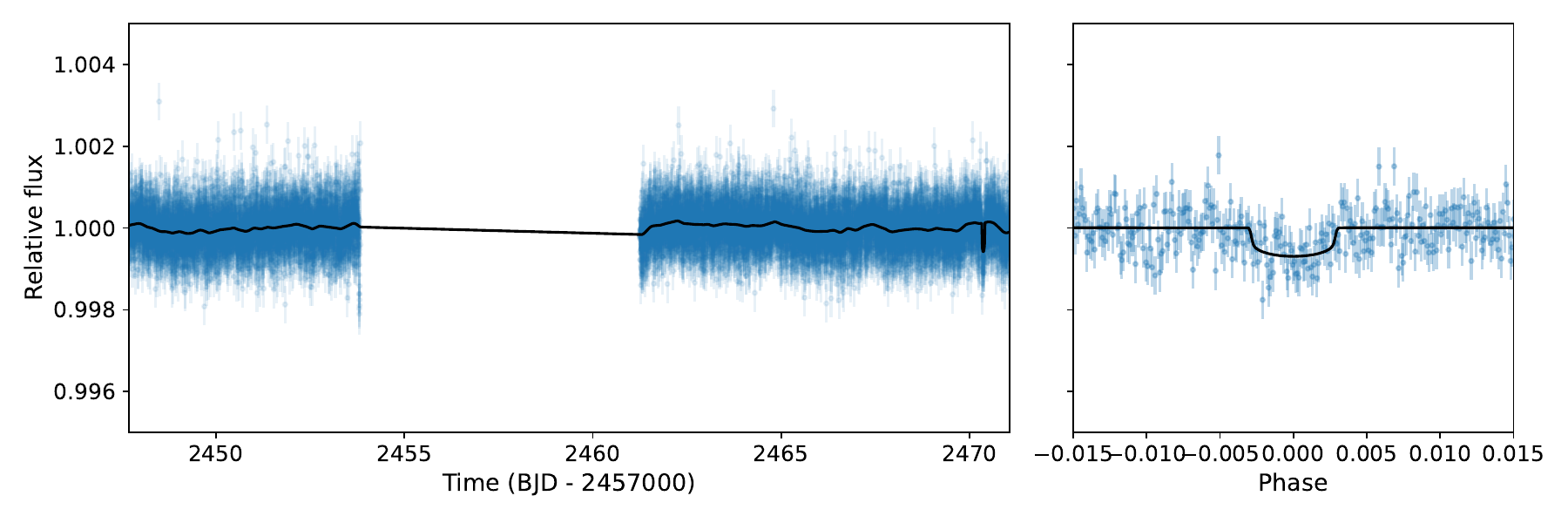}
    \includegraphics[width=.95\textwidth]{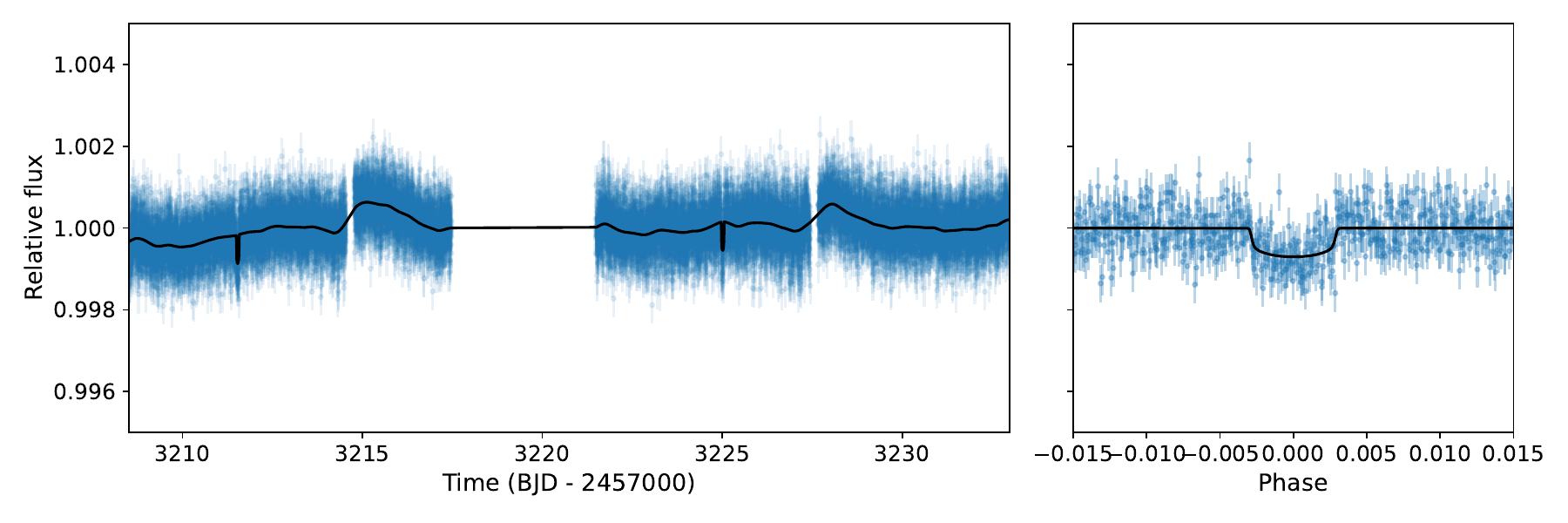}
    \caption{\textit{TESS} data with the median circular model and GP for TOI-260~b (left) and phase-folded transits (right), for sectors 3 (top), 42 (middle), and 70 (bottom).}
    \label{fig:toi-260_TESS_sectors_circ}
\end{figure*}

\begin{figure*}
    \centering
    \includegraphics[width=.95\textwidth]{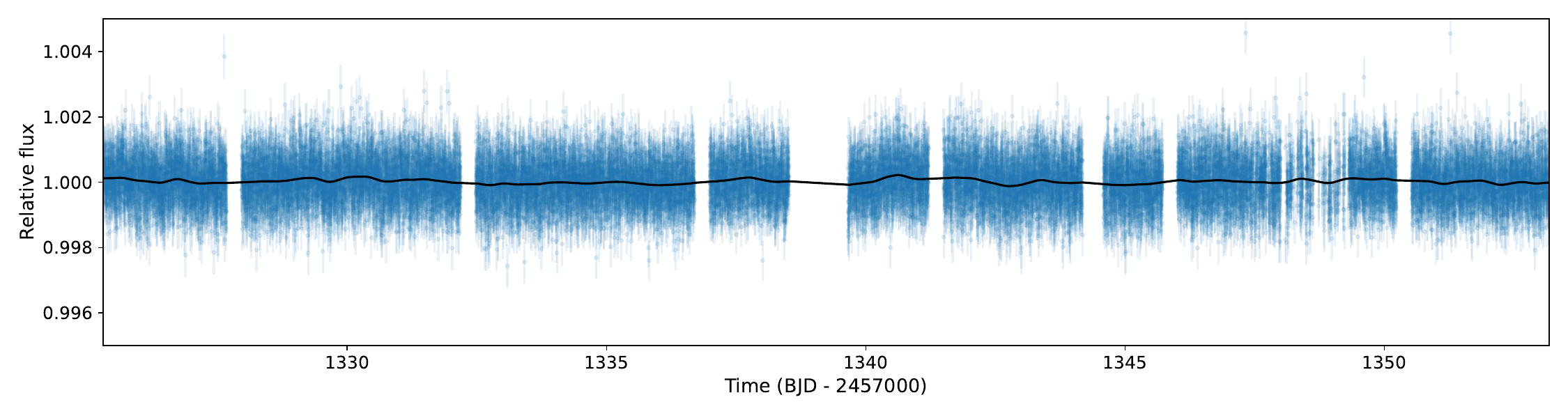}
    \includegraphics[width=.95\textwidth]{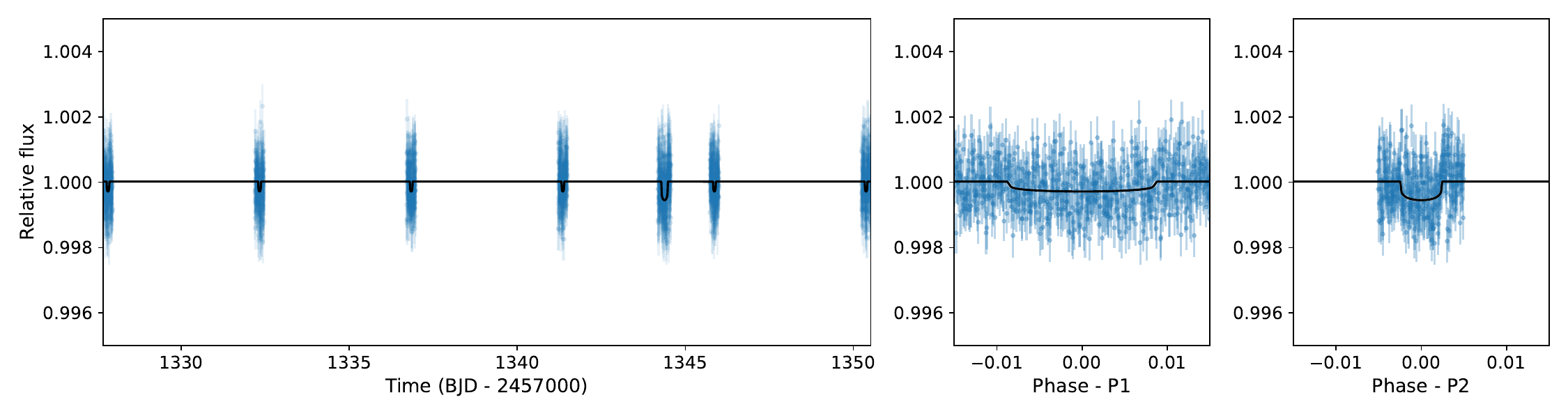}
    \caption{Top: out-of-transit \textit{TESS} sector 1 data for TOI-286 (blue points) and GP model (black line). Bottom: in-transit \textit{TESS} sector 1 data with the median circular model (left) and phase-folded transits for planets b (centre) and c (right).}
    \label{fig:toi-286_TESS_sectors_circ}
\end{figure*}

\begin{figure*}
    \centering
    \includegraphics[width=.95\textwidth]{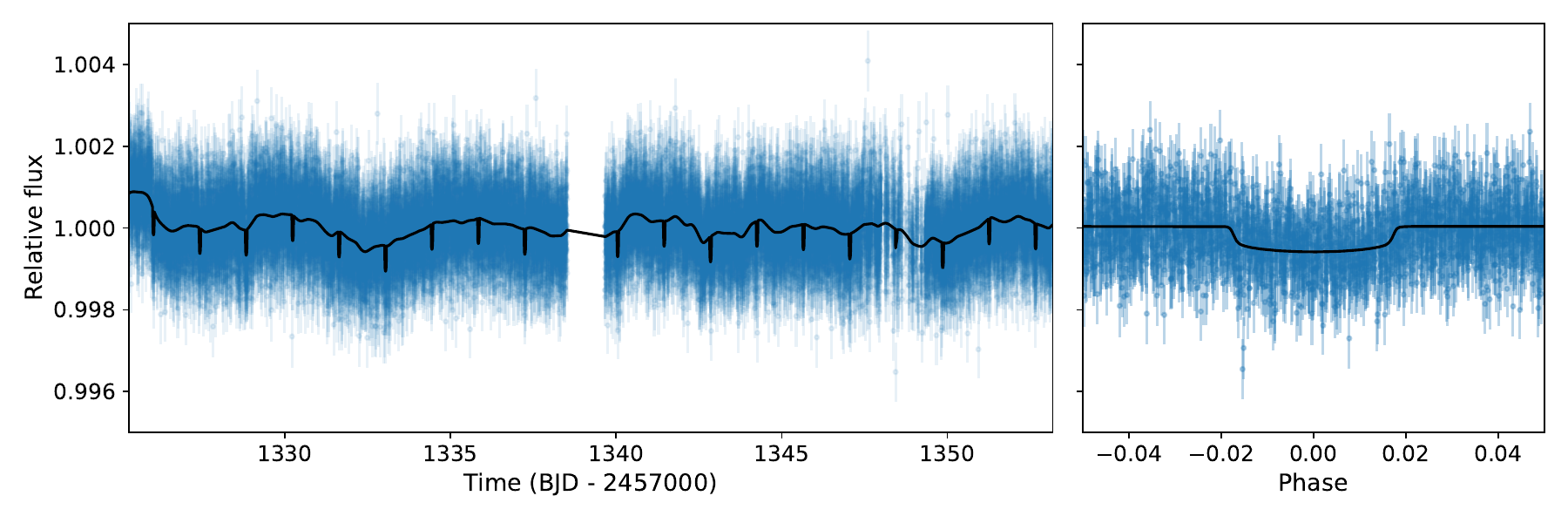}
    \includegraphics[width=.95\textwidth]{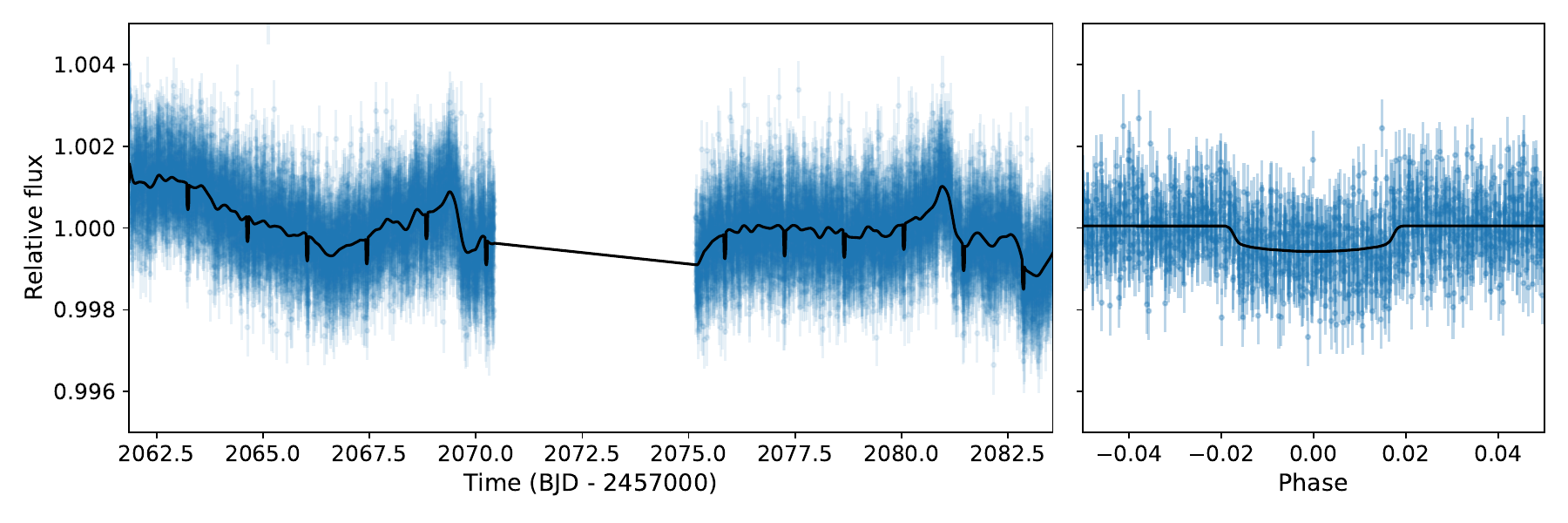}
    \includegraphics[width=.95\textwidth]{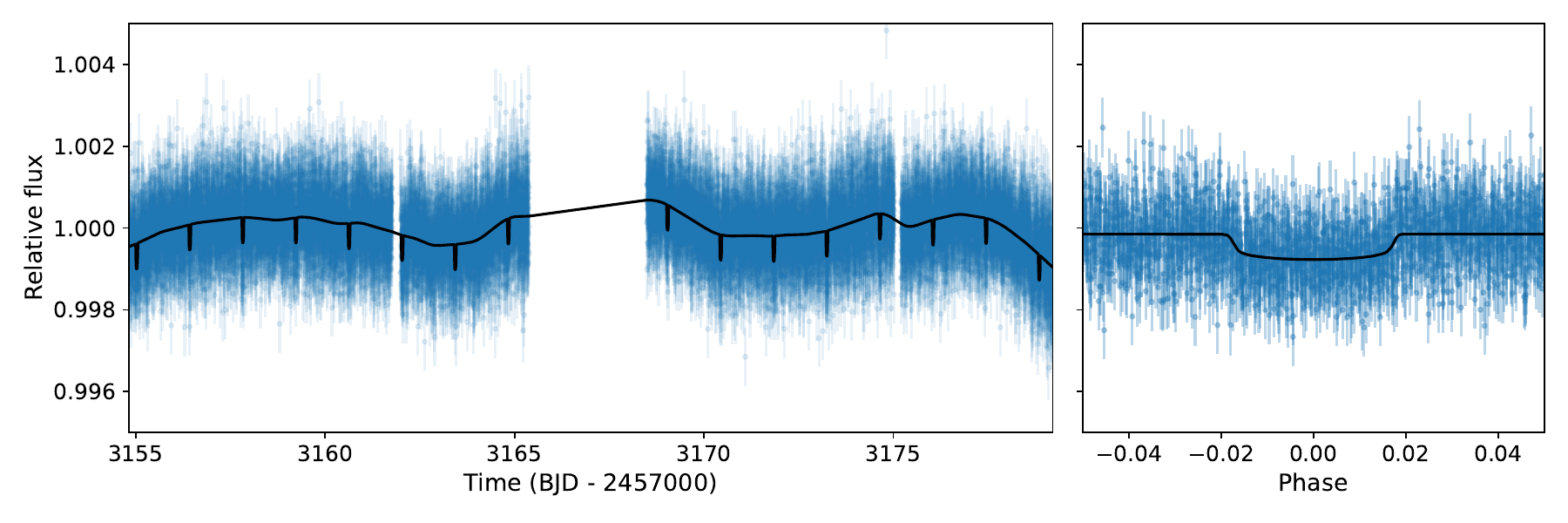}
    \caption{\textit{TESS} data with the median circular model and GP for TOI-134~b (left) and phase-folded transits (right), for sectors 1 (top), 28 (middle), and 68 (bottom).}
    \label{fig:toi-134_TESS_sectors_circ}
\end{figure*}

\clearpage

\section{\textit{TESS} sector-by-sector GPs for TOI-286}\label{ap:TOI286_TESSGP}

In this appendix, we give the priors and posteriors of the GPs fitted to the out-of-transit data for TOI-286 for each sector. For simplicity, we group them into three tables: \ref{tab:TOI286_TESSGP_prime} lists the prime mission sectors, \ref{tab:TOI286_TESSGP_em1} lists the first extended mission sectors, and \ref{tab:TOI286_TESSGP_em2} lists the second extended mission sectors.

\begin{table}[pht] 
\begin{center} 
\caption{Prior and posterior planetary parameter distributions obtained with \texttt{juliet} for the sector-by-sector out-of-transit detrending of \textit{TESS} data for TOI-286, for the prime mission.} 
\label{tab:TOI286_TESSGP_prime} 
\centering 
\resizebox{\columnwidth}{!}{%
\begin{tabular}{lll} 
\hline  \hline 
Parameter & Prior & Posterior \\ 
\hline 
$\mathrm{m_{dilution,TESS1}}$ \dotfill & $\mathrm{fixed}$ & \twoeightysixTESSmdilutionTESSone \\
$\mathrm{m_{flux,TESS1}}$ \dotfill & $\mathcal{N}(0.0,0.1)$ & \twoeightysixTESSmfluxTESSone \\
$\mathrm{\sigma_{w,TESS1}}$ \dotfill & $\mathcal{J}(0.1,1000.0)$ & \twoeightysixTESSsigmawTESSone \\
$\mathrm{\sigma_{GP,TESS1}}$ \dotfill & $\mathcal{J}(1e-06,1000000.0)$ & \twoeightysixTESSGPsigmaTESSone \\
$\mathrm{\rho_{GP,TESS1}}$ \dotfill & $\mathcal{J}(0.001,1000.0)$ & \twoeightysixTESSGPrhoTESSone \\
$\mathrm{m_{dilution,TESS2}}$ \dotfill & $\mathrm{fixed}$ & \twoeightysixTESSmdilutionTESStwo \\
$\mathrm{m_{flux,TESS2}}$ \dotfill & $\mathcal{N}(0.0,0.1)$ & \twoeightysixTESSmfluxTESStwo \\
$\mathrm{\sigma_{w,TESS2}}$ \dotfill & $\mathcal{J}(0.1,1000.0)$ & \twoeightysixTESSsigmawTESStwo \\
$\mathrm{\sigma_{GP,TESS2}}$ \dotfill & $\mathcal{J}(1e-06,1000000.0)$ & \twoeightysixTESSGPsigmaTESStwo \\
$\mathrm{\rho_{GP,TESS2}}$ \dotfill & $\mathcal{J}(0.001,1000.0)$ & \twoeightysixTESSGPrhoTESStwo \\
$\mathrm{m_{dilution,TESS3}}$ \dotfill & $\mathrm{fixed}$ & \twoeightysixTESSmdilutionTESSthree \\
$\mathrm{m_{flux,TESS3}}$ \dotfill & $\mathcal{N}(0.0,0.1)$ & \twoeightysixTESSmfluxTESSthree \\
$\mathrm{\sigma_{w,TESS3}}$ \dotfill & $\mathcal{J}(0.1,1000.0)$ & \twoeightysixTESSsigmawTESSthree \\
$\mathrm{\sigma_{GP,TESS3}}$ \dotfill & $\mathcal{J}(1e-06,1000000.0)$ & \twoeightysixTESSGPsigmaTESSthree \\
$\mathrm{\rho_{GP,TESS3}}$ \dotfill & $\mathcal{J}(0.001,1000.0)$ & \twoeightysixTESSGPrhoTESSthree \\
$\mathrm{m_{dilution,TESS4}}$ \dotfill & $\mathrm{fixed}$ & \twoeightysixTESSmdilutionTESSfour \\
$\mathrm{m_{flux,TESS4}}$ \dotfill & $\mathcal{N}(0.0,0.1)$ & \twoeightysixTESSmfluxTESSfour \\
$\mathrm{\sigma_{w,TESS4}}$ \dotfill & $\mathcal{J}(0.1,1000.0)$ & \twoeightysixTESSsigmawTESSfour \\
$\mathrm{\sigma_{GP,TESS4}}$ \dotfill & $\mathcal{J}(1e-06,1000000.0)$ & \twoeightysixTESSGPsigmaTESSfour \\
$\mathrm{\rho_{GP,TESS4}}$ \dotfill & $\mathcal{J}(0.001,1000.0)$ & \twoeightysixTESSGPrhoTESSfour \\
$\mathrm{m_{dilution,TESS5}}$ \dotfill & $\mathrm{fixed}$ & \twoeightysixTESSmdilutionTESSfive \\
$\mathrm{m_{flux,TESS5}}$ \dotfill & $\mathcal{N}(0.0,0.1)$ & \twoeightysixTESSmfluxTESSfive \\
$\mathrm{\sigma_{w,TESS5}}$ \dotfill & $\mathcal{J}(0.1,1000.0)$ & \twoeightysixTESSsigmawTESSfive \\
$\mathrm{\sigma_{GP,TESS5}}$ \dotfill & $\mathcal{J}(1e-06,1000000.0)$ & \twoeightysixTESSGPsigmaTESSfive \\
$\mathrm{\rho_{GP,TESS5}}$ \dotfill & $\mathcal{J}(0.001,1000.0)$ & \twoeightysixTESSGPrhoTESSfive \\
$\mathrm{m_{dilution,TESS7}}$ \dotfill & $\mathrm{fixed}$ & \twoeightysixTESSmdilutionTESSseven \\
$\mathrm{m_{flux,TESS7}}$ \dotfill & $\mathcal{N}(0.0,0.1)$ & \twoeightysixTESSmfluxTESSseven \\
$\mathrm{\sigma_{w,TESS7}}$ \dotfill & $\mathcal{J}(0.1,1000.0)$ & \twoeightysixTESSsigmawTESSseven \\
$\mathrm{\sigma_{GP,TESS7}}$ \dotfill & $\mathcal{J}(1e-06,1000000.0)$ & \twoeightysixTESSGPsigmaTESSseven \\
$\mathrm{\rho_{GP,TESS7}}$ \dotfill & $\mathcal{J}(0.001,1000.0)$ & \twoeightysixTESSGPrhoTESSseven \\
$\mathrm{m_{dilution,TESS8}}$ \dotfill & $\mathrm{fixed}$ & \twoeightysixTESSmdilutionTESSeight \\
$\mathrm{m_{flux,TESS8}}$ \dotfill & $\mathcal{N}(0.0,0.1)$ & \twoeightysixTESSmfluxTESSeight \\
$\mathrm{\sigma_{w,TESS8}}$ \dotfill & $\mathcal{J}(0.1,1000.0)$ & \twoeightysixTESSsigmawTESSeight \\
$\mathrm{\sigma_{GP,TESS8}}$ \dotfill & $\mathcal{J}(1e-06,1000000.0)$ & \twoeightysixTESSGPsigmaTESSeight \\
$\mathrm{\rho_{GP,TESS8}}$ \dotfill & $\mathcal{J}(0.001,1000.0)$ & \twoeightysixTESSGPrhoTESSeight \\
$\mathrm{m_{dilution,TESS9}}$ \dotfill & $\mathrm{fixed}$ & \twoeightysixTESSmdilutionTESSnine \\
$\mathrm{m_{flux,TESS9}}$ \dotfill & $\mathcal{N}(0.0,0.1)$ & \twoeightysixTESSmfluxTESSnine \\
$\mathrm{\sigma_{w,TESS9}}$ \dotfill & $\mathcal{J}(0.1,1000.0)$ & \twoeightysixTESSsigmawTESSnine \\
$\mathrm{\sigma_{GP,TESS9}}$ \dotfill & $\mathcal{J}(1e-06,1000000.0)$ & \twoeightysixTESSGPsigmaTESSnine \\
$\mathrm{\rho_{GP,TESS9}}$ \dotfill & $\mathcal{J}(0.001,1000.0)$ & \twoeightysixTESSGPrhoTESSnine \\
$\mathrm{m_{dilution,TESS10}}$ \dotfill & $\mathrm{fixed}$ & \twoeightysixTESSmdilutionTESSten \\
$\mathrm{m_{flux,TESS10}}$ \dotfill & $\mathcal{N}(0.0,0.1)$ & \twoeightysixTESSmfluxTESSten \\
$\mathrm{\sigma_{w,TESS10}}$ \dotfill & $\mathcal{J}(0.1,1000.0)$ & \twoeightysixTESSsigmawTESSten \\
$\mathrm{\sigma_{GP,TESS10}}$ \dotfill & $\mathcal{J}(1e-06,1000000.0)$ & \twoeightysixTESSGPsigmaTESSten \\
$\mathrm{\rho_{GP,TESS10}}$ \dotfill & $\mathcal{J}(0.001,1000.0)$ & \twoeightysixTESSGPrhoTESSten \\
$\mathrm{m_{dilution,TESS11}}$ \dotfill & $\mathrm{fixed}$ & \twoeightysixTESSmdilutionTESSeleven \\
$\mathrm{m_{flux,TESS11}}$ \dotfill & $\mathcal{N}(0.0,0.1)$ & \twoeightysixTESSmfluxTESSeleven \\
$\mathrm{\sigma_{w,TESS11}}$ \dotfill & $\mathcal{J}(0.1,1000.0)$ & \twoeightysixTESSsigmawTESSeleven \\
$\mathrm{\sigma_{GP,TESS11}}$ \dotfill & $\mathcal{J}(1e-06,1000000.0)$ & \twoeightysixTESSGPsigmaTESSeleven \\
$\mathrm{\rho_{GP,TESS11}}$ \dotfill & $\mathcal{J}(0.001,1000.0)$ & \twoeightysixTESSGPrhoTESSeleven \\
$\mathrm{m_{dilution,TESS12}}$ \dotfill & $\mathrm{fixed}$ & \twoeightysixTESSmdilutionTESStwelve \\
$\mathrm{m_{flux,TESS12}}$ \dotfill & $\mathcal{N}(0.0,0.1)$ & \twoeightysixTESSmfluxTESStwelve \\
$\mathrm{\sigma_{w,TESS12}}$ \dotfill & $\mathcal{J}(0.1,1000.0)$ & \twoeightysixTESSsigmawTESStwelve \\
$\mathrm{\sigma_{GP,TESS12}}$ \dotfill & $\mathcal{J}(1e-06,1000000.0)$ & \twoeightysixTESSGPsigmaTESStwelve \\
$\mathrm{\rho_{GP,TESS12}}$ \dotfill & $\mathcal{J}(0.001,1000.0)$ & \twoeightysixTESSGPrhoTESStwelve \\
$\mathrm{m_{dilution,TESS13}}$ \dotfill & $\mathrm{fixed}$ & \twoeightysixTESSmdilutionTESSthirteen \\
$\mathrm{m_{flux,TESS13}}$ \dotfill & $\mathcal{N}(0.0,0.1)$ & \twoeightysixTESSmfluxTESSthirteen \\
$\mathrm{\sigma_{w,TESS13}}$ \dotfill & $\mathcal{J}(0.1,1000.0)$ & \twoeightysixTESSsigmawTESSthirteen \\
$\mathrm{\sigma_{GP,TESS13}}$ \dotfill & $\mathcal{J}(1e-06,1000000.0)$ & \twoeightysixTESSGPsigmaTESSthirteen \\
$\mathrm{\rho_{GP,TESS13}}$ \dotfill & $\mathcal{J}(0.001,1000.0)$ & \twoeightysixTESSGPrhoTESSthirteen \\
\hline 
\end{tabular} 
} 
\end{center} 
\end{table} 

\begin{table}[pht] 
\begin{center} 
\caption{Prior and posterior planetary parameter distributions obtained with \texttt{juliet} for the sector-by-sector out-of-transit detrending of TESS data for TOI-286, for the first extended mission.} 
\label{tab:TOI286_TESSGP_em1} 
\centering 
\resizebox{.9\columnwidth}{!}{%
\begin{tabular}{lll} 
\hline  \hline 
Parameter & Prior & Posterior \\ 
\hline 
$\mathrm{m_{dilution,TESS27}}$ \dotfill & $\mathrm{fixed}$ & \twoeightysixTESSmdilutionTESStwentyseven \\
$\mathrm{m_{flux,TESS27}}$ \dotfill & $\mathcal{N}(0.0,0.1)$ & \twoeightysixTESSmfluxTESStwentyseven \\
$\mathrm{\sigma_{w,TESS27}}$ \dotfill & $\mathcal{J}(0.1,1000.0)$ & \twoeightysixTESSsigmawTESStwentyseven \\
$\mathrm{\sigma_{GP,TESS27}}$ \dotfill & $\mathcal{J}(1e-06,1000000.0)$ & \twoeightysixTESSGPsigmaTESStwentyseven \\
$\mathrm{\rho_{GP,TESS27}}$ \dotfill & $\mathcal{J}(0.001,1000.0)$ & \twoeightysixTESSGPrhoTESStwentyseven \\
$\mathrm{m_{dilution,TESS28}}$ \dotfill & $\mathrm{fixed}$ & \twoeightysixTESSmdilutionTESStwentyeight \\
$\mathrm{m_{flux,TESS28}}$ \dotfill & $\mathcal{N}(0.0,0.1)$ & \twoeightysixTESSmfluxTESStwentyeight \\
$\mathrm{\sigma_{w,TESS28}}$ \dotfill & $\mathcal{J}(0.1,1000.0)$ & \twoeightysixTESSsigmawTESStwentyeight \\
$\mathrm{\sigma_{GP,TESS28}}$ \dotfill & $\mathcal{J}(1e-06,1000000.0)$ & \twoeightysixTESSGPsigmaTESStwentyeight \\
$\mathrm{\rho_{GP,TESS28}}$ \dotfill & $\mathcal{J}(0.001,1000.0)$ & \twoeightysixTESSGPrhoTESStwentyeight \\
$\mathrm{m_{dilution,TESS29}}$ \dotfill & $\mathrm{fixed}$ & \twoeightysixTESSmdilutionTESStwentynine \\
$\mathrm{m_{flux,TESS29}}$ \dotfill & $\mathcal{N}(0.0,0.1)$ & \twoeightysixTESSmfluxTESStwentynine \\
$\mathrm{\sigma_{w,TESS29}}$ \dotfill & $\mathcal{J}(0.1,1000.0)$ & \twoeightysixTESSsigmawTESStwentynine \\
$\mathrm{\sigma_{GP,TESS29}}$ \dotfill & $\mathcal{J}(1e-06,1000000.0)$ & \twoeightysixTESSGPsigmaTESStwentynine \\
$\mathrm{\rho_{GP,TESS29}}$ \dotfill & $\mathcal{J}(0.001,1000.0)$ & \twoeightysixTESSGPrhoTESStwentynine \\
$\mathrm{m_{dilution,TESS30}}$ \dotfill & $\mathrm{fixed}$ & \twoeightysixTESSmdilutionTESSthirty \\
$\mathrm{m_{flux,TESS30}}$ \dotfill & $\mathcal{N}(0.0,0.1)$ & \twoeightysixTESSmfluxTESSthirty \\
$\mathrm{\sigma_{w,TESS30}}$ \dotfill & $\mathcal{J}(0.1,1000.0)$ & \twoeightysixTESSsigmawTESSthirty \\
$\mathrm{\sigma_{GP,TESS30}}$ \dotfill & $\mathcal{J}(1e-06,1000000.0)$ & \twoeightysixTESSGPsigmaTESSthirty \\
$\mathrm{\rho_{GP,TESS30}}$ \dotfill & $\mathcal{J}(0.001,1000.0)$ & \twoeightysixTESSGPrhoTESSthirty \\
$\mathrm{m_{dilution,TESS31}}$ \dotfill & $\mathrm{fixed}$ & \twoeightysixTESSmdilutionTESSthirtyone \\
$\mathrm{m_{flux,TESS31}}$ \dotfill & $\mathcal{N}(0.0,0.1)$ & \twoeightysixTESSmfluxTESSthirtyone \\
$\mathrm{\sigma_{w,TESS31}}$ \dotfill & $\mathcal{J}(0.1,1000.0)$ & \twoeightysixTESSsigmawTESSthirtyone \\
$\mathrm{\sigma_{GP,TESS31}}$ \dotfill & $\mathcal{J}(1e-06,1000000.0)$ & \twoeightysixTESSGPsigmaTESSthirtyone \\
$\mathrm{\rho_{GP,TESS31}}$ \dotfill & $\mathcal{J}(0.001,1000.0)$ & \twoeightysixTESSGPrhoTESSthirtyone \\
$\mathrm{m_{dilution,TESS32}}$ \dotfill & $\mathrm{fixed}$ & \twoeightysixTESSmdilutionTESSthirtytwo \\
$\mathrm{m_{flux,TESS32}}$ \dotfill & $\mathcal{N}(0.0,0.1)$ & \twoeightysixTESSmfluxTESSthirtytwo \\
$\mathrm{\sigma_{w,TESS32}}$ \dotfill & $\mathcal{J}(0.1,1000.0)$ & \twoeightysixTESSsigmawTESSthirtytwo \\
$\mathrm{\sigma_{GP,TESS32}}$ \dotfill & $\mathcal{J}(1e-06,1000000.0)$ & \twoeightysixTESSGPsigmaTESSthirtytwo \\
$\mathrm{\rho_{GP,TESS32}}$ \dotfill & $\mathcal{J}(0.001,1000.0)$ & \twoeightysixTESSGPrhoTESSthirtytwo \\
$\mathrm{m_{dilution,TESS33}}$ \dotfill & $\mathrm{fixed}$ & \twoeightysixTESSmdilutionTESSthirtythree \\
$\mathrm{m_{flux,TESS33}}$ \dotfill & $\mathcal{N}(0.0,0.1)$ & \twoeightysixTESSmfluxTESSthirtythree \\
$\mathrm{\sigma_{w,TESS33}}$ \dotfill & $\mathcal{J}(0.1,1000.0)$ & \twoeightysixTESSsigmawTESSthirtythree \\
$\mathrm{\sigma_{GP,TESS33}}$ \dotfill & $\mathcal{J}(1e-06,1000000.0)$ & \twoeightysixTESSGPsigmaTESSthirtythree \\
$\mathrm{\rho_{GP,TESS33}}$ \dotfill & $\mathcal{J}(0.001,1000.0)$ & \twoeightysixTESSGPrhoTESSthirtythree \\
$\mathrm{m_{dilution,TESS34}}$ \dotfill & $\mathrm{fixed}$ & \twoeightysixTESSmdilutionTESSthirtyfour \\
$\mathrm{m_{flux,TESS34}}$ \dotfill & $\mathcal{N}(0.0,0.1)$ & \twoeightysixTESSmfluxTESSthirtyfour \\
$\mathrm{\sigma_{w,TESS34}}$ \dotfill & $\mathcal{J}(0.1,1000.0)$ & \twoeightysixTESSsigmawTESSthirtyfour \\
$\mathrm{\sigma_{GP,TESS34}}$ \dotfill & $\mathcal{J}(1e-06,1000000.0)$ & \twoeightysixTESSGPsigmaTESSthirtyfour \\
$\mathrm{\rho_{GP,TESS34}}$ \dotfill & $\mathcal{J}(0.001,1000.0)$ & \twoeightysixTESSGPrhoTESSthirtyfour \\
$\mathrm{m_{dilution,TESS35}}$ \dotfill & $\mathrm{fixed}$ & \twoeightysixTESSmdilutionTESSthirtyfive \\
$\mathrm{m_{flux,TESS35}}$ \dotfill & $\mathcal{N}(0.0,0.1)$ & \twoeightysixTESSmfluxTESSthirtyfive \\
$\mathrm{\sigma_{w,TESS35}}$ \dotfill & $\mathcal{J}(0.1,1000.0)$ & \twoeightysixTESSsigmawTESSthirtyfive \\
$\mathrm{\sigma_{GP,TESS35}}$ \dotfill & $\mathcal{J}(1e-06,1000000.0)$ & \twoeightysixTESSGPsigmaTESSthirtyfive \\
$\mathrm{\rho_{GP,TESS35}}$ \dotfill & $\mathcal{J}(0.001,1000.0)$ & \twoeightysixTESSGPrhoTESSthirtyfive \\
$\mathrm{m_{dilution,TESS36}}$ \dotfill & $\mathrm{fixed}$ & \twoeightysixTESSmdilutionTESSthirtysix \\
$\mathrm{m_{flux,TESS36}}$ \dotfill & $\mathcal{N}(0.0,0.1)$ & \twoeightysixTESSmfluxTESSthirtysix \\
$\mathrm{\sigma_{w,TESS36}}$ \dotfill & $\mathcal{J}(0.1,1000.0)$ & \twoeightysixTESSsigmawTESSthirtysix \\
$\mathrm{\sigma_{GP,TESS36}}$ \dotfill & $\mathcal{J}(1e-06,1000000.0)$ & \twoeightysixTESSGPsigmaTESSthirtysix \\
$\mathrm{\rho_{GP,TESS36}}$ \dotfill & $\mathcal{J}(0.001,1000.0)$ & \twoeightysixTESSGPrhoTESSthirtysix \\
$\mathrm{m_{dilution,TESS37}}$ \dotfill & $\mathrm{fixed}$ & \twoeightysixTESSmdilutionTESSthirtyseven \\
$\mathrm{m_{flux,TESS37}}$ \dotfill & $\mathcal{N}(0.0,0.1)$ & \twoeightysixTESSmfluxTESSthirtyseven \\
$\mathrm{\sigma_{w,TESS37}}$ \dotfill & $\mathcal{J}(0.1,1000.0)$ & \twoeightysixTESSsigmawTESSthirtyseven \\
$\mathrm{\sigma_{GP,TESS37}}$ \dotfill & $\mathcal{J}(1e-06,1000000.0)$ & \twoeightysixTESSGPsigmaTESSthirtyseven \\
$\mathrm{\rho_{GP,TESS37}}$ \dotfill & $\mathcal{J}(0.001,1000.0)$ & \twoeightysixTESSGPrhoTESSthirtyseven \\
$\mathrm{m_{dilution,TESS38}}$ \dotfill & $\mathrm{fixed}$ & \twoeightysixTESSmdilutionTESSthirtyeight \\
$\mathrm{m_{flux,TESS38}}$ \dotfill & $\mathcal{N}(0.0,0.1)$ & \twoeightysixTESSmfluxTESSthirtyeight \\
$\mathrm{\sigma_{w,TESS38}}$ \dotfill & $\mathcal{J}(0.1,1000.0)$ & \twoeightysixTESSsigmawTESSthirtyeight \\
$\mathrm{\sigma_{GP,TESS38}}$ \dotfill & $\mathcal{J}(1e-06,1000000.0)$ & \twoeightysixTESSGPsigmaTESSthirtyeight \\
$\mathrm{\rho_{GP,TESS38}}$ \dotfill & $\mathcal{J}(0.001,1000.0)$ & \twoeightysixTESSGPrhoTESSthirtyeight \\
$\mathrm{m_{dilution,TESS39}}$ \dotfill & $\mathrm{fixed}$ & \twoeightysixTESSmdilutionTESSthirtynine \\
$\mathrm{m_{flux,TESS39}}$ \dotfill & $\mathcal{N}(0.0,0.1)$ & \twoeightysixTESSmfluxTESSthirtynine \\
$\mathrm{\sigma_{w,TESS39}}$ \dotfill & $\mathcal{J}(0.1,1000.0)$ & \twoeightysixTESSsigmawTESSthirtynine \\
$\mathrm{\sigma_{GP,TESS39}}$ \dotfill & $\mathcal{J}(1e-06,1000000.0)$ & \twoeightysixTESSGPsigmaTESSthirtynine \\
$\mathrm{\rho_{GP,TESS39}}$ \dotfill & $\mathcal{J}(0.001,1000.0)$ & \twoeightysixTESSGPrhoTESSthirtynine \\
\hline 
\end{tabular} 
} 
\end{center} 
\end{table} 

\begin{table}[pht] 
\begin{center} 
\caption{Prior and posterior planetary parameter distributions obtained with \texttt{juliet} for the sector-by-sector out-of-transit detrending of TESS data for TOI-286, for the second extended mission.} 
\label{tab:TOI286_TESSGP_em2} 
\centering 
\resizebox{\columnwidth}{!}{%
\begin{tabular}{lll} 
\hline  \hline 
Parameter & Prior & Posterior \\ 
\hline 
$\mathrm{m_{dilution,TESS61}}$ \dotfill & $\mathrm{fixed}$ & \twoeightysixTESSmdilutionTESSsixtyone \\
$\mathrm{m_{flux,TESS61}}$ \dotfill & $\mathcal{N}(0.0,0.1)$ & \twoeightysixTESSmfluxTESSsixtyone \\
$\mathrm{\sigma_{w,TESS61}}$ \dotfill & $\mathcal{J}(0.1,1000.0)$ & \twoeightysixTESSsigmawTESSsixtyone \\
$\mathrm{\sigma_{GP,TESS61}}$ \dotfill & $\mathcal{J}(1e-06,1000000.0)$ & \twoeightysixTESSGPsigmaTESSsixtyone \\
$\mathrm{\rho_{GP,TESS61}}$ \dotfill & $\mathcal{J}(0.001,1000.0)$ & \twoeightysixTESSGPrhoTESSsixtyone \\
$\mathrm{m_{dilution,TESS62}}$ \dotfill & $\mathrm{fixed}$ & \twoeightysixTESSmdilutionTESSsixtytwo \\
$\mathrm{m_{flux,TESS62}}$ \dotfill & $\mathcal{N}(0.0,0.1)$ & \twoeightysixTESSmfluxTESSsixtytwo \\
$\mathrm{\sigma_{w,TESS62}}$ \dotfill & $\mathcal{J}(0.1,1000.0)$ & \twoeightysixTESSsigmawTESSsixtytwo \\
$\mathrm{\sigma_{GP,TESS62}}$ \dotfill & $\mathcal{J}(1e-06,1000000.0)$ & \twoeightysixTESSGPsigmaTESSsixtytwo \\
$\mathrm{\rho_{GP,TESS62}}$ \dotfill & $\mathcal{J}(0.001,1000.0)$ & \twoeightysixTESSGPrhoTESSsixtytwo \\
$\mathrm{m_{dilution,TESS63}}$ \dotfill & $\mathrm{fixed}$ & \twoeightysixTESSmdilutionTESSsixtythree \\
$\mathrm{m_{flux,TESS63}}$ \dotfill & $\mathcal{N}(0.0,0.1)$ & \twoeightysixTESSmfluxTESSsixtythree \\
$\mathrm{\sigma_{w,TESS63}}$ \dotfill & $\mathcal{J}(0.1,1000.0)$ & \twoeightysixTESSsigmawTESSsixtythree \\
$\mathrm{\sigma_{GP,TESS63}}$ \dotfill & $\mathcal{J}(1e-06,1000000.0)$ & \twoeightysixTESSGPsigmaTESSsixtythree \\
$\mathrm{\rho_{GP,TESS63}}$ \dotfill & $\mathcal{J}(0.001,1000.0)$ & \twoeightysixTESSGPrhoTESSsixtythree \\
$\mathrm{m_{dilution,TESS64}}$ \dotfill & $\mathrm{fixed}$ & \twoeightysixTESSmdilutionTESSsixtyfour \\
$\mathrm{m_{flux,TESS64}}$ \dotfill & $\mathcal{N}(0.0,0.1)$ & \twoeightysixTESSmfluxTESSsixtyfour \\
$\mathrm{\sigma_{w,TESS64}}$ \dotfill & $\mathcal{J}(0.1,1000.0)$ & \twoeightysixTESSsigmawTESSsixtyfour \\
$\mathrm{\sigma_{GP,TESS64}}$ \dotfill & $\mathcal{J}(1e-06,1000000.0)$ & \twoeightysixTESSGPsigmaTESSsixtyfour \\
$\mathrm{\rho_{GP,TESS64}}$ \dotfill & $\mathcal{J}(0.001,1000.0)$ & \twoeightysixTESSGPrhoTESSsixtyfour \\
$\mathrm{m_{dilution,TESS65}}$ \dotfill & $\mathrm{fixed}$ & \twoeightysixTESSmdilutionTESSsixtyfive \\
$\mathrm{m_{flux,TESS65}}$ \dotfill & $\mathcal{N}(0.0,0.1)$ & \twoeightysixTESSmfluxTESSsixtyfive \\
$\mathrm{\sigma_{w,TESS65}}$ \dotfill & $\mathcal{J}(0.1,1000.0)$ & \twoeightysixTESSsigmawTESSsixtyfive \\
$\mathrm{\sigma_{GP,TESS65}}$ \dotfill & $\mathcal{J}(1e-06,1000000.0)$ & \twoeightysixTESSGPsigmaTESSsixtyfive \\
$\mathrm{\rho_{GP,TESS65}}$ \dotfill & $\mathcal{J}(0.001,1000.0)$ & \twoeightysixTESSGPrhoTESSsixtyfive \\
$\mathrm{m_{dilution,TESS66}}$ \dotfill & $\mathrm{fixed}$ & \twoeightysixTESSmdilutionTESSsixtysix \\
$\mathrm{m_{flux,TESS66}}$ \dotfill & $\mathcal{N}(0.0,0.1)$ & \twoeightysixTESSmfluxTESSsixtysix \\
$\mathrm{\sigma_{w,TESS66}}$ \dotfill & $\mathcal{J}(0.1,1000.0)$ & \twoeightysixTESSsigmawTESSsixtysix \\
$\mathrm{\sigma_{GP,TESS66}}$ \dotfill & $\mathcal{J}(1e-06,1000000.0)$ & \twoeightysixTESSGPsigmaTESSsixtysix \\
$\mathrm{\rho_{GP,TESS66}}$ \dotfill & $\mathcal{J}(0.001,1000.0)$ & \twoeightysixTESSGPrhoTESSsixtysix \\
$\mathrm{m_{dilution,TESS67}}$ \dotfill & $\mathrm{fixed}$ & \twoeightysixTESSmdilutionTESSsixtyseven \\
$\mathrm{m_{flux,TESS67}}$ \dotfill & $\mathcal{N}(0.0,0.1)$ & \twoeightysixTESSmfluxTESSsixtyseven \\
$\mathrm{\sigma_{w,TESS67}}$ \dotfill & $\mathcal{J}(0.1,1000.0)$ & \twoeightysixTESSsigmawTESSsixtyseven \\
$\mathrm{\sigma_{GP,TESS67}}$ \dotfill & $\mathcal{J}(1e-06,1000000.0)$ & \twoeightysixTESSGPsigmaTESSsixtyseven \\
$\mathrm{\rho_{GP,TESS67}}$ \dotfill & $\mathcal{J}(0.001,1000.0)$ & \twoeightysixTESSGPrhoTESSsixtyseven \\
$\mathrm{m_{dilution,TESS68}}$ \dotfill & $\mathrm{fixed}$ & \twoeightysixTESSmdilutionTESSsixtyeight \\
$\mathrm{m_{flux,TESS68}}$ \dotfill & $\mathcal{N}(0.0,0.1)$ & \twoeightysixTESSmfluxTESSsixtyeight \\
$\mathrm{\sigma_{w,TESS68}}$ \dotfill & $\mathcal{J}(0.1,1000.0)$ & \twoeightysixTESSsigmawTESSsixtyeight \\
$\mathrm{\sigma_{GP,TESS68}}$ \dotfill & $\mathcal{J}(1e-06,1000000.0)$ & \twoeightysixTESSGPsigmaTESSsixtyeight \\
$\mathrm{\rho_{GP,TESS68}}$ \dotfill & $\mathcal{J}(0.001,1000.0)$ & \twoeightysixTESSGPrhoTESSsixtyeight \\
$\mathrm{m_{dilution,TESS69}}$ \dotfill & $\mathrm{fixed}$ & \twoeightysixTESSmdilutionTESSsixtynine \\
$\mathrm{m_{flux,TESS69}}$ \dotfill & $\mathcal{N}(0.0,0.1)$ & \twoeightysixTESSmfluxTESSsixtynine \\
$\mathrm{\sigma_{w,TESS69}}$ \dotfill & $\mathcal{J}(0.1,1000.0)$ & \twoeightysixTESSsigmawTESSsixtynine \\
$\mathrm{\sigma_{GP,TESS69}}$ \dotfill & $\mathcal{J}(1e-06,1000000.0)$ & \twoeightysixTESSGPsigmaTESSsixtynine \\
$\mathrm{\rho_{GP,TESS69}}$ \dotfill & $\mathcal{J}(0.001,1000.0)$ & \twoeightysixTESSGPrhoTESSsixtynine \\
\hline 
\end{tabular} 
} 
\end{center} 
\end{table} 

\clearpage

\section{Cornerplots of the juliet fits}\label{ap:cornerplots}

In this appendix, we show the cornerplots of the posterior parameter distributions obtained from the  \texttt{juliet} fits. For each fit for TOI-260 and TOI-134 we split the plots into orbital parameters, instrumental RV parameters, and photometric instrumental parameters, for ease of viewing. For the TOI-286 fits, as there are two planets and most of the photometric parameters are fixed, we instead split them into orbital parameters for the inner planet, orbital parameters for the outer planet (when applicable), and instrumental parameters. As the variations of the period and $\mathrm{t_0}$ tend to be very small, we subtract the median in order to have more legible axes for these parameters.

\begin{figure*}
    \centering
    \includegraphics[width=\textwidth]{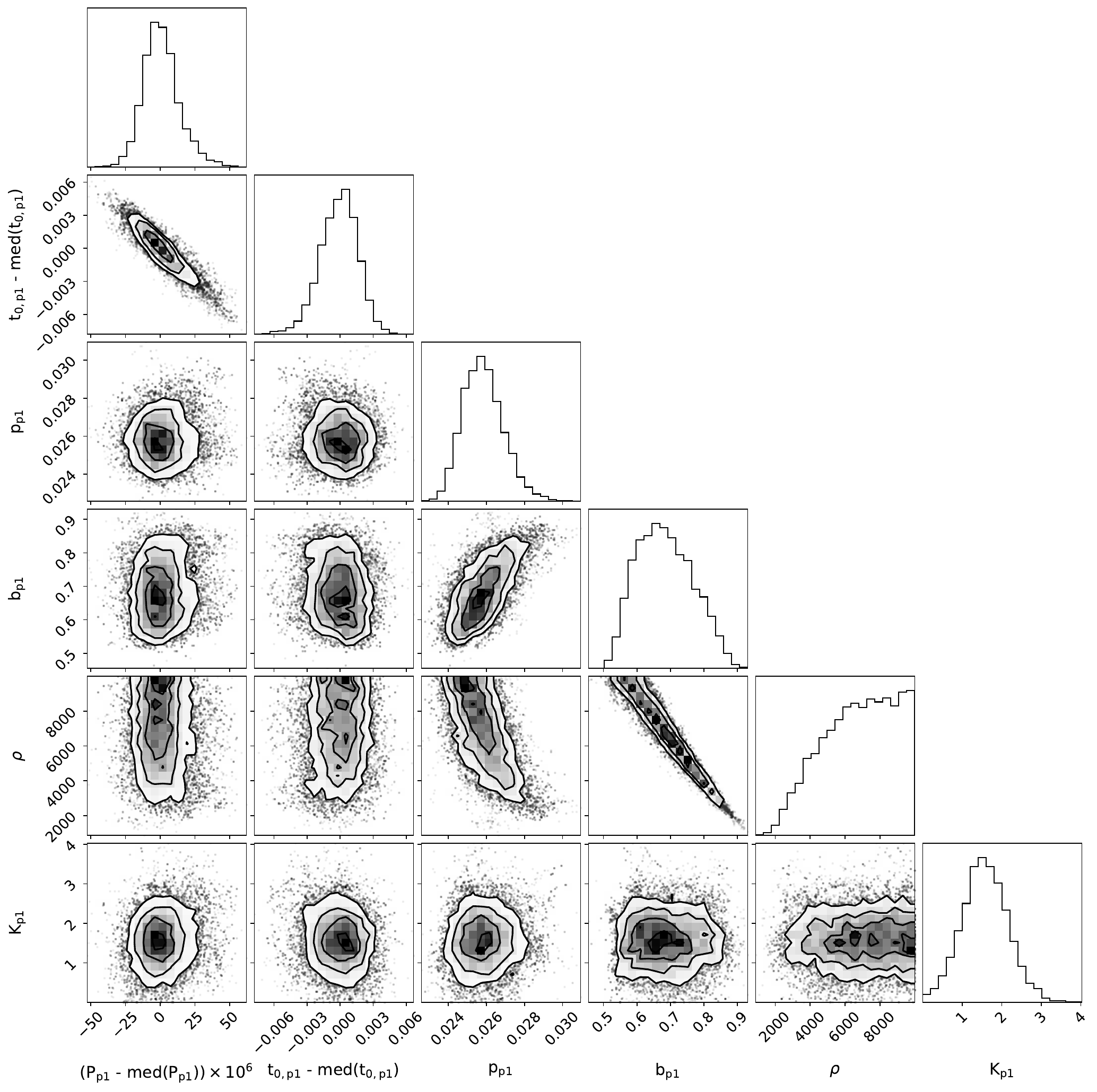}
    \caption{Cornerplot of the orbital posterior planetary parameter distributions obtained
with \texttt{juliet} for TOI-260, for the fit with eccentricity fixed to 0.}
    \label{fig:toi-260_cornerplot_circ_orbital}
\end{figure*}

\begin{figure*}
    \centering
    \includegraphics[width=\textwidth]{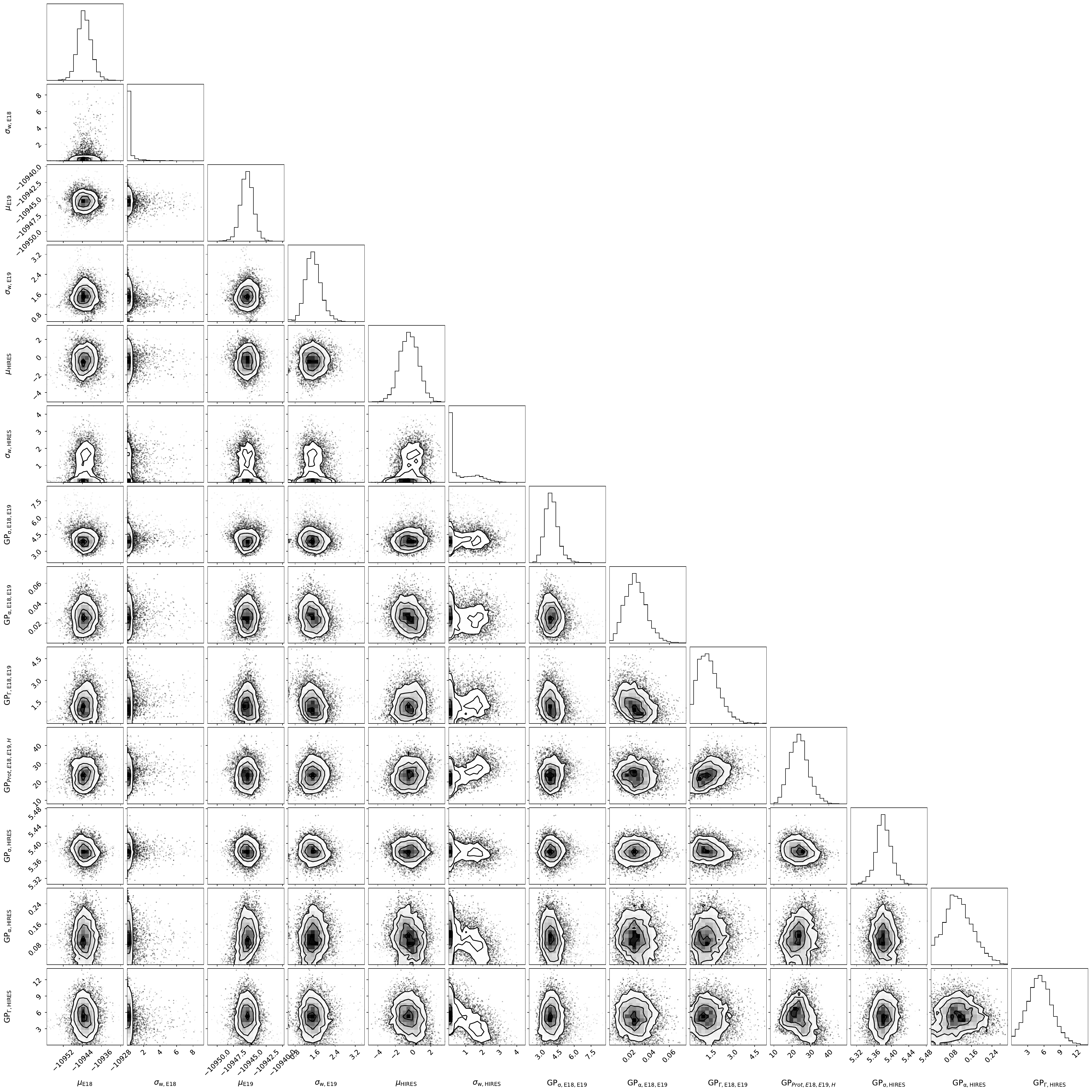}
    \caption{Cornerplot of the RV instrumental posterior parameter distributions obtained
with \texttt{juliet} for TOI-260, for the fit with eccentricity fixed to 0.}
    \label{fig:toi-260_cornerplot_circ_rv}
\end{figure*}

\begin{figure*}
    \centering
    \includegraphics[width=\textwidth]{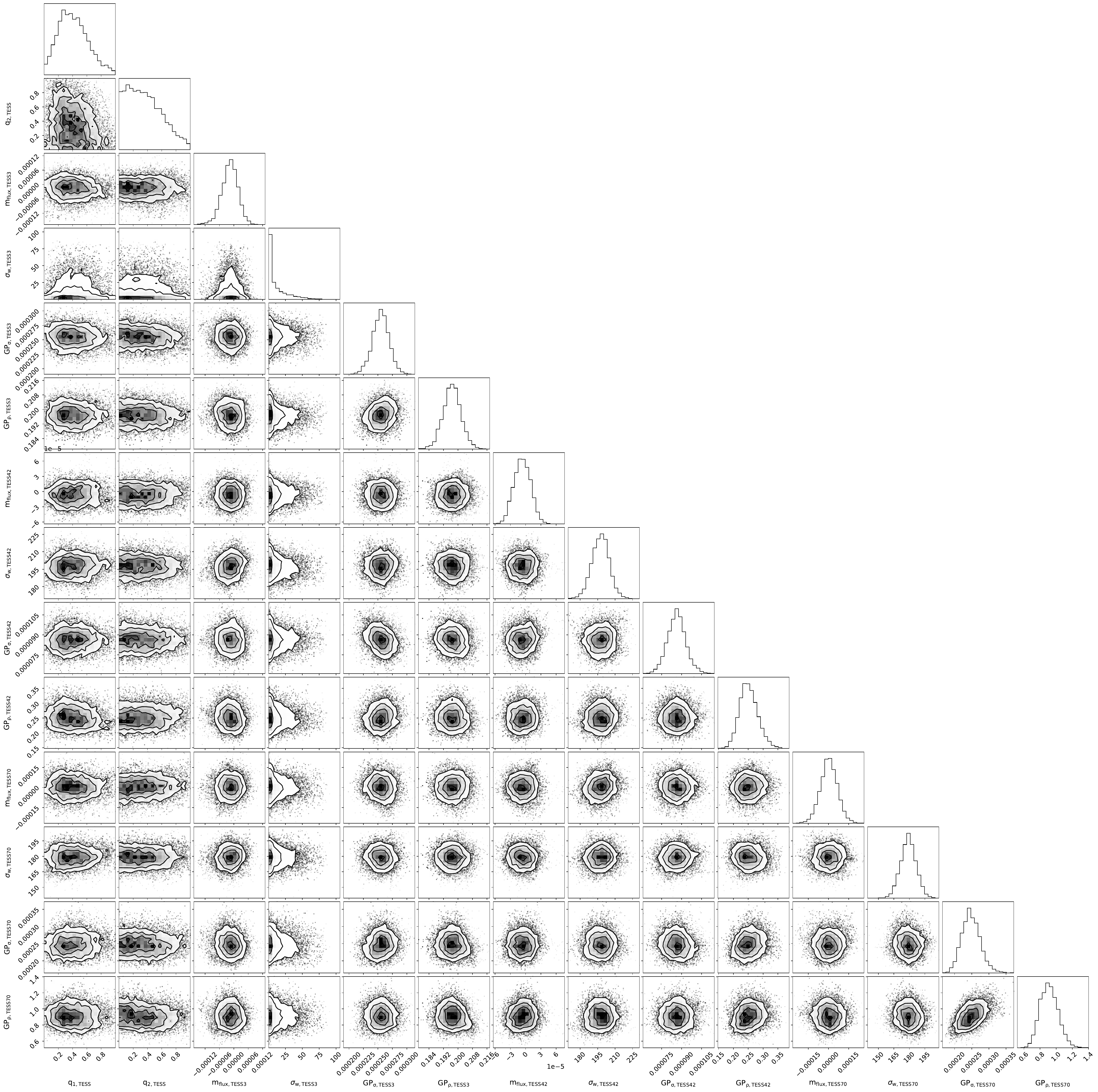}
    \caption{Cornerplot of the photometric instrumental posterior parameter distributions obtained
with \texttt{juliet} for TOI-260, for the fit with eccentricity fixed to 0.}
    \label{fig:toi-260_cornerplot_circ_transit}
\end{figure*}

\begin{figure*}
    \centering
    \includegraphics[width=\textwidth]{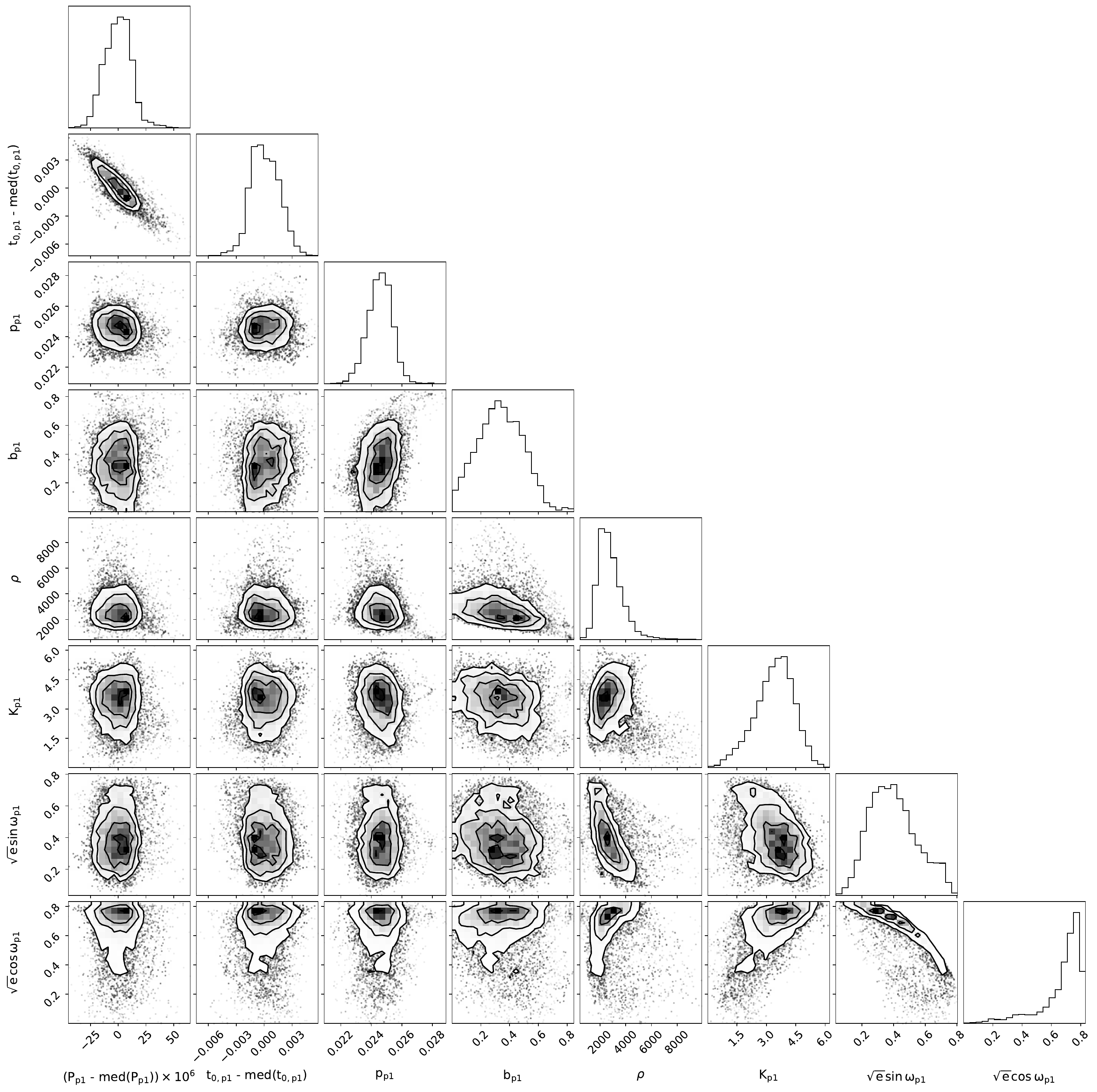}
    \caption{Cornerplot of the orbital posterior planetary parameter distributions obtained
with \texttt{juliet} for TOI-260, for the fit with free eccentricity, ($\mathrm{\sqrt{e} \sin \omega, \sqrt{e} \cos \omega}$) parametrization, and an upper limit on the eccentricity of 0.7.}
    \label{fig:toi-260_cornerplot_e07_orbital}
\end{figure*}

\begin{figure*}
    \centering
    \includegraphics[width=\textwidth]{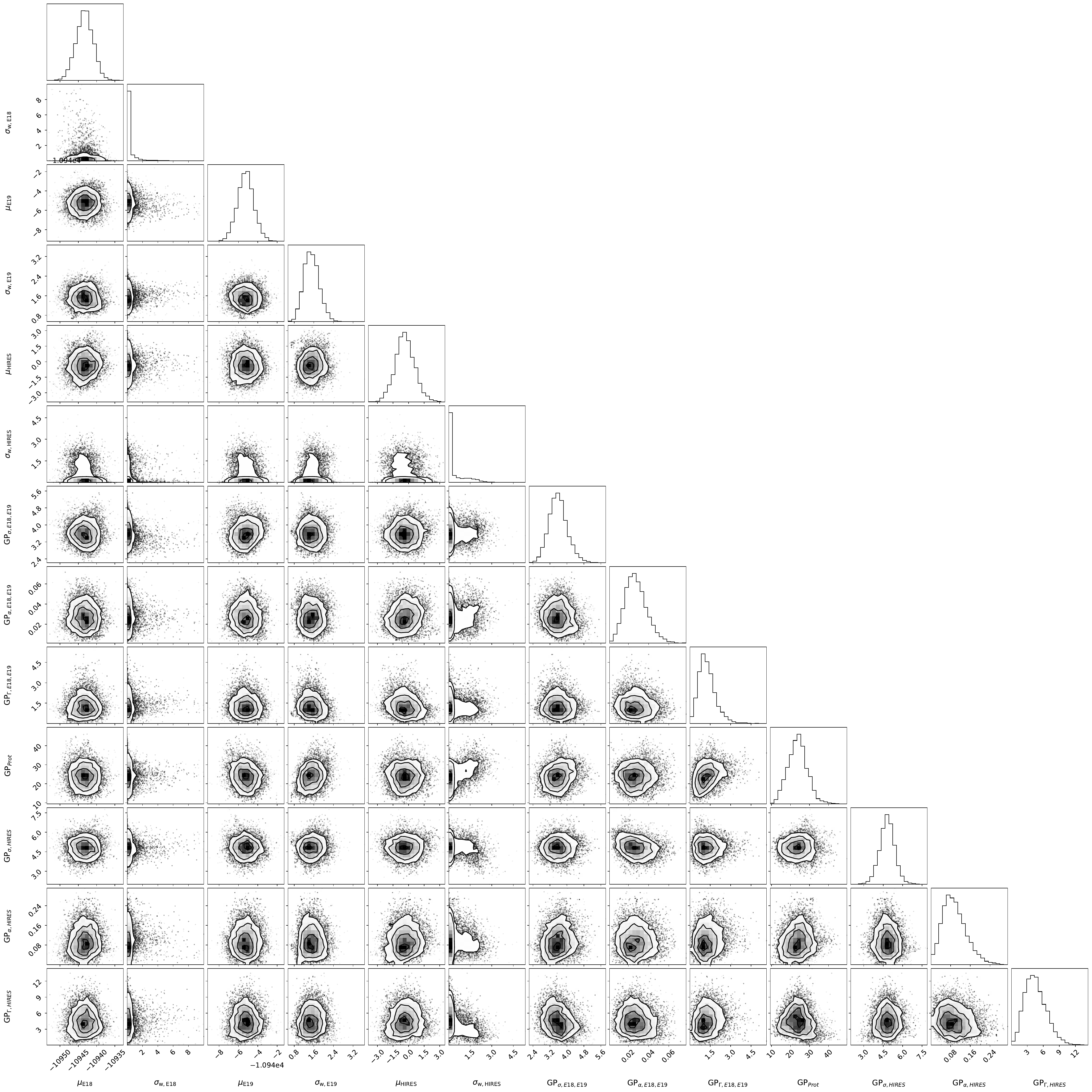}
    \caption{Cornerplot of the RV instrumental posterior parameter distributions obtained
with \texttt{juliet} for TOI-260, for the fit with free eccentricity, ($\mathrm{\sqrt{e} \sin \omega, \sqrt{e} \cos \omega}$) parametrization, and an upper limit on the eccentricity of 0.7.}
    \label{fig:toi-260_cornerplot_e07_rv}
\end{figure*}

\begin{figure*}
    \centering
    \includegraphics[width=\textwidth]{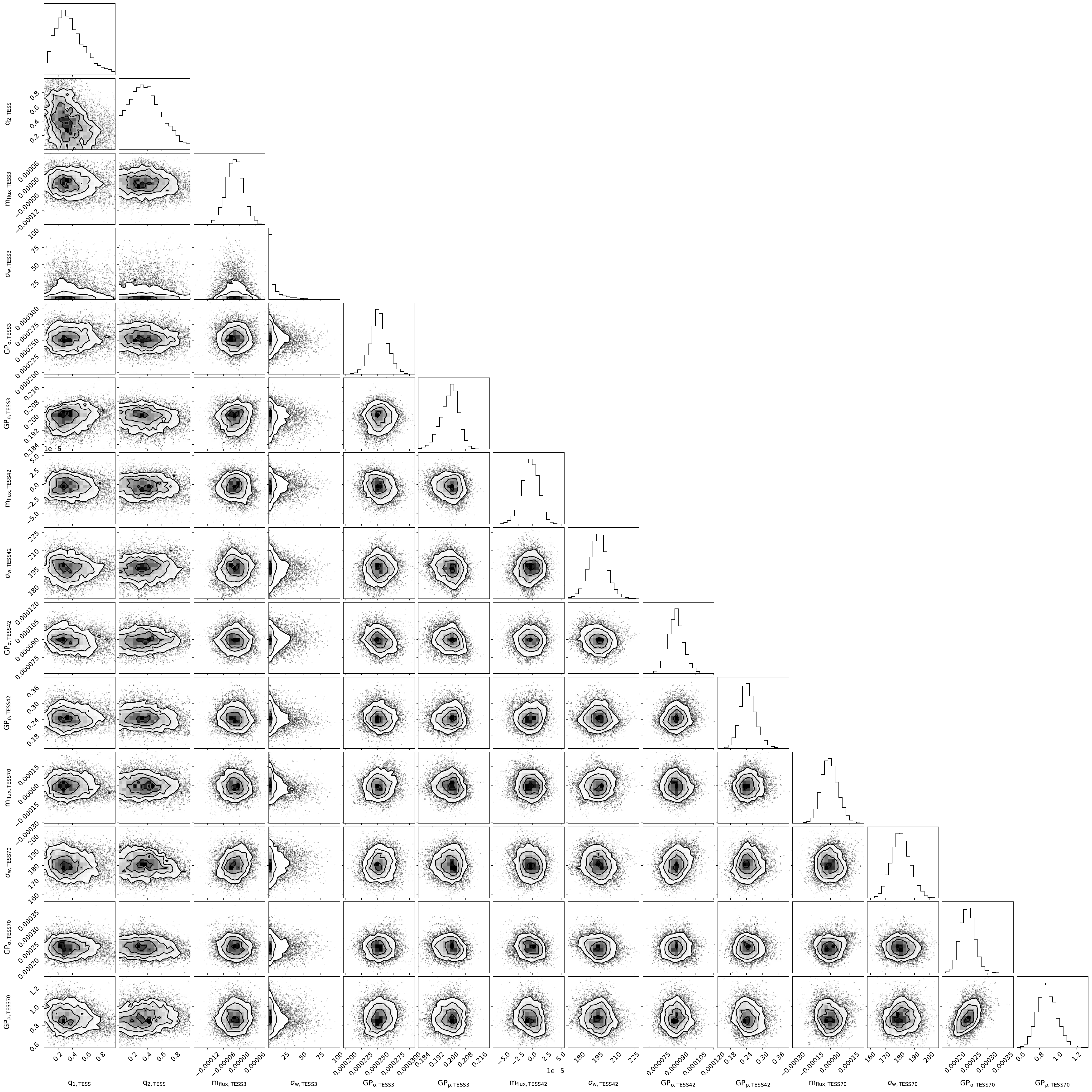}
    \caption{Cornerplot of the photometric instrumental posterior parameter distributions obtained
with \texttt{juliet} for TOI-260, for the fit with free eccentricity, ($\mathrm{\sqrt{e} \sin \omega, \sqrt{e} \cos \omega}$) parametrization, and an upper limit on the eccentricity of 0.7.}
    \label{fig:toi-260_cornerplot_e07_transit}
\end{figure*}


\begin{figure*}
    \centering
    \includegraphics[width=\textwidth]{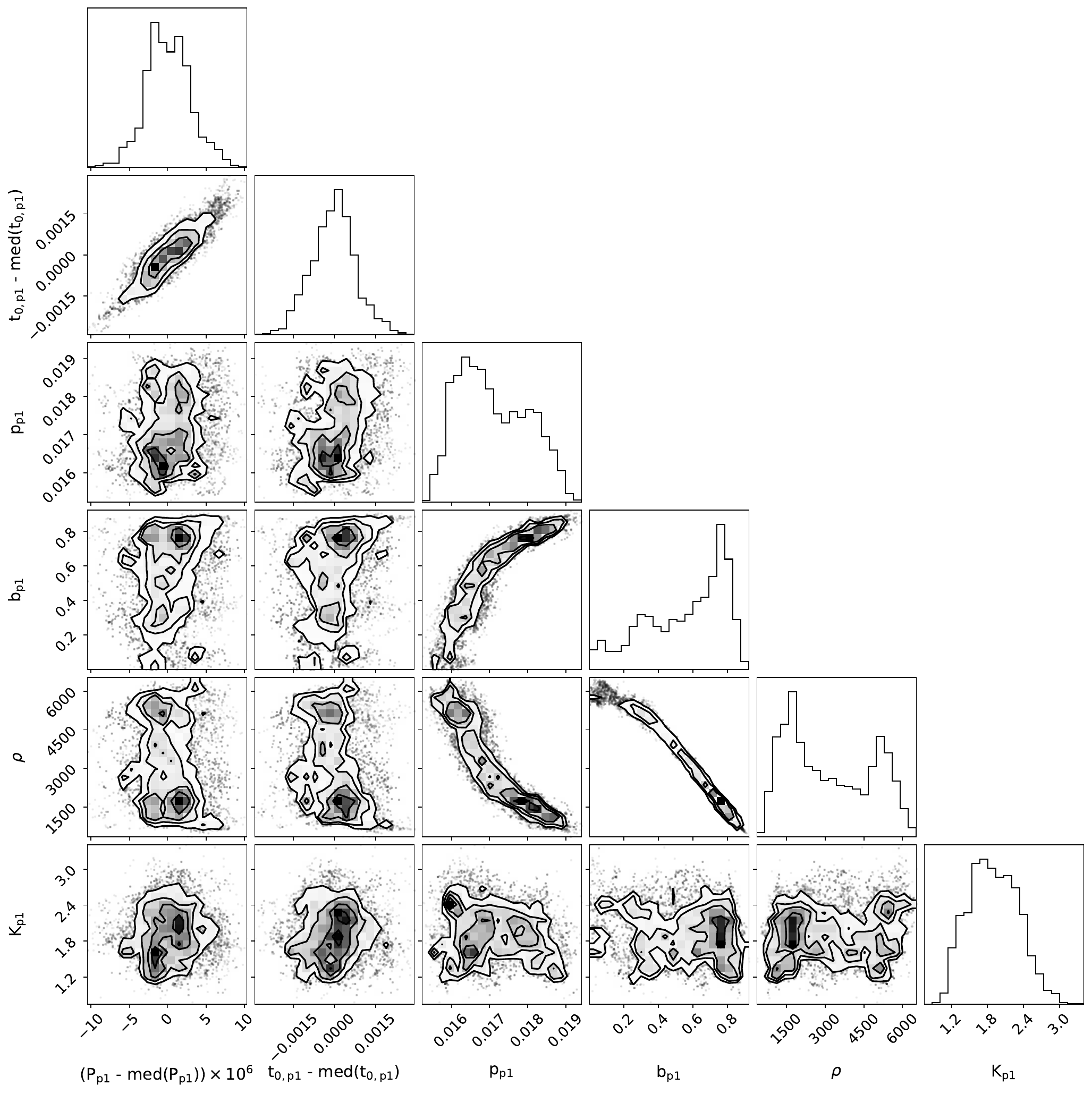}
    \caption{Cornerplot of the orbital posterior planetary parameter distributions obtained
with \texttt{juliet} for TOI-286~b, for the one-planet fit.}
    \label{fig:toi-286_cornerplot_1pl_p1}
\end{figure*}

\begin{figure*}
    \centering
    \includegraphics[width=	\textwidth]{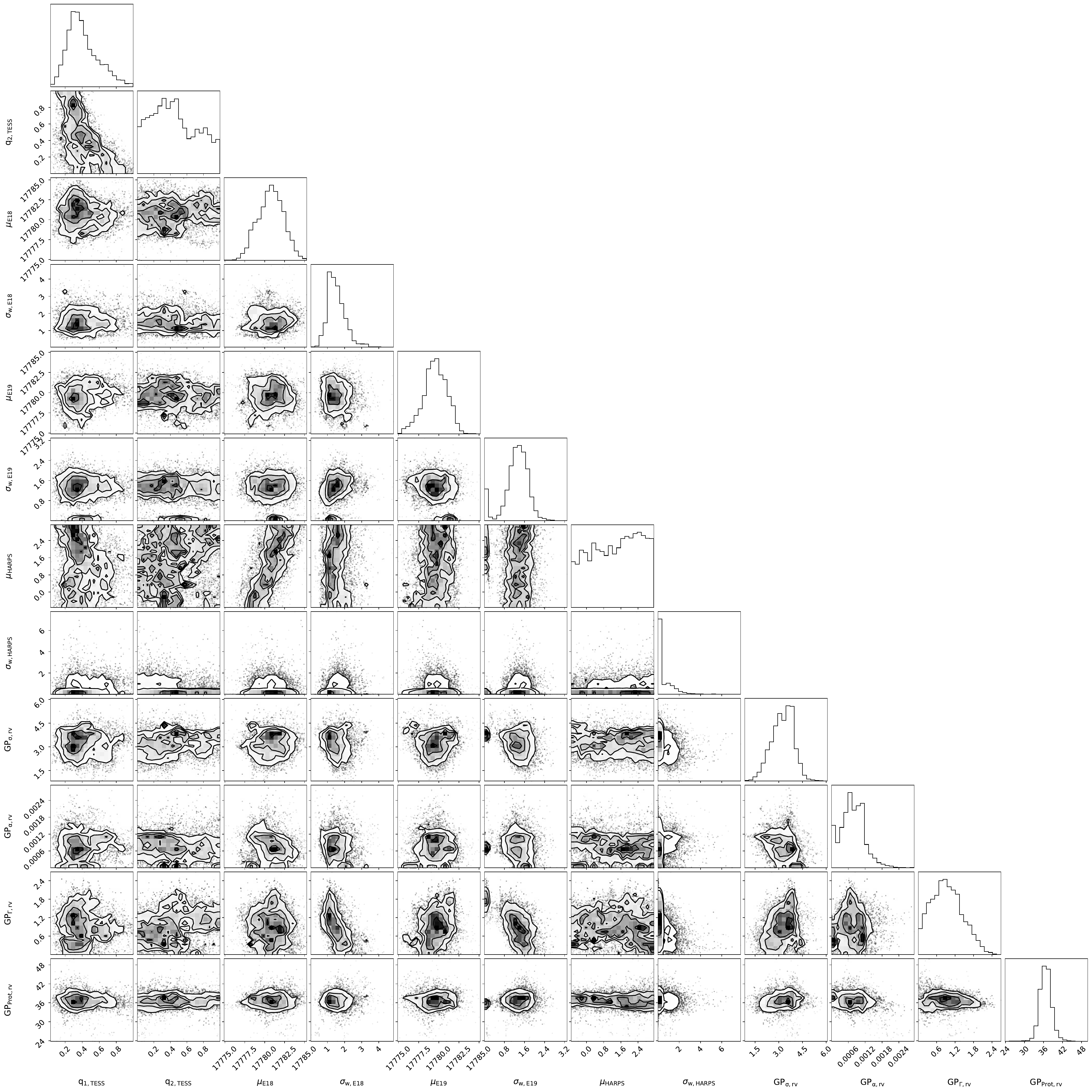}
    \caption{Cornerplot of the instrumental posterior parameter distributions obtained
with \texttt{juliet} for TOI-286, for the one-planet fit.}
    \label{fig:toi-286_cornerplot_1pl_instr}
\end{figure*}

\begin{figure*}
    \centering
    \includegraphics[width=\textwidth]{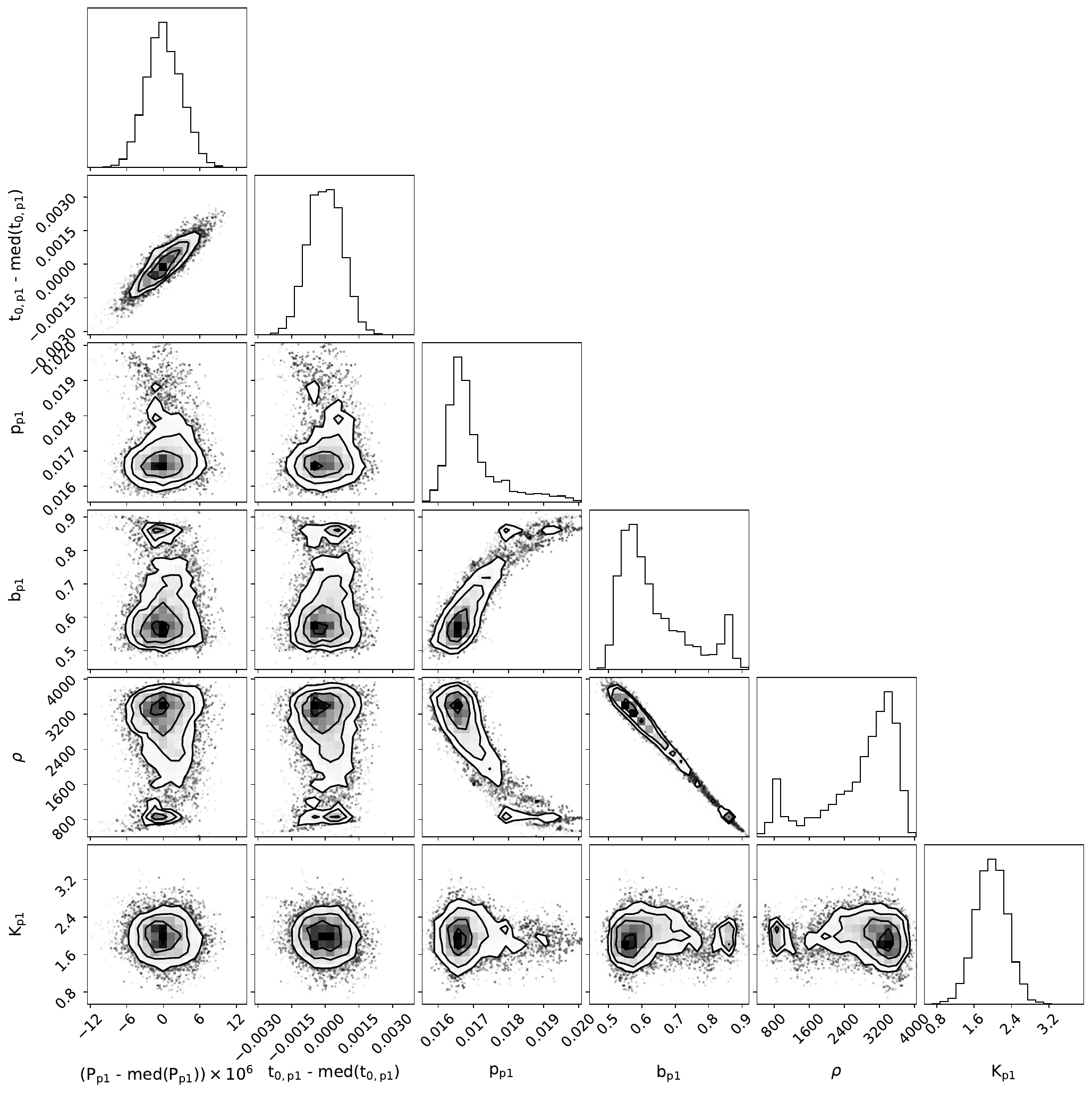}
    \caption{Cornerplot of the orbital posterior planetary parameter distributions obtained
with \texttt{juliet} for TOI-286~b, for the two-planet fit with eccentricities fixed to 0.}
    \label{fig:toi-286_cornerplot_circ_p1}
\end{figure*}

\begin{figure*}
    \centering
    \includegraphics[width=\textwidth]{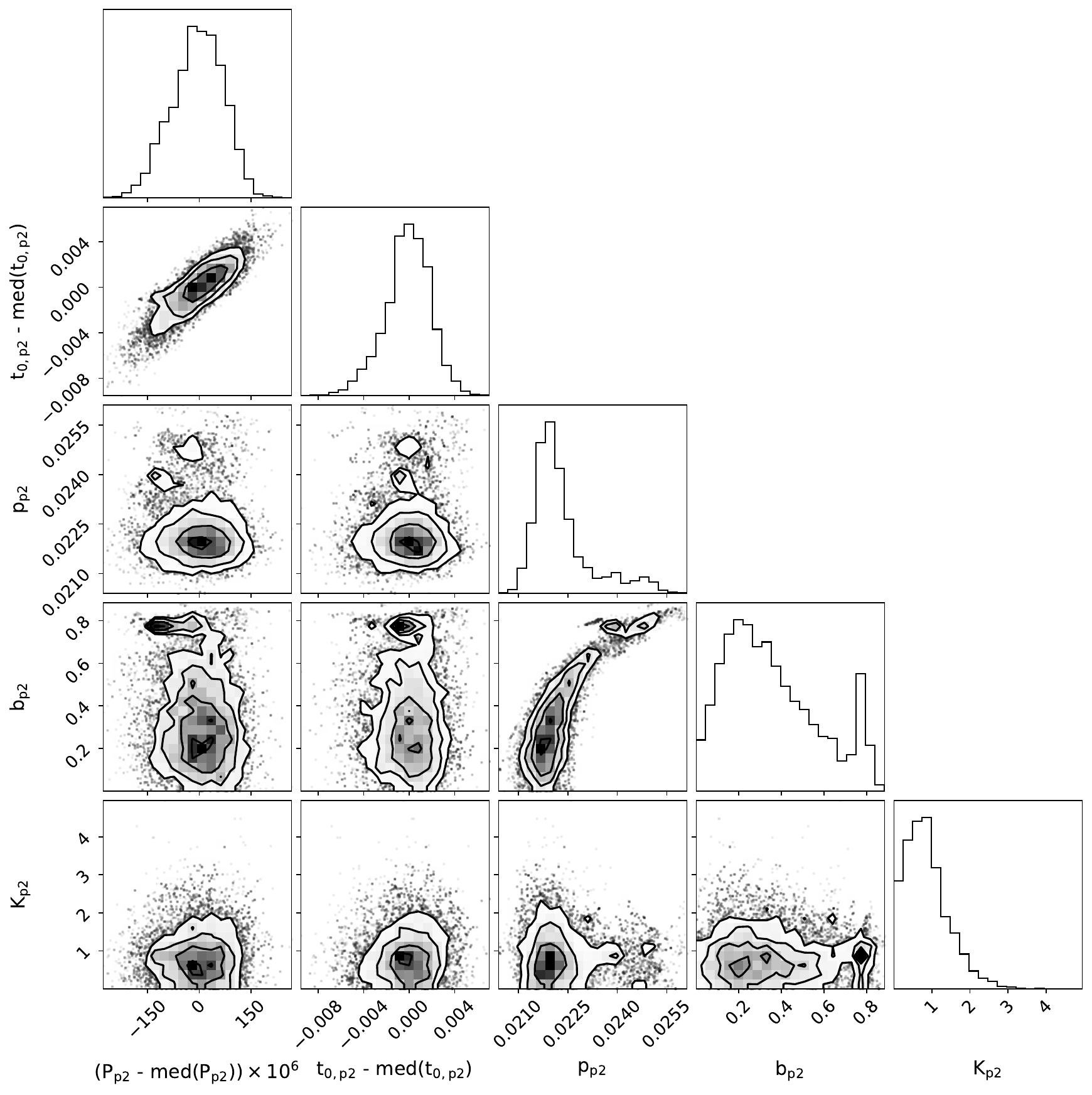}
    \caption{Cornerplot of the orbital posterior planetary parameter distributions obtained
with \texttt{juliet} for TOI-286~c, for the two-planet fit with eccentricities fixed to 0.}
    \label{fig:toi-286_cornerplot_circ_p2}
\end{figure*}

\begin{figure*}
    \centering
    \includegraphics[width=\textwidth]{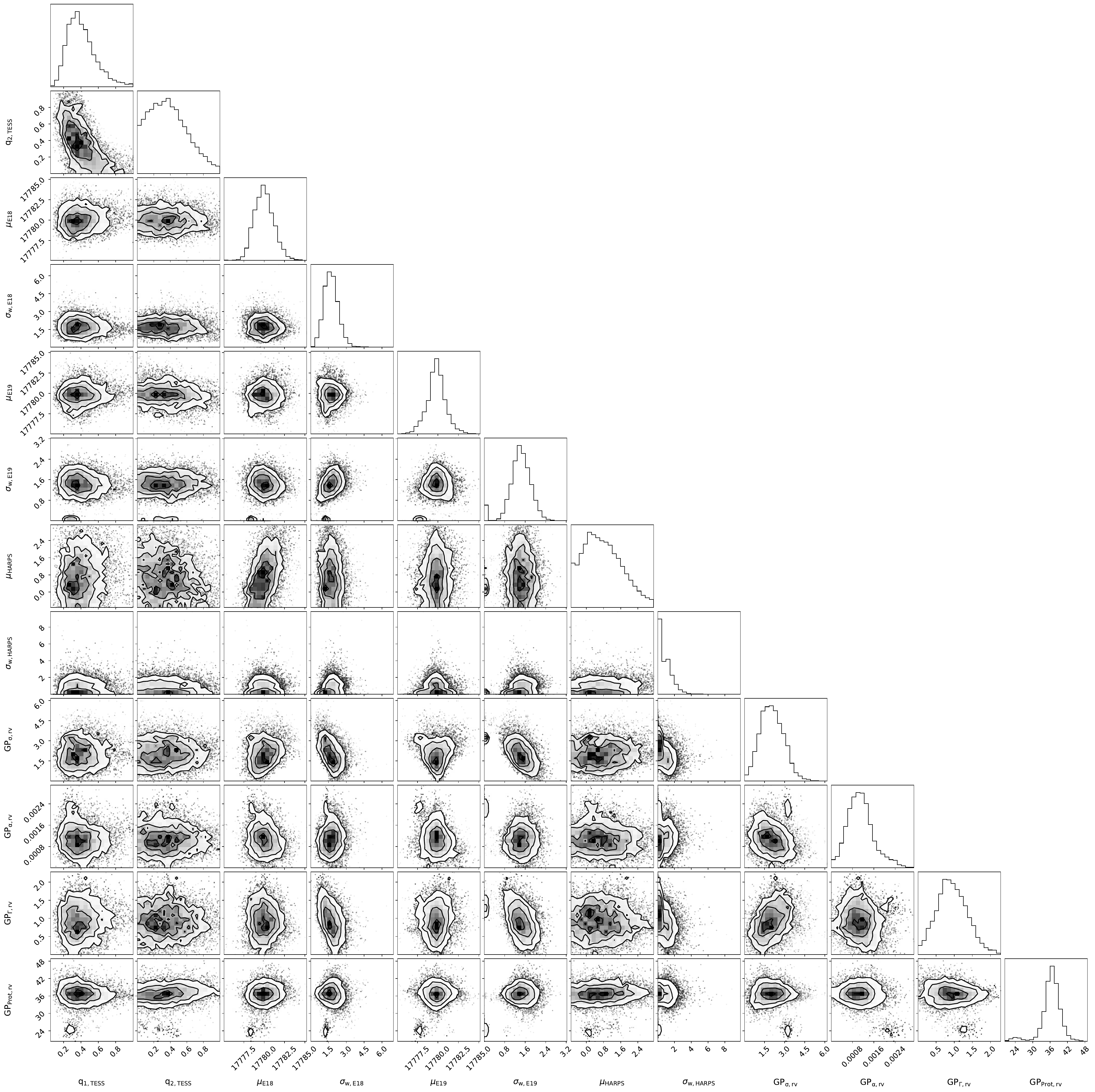}
    \caption{Cornerplot of the instrumental posterior parameter distributions obtained
with \texttt{juliet} for TOI-286, for the two-planet fit with eccentricities fixed to 0.}
    \label{fig:toi-286_cornerplot_circ_instr}
\end{figure*}

\begin{figure*}
    \centering
    \includegraphics[width=\textwidth]{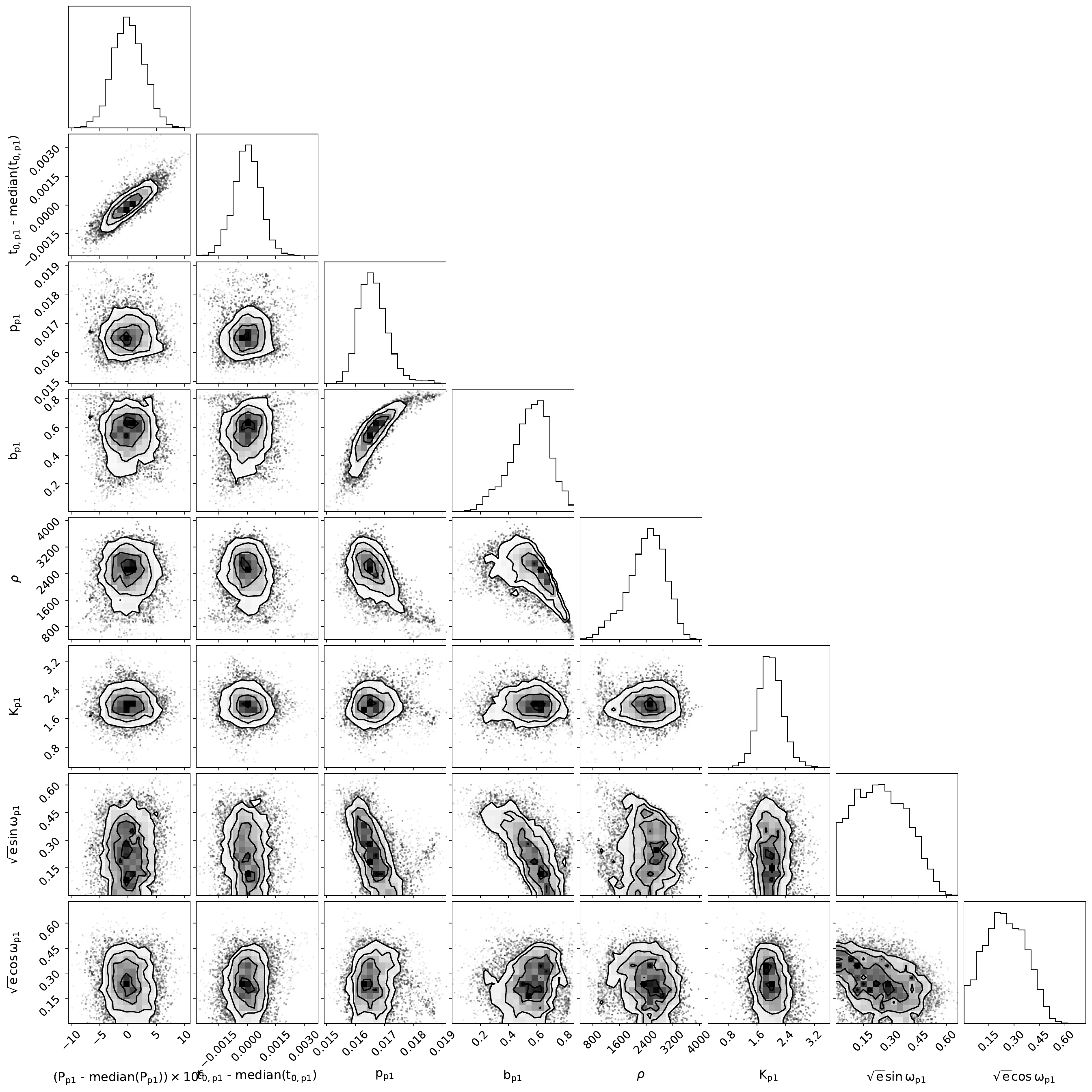}
    \caption{Cornerplot of the orbital posterior planetary parameter distributions obtained
with \texttt{juliet} for TOI-286~b, for the fit with free eccentricity, ($\mathrm{\sqrt{e} \sin \omega, \sqrt{e} \cos \omega}$) parametrization, and an upper limit on the eccentricity of 0.7.}
    \label{fig:toi-286_cornerplot_e07_p1}
\end{figure*}

\begin{figure*}
    \centering
    \includegraphics[width=\textwidth]{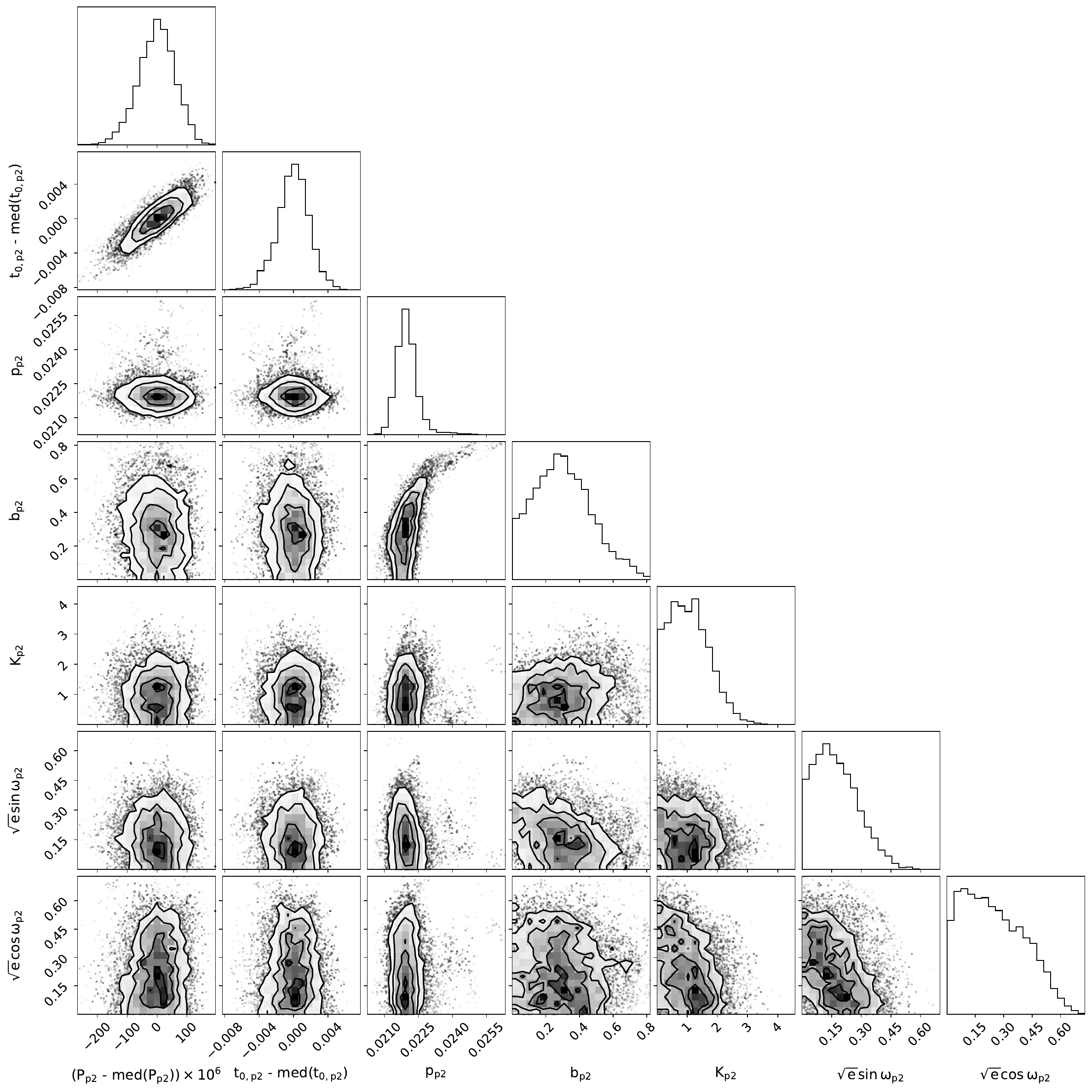}
    \caption{Cornerplot of the orbital posterior planetary parameter distributions obtained
with \texttt{juliet} for TOI-286~c, for the fit with free eccentricity, ($\mathrm{\sqrt{e} \sin \omega, \sqrt{e} \cos \omega}$) parametrization, and an upper limit on the eccentricity of 0.7.}
    \label{fig:toi-286_cornerplot_e07_p2}
\end{figure*}

\begin{figure*}
    \centering
    \includegraphics[width=\textwidth]{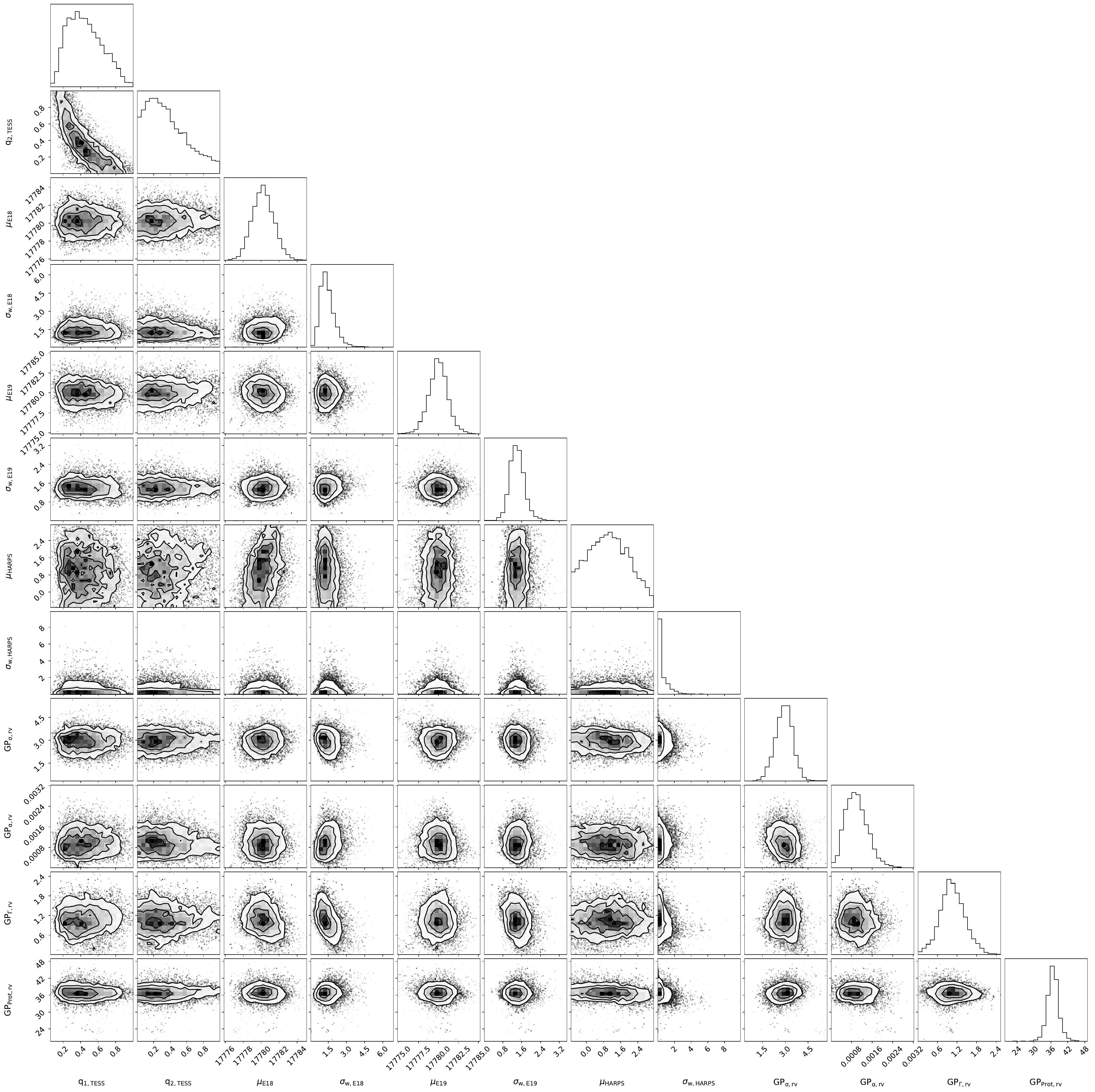}
    \caption{Cornerplot of the instrumental posterior parameter distributions obtained
with \texttt{juliet} for TOI-286, for the fit with free eccentricity, ($\mathrm{\sqrt{e} \sin \omega, \sqrt{e} \cos \omega}$) parametrization, and an upper limit on the eccentricity of 0.7.}
    \label{fig:toi-286_cornerplot_e07_instr}
\end{figure*}

\begin{figure*}
    \centering
    \includegraphics[width=\textwidth]{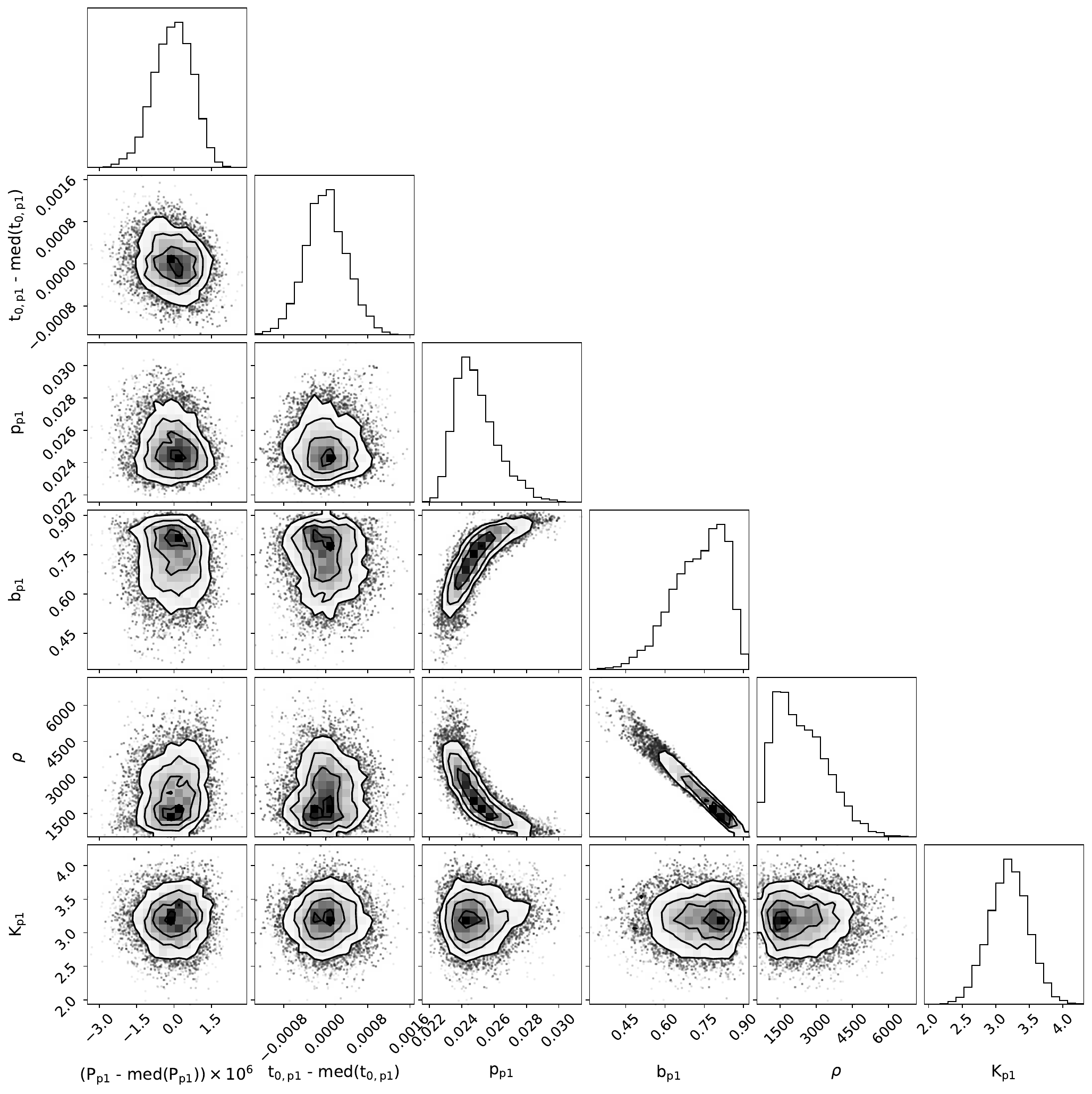}
    \caption{Cornerplot of the orbital posterior planetary parameter distributions obtained
with \texttt{juliet} for TOI-134, for the fit with eccentricity fixed to 0.}
    \label{fig:toi-134_cornerplot_circ_orbital}
\end{figure*}

\begin{figure*}
    \centering
    \includegraphics[width=\textwidth]{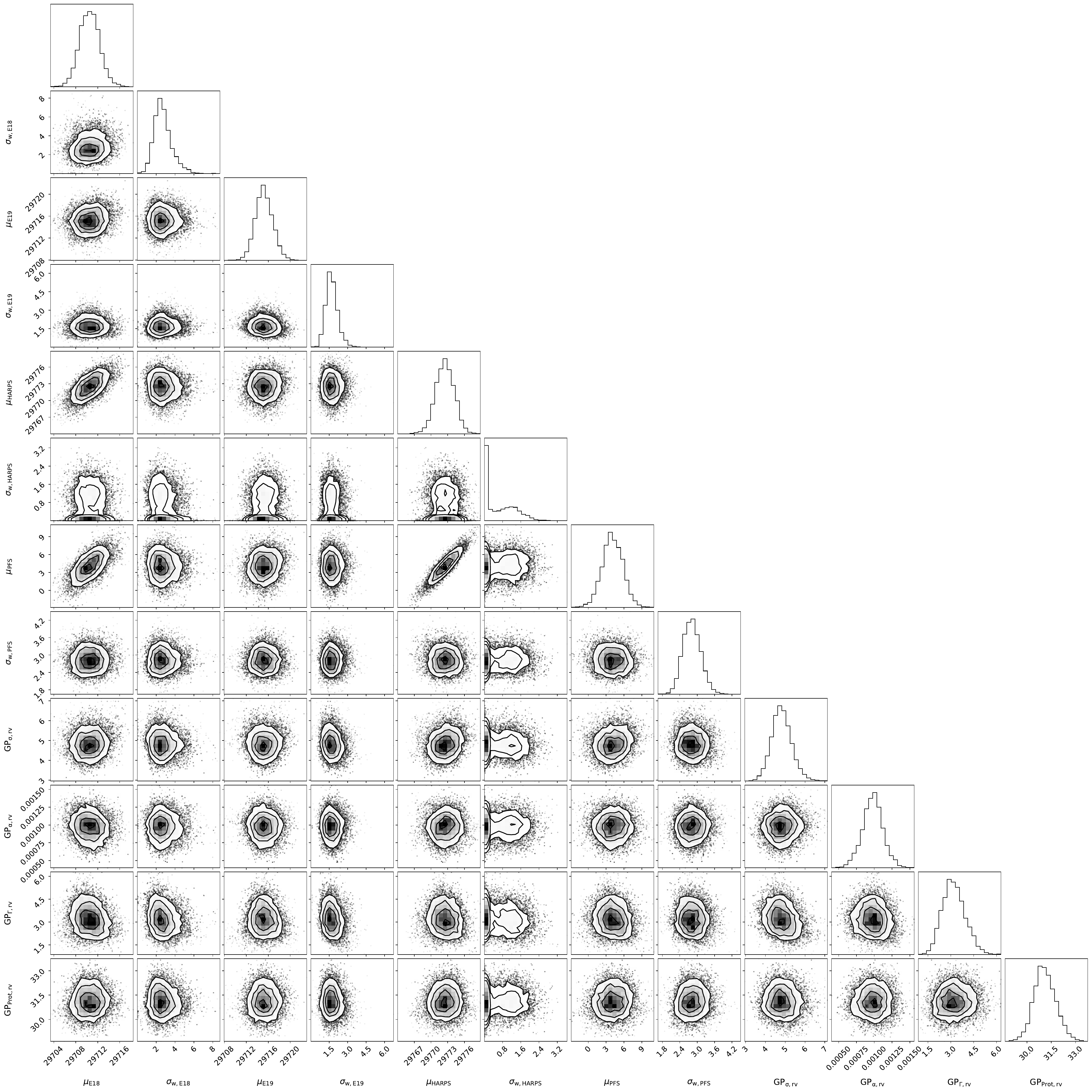}
    \caption{Cornerplot of the RV instrumental posterior parameter distributions obtained
with \texttt{juliet} for TOI-134, for the fit with eccentricity fixed to 0.}
    \label{fig:toi-134_cornerplot_circ_rv}
\end{figure*}

\begin{figure*}
    \centering
    \includegraphics[width=\textwidth]{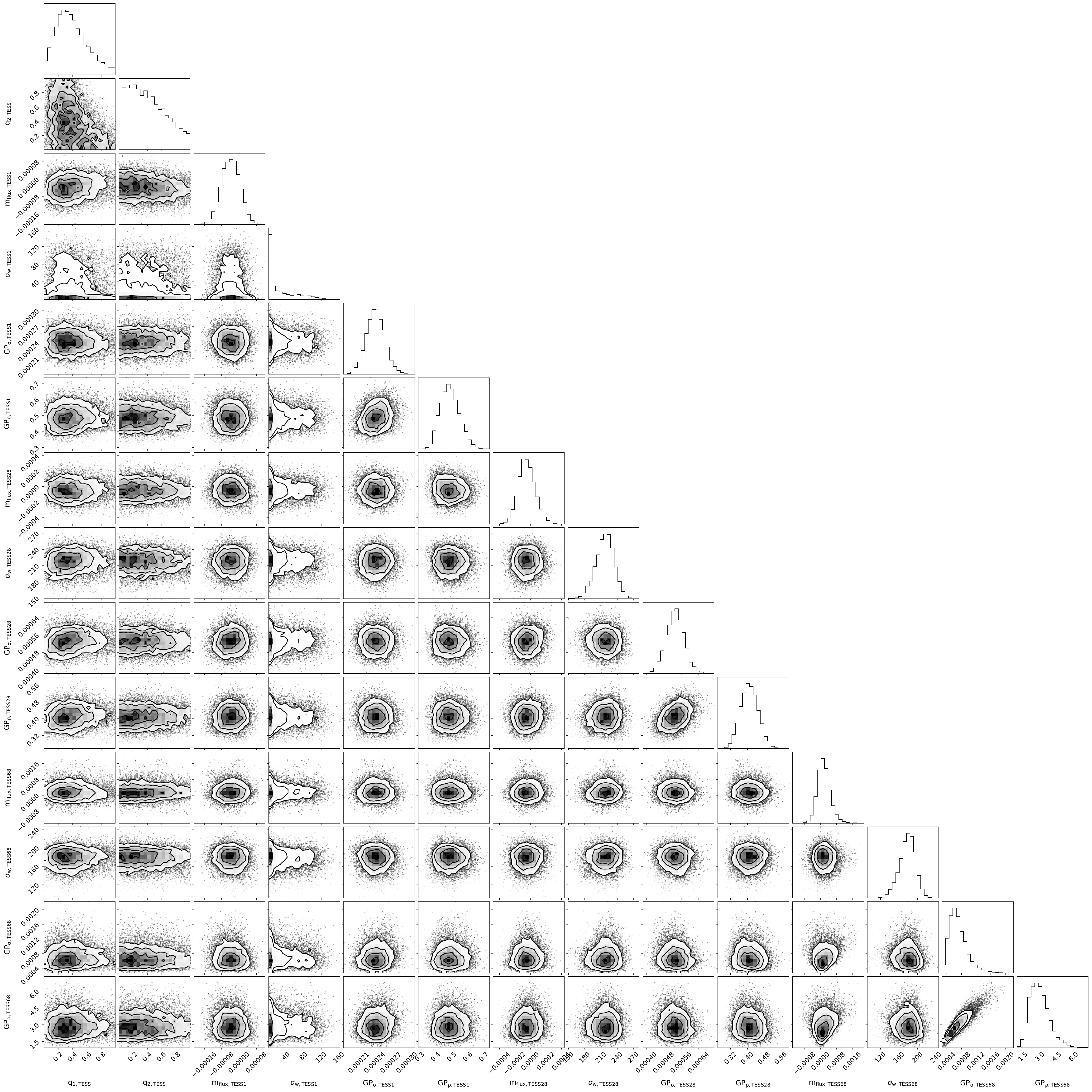}
    \caption{Cornerplot of the photometric instrumental posterior parameter distributions obtained
with \texttt{juliet} for TOI-134, for the fit with eccentricity fixed to 0.}
    \label{fig:toi-134_cornerplot_circ_transit}
\end{figure*}

\begin{figure*}
    \centering
    \includegraphics[width=\textwidth]{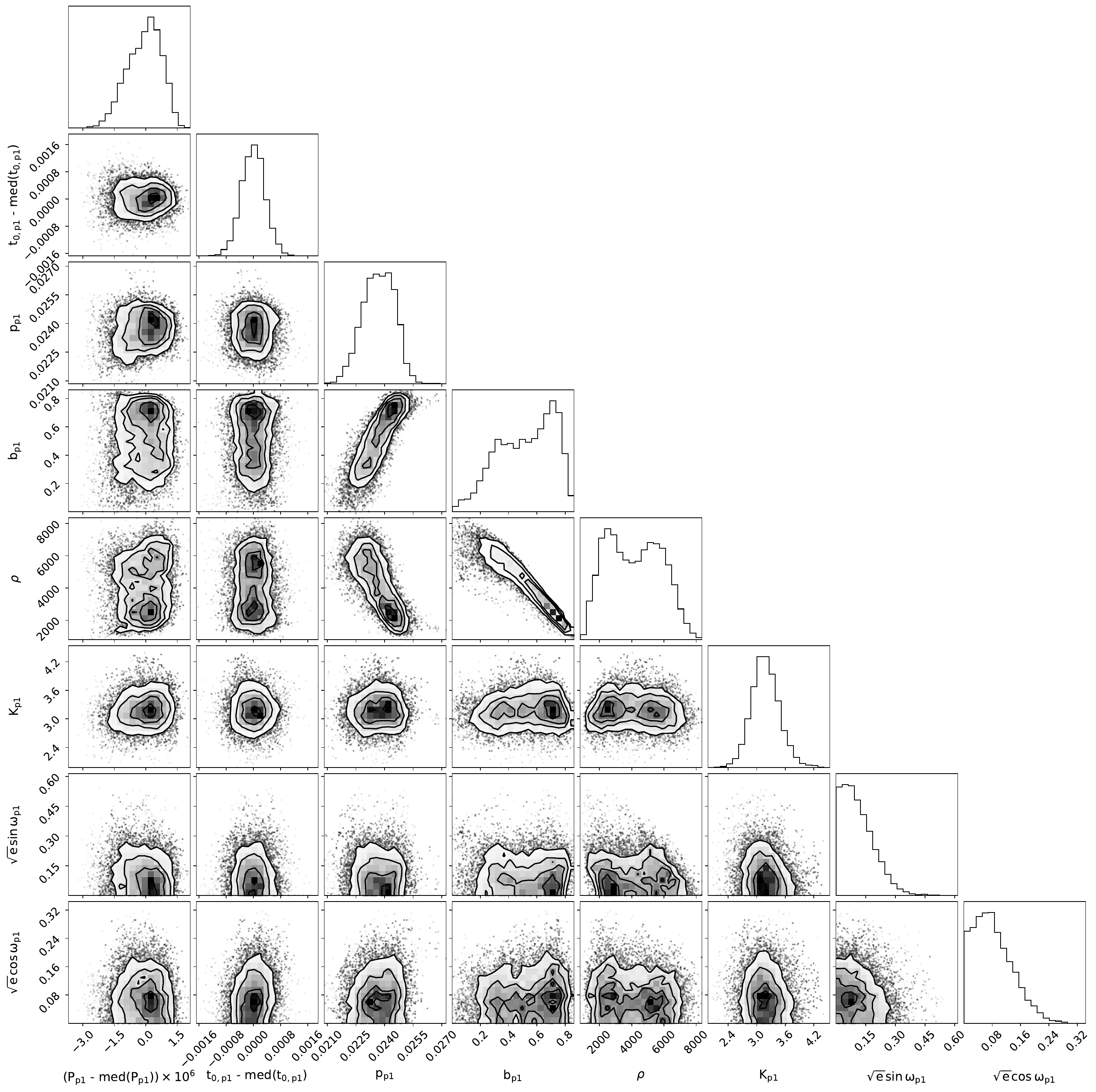}
    \caption{Cornerplot of the orbital posterior planetary parameter distributions obtained
with \texttt{juliet} for TOI-134, for the fit with free eccentricity, ($\mathrm{\sqrt{e} \sin \omega, \sqrt{e} \cos \omega}$) parametrization, and an upper limit on the eccentricity of 0.7.}
    \label{fig:toi-134_cornerplot_efree_orbital}
\end{figure*}

\begin{figure*}
    \centering
    \includegraphics[width=\textwidth]{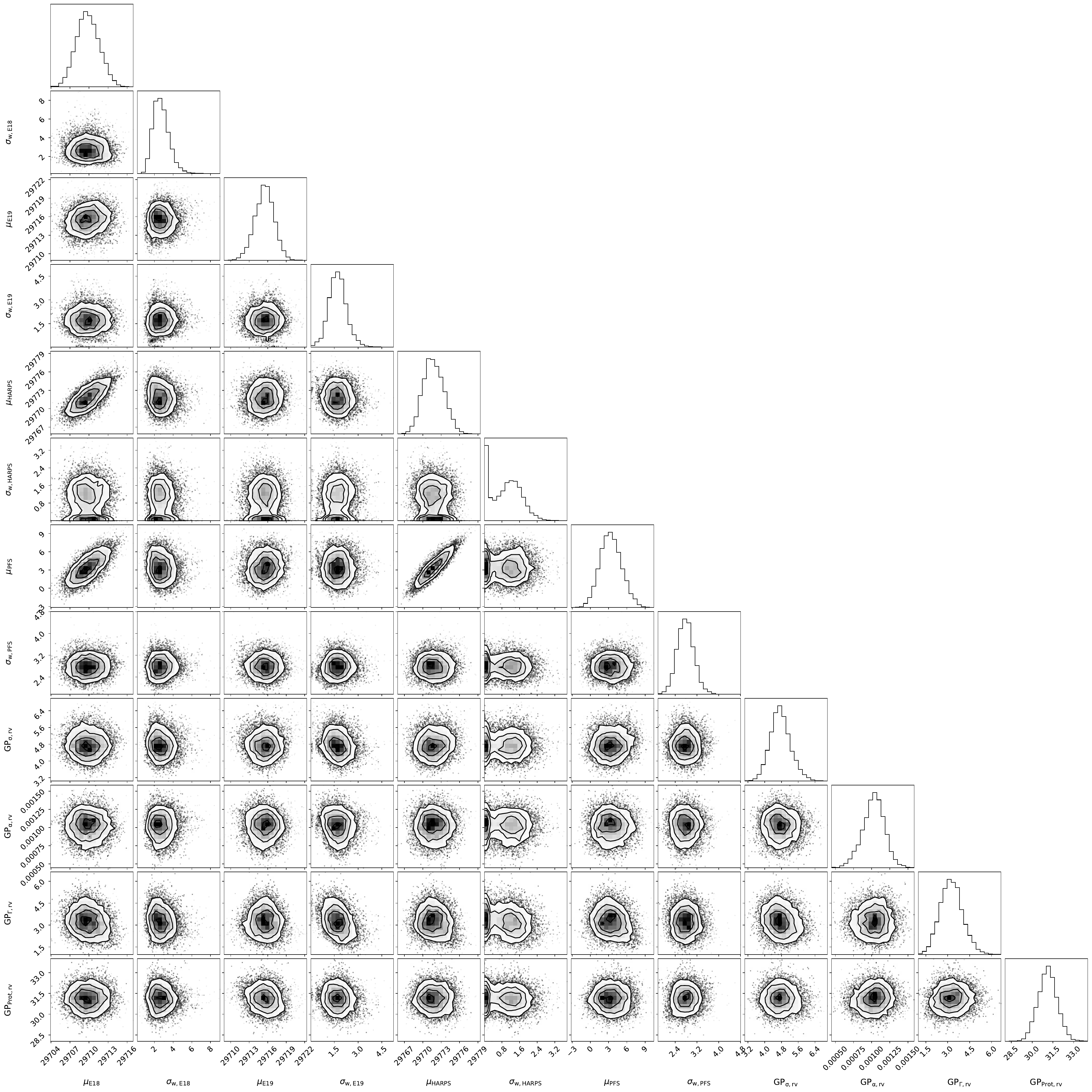}
    \caption{Cornerplot of the RV instrumental posterior parameter distributions obtained
with \texttt{juliet} for TOI-134, for the fit with free eccentricity, ($\mathrm{\sqrt{e} \sin \omega, \sqrt{e} \cos \omega}$) parametrization, and an upper limit on the eccentricity of 0.7.}
    \label{fig:toi-134_cornerplot_efree_rv}
\end{figure*}

\begin{figure*}
    \centering
    \includegraphics[width=\textwidth]{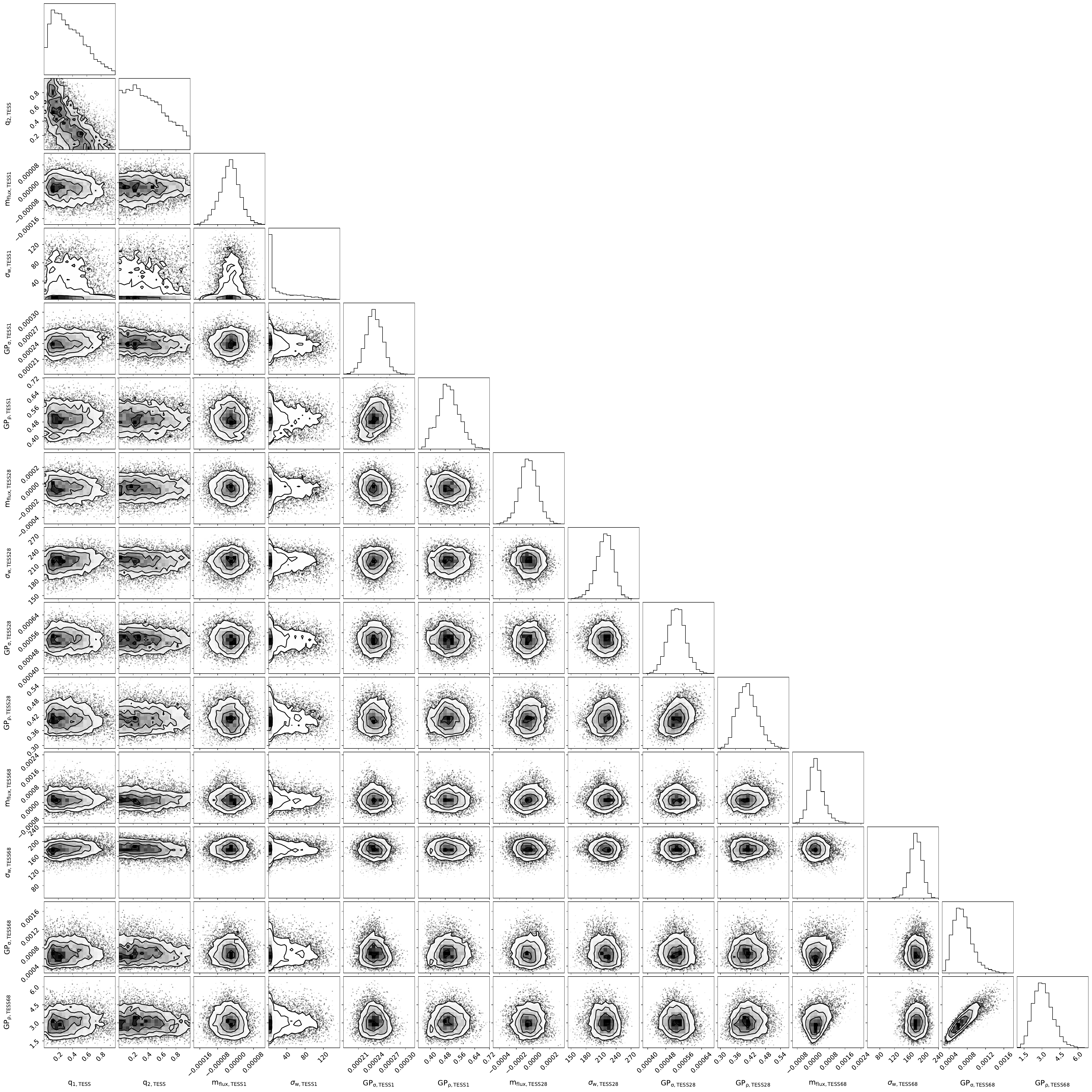}
    \caption{Cornerplot of the photometric instrumental posterior parameter distributions obtained
with \texttt{juliet} for TOI-134, for the fit with free eccentricity, ($\mathrm{\sqrt{e} \sin \omega, \sqrt{e} \cos \omega}$) parametrization, and an upper limit on the eccentricity of 0.7.}
    \label{fig:toi-134_cornerplot_efree_transit}
\end{figure*}

%
%

\end{document}